\documentclass[a4paper,fleqn,colorlinks,unicode,hypertexnames=false]{cas-dc}

\usepackage{amsmath}
\usepackage{amssymb}
\usepackage{soul}      
\usepackage{multirow}  
\usepackage{tikz}
\usepackage{array}     
\usepackage{pdflscape} 
\usepackage{longtable} 
\usepackage{rotating}  
\usepackage{color}     
\usepackage{pdftexcmds} 

\makeatletter
\let\bibcop@truebibliography\bibliography
\ifnum\pdf@shellescape=0
\else
\fi
\makeatother

\usepackage[title,titletoc]{appendix} 

\usepackage{stmaryrd}  
\SetSymbolFont{stmry}{bold}{U}{stmry}{m}{n} 

\usepackage{verbatim}  
\usepackage{textcomp}  

\ifdefined\DISSERTATION
  \usepackage[numbers,square]{natbib}    
\else
\fi
\usepackage[unicode,hypertexnames=false,colorlinks=true]{hyperref}
\usepackage{cleveref}

\newcommand*\rot{\rotatebox{90}} 

\def\fillandplacepagenumber{
 \par\thispagestyle{empty}%
 \vbox to 0pt{\vss}\vfill
 \vbox to 0pt{\baselineskip0pt
   \hbox to\linewidth{\hss}%
   \baselineskip\footskip
   \hbox to\linewidth{%
     \hfil\thepage\hfil}\vss}}

\makeatletter

\AddToHook{cmd/appendices/before}{%
  \crefalias{section}{appendix}%
  \crefalias{subsection}{appendix}
  \crefalias{subsubsection}{appendix}
  \crefalias{chapter}{appendix}
}
\makeatother

\usepackage{microtype}
\usepackage{subcaption} 
\usepackage{enumitem} 
\usepackage{tikz}

\newcolumntype{P}[1]{>{\centering\arraybackslash}p{#1}}
\newcolumntype{M}[1]{>{\centering\arraybackslash}m{#1}}

\newcommand{\singleColMacro}[1]{
    \ifdefined\DISSERATION 
    \else
        \onecolumn
    \fi
    {#1}
    \ifdefined\DISSERATION 
    \else
        \twocolumn
    \fi
}
\newcommand{\journal}{Applied Ocean Research}

\usetikzlibrary{shapes.geometric}
\newcommand{\Stars}[2][fill=black,draw=black]{
    \begin{tikzpicture}[baseline=-0.35em,#1]
    \foreach \X in {1,...,3}
    {\pgfmathsetmacro{\xfill}{min(1,max(1+#2-\X,0))}
    \path (\X*1.1em,0) 
    node[star,draw,star point height=0.25em,minimum size=1em,inner sep=0pt,
    path picture={\fill (path picture bounding box.south west) 
    rectangle  ([xshift=\xfill*1em]path picture bounding box.north west);}]{};
    }
    \end{tikzpicture}
}

\input{numeric-results.tex}

\usepackage[authoryear]{natbib}

\def\tsc#1{\csdef{#1}{\textsc{\lowercase{#1}}\xspace}}
\tsc{WGM}
\tsc{QE}

\begin{document}

\let\WriteBookmarks\relax
\def\floatpagepagefraction{1}
\def\textpagefraction{.001}

\shorttitle{Development, Validation, and Benchmarking of a Multidisciplinary Semi-Analytical Model for Wave Energy Converters}
\shortauthors{R. McCabe et~al.}

\title[mode=title]
{Development, Validation, and Benchmarking of a Multidisciplinary Semi-Analytical Model for Wave Energy Converters}

%

\author[MAE]{Rebecca McCabe}[
orcid=0000-0001-5108-998X,
auid=57980095200,
linkedin=rebecca-mccabe]
\cormark[1]
\fnmark[1]
\ead{rgm222@cornell.edu}
\ead[url]{http://sea.mae.cornell.edu}

\credit{Conceptualization, Methodology, Software, Validation, Writing - Original draft}

\author[MAE]{Madison Dietrich}

\credit{Software, Visualization,  Writing - Review \& Editing}

\author[mich]{Maha Haji}

\credit{Supervision, Writing - Review \& Editing}

\affiliation[MAE]{
    organization={Sibley School of Mechanical and Aerospace Engineering, Cornell University},
    addressline={124 Hoy Rd.},
    city={Ithaca},
    postcode={14853},
    state={NY},
    country={USA}
}

\affiliation[mich]{
    organization={Department of Mechanical Engineering, University of Michigan},
    addressline={G.G. Brown Laboratory, 2350 Hayward},
    city={Ann Arbor},
    postcode={48109},
    state={MI},
    country={USA}
}

\cortext[1]{Corresponding author}

\begin{abstract}
Wave energy converters (WECs) require system-level techno-economic analysis to balance power production, cost, and survivability, yet existing simulation tools are often either too computationally intensive for large-scale optimization or too narrow in disciplinary scope to support integrated design studies.
This work presents MDOcean, a novel modular WEC simulation framework developed for rapid early-stage design exploration, parametric analysis, and multidisciplinary optimization.
MDOcean integrates hydrodynamics, dynamics, structures, and economics within a computationally efficient architecture based on analytical and semi-analytical methods that substantially reduce runtime while maintaining near-numerical accuracy. 

The framework includes a mesh-free eigenfunction-based linear hydrodynamic solver, a quasi-linearized frequency-domain dynamics engine capable of modeling drag and saturation nonlinearities, a structural sizing module incorporating realistic yield, ultimate, buckling, storm, and fatigue design criteria, and a simplified cost model for techno-economic assessment.
Particular emphasis is placed on the linearized pseudo-spectral dynamics formulation, which extends prior frequency-domain constraint-handling approaches through a unified describing-function and analytical quadratically-constrained quadratic program (QCQP) framework.
This formulation enables efficient treatment of nonlinear constraints while preserving compatibility with optimization workflows and frequency-domain analysis techniques commonly used in WEC design.

Validation and benchmarking demonstrate that MDOcean's \resultsAOR[simRuntime] runtime is orders of magnitude faster than leading WEC simulation tools while maintaining agreement with higher-fidelity baselines to within a few percent in most cases.
Beyond computational performance, the framework also provides insight into limiting behaviors, scaling laws, subsystem interactions, and key tradeoffs governing WEC design and techno-economic performance.
MDOcean is released as open-source software to support accelerated WEC research, design, and optimization.

\end{abstract}

\begin{graphicalabstract}
\includegraphics[width=\linewidth]{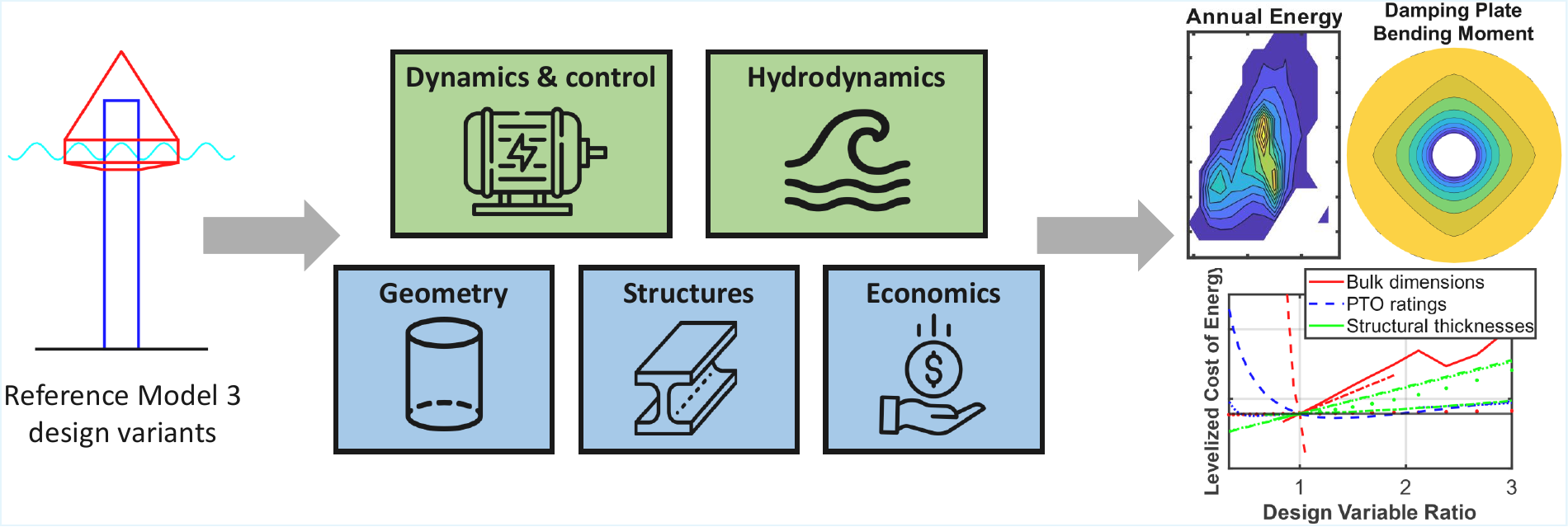}
\end{graphicalabstract}

\begin{highlights}
\item Presents MDOcean, an open-source semi-analytical wave energy converter simulation.
\item Integrates hydrodynamics, powertrain, structures, \& economics for coupling intuition.

\ifdefined\DISSERTATION
    \item Quasi-linear frequency domain dynamics model captures nonlinearities such as generator force saturation and drag.
    \item Derives analytical constrained optimal control solution for underactuated wave-to-wire multi-port model respecting effort, flow, and power limits.
    \item Analytical stiffened-plate structural model assesses stress and stability failure modes under fatigue and storm survival loadcases.
\else
    \item Derives analytical frequency-domain constrained optimal control with nonlinearities.
\fi

\item Achieves \resultsAOR[simRuntime] runtime, 10-1000x faster than existing tools.
\item Verifies <10\% power \& amplitude error for standard reference model 3 design.

\end{highlights}


\maketitle


\section{Introduction}
\label{sec:intro}
\subsection{Wave Energy Overview}

The global climate crisis requires a transition to carbon-free energy sources such as ocean wave energy.
Ocean waves have higher consistency, predictability, and energy density than other renewable energy sources such as wind and solar, while the temporal complementarity of waves with other resources can improve grid resilience and capacity adequacy, decrease energy prices, and decrease requirements for energy storage and balancing power \citep{akdemir_opportunities_2023,bhattacharya_timing_2021,pennock_temporal_2022}.
Wave energy converters (WECs), the devices that harness and convert this energy, could provide both electricity for the grid at large scale and power for smaller offshore technologies like aquaculture, desalination, and marine sensing \citep{livecchi_powering_2019}.
\ifdefined\DISSERTATION
Despite the potential advantages of WECs, a significant decrease in the cost per unit energy production is required before the technology can be deployed at large scales. 
Thus, a design process that emphasizes techno-economic viability from an early stage is necessary.
\else
    However, WECs require substantial cost reduction before large-scale deployment, motivating a design process that emphasizes techno-economic viability from an early stage.
\fi

\ifdefined\DISSERTATION
    However, the presence of strong interdisciplinary coupling complicates the application of typical design and optimization techniques to WECs.
    The powertrain, controller, bulk dimensions, hydrodynamic shape, and structural thicknesses must be selected to balance the complex tradeoff between power production and cost while ensuring survivability and practical feasibility.
    Concentrating and automating the design decision-making efforts of structural, control, hydrodynamics, and electrical engineers while accurately capturing all major relationships and requirements is challenging, not to mention computationally prohibitive.
    Nonetheless, leveraging the coupling between subsystems and across technical domains can produce large gains in techno-economic viability.
    Automating this process through optimization allows WEC designers to more quickly and systematically explore tradeoffs of major design decisions, and could eventually allow standardized comparisons between different WEC architectures.
    Thus, performing a system-level WEC optimization that simultaneously considers the coupled techno-economic goals, design decisions, and requirements is highly advantageous.
\else
    The strong interdisciplinary coupling of between powertrain, controller, bulk dimensions, hydrodynamic shape, and structural thicknesses complicates conventional WEC design techniques.
    Systematically exploring these coupled tradeoffs requires a fast multidisciplinary simulation that can be embedded in an optimization loop.
\fi

\ifdefined\DISSERTATION
    An initial attempt to consolidate a multidisciplinary WEC design process, though without optimization, was the Reference Model Project.
    From 2010 to 2014, the U.S.\ Department of Energy funded the creation of six benchmark designs for marine energy devices, published a report \citep{RM3} describing the design considerations and performance calculations, and released design artifacts and other documentation on its website.\footnote{\url{https://openei.org/wiki/PRIMRE/Signature_Projects/Reference_Model}} Three designs are for tidal and current energy turbines, and three are WECs.
    The third reference model, known as RM3, is a two-body point absorber WEC designed for a reference site located in Humboldt Bay, California.
    The major structural components of the device are a surface float and a spar consisting of a vertical column and a subsurface damping plate, with labeled dimensions shown in \Cref{fig:dims}.
    The float oscillates up and down with oncoming incident waves while the spar stays mostly stationary, and electrical power is produced from this relative heave motion.
    RM3 has received considerable research attention in the decade since the original report was published.
    The present study builds on that work and presents a multidisciplinary techno-economic simulation of the RM3.
\else
    This paper builds on the U.S.\ Department of Energy's Reference Model Project, specifically the third reference model (RM3) \citep{RM3}: a two-body point-absorber WEC designed for Humboldt Bay, California, comprising a surface float and a spar with a subsurface damping plate (\Cref{fig:dims}).
    Electrical power is produced from the relative heave motion of the float against the spar.
\fi

\ifdefined\DISSERTATION
    The remainder of the introduction will review WEC system modeling literature, articulate the appeal of semi-analytical modeling to address gaps in the field, and identify the paper's key contributions and structure.
\else
    The remainder of this introduction reviews WEC modeling literature, identifies the gap motivating MDOcean, and summarizes the paper's contributions.
\fi

\subsection{WEC System Modeling: State of the Art and Gaps}
\label{sec:lit}
\ifdefined\DISSERTATION
    Existing models of WECs in the literature vary widely in their disciplinary scope, fidelity, and suitability for optimization.
    In the wave energy industry, multidisciplinary modeling is common due to the imperative of considering all design elements, although different disciplines are often modeled separately and sequentially rather than concurrently.
    \citet{trueworthy_wave_2020} provides a critical summary of the WEC design process and shares survey results from 25 WEC designers and developers.
    50\% of respondents design all subsystems concurrently.
    The others tend to start with the WEC shape and PTO, then move to the controller and moorings, and finish with the power transportation system.
    \citet{trueworthy_wave_2020} concludes that concurrent design has promise to take advantage of subsystem interactions and is an under-utilized technique worthy of further study, emphasizing the importance of incorporating all design requirements at an early stage.
\else
    Existing WEC models vary widely in disciplinary scope, fidelity, and suitability for optimization.
In industry, multidisciplinary modeling is common but typically sequential rather than concurrent; a survey of 25 WEC designers \citep{trueworthy_wave_2020} found that only 50\% design all subsystems concurrently, and identified concurrent design as an under-utilized technique worthy of further study.
\fi

The academic state-of-the-art typically focuses on modeling and optimizing one or a few disciplines at a time, with the most common being hydrodynamics, dynamics, and controls.
Major modeling choices include the representation (or lack thereof) of the WEC's hydrodynamics, dynamics, controls, mooring, powertrain, structures, and economic viability, as well as the sea states considered.

\Cref{fig:model-taxonomy} provides a simplified taxonomy of the various WEC system modeling approaches used in the literature, and \Cref{tab:lit} compares the most relevant studies along this taxonomy.

\ifdefined\DISSERTATION
    Major modeling choices include the representation (or lack thereof) of the WEC's hydrodynamics, dynamics, controls, mooring, powertrain, structures, and economic viability, as well as the sea states considered.
    After examining the overall trends in \Cref{tab:lit}, each table entry will be described in more detail.
    Point absorbers (PA) seem to be the most common device architecture, though most major device types are represented, and several studies optimize arrays of multiple devices.
    Nearly all model hydrodynamics and power production in some way.
    Hydrodynamic modeling is typically performed with the boundary element method (BEM), occasionally with semi-analytical models like matched eigenfunction expansion method (MEEM), and rarely with high fidelity computational fluid dynamics (CFD), fitting a model to numerical or experimental data (FIT), or algebraic approximations (ALG).
    Nonlinear forces such as drag are most often linearized and included as an additional linear damping (LD), though other methods exist including quadratic representation (Q), describing function (DF), strip-theory integration of Morison's equation (M), and CFD.
    Dynamic modeling is frequently performed both in the frequency domain (F) and time domain (T), with a handful of studies using extensions of frequency domain techniques (F+) to handle constraints and nonlinearities that are typically only handled in the time domain, and a few using the newer pseudo-spectral method (PS).
    A range of controllers are considered, including proportional-integral / reactive (PI), proportional / pure damping (P), their extensions to incorporate constraints (P+/PI+), nonlinear structured (NL), and unstructured (UNS).

    Optimizations that incorporate powertrain, structures, and economics are less ubiquitous but still common.
    When a discipline is included, its fidelity can vary widely.
    For instance, some studies model the PTO only with an efficiency (EF), while others model the PTO dynamics (DY).
    Structural models are split relatively evenly between analytical models for stress as a function of dimensions (AN), calculating loads without considering dimension-dependent stress (LO), and finite element analysis (FEA).
    Cost models most often use a geometric cost proxy (GEO) such as hull volume or area.
    Some estimate costs of the structural material (STR), power take-off (PTO), mooring (MOOR), or include a comprehensive cost model (FULL).
    The use of a mooring model is relatively rare, with four studies modeling the mooring system physical design (DE) and one modeling its dynamic properties (DY).
    It is most common to simulate irregular waves (IRR), though several studies assume regular waves (REG).
    Others incorporate multiple sea states via a joint probability density matrix (IRR+/REG+).
    Only a few consider storm sea states (STO).
\else
    The dominant trends across these studies are: point absorbers as the most common architecture; boundary element method (BEM) hydrodynamics with occasional semi-analytical (MEEM) treatments; drag handled via linear damping or describing functions; dynamics in either frequency or time domain (with extensions of either to handle constraints); and limited inclusion of structures, mooring, and cost modeling. 
\fi

\ifdefined\DISSERTATION
    \paragraph{Notable large-scale optimization studies}
    More specifically, a few studies \citep{gaudin_single_2021,khanal_multi-objective_2024,edwards_optimisation_2022,garcia-teruel_reliability-based_2021} perform WEC optimization with several disciplines.
    These studies are notable for their large scales of 120 degrees of freedom and 14, 15, and 66 design variables respectively.
    \citet{gaudin_single_2021} optimizes an array of 20 submerged cylinders, holding the WEC design constant but optimizing site selection, array layout, and mooring design while considering cable and pile cost as well as pile load capacity.
    \citet{khanal_multi-objective_2024} optimizes an array of four cylindrical point absorbers, with 4 modeled disciplines of array layout, hydrodynamics, controls, and economics.
    \citet{edwards_optimisation_2022} is a point absorber shape optimization study that minimizes surface area as a cost proxy, accounts for stability constraints, and considers only geometries with power extraction equal to theoretical radiation, amplitude, and steepness limits.
    \citet{garcia-teruel_reliability-based_2021}, along with the smaller scale study \citep{garcia-teruel_design_2022} by the same authors, apply multi-objective optimization to trade off power with fatigue load and device volume, respectively.
    \citet{cotten_multi-objective_2022} similarly optimizes fatigue load for an attenuator-style WEC with many more degrees of freedom but fewer (8) design variables.
    A different shape optimization \citep{abdulkadir_control_2024} notably calculates hydrodynamic coefficients via the semi-analytical matched eigenfunction expansion method (MEEM), rather than from boundary element method as is typical, but does not include disciplines beyond hydrodynamics and dynamics/controls.
    A team at NREL attempted local optimization for a terminator shape study but found they could only reasonably optimize 3 of the intended 13 design variables at a time due to the high computational cost of their time-domain simulation \citep{housner_numerical_2024}.
    The previous studies \citet{garcia-teruel_reliability-based_2021,garcia-teruel_design_2022,cotten_multi-objective_2022} avoided this issue by simulating primarily in the frequency domain and using the time domain only to address PTO constraints.

    \paragraph{Structural modeling and optimization}
    Structural optimization of WECs faces a similar challenge in the high compute time of finite element analysis (FEA).
    References \citep{coe_survey_2018, ove_arup__partners_ltd_structural_2016,paduano_towards_2024,giannini_wave_2022} provide guidelines for WEC survivability modeling and design, including the prediction of both extreme loads and structural failure.
    Fatigue modeling methods include spectral analysis, rainflow counting, damage equivalent loads, and Miner's law, augmented with standard stress concentration factors or FEA for local hotspots \citep{ove_arup__partners_ltd_structural_2016}.
    Strength modeling methods consist primarily of various fidelities of FEA, although hand and spreadsheet calculations permit sizing to a specified stress early in the design process \citep{ove_arup__partners_ltd_structural_2016}.
    FEA of WECs is most often performed with manual design iterations, as in the studies \citep{RM3,mi_multi-scale_2025}.
    These two studies, while not optimization, are extremely multidisciplinary, encompassing mooring design and detailed economic evaluation in addition to the more common hydrodynamics, powertrain, controls, and structures.
    Automating the design process reported there remains a long-term vision for WEC multidisciplinary design.

    Notably, one study \citep{an_optimal_2024} performs true optimization of WEC structural FEA using the commercial computer aided engineering software Altair Inspire.
    The study examines four design variables, although it keeps the power calculations entirely separate from the structural optimization, neglecting the interdisciplinary coupling.
    Another structural optimization does not consider power at all, but is able to perform a brute-force sweep of 3 design variables by using analytical equations for stress and other failure modes \citep{ambuhl_reliability-based_2014}.
    The new Project SEA Stack modeling suite adds hydrodynamics to Project Chrono, a multiphysics software with the ability to perform finite element analysis using beam, shell, or solid elements, although it appears that so far the structural modeling capabilities have only been used for offshore wind turbines rather than WECs \citep{ogden_hydrochrono_2023}.
    Finally, several studies consider structures indirectly by examining the load on the device rather than the stress, e.g.\ \citep{nguyen_theoretical_2024, ferri_balancing_2014}, or equivalently assuming that stress scales with load and does not depend on any other dimensional design variable, e.g.\ \citep{garcia-teruel_reliability-based_2021, cotten_multi-objective_2022}.
    Like the earlier MEEM study, \citet{nguyen_theoretical_2024} uses the equivalent analytical hydrodynamics in elliptical coordinates for an oscillating surge WEC.

    \paragraph{Controls and powertrain co-design}
    The study \citep{ferri_balancing_2014} is notable because in addition to structural loads, it compares various control schemes and simulates second-order drivetrain dynamics.
    The control co-design (CCD) methodology \citep{garcia-sanz_control_2019} emphasizes the importance of considering drivetrain dynamics and generator efficiency due to their strong effect on electrical power generation \citep{coe_useful_2023}.
    A number of recent studies \citep{rosati_control_2023,son_performance_2016,anderson_re-imagining_2024,devin_high-dimensional_2024,grasberger_control_2024} simultaneously optimize the PTO with the controller.
    \citet{rosati_control_2023} optimizes an OWC and is notable for its comparatively detailed cost modeling and its optimization of both economic and power variability metrics.
    \citet{mccabe_multidisciplinary_2022} also optimized power variation, although that study's use of regular waves makes the variation formulation less useful.
    Meanwhile, one point absorber PTO optimization \citep{son_performance_2016} is additionally notable for its use of the MEEM method.
    \citet{gaebele_tpl_2025} is relevant because it optimizes the dimensional scale and PTO of the RM3 WEC directly using a control co-design approach.
    It also efficiently incorporates many time-domain constraints (maximum force, speed, average power, RMS torque, and stroke) while maintaining frequency-domain hydrodynamics via the pseudo-spectral (PS) modeling method, which \citep{devin_high-dimensional_2024,grasberger_control_2024} also use.

    Other CCD papers optimize WEC dimensions without modeling the PTO.
    \citet{herber_dynamic_2014} performs a direct transcription co-optimization of cylinder draft and radius with control.
    \citet{lin_fast_2025} introduces a new CCD formulation for constrained control that is even faster than PS.
    The method leverages the analytical Pontryagin Maximum Principle (PMP) and performs a case study to optimize a single geometric variable.
    A different PMP formulation \citep{abdulkadir_optimal_2024} incorporates PTO-constrained array dynamics, although it has not yet been used for design optimization.
\else
    Across these studies, the most directly relevant work falls into three groups.
    First, several large-scale optimization studies \citep{gaudin_single_2021,khanal_multi-objective_2024,edwards_optimisation_2022,garcia-teruel_reliability-based_2021,garcia-teruel_design_2022,cotten_multi-objective_2022} address one to four disciplines but invariably exclude either structures or powertrain.
    Second, structural optimization of WECs is hampered by the cost of FEA \citep{coe_survey_2018,ove_arup__partners_ltd_structural_2016,paduano_towards_2024,giannini_wave_2022}; the few WEC studies that perform structural optimization \citep{an_optimal_2024,ambuhl_reliability-based_2014} decouple it from power calculations.
    Third, recent control co-design studies \citep{rosati_control_2023,son_performance_2016,gaebele_tpl_2025,devin_high-dimensional_2024,grasberger_control_2024,herber_dynamic_2014,lin_fast_2025,abdulkadir_optimal_2024} efficiently handle constraints via pseudo-spectral methods or analytical Pontryagin Maximum Principle, but none integrate structures or full economic costs.
    The directly comparable RM3 study \citep{gaebele_tpl_2025} optimizes dimensional scale and PTO with control co-design but excludes structures.
\fi

\paragraph{Software tools and gaps}
\ifdefined\DISSERTATION
    Beyond models used in individual studies, it is also helpful to analyze the capabilities and gaps of available simulation tools.
    The popular time-domain hydrodynamics tool WEC-Sim \citep{ruehl_wec-simwec-sim_2024} can become multidisciplinary through the use of integrations like NEMOH/Capytaine (hydrodynamic coefficients), PTO-Sim (electric and hydraulic components), WEC-Sim Applications (optimal control and other useful tools), and MoorDyn (mooring).
    However, as \citep{housner_numerical_2024} reveals, it is too slow to use for large-scale optimization.
    Meanwhile, the newer pseudospectral tool WecOptTool \citep{coe_initial_2020} is faster and suitable for control co-design, but its disciplines are currently limited to hydrodynamics (via Capytaine) and control, with the ability to easily add custom dynamics like PTO and mooring as needed, and it still can take considerable time for large optimizations.

    A 2023 marine energy software assessment \citep{ruehl_next-generation_2023} identified that industry-wide needs broadly include using state-of-the-art computation; overcoming limitations of existing software; ensuring appropriate language, architecture, and license of new software; accessing data; and ensuring usability and accuracy of free open source software.
    Specifically, relevant gaps include tools for powertrain and controls modeling, co-design, and optimization; high-fidelity models with faster runtime and low/mid-fidelity models with higher accuracy; inter-operabilty of models across disciplines and fidelities; and data for verification and validation \citep{ruehl_next-generation_2023}.
    This lack of a multidisciplinary WEC simulation platform that is sufficiently fast for optimization helps explain why no optimization study in \Cref{tab:lit} simultaneously optimizes the device geometry, structures, and powertrain.
\else
    The two leading open-source WEC simulation tools illustrate the gap.
    WEC-Sim \citep{ruehl_wec-simwec-sim_2024} is a time-domain simulator extensible to multidisciplinary studies via integrations (NEMOH/Capytaine for hydrodynamics, PTO-Sim for powertrain, MoorDyn for mooring), but is too slow for large-scale optimization \citep{housner_numerical_2024}.
    WecOptTool \citep{coe_initial_2020} is a faster pseudospectral tool suitable for control co-design but currently limited to hydrodynamics and control.
    A 2023 marine energy software assessment \citep{ruehl_next-generation_2023} explicitly identifies the absence of a fast multidisciplinary optimization platform as an industry-wide gap, which explains why no study in \Cref{tab:lit} simultaneously optimizes geometry, structures, and powertrain. 
    
    Existing studies therefore tend to sacrifice either disciplinary breadth or computational tractability: high-fidelity multidisciplinary models are generally too expensive for large-scale optimization, while optimization-oriented models often omit structures, powertrain dynamics, or economic assessment.
    This tradeoff motivates the development of a computationally efficient framework capable of concurrently capturing the major techno-economic couplings governing WEC performance.
\fi
\begin{landscape}
\begingroup
\begin{figure}[htbp]
\centering
\includegraphics[width=1.05\linewidth]{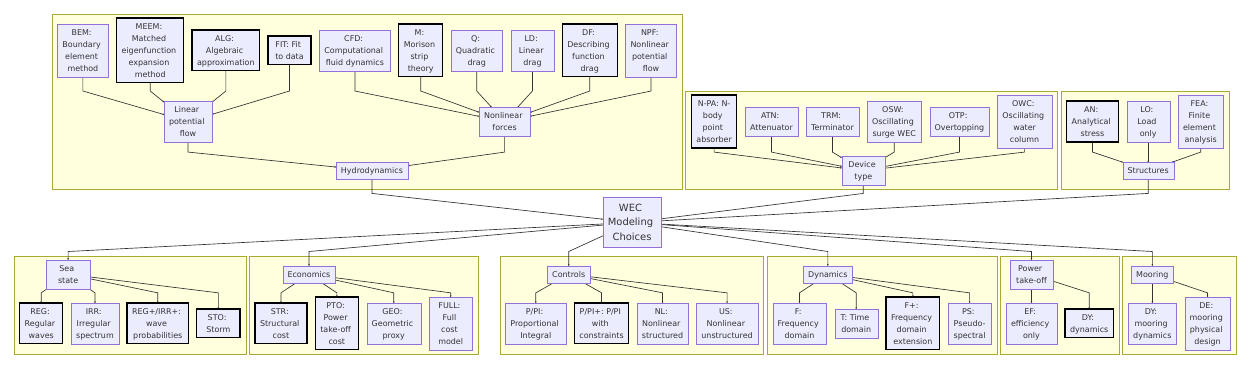}
\caption{WEC system modeling approaches, with MDOcean's capabilities outlined in black.}\label{fig:model-taxonomy}
\fillandplacepagenumber
\end{figure}
\endgroup
\end{landscape}
\singleColMacro{
\begin{longtable}{
    P{0.13\linewidth}| 
    P{0.055\linewidth}| 
    P{0.08\linewidth}| 
    P{0.04\linewidth}| 
    P{0.025\linewidth}| 
    P{0.05\linewidth}| 
    P{0.04\linewidth}| 
    P{0.035\linewidth}| 
    P{0.05\linewidth}| 
    P{0.06\linewidth}| 
    P{0.065\linewidth}} 
\rot{\textbf{Ref}} & \rot{\textbf{Type}} & \rot{\textbf{Hydro}}& \rot{\textbf{Drag}} & \rot{\textbf{Dyn}} & \rot{\textbf{Ctrl}} & \rot{\textbf{Moor}} & \rot{\textbf{PTO}} & \rot{\textbf{Struct}} & \rot{\textbf{Econ}} & \rot{\textbf{Waves}} \\
\hline
This work & 2-PA & MEEM, ALG, FIT & DF, M & F+ & PI+ & - & DY & AN & STR, PTO & REG+, STO \\

\cite{mccabe_multidisciplinary_2022} & 2-PA & ALG & - & F+ & PI+ & - & EF & AN & STR & REG+, STO \\

\cite{khanal_multi-objective_2024} & 1-PA* & BEM & - & F & PI+ & - & - & - & GEO & REG \\

\cite{gaudin_single_2021} & 1-PA* & MEEM & - & F & PI & DE & - & AN & MOOR & IRR+, STO \\

\cite{edwards_optimisation_2022} & 1-PA & BEM & - & F & P & - & - & - & GEO & REG \\

\cite{garcia-teruel_reliability-based_2021}& 1-PA & BEM & LD & F+ & PI & - & - & LD & - & IRR+ \\

\cite{garcia-teruel_design_2022}& 1-PA & BEM & - & F+ & PI & - & - & - & - & IRR \\

\cite{cotten_multi-objective_2022} & ATN & BEM & - & F+ & P & - & - & LD & - & IRR+ \\

\cite{abdulkadir_control_2024} & 1-PA & MEEM & LD & T & NL P & - & - & - & - & IRR \\

\cite{housner_numerical_2024} & TRM & BEM & - & T & PI & DY & - & - & - & REG \\

\cite{al_shami_parameter_2019} & 2-PA & BEM & DF & F & PI & - & - & - & - & REG \\

\cite{RM3} & 2-PA & BEM & Q & T & P & DE & - & FEA & FULL & IRR+, STO \\

\cite{mi_multi-scale_2025} & OSW & BEM & \small{Various} & T & PI & DE & DY & FEA & FULL & IRR+, STO \\

\cite{an_optimal_2024} & OTP & CFD & CFD & T & - & - & - & FEA & - & STO \\

\cite{ambuhl_reliability-based_2014} & 1-PA & - & Q & - & - & DE & - & AN & STR & STO \\

\cite{nguyen_theoretical_2024} & OSW & MEEM & LD & F & PI & - & - & LD & - & IRR \\

\cite{ferri_balancing_2014} & 1-PA & BEM & LD & T & P,PI+, US & - & DY & LD & STR & IRR+ \\

\cite{rosati_control_2023} & OWC & BEM & - & T & NL P+ & - & EF & - & PTO & IRR+ \\

\cite{son_performance_2016} & 2-PA & MEEM & LD & F & P & - & EF & - & - & REG \\

\cite{gaebele_tpl_2025} & 2-PA & BEM & - & PS & PI & - & DY & - & PTO, GEO & IRR \\

\cite{devin_high-dimensional_2024} & 1-PA & FIT & - & PS & US & - & DY & - & - & REG \\

\cite{grasberger_control_2024} & OSW & BEM & - & PS & PI & - & DY & - & GEO & IRR \\

\cite{herber_dynamic_2014} & 1-PA & FIT & LD & T & US & - & - & - & - & IRR \\

\cite{lin_fast_2025} & 1-PA & BEM & LD & F+ & US & - & - & - & - & IRR+ \\

\cite{abdulkadir_optimal_2024} & 1-PA* & BEM & - & T & US & - & - & - & - & IRR \\



\hline 
\caption{Comparison of the disciplinary scope and fidelity of optimization-relevant WEC models.
See \Cref{fig:model-taxonomy} for abbreviations.
An asterisk * in type column means array effects from multiple interacting devices are modeled.}
\label{tab:lit}
\end{longtable}
}

\subsection{Relevance of Semi-Analytical Modeling}   
An optimization study can use any kind of simulation model to assess the design's performance, including simple algebraic models, theory-intensive semi-analytical models, and standard numerical models.
\Cref{tab:model-types} summarizes the tradeoffs.

\begin{table}[htbp]
    \centering
    \caption{Comparison of model types}
    \label{tab:model-types}
\begin{tabular}{cM{0.14\linewidth}cM{0.13\linewidth}}
         \textbf{Model type}&  \textbf{Expertise required}&  \textbf{Accuracy}&  \textbf{Computation cost}\\ \hline
         \textbf{Algebraic}&  Low&  Low&  Low\\
         \textbf{Semi-analytical}&  High&  Med&  Low\\
         \textbf{Numerical}&  Med&  Med/High&  High\\
    \end{tabular}
    \end{table}

For early-stage WEC design optimization, where the model must be run many thousands of times, semi-analytical models offer the most beneficial combination of speed and accuracy.

\ifdefined\DISSERTATION
    Semi-analytical models can be orders of magnitude faster than numerical models, often with minimal sacrifices in accuracy.
    Additionally, semi-analytical models typically vary smoothly with their inputs and can make it easier to obtain accurate gradients, a feature that is often not the case for fully numerical models that may contain non-differentiable artifacts.
    Accurate gradients can also speed up convergence of the optimization and allow it to navigate high-dimension design spaces that would be intractable with heuristic algorithms.

    Unfortunately, semi-analytical models are less likely to be available commercially or open-source as numerical models are, and developing them from scratch requires substantial background in mathematics, numerical methods, and the technical domain of the model (for example, hydrodynamics or constrained optimal control).
    Often, semi-analytical models remain in the realm of theoretical academics, and ``application'' studies that use them to perform design/optimization usually consider just one discipline and have low practical realism.
    This is especially true if the models are closed-source, preventing others besides the model authors from using them.
    In the context of WECs, low practical realism usually means studies are limited to mechanical power production, without considering cost and realistic constraints, or perhaps using volume or surface area as a cost proxy.
    The design studies with high practical realism tend to use numerical models instead.
    As the literature review in \Cref{tab:lit} revealed, the high computational cost of the numerical models means that often a study is either multidisciplinary without optimization \citep{RM3,mi_multi-scale_2025}, or performs optimization for a single discipline, such as controls, powertrain, or structures, not all at once.

    Of the few studies that perform multidisciplinary optimization, those that use numerical models take many hours to run, especially when using heuristic optimizers, preventing designers from iterating quickly.
    For example, the BEM-based 4-WEC array optimization in the study \citep{khanal_multi-objective_2024} took 36 hours to run on a high-end workstation for only a single frequency.
    This type of model is too slow for use in realistic large-scale optimizations.
    On the other extreme, \citet{mccabe_multidisciplinary_2022} uses an algebraic hydrodynamic model that runs quickly but lacks sufficient accuracy to draw meaningful design conclusions.
    The present paper introduces a set of interoperable semi-analytical models that are designed to have the speed and robustness required for multidisciplinary optimization.
\else
    The directly comparable studies above either use slow numerical models that limit optimization scope (e.g.\ a BEM-based 4-WEC array optimization that took 36 hours per frequency on a high-end workstation \citep{khanal_multi-objective_2024}) or fast algebraic models that lack accuracy for meaningful design conclusions (e.g.\ \citep{mccabe_multidisciplinary_2022}).
    Semi-analytical models offer a useful intermediate that has not yet been combined with multidisciplinary scope in an open-source WEC tool, which is the gap MDOcean addresses.
\fi

\subsection{Paper Contributions and Roadmap}
\ifdefined\DISSERTATION
    To address the gaps just described, the authors of this paper create a fast, multidisciplinary, and open-source WEC simulation software.
    The developed software, called MDOcean \citep{mccabe_mdocean_2024}, implements relevant semi-analytical models from various fields, integrates them in a scalable framework for WEC simulation and concurrent optimization, and validates them in a WEC system context.
    The modeling framework articulated here is scalable to more disciplines, and enables systematic WEC design optimization.
\else
    This paper presents MDOcean \citep{mccabe_mdocean_2024}, an open-source semi-analytical WEC simulation framework addressing the gap identified above.
    The work provides the following contributions:
    \begin{itemize}[leftmargin=*]
        \item \textbf{Integrates semi-analytical models} from hydrodynamics, dynamics, control, structures, and economics in a single framework suitable for optimization
        \item Formulates an underactuated optimal control problem with coupled nonlinear multi-port dynamics, and introduces the \textbf{linearized pseudo-spectral (LPS) method} to approximate time-domain constraints and nonlinearities
        \item \textbf{Analytically solves} the LPS optimal control problem as a univariate quadratically-constrained quadratic program (QCQP) with a geometric interpretation on the complex reflection coefficient plane
        \item \textbf{Validates} against WEC-Sim and the RM3 reference design across power, structural, and economic outputs
        \item Demonstrates \textbf{computational performance} (\resultsAOR[simRuntime] simulation runtime) enabling optimization studies that are otherwise infeasible
        \item Illustrates the \textbf{mathematical structure and effect magnitude} of subsystem interactions and techno-economic scaling relationships, providing intuition to facilitate early-stage design tradeoffs
    \end{itemize}
\fi
\ifdefined\DISSERTATION
    In \Cref{sec:model-structure}, the research objective and design scope are first used to define a broad problem formulation and module decomposition.
    Next, \Cref{sec:modules} describes the construction of the simulation model.
    It develops the assumptions, analysis method, and implementation details of each module based on the appropriate balance of accuracy and speed.
    \Cref{sec:validation-benchmarking} discusses model validation and runtime benchmarking.
    Finally, \Cref{sec:discussion} showcases results from model sweeps and insights from the model structure.
\else
    \Cref{sec:model-structure} introduces MDOcean's modular structure and the rationale behind its analysis architecture.
    \Cref{sec:modules} documents the assumptions, analysis methods, and implementation choices of each module -- geometry, hydrodynamics, dynamics and control, structures, and economics.
    \Cref{sec:validation-benchmarking} compares MDOcean against WEC-Sim and the RM3 reference design, demonstrates the runtime gain over established tools, and discusses where the model is reliable.
    \Cref{sec:discussion} presents results from variable sweeps and model-derived insights, along with limitations and future work.
    Extended derivations, validation studies, and implementation details are provided in the appendices to preserve readability of the primary narrative while maintaining reproducibility.
    The companion paper \citep{mccabe_leveraging_2026} applies MDOcean to a multidisciplinary techno-economic optimization of the RM3.
\fi
\section{Model Structure}
\label{sec:model-structure}

The simulation is composed of five coupled modules: geometry, hydrodynamics, dynamics and control, structures, and economics.
The geometry module computes derived properties such as areas, volumes, masses, centers of gravity, hydrostatics, and stability margins from bulk dimensions and structural thicknesses.
The hydrodynamics module evaluates frequency-domain hydrodynamic coefficients using a semi-analytical formulation based on the device geometry.
The dynamics and control module combines hydrodynamic coefficients, mass properties, and generator ratings to compute system response, loads, and power production using a quasi-linear frequency-domain model with nonlinear corrections for drag and powertrain saturation.
Both operational and extreme (storm) conditions are considered.
The structures module evaluates stress and factor-of-safety constraints under yield, ultimate, buckling, and fatigue criteria.
Finally, the economics module estimates capital cost (PTO and structure) and levelized cost of energy using power production and material usage.

\begin{figure*}[htbp]
\centering
\includegraphics[width=\linewidth]{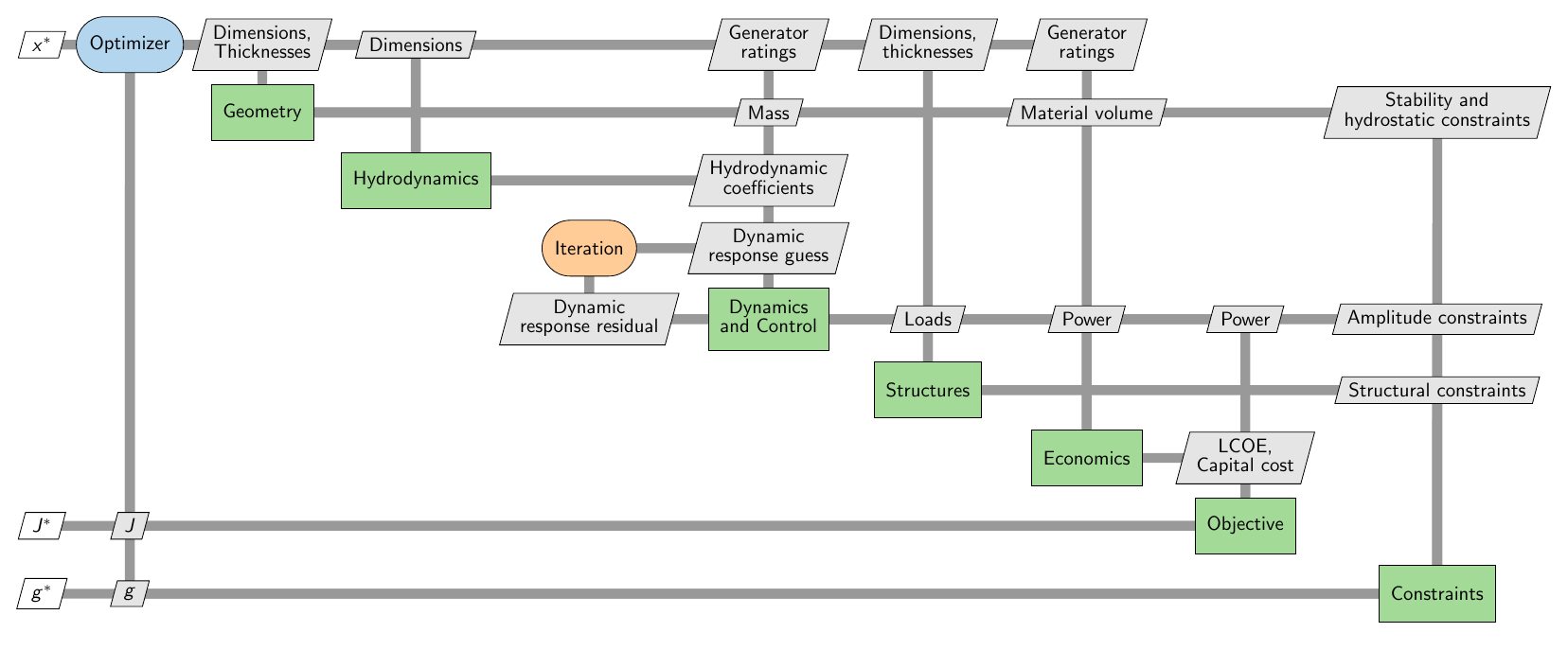}
\caption{Simplified XDSM diagram}\label{fig:n2}
\end{figure*}

Although this work does not perform optimization, the model is developed to support system-level design optimization.
The module coupling structure is illustrated in the XDSM diagram in \Cref{fig:n2}.
Inputs and outputs are organized according to standard MDO conventions \citep{lambe_extensions_2012}, and an optimizer would update design variables $x$ using objective $J$ and constraints $g$ until convergence to $x^*$.

\begin{figure*}[b!]
\centering
\includegraphics[width=1\linewidth]{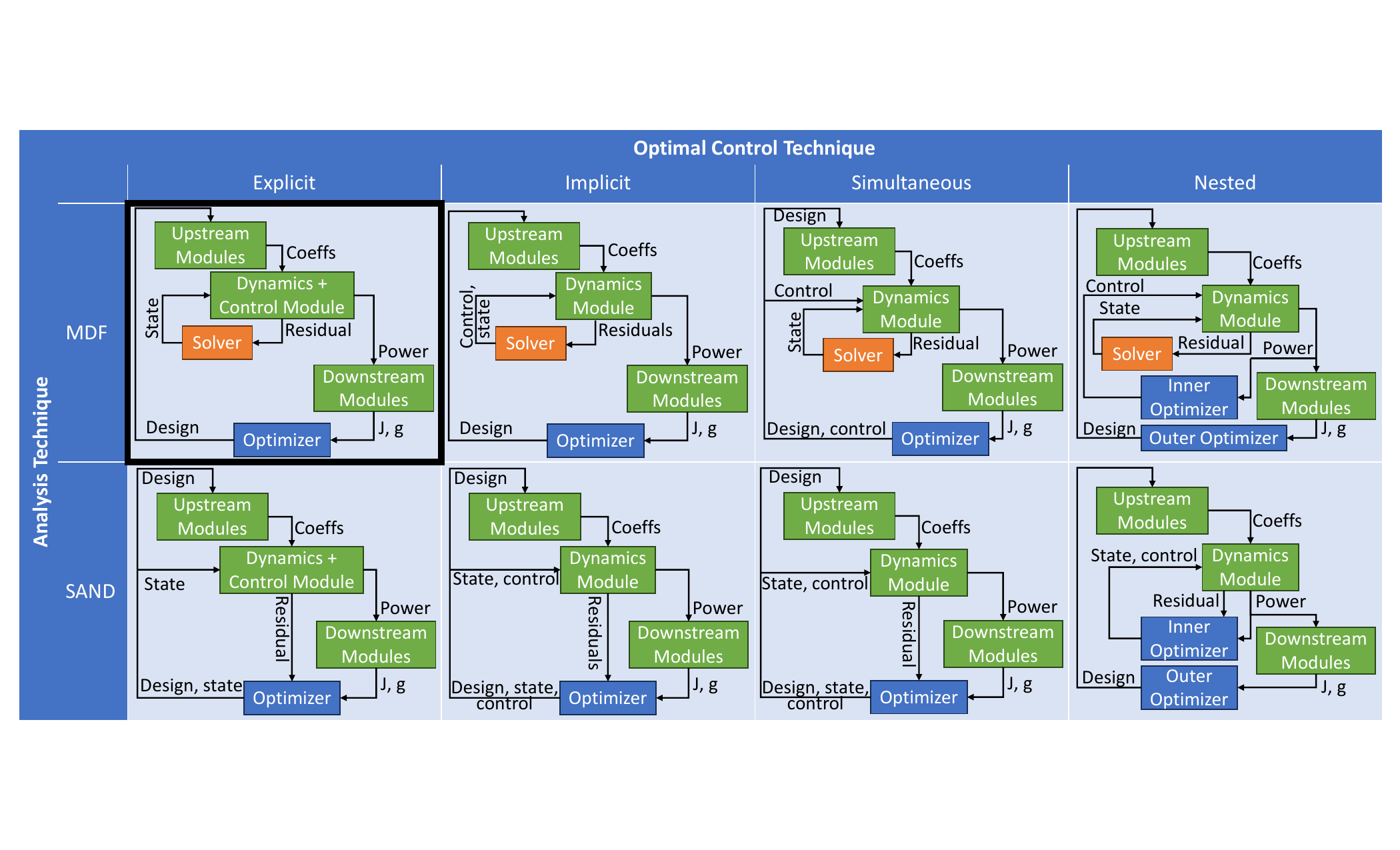}
\caption{Optimization architectures organized by analysis method (MDF and SAND) and control method (explicit, implicit, simultaneous, and nested).}
\label{fig:control-arch}
\end{figure*}

\ifdefined\DISSERTATION
    \paragraph{Absence of Feedback Coupling}
    The lack of variables in the lower left portion of the diagram indicates that there is no feedback coupling between modules, so an external solver to enforce consistency is not necessary.
    This helps decrease the simulation computation time and is appealing for optimization.
    \paragraph{Presence of Design Variable and Feed-Forward Coupling}
    The fact that the same design variable group is an input to multiple modules indicates the presence of design variable coupling.
    Most heavily coupled are the bulk dimensions, which are input to the geometry, hydrodynamics, and structural modules.
    Thicknesses and generator ratings are each input to two modules, with thicknesses coupling geometry and structures and generator ratings coupling dynamics and economics.
    Moreover, the presence of variables in the upper right of the diagram indicates feed-forward coupling in which the output of one module directly affects subsequent modules.
    
    Analysis of coupling is important because under certain conditions, the optimization can be decomposed into separate problems to be solved in sequence or in parallel without sacrificing performance.
    For example, techniques like block coordinate descent or alternating direction method of multipliers apply if coupling is convex \citep{wright_coordinate_2015}, and monotonicity-based decomposition and constraint elimination apply if coupling is monotonic \citep{azarm_monotonicity-based_1988,papalambros_principles_2017}.
    The present coupling of WEC disciplines is non-convex and non-monotonic, so concurrent optimization is required to obtain the system optimal design.
    Non-monotonicity means that perturbing a variable from one module in a certain direction does not necessarily determine the direction of the propagated change in coupling variables, objective, and constraints computed in other modules.
    Non-convexity arises from the presence of both positive and negative curvature in the coupled variable space, leading to multiple local optima.
    With an objective $J$ of LCOE, for example, while an increase in hydrodynamic damping generally increases power production (beneficial to $J$), it also increases structural loading (detrimental to $g$), and the larger bulk dimensions required to achieve that higher damping may produce higher stresses (detrimental to $g$) and increase the required structural material (detrimental to $J$).
    Thus, it would be sub-optimal to optimize the hydrodynamics module strictly for damping or power production followed by a separate optimization considering structures and economics.
    Furthermore, while even in a unified optimization it may be tempting to calculate the structural thicknesses within the structures module as the minimum required to sustain loads without failure, it is important to keep structural thicknesses as design variables because they also contribute to the hydrostatic constraints computed in the geometry module.
    If active, these constraints would make the system-level optimum material thickness larger than what is structurally necessary.
    Using a single optimizer to drive the design of all modules and consider all constraints addresses this non-monotonic coupling. 
    
    \paragraph{Multiple-Discipline-Feasible Analysis Architecture}
    All modules except the dynamics module are explicit, meaning they require no internal iteration to converge.
    The dynamics module requires iteration to incorporate nonlinearities into the quasi-linear frequency domain model, shown in the XDSM diagram as the orange box.
    The dynamics iteration occurs separately for each outer iteration of the optimizer.
    This structure is known as the multiple discipline feasible (MDF) architecture.
    In principle, it is possible to remove the dynamics iteration and instead incorporate the nonlinear dynamics as a residual equality constraint within the optimization, which is known as the simultaneous analysis and design (SAND) architecture \citep{martins_multidisciplinary_2013}.
    SAND generally decreases the runtime of the simulation (analysis) but requires more optimization iterations, since it is the optimization rather than the simulation that must converge residuals. 
    The WecOptTool control co-optimization software pursues the SAND strategy by collocating the dynamics constraints with the pseudo-spectral method \citep{coe_initial_2020}.
    
    Readers familiar with trajectory optimization should note that the choice of MDF versus SAND as an MDO architecture parallels that of shooting versus collocation as a transcription method, in the sense that one chooses to enforce governing equations with simulation (MDF and shooting) versus optimization (SAND and collocation) \citep{tedrake_underactuated_2024}.
    In MDO, the appropriate decision depends on characteristics of all modules, not merely the module whose governing equations are in question.
    As \Cref{sec:sim-runtime} will demonstrate, the hydrodynamics module, not dynamics and controls, is the dominant computational cost of the simulation.
    This means that even a substantial dynamics speedup represents only a small speedup of the full simulation and is unlikely to outweigh the slowdown of more (hydrodynamics-dominated) simulation evaluations.
    For this reason, SAND would likely increase the runtime of the full optimization, motivating MDOcean's selection of MDF as the more suitable architecture.
    Another advantage of MDF over SAND is that the latter invites the possibility of inconsistent dynamics if the optimization terminates unsuccessfully \citep{martins_multidisciplinary_2013} and requires optimizing a dummy constant objective to perform simulation without optimization.
    
    \paragraph{Control Optimization Architecture}
    In control co-design, the nested and simultaneous architectures are often contrasted \citep{sundarrajan_towards_2021,herber_nested_2018}.
    The nested architecture, known as ``asymmetric subspace optimization'' in MDO literature \citep{chittick_asymmetric_2009}, optimizes control variables in an inner loop for each iteration of the outer design optimization.
    This takes advantage of the fact that the inner optimization is often convex and can be solved quickly.
    On the other hand, the simultaneous architecture uses a single optimizer to optimize design and control variables concurrently, creating a larger (typically non-convex) optimization problem and avoiding the difficulty of inner loop infeasibility that may arise in the nested approach \citep{herber_nested_2018}. 
    We presently distinguish two additional types of control co-design architectures in which the controller is not numerically optimized and instead obtained through an optimal control law.
    The optimal control law can be either explicitly derived and applied in the simulation (``explicit'' or analytical), or implicitly solved for by numerically zeroing the residual of an analytical optimal control condition (``implicit'' or semi-analytical).
    
    Each of these four control architectures can coexist with both MDF and SAND analysis architectures, shown in \Cref{fig:control-arch}.
    The analysis architecture determines whether it is a solver or an optimization constraint which handles the dynamic residuals to converge the state, while the control architecture determines whether control variables are determined via simulation, residual solve, or simultaneous/nested optimization.
    In the ``implicit'' control diagram, the plural ``residuals'' is used rather than ``residual'' to convey that there exist residuals on the optimal control conditions to determine the control variables, in addition to those on the dynamic equations to determine the state variables.
    While the implicit-SAND and simultaneous-SAND block diagrams look nearly identical, the former determines the control variables via residuals of pre-determined optimality conditions expressed as constraints to the optimizer, while the latter determines them to maximize the optimization objective without a-priori knowledge of the optimal control conditions.
    
    In WecOptTool, the pseudo-spectral optimizer determines the state and control variables for a fixed design without using previously-known optimality conditions.
    Depending on whether a user adds additional design variables to the existing pseudo-spectral optimizer or in a separate loop outside, WecOptTool would be considered either simultaneous-SAND or nested-SAND respectively \citep{coe_initial_2020}.
    
    For MDOcean, we choose the explicit-MDF architecture.
    Like the implicit-MDF architecture, it minimizes computational cost by avoiding the need to include control variables in an optimization and lowering the number of dynamic constraints which the optimizer must consider.
    This is possible due to the semi-analytical dynamics and control module, where the optimal constrained control conditions are derived analytically.
    Explicit-MDF is chosen over implicit-MDF because it avoids the latter's numerical root-finding and is thus faster, as \Cref{sec:mod-freq-domain} will detail.
    An early version of MDOcean used the implicit-MDF architecture, which is described in \Cref{sec:appendix-qp-numerical} for completeness.
\else
    The model exhibits strong design-variable coupling but no direct feedback coupling between modules, eliminating the need for a system-level consistency iteration.
    This improves computational efficiency and is advantageous for optimization.
    Coupling occurs primarily through shared design variables, particularly bulk dimensions, which affect geometry, hydrodynamics, and structures, as well as generator ratings and structural thicknesses, which couple multiple subsystems.
    Feed-forward coupling is also present through inter-module dependencies.
    
    This coupling is both non-convex (e.g., larger bulk dimensions raise power but also raise structural loads and material requirements) and non-monotonic, motivating concurrent rather than sequential optimization.
    
    \paragraph{MDO Architecture}
    All modules except dynamics are explicit.
    The dynamics module requires an internal fixed-point iteration to resolve nonlinearities in the quasi-linear frequency-domain formulation.
    This leads to a multiple-discipline feasible (MDF) architecture, where dynamics convergence is handled within each simulation call.
    
    An alternative simultaneous analysis and design (SAND) formulation \citep{martins_multidisciplinary_2013} would enforce dynamic residuals directly within the optimization, potentially reducing simulation cost but increasing optimization complexity.
    However, since hydrodynamics dominates computational cost in this framework (\Cref{sec:sim-runtime}), the benefit of removing dynamics iteration is limited.
    For this reason, MDF is adopted.
    
    \paragraph{Control Co-Design Architecture}
    Control co-design problems are commonly classified into nested and simultaneous formulations \citep{sundarrajan_towards_2021,herber_nested_2018}.
    In the nested approach, control is optimized in an inner loop, while in the simultaneous approach control and design variables are optimized together.
    In addition, optimal control-based formulations may determine control either explicitly or implicitly via optimality conditions. 
    
    These control formulations can be combined with both MDF and SAND analysis architectures.
    MDOcean uses an implicit-MDF architecture: the dynamics module iterates internally to converge the quasi-linear drag and saturation describing functions, while all other modules are explicit.
    The implicit formulation is preferred over explicit alternatives due to the structure of the derived optimal control conditions (\Cref{sec:mod-freq-domain}).
    It also avoids introducing control variables into the outer optimization while preserving optimal control behavior, minimizing total runtime given that hydrodynamics dominates computational cost (\Cref{sec:sim-runtime}).
    Detailed architecture rationale, coupling analysis, and the relation to MDO literature are provided in \cite{mccabe_dissertation_2026}.
    
    \Cref{fig:control-arch} summarizes the relationship between analysis and control architectures, with the explicit-MDF formulation used in this study highlighted.
\fi

\section{Module Details}
\label{sec:modules}
\ifdefined\DISSERTATION
    The five modules (geometry, hydrodynamics, dynamics and controls, structures, and economics) will now be described one at a time.
    For notation, unbolded unarrowed variables refer to scalars (e.g. $x$ or $X$), arrowed variables refer to vectors ($\vec{x}$ or $\vec{X}$), and bold variables refer to matrices ($\mathbf{x}$ or $\mathbf{X}$).
    For dynamic (time-varying) quantities, the time domain representation is denoted $x(t)$, and can also be represented as the real part of a complex signal $x(t)=\Re\left[\tilde{x}(t)\right]$.
    Assuming the signal is sinusoidal, the complex signal can be expressed as $\tilde{x}(t)=\hat{X}e^{i\omega t}$ with complex phasor $\hat{X}=|\hat{X}|e^{i\angle \hat{X}}$ and angular frequency $\omega$.
    Altogether, this gives $x(t)=|\hat{X}|\cos(\omega t+\angle\hat{X})$. 
    Dots ($\dot{x}(t)=\Re(\hat{\dot{X}}e^{i\omega t})=\Re(i\omega\hat{X}e^{i\omega t})$) indicate time-derivatives.
    For static quantities, no special notation differentiates complex from real variables, as this will be indicated directly in the text or with $\mathbb{R}$ or $\mathbb{C}$ if clarification is needed.
\else
    The five modules (geometry, hydrodynamics, dynamics and controls, structures, and economics) will now be described one at a time.
    For notation, unbolded unarrowed variables refer to scalars, arrowed variables refer to vectors, and bold variables refer to matrices.
    Time-varying quantities $x(t)$ are represented by complex phasor $\hat{X}$ such that $x(t)=\Re[\hat{X}e^{i\omega t}]$.
\fi
\subsection{Geometry}\label{sec:geom}
\begin{figure}[htbp]
\centering
    \includegraphics[width=\linewidth]{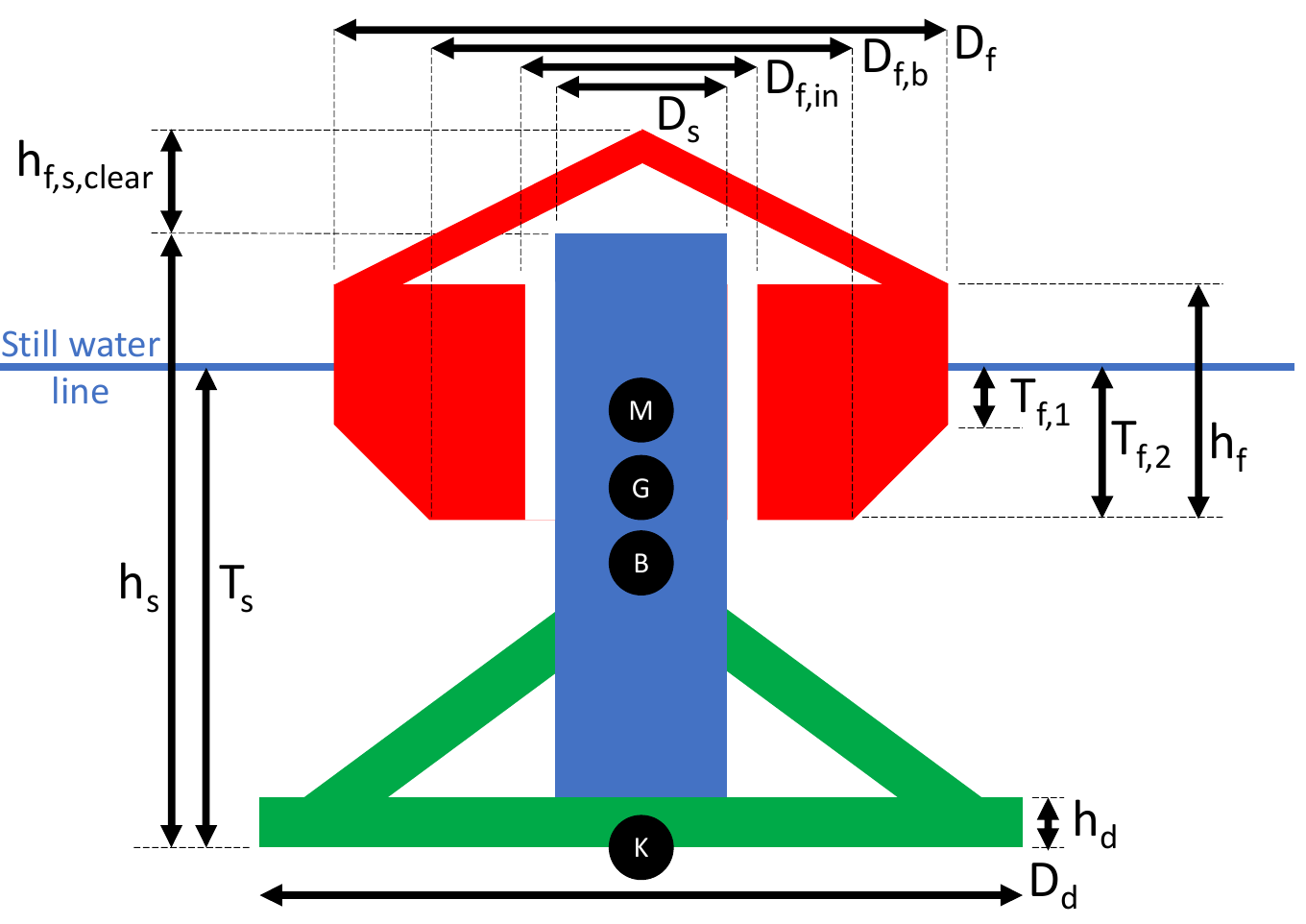}
\caption{Dimension labeling of system}\label{fig:dims}
\end{figure}

The geometry module computes submerged volume $V_{\text{sub}}$, structural volume $V_{\text{struct}}$, and associated hydrostatic properties from bulk dimensions.
\Cref{fig:dims} defines the principal geometric variables, where $T$ denotes drafts, $h$ heights, and $D$ diameters.
Subscripts $f$, $s$, and $d$ refer to the float, spar, and damping plate, respectively.
Key points include the center of buoyancy $B$, center of gravity $G$, the keel $K$, and metacenter $M$ for the combined system.
Additional structural dimensions and thickness definitions are provided in \Cref{sec:structures}.

Static pitch stability is enforced by requiring:
\begin{equation}\label{eq:GM}
    \overline{GM} = \overline{KB} + \overline{BM} - \overline{KG} > 0,
\end{equation}
\noindent where $\overline{BM} = \pi D_f^4/(64V_{\text{sub}})$ \citep{newman}.

Hydrostatic equilibrium is enforced by matching displaced water mass to total system mass, with any mismatch compensated using ballast with mass $m_{bal}$:
\begin{equation}
    m_{bal} = \rho_w V_{sub} - \rho_M V_{struct},
\end{equation}
where $\rho_w$ and $\rho_M$ are water and structural material densities.
To ensure feasible buoyancy, $m_{bal}$ must be non-negative and physically storable within the hull.
Assuming seawater ballast, this imposes a volume constraint:
\begin{equation}\label{eq:vol-constraint}
V_{struct} \le V_{sub} + V_{surf} - V_{pto},
\end{equation}
where $V_{\text{surf}}$ (design-dependent) and $V_{pto}$ (fixed) are volumes of the above-water structures and PTO respectively.

\subsection{Hydrodynamic Coefficients}\label{sec:hydro}
The first-order hydrodynamic force phasors $\vec{\hat{F}}$ for interacting floating bodies are expressed in the frequency domain as the sum of wave excitation $\vec{\hat{F}}_{e}$, hydrodynamic radiation $\vec{\hat{F}}_{rad}$, and hydrostatic restoring $\vec{\hat{F}}_{res}$ contributions:
\begin{equation}\label{eq:hydro-forces}
    \vec{\hat{F}} = \vec{\hat{F}}_{e} + \vec{\hat{F}}_{rad} + \vec{\hat{F}}_{res}
\end{equation}
where, under monochromatic forcing,
\begin{equation}\label{eq:hydro-forces-expanded}
\begin{aligned}
    \vec{\hat{F}}_{e} &= \vec{\gamma}\,\hat{\zeta},&~~
    \vec{\hat{F}}_{rad} &= -\mathbf{A}_h\vec{\hat{\ddot{\xi}}} - \mathbf{B}_h\vec{\hat{\dot{\xi}}},&~~
    \vec{\hat{F}}_{res} &= -\mathbf{K}_h\vec{\hat{\xi}}.
\end{aligned}
\end{equation}
Here $\vec{\hat{\xi}}$ is the body displacement phasor vector from the origin at the center of the float at the still waterline.
The hydrodynamic coefficients $\vec{\gamma}$, $\mathbf{A}_h$, $\mathbf{B}_h$, and $\mathbf{K}_h$ are the wave excitation vector, added mass matrix, radiation damping matrix, and hydrostatic stiffness matrix, respectively.

The displacement vector $\vec{\hat{\xi}}$ contains a state for each degree of freedom.
We define separate displacement and excitation vectors for the operational ($op$) and storm ($st$) cases.
A two-body heaving point absorber that rigidly locks the bodies in a storm following \cite{RM3} has vectors:
\begin{equation}\label{eq:define-vectors}
    \vec{\hat{\xi}}_{op} = \begin{bmatrix}
        \hat{\xi}_f \\ \hat{\xi}_s
    \end{bmatrix}, ~~
    \vec{\hat{\xi}}_{st}=\begin{bmatrix}\hat{\xi}_m 
    \end{bmatrix}, ~~
     \vec{\gamma}_{op} = \begin{bmatrix}
        \gamma_f \\ \gamma_s
    \end{bmatrix},~~
     \vec{\gamma}_{st} = \begin{bmatrix}
        \gamma_m
    \end{bmatrix}.
\end{equation}
with subscripts for the float ($f$), spar ($s$), their coupling ($c$), and the merged float-spar ($m$).
Operational hydrodynamic matrices are symmetric $2\times 2$:
\begin{equation}\label{eq:op-hydro-coeffs}
\begin{aligned}
    \mathbf{A}_{h,op} = \begin{bmatrix}
        A_f & A_c \\ A_c & A_s
    \end{bmatrix},~ 
    \mathbf{B}_{h,op} &= \begin{bmatrix}
        B_f & B_c \\ B_c & B_s
    \end{bmatrix},\\
    \mathbf{K}_{h,op} &= \begin{bmatrix}
        K_f & 0\\ 0 & K_s
    \end{bmatrix}
\end{aligned}
\end{equation}
while storm (merged-body) coefficients are scalar:
\begin{equation}\label{eq:st-hydro-coeffs}
\begin{aligned}
\vec{\gamma}_{st}&=\gamma_{m}=\Sigma~\vec{\gamma}_{op},&
~\mathbf{A}_{h,st}&=A_m=\Sigma~\mathbf{A}_{h,op}, \\
~\mathbf{B}_{h,st}&=B_m=\Sigma~\mathbf{B}_{h,op},&
~\mathbf{K}_{h,st}&=K_m=\Sigma~\mathbf{K}_{h,op}.
\end{aligned}
\end{equation}
with $\Sigma$ denoting the sum of all elements in the matrix or vector (``grand sum'').

The remaining task is to compute the float, spar, and coupling coefficients, with operational and storm implications discussed in \Cref{sec:design-load-cases}.
The hydrostatic stiffness terms ($K_f$, $K_s$) are frequency-independent and are computed directly from geometry (see \Cref{eq:gamma-K} in \Cref{sec:appendix-meem-details}), while the remaining coefficients require solving the frequency-domain radiation boundary value problem for the velocity potential.
Traditionally, this is performed using a BEM solver, which discretizes the body surfaces and solves a large linear system.
In MDOcean, float coefficients are instead computed using the semi-analytical Matched Eigenfunction Expansion Method (MEEM), exploiting cylindrical symmetry for improved efficiency.
Spar and coupling terms are obtained via algebraic approximations and interpolation of pre-computed BEM data, and could be extended to MEEM in future work.
\Cref{tab:hydro-methods} summarizes the methods used for each coefficient.

\subsubsection{Float Coefficients: Matched Eigenfunction Expansion Method}\label{sec:hydro-meem}
MEEM solves the radiation and excitation problems by expanding the velocity potential in eigenfunctions within each fluid region and enforcing matching across region boundaries.
The MEEM radiation solution for two concentric heaving cylinders was first presented by \citet{mavrakos_hydrodynamic_2004} and detailed by \citet{chau_inertia_2012}.
In this approach, the fluid domain is partitioned into regions with separable analytical solutions, and continuity conditions are enforced at region interfaces to determine the unknown expansion coefficients.
The open-source implementation used here, summarized in \Cref{sec:appendix-meem-details}, follows the formulation by \citet{mccabe_open-source_2024}.
Forthcoming companion papers \citet{bimali_matrix_2026,best_openflash_2026} detail the numerical properties and code implementation respectively of an expanded implementation. 

The primary advantage of MEEM is computational efficiency.
Timing comparisons (\Cref{sec:sim-runtime}) show more than an order-of-magnitude speedup over Capytaine BEM at comparable convergence, significantly reducing the cost of multidisciplinary analysis.
Additional benefits include lower memory usage, elimination of meshing, and avoidance of numerical Green’s function approximation.

\begin{figure}[b!]
    \centering
     \includegraphics[width=\linewidth]{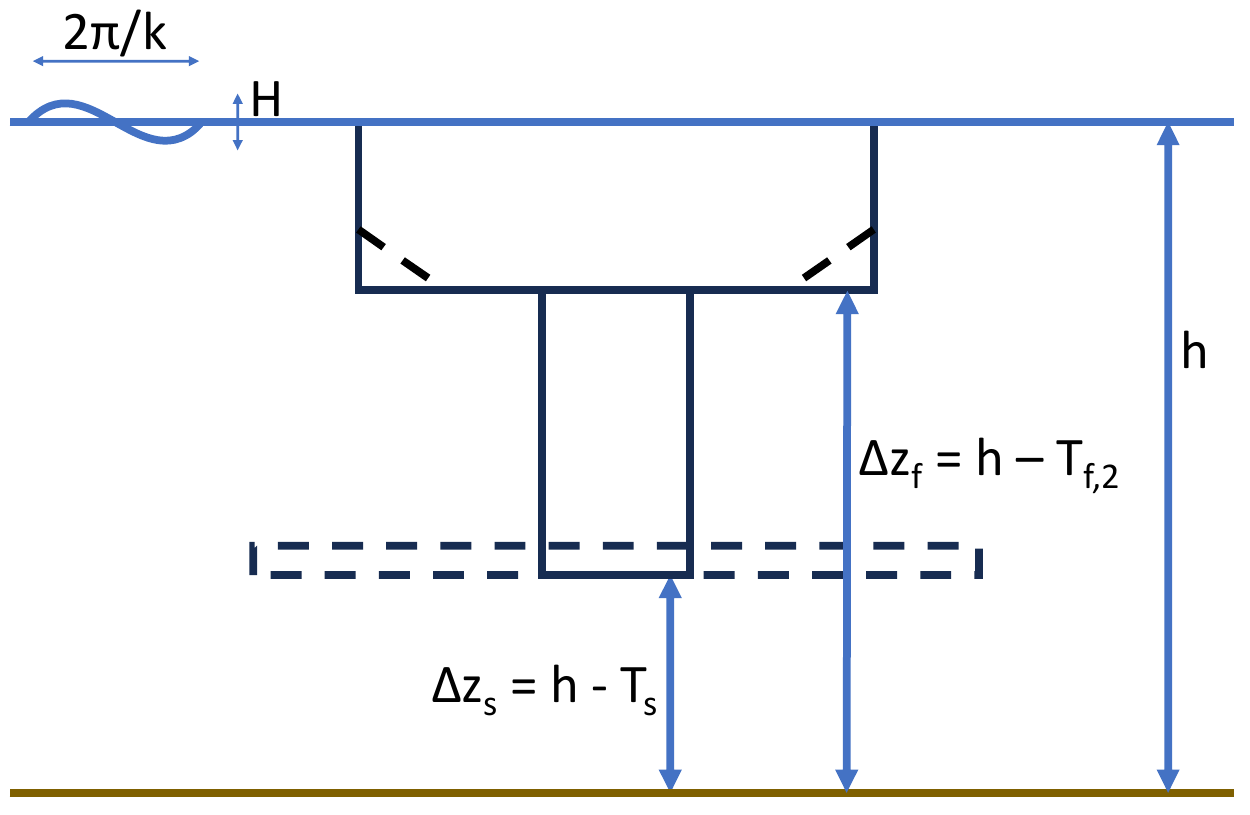}
    \caption{MEEM geometry.
Dashed lines represent geometry that is neglected in the MEEM calculations}
   \label{fig:meem-geom}
\end{figure}

Limitations include geometric restrictions and numerical overflow (discussed in \Cref{sec:appendix-meem-details}).
As a semi-analytical method, MEEM is restricted to simplified geometries; the formulation used here assumes a dual concentric cylinder (\Cref{fig:meem-geom}).
Consequently, float coefficients neglect the damping plate and approximate the float bottom as flat rather than slanted.
The damping plate enters through the spar coefficients (next subsection).
These approximations have minor impact and could be relaxed in future extensions using more general MEEM formulations \citep{olaya_hydrodynamic_2015,kokkinowrachos_behaviour_1986,bimali_matrix_2026}.

\subsubsection{Other coefficients: approximations}\label{sec:hydro-other}
Because the damping plate affects the spar coefficients more strongly than the float, the former are obtained by scaling existing solutions that include the damping plate rather than using MEEM.

\begin{table}[htbp]
    \centering
    \caption{Method of computing hydrodynamic coefficients in MDOcean}
    \label{tab:hydro-methods}
    \begin{tabular}
    {M{0.25\linewidth}M{0.63\linewidth}}
         Coefficient& Method\\ \hline
         $A_f$, $B_f$, $\gamma_f$ & MEEM (\Cref{eq:hydro,eq:gamma-K} in \Cref{sec:appendix-meem-details})\\
         $A_s$& Approximate $\omega\rightarrow\infty$ solution (\Cref{eq:As} in \Cref{sec:appendix-additional-hydro})\\
         $A_c$, $B_s$, $B_c$, $\gamma_s$& Nominal BEM solution, scaled with $D_d$ and $T_s$ (\Cref{eq:scale-wamit} in \Cref{sec:appendix-additional-hydro})\\
    \end{tabular}
\end{table}
The spar added mass coefficient $A_s$ is approximated as a frequency-independent quantity based on the displaced water volume in the spherical projection of the damping plate, with the formula given in \Cref{sec:appendix-additional-hydro}.

The spar damping ($B_s$), excitation ($\gamma_s$), and coupling terms ($A_c$, $B_c$) are obtained by interpolating a pre-computed WAMIT BEM solution over the nondimensional parameter $kD_d$ (see \Cref{sec:appendix-additional-hydro}).
The excitation term is additionally scaled as $\exp(-k(T_s - T_{s,nom}))$ to account for depth-dependent attenuation.


Because float and spar coefficients are computed using different approximations, positive definiteness of the hydrodynamic matrices is not guaranteed (see \Cref{sec:appendix-additional-hydro}).
To ensure physicality, the added mass and damping matrices are required to remain positive definite.
Any violations are corrected by adjusting the coupling terms $A_c$ and $B_c$.

\subsection{Dynamics and Control}\label{sec:dynamics}
This subsection introduces the dynamics and control modeling assumptions before detailing the frequency-domain methods and control formulations.

\subsubsection{Modified frequency-domain methods}
\label{sec:mod-freq-domain}

\begin{figure}[htbp]
    \centering
    \includegraphics[width=.9\linewidth]{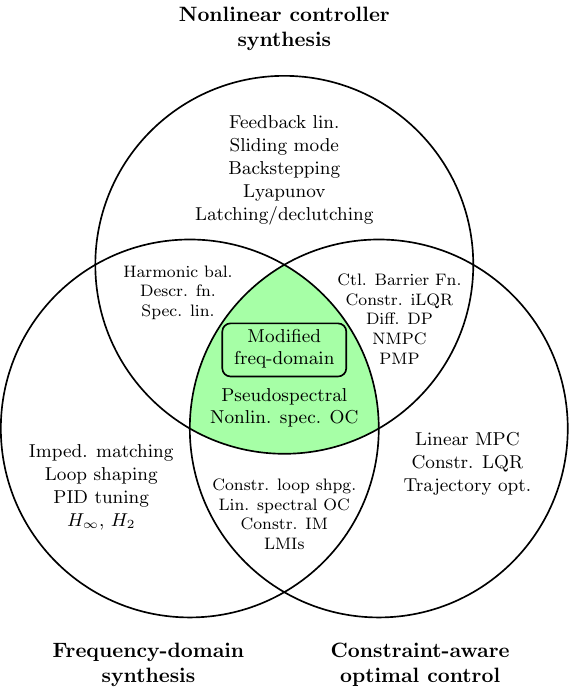}
    \caption{Venn diagram comparing different methods for control synthesis, with MDOcean's quasi-linearized pseudo-spectral approach at the intersection of frequency domain, nonlinear, and constrained optimal control.
    See text for acronym descriptions.}
    \label{fig:venn-diagram}
\end{figure}

\ifdefined\DISSERTATION
    Methods for WEC dynamics and control simulation in an optimization context were reviewed in \Cref{sec:lit} and include time-stepping (nonlinear, numerical, high computation cost), frequency domain (linear, analytical, low computation cost), pseudo-spectral (nonlinear, numerical, medium computation cost), quasi-linearized pseudo-spectral techniques (approximate nonlinear, semi-analytical, low computational cost), and Pontryagin's maximum principle (nonlinear, analytical, low computational cost).
    In wave energy, both spectral-domain numerical methods and analytical methods have generated recent research interest, with studies \cite{bacelli_numerical_2015,coe_initial_2020,tan_computationally_2026,nie_optimal_2016,da_silva_stochastic_2023} and studies \cite{abdulkadir_optimal_2024,lin_fast_2025} serving as illustrative examples of each case.
    This section describes both the technical concept and the practical rationale for the choice of a quasi-linearized pseudo-spectral method, and contextualizes it with respect to other constrained optimal control methods in more detail than was possible in the literature review.

    MDOcean requires the ability to incorporate nonlinearities, at least approximately.
    Inherent nonlinearities such as drag significantly affect the amplitude and power production at resonance.
    Even with linearized intrinsic dynamics, maintaining optimality subject to dynamic constraints also requires a nonlinear controller.
    Dynamic constraints on amplitude, force, and peak power significantly influence the trade-off of power and cost, and amplitude limits are required to obtain a physically meaningful result.
    Besides nonlinearities, MDOcean also requires a low computation cost, since the dynamics will be evaluated many thousands of times during optimization.
    Of the methods mentioned, Pontryagin's maximum principle (PMP) and quasi-linearized pseudo-spectral techniques meet these requirements.
    PMP requires an analytically difficult derivation unique to the specific nonlinearities in question because wave energy dynamics have a particular form requiring so-called singular control \cite{zou_optimal_2017}.
    On the other hand, quasi-linearized pseudo-spectral techniques are simple and intuitive.
    Therefore, MDOcean pursues a semi-analytical quasi-linearized pseudo-spectral method.
\else
    Standard linear frequency-domain WEC analysis assumes unconstrained linear dynamics, enabling closed-form impedance-matched optimal control.
    However, real WECs are subject to (i) dynamic constraints (e.g., generator force, power, and stroke limits) and (ii) nonlinear forces, particularly viscous drag.
    Time-domain simulation handles both naturally but is computationally expensive, especially in an optimization loop.
    Pseudo-spectral methods are nonlinear and handle constraints by collocating the dynamics at specific time-steps, but require an outer numerical optimizer for both simulation and control synthesis.
\fi
MDOcean adopts a quasi-linearized pseudo-spectral (QLPS) approach that retains the speed of frequency-domain analysis while handling both temporal constraints and drag nonlinearities.
This formulation seeks to preserve the computational advantages and simplicity of classical linear analysis while extending its applicability to practical WEC operating regimes that include saturation, drag, and dynamic constraints.
QLPS combines two techniques: (a) describing functions, which quasi-linearize a nonlinear waveform by retaining only its fundamental harmonic,
and (b) constrained spectral optimal control, which solves the optimal control problem as a quadratically-constrained quadratic program (QCQP) assuming linear dynamics and spectral-domain constraints.
\Cref{fig:venn-diagram} positions QLPS relative to existing methods.
Augmenting spectral optimal control with describing functions extends its scope to approximate time-domain constraints and nonlinear dynamics.
Furthermore, because QLPS assumptions under monochromatic forcing lead to a monochromatic problem, the QCQP becomes univariate for systems with a single control degree of freedom.
The optimal constrained linear controller can then be found analytically through a geometric interpretation that exploits the low dimensionality. 
Its nonlinear counterpart can be reconstructed semi-analytically with inverse describing functions.
Avoiding iterative numerical optimization substantially reduces computational cost in large parametric studies and produces insight into the mathematical structure of dynamic tradeoffs to inform PTO sizing.

\ifdefined\DISSERTATION
    \Cref{fig:venn-diagram} shows a Venn diagram comparing different methods for optimal control synthesis in greater detail.
    Typical frequency domain methods for control design, such as loop shaping, PID tuning, and $H_\infty$/$H_2$ control for general applications, along with complex-conjugate impedance matching for power-maximizing applications such as wave energy, cannot take into account constraints and nonlinearities.
    Meanwhile, typical methods for nonlinear control design, such as feedback linearization, sliding mode control, backstepping, and Lyapunov-based control, are performed in the time domain and do not incorporate constraints.
    Latching and declutching control are wave-energy-specific passive nonlinear bang-bang control methods that utilize the frequency-domain concept of phase-matching the fundamental harmonic of the response to the wave excitation, but they are still primarily time-domain methods and do not incorporate constraints \cite{hals_comparison_2011}.
    Methods for constrained optimal control, meanwhile, such as model predictive control (MPC), constrained linear-quadratic regulation (LQR), and trajectory optimization are likewise time-domain and were originally derived for linear control problems, although nonlinear extensions exist including nonlinear MPC (NMPC), constrained iterative LQR (iLQR), and differential dynamic programming (DP).
    Other methods for nonlinear constrained optimal control in the time domain include the already-discussed PMP and the control barrier function method.
    Methods to incorporate nonlinearities in the frequency domain include harmonic balance synthesis, which retains full nonlinearity in a sinusoidal-steady state analysis so technically falls in the spectral domain rather than the frequency domain; its linearized counterpart, describing function (DF)-based synthesis \cite{atherton_nonlinear_1982,gelb_multiple-input_1968}; and the extension of the latter to broadband inputs, spectral linearization (also called stochastic linearization, Gaussian closure, and quasi-linear control) \cite{roberts_random_2003,ching_quasilinear_2010}.
    These methods are not optimal control methods and do not incorporate constraints, although it is possible to sub-optimally enforce constraints by applying a saturation nonlinearity to the constrained quantity.
    Finally, methods which combine frequency domain synthesis with constrained optimal control include constrained loop shaping (often enforcing performance constraints while minimizing the distance to an optimal loop shape, in this case the complex-conjugate impedance); linear spectral optimal control (OC), which poses a trajectory optimization problem for a linear system represented with global basis functions in the spectral domain; constrained impedance matching (IM), the narrow-band, frequency-domain equivalent of spectral optimal control; and convex control formulated using linear matrix inequalities (LMIs), typically in the context of robust control \cite{vanantwerp_tutorial_2000}.

    At the intersection of all three categories lies nonlinear spectral OC, pseudo-spectral OC, and the method used in MDOcean, which we call ``quasi-linearized pseudo-spectral OC.''
    The pseudospectral method uses the same global basis function representation as spectral OC, but instead of enforcing the dynamics at all points in the trajectory, it does so only at specific collocation points, which enables the representation of time-domain constraints and nonlinearities as opposed to harmonic constraints and nonlinearities \cite{elnagar_pseudospectral_1995}.
    Pseudo-spectral methods often employ Legendre or Chebyshev polynomial bases, although in wave energy Fourier bases are more typical.
    In this case, the method is equivalent to a nonlinear trajectory optimization problem that enforces both spectral-domain harmonic balance and time-domain constraints.
    The quasi-linearized pseudo-spectral method used in MDOcean borrows the pseudo-spectral concept of incorporating time-domain constraints in the spectral domain, but retains only the fundamental harmonic to collapse to the frequency domain.
    This is exactly analogous to the way that describing functions linearize the harmonic balance method.
    It can also be thought of as the combination of describing functions with constrained impedance matching.
\fi 

\begin{figure*}[htbp]
\centering
\caption{Flowchart depicting the QLPS method (M1). The describing functions (DFs) for response/constraint, controller, and plant are shown in yellow, pink, and blue respectively.}
\begin{subfigure}[t]{.9\linewidth}
    \includegraphics[width=\linewidth]{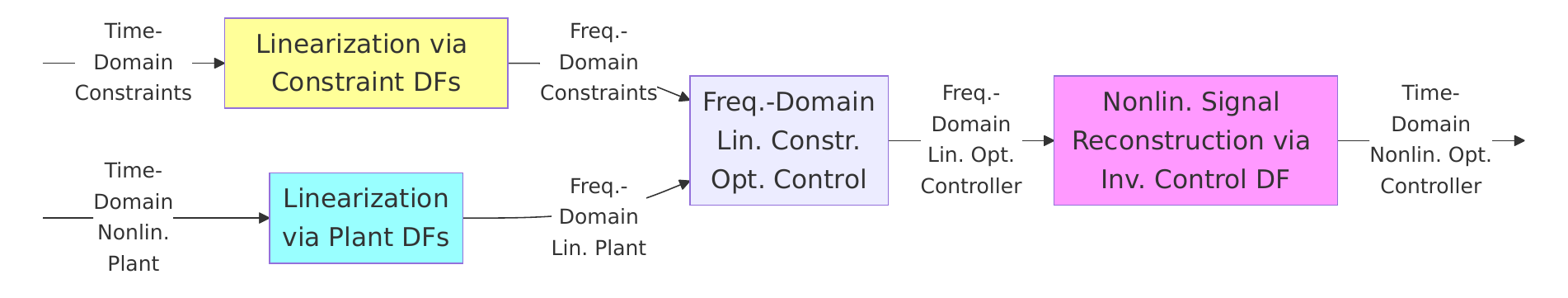}
    \caption{Control synthesis - applies to (M1.2) only.}
    \label{fig:mod-freq-domain-synthesis}
\end{subfigure}
\vfill
\begin{subfigure}[t]{.9\linewidth}
    \includegraphics[width=\linewidth]{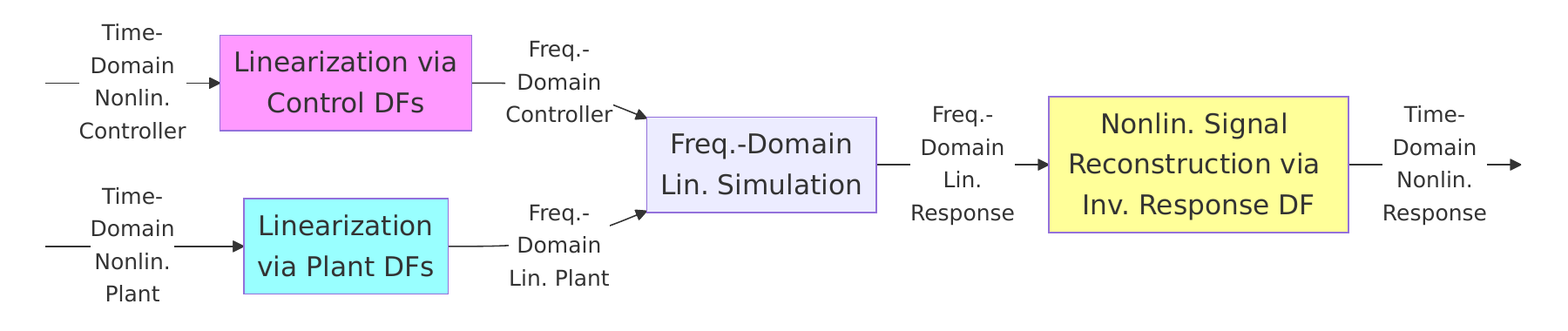}
    \caption{Control evaluation - applies to both (M1.1) and (M1.2).}
    \label{fig:mod-freq-domain-evaluation}
\end{subfigure}
\vfill
\begin{subfigure}[t]{.9\linewidth}
    \includegraphics[width=\linewidth]{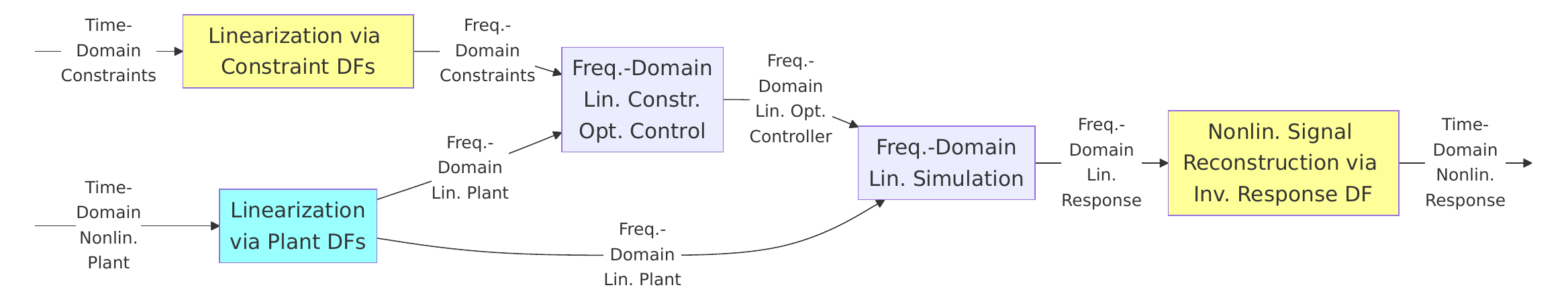}
    \caption{Combined synthesis and evaluation.}
    \label{fig:mod-freq-domain-combined}
\end{subfigure}
\end{figure*}

\Cref{fig:mod-freq-domain-synthesis,fig:mod-freq-domain-evaluation,fig:mod-freq-domain-combined} distinguish three application workflows for the QLPS method: control synthesis (the design of a potentially nonlinear controller to obey constraints and maximize performance of the quasi-linearized system), evaluation (simulation of an arbitrary controller's power performance, where the controller, plant, or both are nonlinear), and both.

MDOcean uses QLPS for both control synthesis and evaluation.
In prior wave energy studies, related approaches have been used for evaluation, with simplified treatments of constraints (see F+ and PI+ entries in \Cref{tab:lit}).
We refer to these as the modified frequency-domain family of methods, of which QLPS is the most accurate and optimal.
In contrast, the present work applies QLPS and other modified frequency-domain methods to control synthesis, enabling nonlinear controller design within a frequency-domain formulation.
This differs from linearization approaches such as stochastic linearization, which are generally limited to linear controller design.

When handling dynamic constraints with a modified frequency-domain method, first the system is (quasi-) linearized and the unconstrained optimal controller (and the response signals corresponding to that controller) are determined in the frequency domain.
Then the unconstrained response is checked for constraint violations.
This check can be performed in the frequency domain if the time-domain constraint can be adequately linearized (via describing functions) as a harmonic constraint, as shown in the top of \Cref{fig:mod-freq-domain-synthesis}, or the response can be reconstructed in the time domain via the inverse describing function if the constraint is not easily representable via harmonics.
If the response for a given sea state does not violate any constraints, then the unconstrained controller and corresponding power is used for that sea state as the frequency-domain linear optimal controller.
Otherwise, a variety of approaches are possible to address the violated constraint, summarized in \Cref{tab:constraint-approaches} and described below.
Methods M1--M2 can be used for both control synthesis and evaluation, while methods M4--M5 can only be used for evaluation since they do not attempt to find a controller that satisfies the constraint.
Method M3 can, in general, only be used for evaluation.
However, in the special case where the only time-domain constraint is on the control force, the method can be used for control synthesis.

\begin{table}[htbp]
\centering
\caption{Constraint satisfaction approaches for modified frequency domain methods. Filled stars indicate better performance.}
\label{tab:constraint-approaches}
\begin{tabular}{M{.03\linewidth} >{\raggedright\arraybackslash}m{0.45\linewidth}l l}
&\textbf{Method} & \textbf{Accuracy} & \textbf{Optimality} \\ \toprule
M1 & Nonlinear optimal (M1.1) or near-optimal (M1.2) controller, quasi-linearized via describing functions& \Stars{3} &\Stars{3} \\ \hline
M2 & Linear optimal controller& \Stars{3} & \Stars{2} \\ \hline
M3 & Saturate or zero the times with constraint violations, neglecting the effect on other signals& \Stars{2} & \Stars{1}\\ \hline
M4 & Saturate or zero entire sea state& \Stars{1} & \Stars{0.5} \\ \hline
M5 &Mark design as infeasible& N/A& \Stars{0} \\
\end{tabular}
\end{table}

\paragraph{M1: QLPS with nonlinear control}
The modified frequency-domain approach with the highest accuracy and optimality (M1) is to synthesize a new nonlinear optimal (or near-optimal) time-domain controller that obeys the constraint, taking into account the linearized effect of that nonlinear controller on the frequency domain response.
This synthesis can occur in a number of ways.
If a time-domain optimal control law (e.g. those that \citet{abdulkadir_optimal_2024,lin_fast_2025} develop via the Pontryagin Maximum Principle (PMP)) is available for the constraint under consideration (M1.1), it can be used as the nonlinear optimal controller directly, essentially bypassing the use of QLPS in the control synthesis step (\Cref{fig:mod-freq-domain-synthesis}) and using it only for evaluation (\Cref{fig:mod-freq-domain-evaluation}).
\ifdefined\DISSERTATION
    This differs from the standard PMP method in that the evaluation of the response and power corresponding to the time-domain PMP controller is determined in the frequency domain, rather than via numerical time-integration (as in the method of \cite{zou_optimal_2017}) or analytical time-integration (as in the method of \cite{lin_fast_2025}).
    This is advantageous if the computational cost or mathematical labor of the numerical or analytical PMP evaluation methods, respectively, are of concern.
\fi
For situations without an explicit PMP control law (M1.2), a nonlinear near-optimal controller can be constructed by first deriving the optimal linear controller via constrained optimization
(M2) and then adjusting it to be nonlinear using insights from signal saturation and filtering.
This is the approach pursued here for the force limit.
M1.2 should be considered the primary formulation of QLPS, and the other modified frequency-domain methods (M2)--(M5) in \Cref{tab:constraint-approaches} act as approximations of QLPS with varying levels of accuracy and optimality.
M1.1 and M1.2 both linearize the nonlinear controller via the describing function method.
In addition to the control nonlinearity, a describing function is also utilized to linearize the plant drag nonlinearity.

\paragraph{M2: QLPS with linear control}
If a nonlinear controller or the corresponding describing function to linearize it are not available, the most optimal linear controller that obeys the constraint can be used instead (M2).
This approach accurately evaluates the maximum power that a linear controller could produce, although this power will be lower (less optimal) than that of the optimal nonlinear controller.
MDOcean pursues this approach for amplitude limits in the operational design load case.

\ifdefined\DISSERTATION
    The constrained optimization to identify the controller in (M1.2) and (M2) can be performed numerically or analytically by analyzing an impedance-mismatched linear system in the frequency domain and will be discussed in detail in \Cref{sec:optimal-control}.
    The current MDOcean software uses the numerical approach, which requires more computation time but is simpler to implement and equally accurate.
    The analytical approach is nonetheless derived in \Cref{sec:optimal-control} (with further details and graphical visualization of the constraints in \Cref{sec:appendix-qp-solution} and reference \cite{mccabe_force-limited_2024}) and recommended for future implementation.
\else
    The constrained optimization used to identify the controller in (M1.2) and (M2) can be performed numerically or analytically by analyzing an impedance-mismatched linear system in the frequency domain;
    the analytical derivation is provided in \Cref{sec:optimal-control}, with additional details in \Cref{sec:appendix-qp-solution}.
\fi

\paragraph{M3--M5: Constrained frequency-domain estimates}
If neither a numerical nor an analytical implementation is available to identify the optimal constrained linear controller, then the violating portion of the time-domain signals can be saturated or zeroed for the sake of power calculation (M3).
This approach neglects the corresponding effect of the saturation/zeroing on other quantities in the system, which lowers the accuracy, and the controller does not have the opportunity to even approximately consider the constraint, resulting in a power production potentially much lower than the constrained optimal.
\ifdefined\DISSERTATION
    This simple approach does not require any derivations to implement.
    It has been used in several previous WEC optimization studies for various constraints \cite{garcia-teruel_design_2022,garcia-teruel_reliability-based_2021,cotten_multi-objective_2022,mccabe_constrained_2013} and is evidently the most common quasi-linearized pseudo-spectral method in the marine energy field.
\else
    This simple approach has been used in several previous WEC optimization studies for various constraints in the studies \citep{garcia-teruel_design_2022,garcia-teruel_reliability-based_2021,cotten_multi-objective_2022,mccabe_constrained_2013}
    and is the most common quasi-linearized pseudo-spectral method in the marine energy field.
\fi
It is utilized here for the power limit.

As a last resort, the energy in that sea state can be zeroed (M4),
essentially assuming that the device enters survival mode in those cases to avoid constraint violation, or the entire design can be marked infeasible with a constraint in the optimization (M5), which reference~\cite{mccabe_constrained_2013} pursues for a radiation limit constraint.
These approaches do not attempt to find dynamic solutions which satisfy the limits, and essentially deal with violations by penalizing the optimization through either the objective or constraints.
The latter approach (M5) is actually the only option for evaluation of slamming amplitude limits in the storm design load case.
\ifdefined\DISSERTATION
    This is because the device is already in survival mode (preventing method M4), is not producing power (preventing methods M3 and M4), and is not applying a control input (preventing methods M1 and M2).
\else
    This is because the device is in survival mode and not applying a control input.
\fi


\Cref{tab:nonlinearities} summarizes the dynamic nonlinearities and limits and the method MDOcean currently uses to handle each.
The describing function for control force is prioritized due to its simpler implementation as well as the force constraint's frequent activity and strong impact on results observed in prior RM3 optimizations \citep{mccabe_multidisciplinary_2022,gaebele_tpl_2025,mcgilton_optimal_2024}.
\ifdefined\DISSERTATION
    In future work, MDOcean could instead utilize describing functions (M1.2) or the optimal constrained linear controller (M2) for the power and operational amplitude limits respectively to improve accuracy and optimality.
\fi

\begin{table}[htbp]
    \centering
    \caption{Dynamic nonlinearities and approach for each}
    \label{tab:nonlinearities}
\begin{tabular}{>{\raggedright\arraybackslash}p{0.25\linewidth}>{\raggedright\arraybackslash}p{0.7\linewidth}}
         \textbf{Dynamic nonlinearity/limit}& \parbox{\linewidth}{\centering\textbf{Method}}\\ \hline
         Drag& Describing function (M1.2) for $\sin(\omega t)|\sin(\omega t)|$ with iteration to find effect on response and optimal controller\\
         Force limit& Describing function (M1.2) for $\text{sat}(\sin(\omega t))$ with iteration to find effect on response and optimal controller\\
        Operational amplitude limit& Optimal linear controller (M2)\\
         Power limit& Average value of $\text{sat}(\sin(\omega t)+C)$, neglecting any effect on response and optimal controller (M3)\\
        Storm amplitude limit&Optimizer considers limit violations as infeasible designs (M5)\\
    \end{tabular}
    \end{table}

Describing functions assume a certain time-domain nonlinear signal shape and then calculate the fundamental amplitude of that signal for use in a typical frequency-domain simulation.
\ifdefined\DISSERTATION
    The two signals where describing functions are used are both force signals.
    For a force signal, the fundamental amplitude is the most relevant harmonic because not only is it typically the largest amplitude, but it also undergoes the least amount of filtering by the low-pass plant and therefore contributes the most to power production.
\fi
Because the fundamental amplitude depends on the response, the frequency domain simulation is only quasi-linear and must either be iterated numerically or solved analytically.
MDOcean uses iteration to handle both the drag force and control force nonlinearities. 
The former encodes the describing function formula directly into the simulation (see derivation in \Cref{sec:drag}), while the latter implements it indirectly using knowledge of the limit cases of describing functions for multiple nonlinear control laws (see discussion in \Cref{sec:appendix-force-sat}).

\ifdefined\DISSERTATION
    This is possible because the describing function quasi-linearized pseudo-spectral method is used for both controller synthesis and evaluation, so the time-domain nonlinear optimal controller that is output from the right of \Cref{fig:mod-freq-domain-synthesis} is fed into the top left of \Cref{fig:mod-freq-domain-evaluation}, resulting in a chaining of the inverse control describing function with the forward control describing function, which cancel out.
    Unless intentionally made different to study robustness, the linearized plant model is identical for the synthesis and evaluation steps.
    In this case, two workflows essentially become one, with no further computation required after the optimal control step.
    Reconstruction of the nonlinear controller is not necessary for the evaluation of the power corresponding to that controller.
    The power can be calculated directly from the frequency-domain response and the describing function of the nonlinear controller.
    The time-domain representation of the nonlinear controller is only necessary for hardware implementation, or if the synthesized controller is to be evaluated with a different method such as time-stepping.
\else
    Because the same describing function is applied in both synthesis and evaluation, the inverse and forward describing functions cancel; the time-domain nonlinear controller need not be reconstructed for power evaluation, only for hardware implementation.
\fi
While the describing function method used here has been known for over half a century, MDOcean represents its first application in a wave energy optimization or open-source simulation, to the authors' knowledge.
The RAFT open-source Python package for offshore wind turbines uses the drag describing function with iteration but does not include describing functions for saturation \cite{hall_open-source_2022}.
MDOcean therefore extends semi-analytical dynamics to wave energy systems, enabling efficient simulation while retaining model fidelity.

\subsubsection{Equation of Motion}\label{sec:eom}
\ifdefined\DISSERTATION
    Before applying the quasi-linearized pseudo-spectral method described above, we first present the equations of motion for a standard linear WEC model, starting in the time domain and quickly moving to the frequency domain.
    Specifically, we combine the bi-conjugate network model of \cite{coe_co-design_2025} with the underactuated multi-degree of freedom model of \cite{faedo_principle_2022} to arrive at a formulation suitable for the impedance matching of a multibody WEC subject to wave forcing, powertrain kinematics and dynamics, and drag.
\else
    The MDOcean dynamic model combines the bi-conjugate network model of \cite{coe_co-design_2025} with the underactuated multi-degree of freedom model of \cite{faedo_principle_2022} to express the equations of motion for a multibody WEC subject to wave forcing, powertrain kinematics and dynamics, and drag.
\fi

The float and spar are each modeled as a floating rigid body, and the two bodies are kinematically coupled to transmit force between them.
Forces include wave excitation $\vec{F}_e$, hydrodynamic radiation $\vec{F}_{rad}$, hydrostatic restoring $\vec{F}_{res}$, drag $\vec{F}_d$, and power takeoff $\vec{F}_p$.

\ifdefined\DISSERTATION
    The MDOcean dynamic model assumes regular waves for reasons to be detailed in \Cref{sec:irregular-waves}, so the system experiences monochromatic wave forcing.
    Combined with the linearity of the model, this implies that all signals are sinusoids.
    Therefore, the time-domain radiation force is a simple product rather than a convolution integral.
    With these considerations, the equation of motion can be expressed in the time domain as:
\else
    Under the regular-wave assumption justified in \Cref{sec:irregular-waves}, all signals are sinusoidal
    and the radiation force reduces to a simple product rather than a convolution integral, yielding the time-domain equation of motion:
\fi

\begin{equation}\label{eq:eom}
    \mathbf{M}\ddot{\vec\xi}(t)
    = \vec{F}_{e}(t) + \vec{F}_{rad}(t) + \vec{F}_{res}(t) + \vec{F}_{d}(t) + \vec{F}_{p}(t)
\end{equation}
where, in the frequency domain under monochromatic forcing,
\begin{equation}\label{eq:eom-forces}
\begin{aligned}
    \vec{F}_{e}(t)   &= \Re\!\left(\vec{\gamma}\,\tilde{\zeta}(t)\right),
    & \vec{F}_{rad}(t) &= -\mathbf{A}_h\ddot{\vec{\xi}}(t) - \mathbf{B}_h\dot{\vec{\xi}}(t), \\
    \vec{F}_{res}(t) &= -\mathbf{K}_h\vec{\xi}(t),
    & \vec{F}_{d}(t)   &= -\mathbf{B}_d\dot{\vec{\xi}}(t) + \Re\!\left(\vec{\gamma}_d\,\tilde{\zeta}(t)\right), \\
    \vec{F}_{p}(t)   &= -\mathbf{B}_p\dot{\vec{\xi}}(t) - \mathbf{K}_p\vec{\xi}(t).
\end{aligned}
\end{equation}
for both operational and storm conditions, where $\vec{\xi}(t)$ (body position) and $\zeta(t)$ (wave elevation) were defined in \Cref{sec:hydro}.
The mass matrix is:
\begin{equation}
\mathbf{M_{op}}=\begin{bmatrix}m_f & 0 \\ 0 & m_s\end{bmatrix}, \quad  \mathbf{M_{storm}}=\begin{bmatrix}m_f + m_s \end{bmatrix}
\end{equation}

\ifdefined\DISSERTATION
    For each force component, a linear or quasi-linear model is adopted, with coefficients $\vec{\gamma}$, $\mathbf{A_h}$, $\mathbf{B_h}$, $\mathbf{K_h}$ from the hydrodynamics module and $\mathbf{B_d}$, $\vec{\gamma}_d$, $\mathbf{B_p}$, and $\mathbf{K_p}$ to be derived in the following sections.
    The excitation coefficients $\vec{\gamma}$ and $\vec{\gamma}_d$ are complex, and all other coefficients are real.
\else
    Hydrodynamic coefficients $\vec{\gamma}$, $\mathbf{A_h}$, $\mathbf{B_h}$, $\mathbf{K_h}$ come from \Cref{sec:hydro};
    drag coefficients $\mathbf{B_d}$, $\vec{\gamma}_d$ are derived in \Cref{sec:drag};
    PTO coefficients $\mathbf{B_p}$, $\mathbf{K_p}$ are derived in \Cref{sec:appendix-pto-model,sec:optimal-control}.
    The excitation coefficients $\vec{\gamma}$ and $\vec{\gamma}_d$ are complex;
    all others are real.
\fi

In the frequency domain, the complex transfer function from real regular wave height $H$ to the vector of complex body velocity phasors $\vec{\hat{\dot{\xi}}}$ is then: 
\begin{equation}\label{eq:eom-freq-domain}
    \frac{\vec{\hat{\dot{\xi}}}}{H} = \frac{1}{2}\left[\mathbf{Z}_i+\mathbf{Z}_p\right]^{-1}(\vec{\gamma}+\vec{\gamma}_d)
\end{equation}
where the inertial, hydrodynamic radiation, hydrostatic restoring, and drag damping terms have been combined into a single intrinsic impedance matrix $\mathbf{Z}_{i}$ for compactness:
\begin{equation}\label{eq:intrinsic-impedance}
\mathbf{Z}_i = -i\omega(\mathbf{M}+\mathbf{A}_h)+(\mathbf{B}_h+\mathbf{B}_d)+\frac{1}{i\omega}(\mathbf{K}_h)
\end{equation}
The PTO kinematics and dynamics are yet to be specified, currently represented by generic impedance matrix $\mathbf{Z}_p = \mathbf{B}_p+\frac{\mathbf{K}_p}{i\omega}$.

\begin{figure}[htbp]
\centering
    \includegraphics[width=\linewidth]{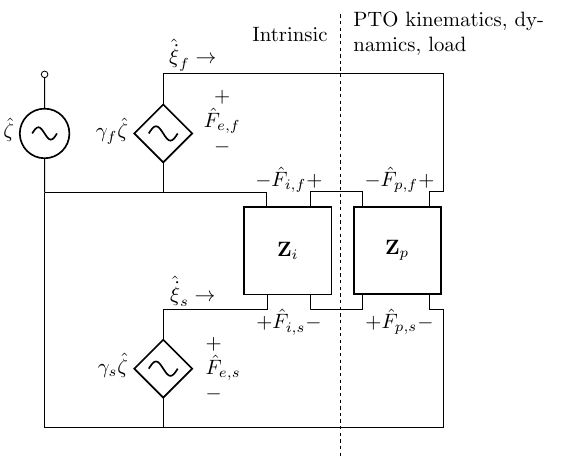}
\caption{Multiport circuit for two hydrodynamic degrees of freedom with intrinsic and powertrain impedances $\mathbf{Z}_i$ and $\mathbf{Z}_p$}
\label{fig:multiport-circuit-intrinsic}
\end{figure}

Equations \eqref{eq:eom-freq-domain}-\eqref{eq:intrinsic-impedance} are represented as a multiport circuit (\Cref{fig:multiport-circuit-intrinsic}), with effort variables $\vec{e}$ and flow variables $\vec{q}$ at each port.
Three matrix forms describe the port relationships:
impedance ($\mathbf{Z}$) groups all ports together;
cascade ($\mathbf{a}$/ABCD) transmits efforts and flows sequentially between port groups and is used for the PTO dynamics (\Cref{sec:appendix-pto-model});
hybrid ($\mathbf{h}$) mixes the two and is used for the PTO kinematics.
Full conventions, conversion formulas, and the cascade inversion rule are given in \Cref{sec:appendix-multiport}, following \cite{reveyrand_multiport_2018}.

Excitation forces are represented as dependent voltage sources (diamonds) driven by a single wave elevation source (circle), and the intrinsic and powertrain dynamics are each represented as a two-port impedance matrix.
The dependent sources and intrinsic impedance matrix are inherently linked through the Haskind relation, which relates excitation coefficient magnitudes $|\gamma_i|$ to diagonal elements of the radiation damping $B_{h,ii}$.
This bounds the maximum absorbed power for any geometry.

\subsubsection{PTO Model}\label{sec:pto-dynamics-overview}
The PTO model specifies the generic powertrain impedance matrix $\mathbf{Z}_p$ in \Cref{eq:eom-freq-domain}.
It is a linear multiport effort-flow model that is valid for multiple degree of freedom devices across various energy domains.
It consists of PTO kinematics (a hybrid matrix relating body degrees of freedom to the PTO port, accommodating underactuation) and PTO dynamics (cascade matrices for a mechanical drivetrain and permanent magnet generator, relating the PTO port to the generator electrical port, with generator voltage $\hat{V}$ and current $\hat{I}$ as the effort and flow variables).
The Thévenin equivalent circuit method is used to find the effective intrinsic impedance $Z_{s,th}$ and open-circuit source voltage $\hat{V}_{s,th}$ viewed from the generator electrical port, as required for impedance matching.
Details and circuit diagrams are provided in \Cref{sec:appendix-pto-dynamics}.

\subsubsection{Power Production}\label{sec:power}
The instantaneous time-domain power at any port is simply the product of the port effort and flow variables, $p(t) = e(t)~q(t)$.
\ifdefined\DISSERTATION
    Substituting $\{e,q\}=\{F_{p},\dot{\xi}\}$ yields the floating body mechanical power,
    $\{e,q\}=\{F_{PTO},\dot{X}_{PTO}\}$ yields the mechanical power at the PTO port,
    $\{e,q\}=\{\tau,\Omega\}$ yields the generator mechanical power,
    and $\{e,q\}=\{V,I\}$ yields the generator electrical power.
\else
    The same expression is evaluated at the floating body, PTO, generator mechanical, and generator electrical ports by substituting the corresponding effort/flow pair $\{e,q\}$.
\fi

In the frequency domain assuming sinusoidal waveforms, the active power $P$, reactive power $Q$, complex power $S$, and apparent power $|S|$ from electrical engineering \citep{saadat_power_1999} are
\begin{equation}\label{eq:power-PQS}
\begin{aligned}
    P &= \tfrac{1}{2}\Re(\hat{e}\hat{q}^*) = \tfrac{1}{2}\Re(Z)|\hat{q}|^2,\\
    Q &= \tfrac{1}{2}\Im(\hat{e}\hat{q}^*) = \tfrac{1}{2}\Im(Z)|\hat{q}|^2, \\
    S &= P + iQ = \tfrac{1}{2}\hat{e}\hat{q}^* = \tfrac{1}{2}Z|\hat{q}|^2, \\
    |S| &= \tfrac{1}{2}|Z||\hat{q}|^2,
\end{aligned}
\end{equation}
where the port is terminated with complex impedance $Z$ and $(\,)^*$ denotes the conjugate transpose.

The average and peak powers over a wave period, derived in \Cref{sec:appendix-power-time}, are
\begin{equation}\label{eq:power-avg-peak}
    p_{avg} = P, \qquad p_{pk} = P + |S|.
\end{equation}
The corresponding minimum power is $p_{\text{min}} = P - |S|$, negative when $|S|>P$ (i.e., reactive control).

To capture non-dynamic power losses such as those in an electrical drive, a constant efficiency $\eta$ describes the gain between average generator electrical power and average absorbed electrical power.
\ifdefined\DISSERTATION
    If the generator cascade matrix is not used (i.e.\ $[\mathbf{a}]_{\tau\Omega\leftarrow VI} = \mathbf{I}$), this efficiency can also approximate losses in the generator itself.
    Note that this is not strictly equivalent to assuming the generator has constant efficiency, since controllers with reactive terms ($K_l\neq 0$) spend some time expending energy, which reverses the direction of the generator efficiency.
    The average efficiency is assumed to also apply to the peak powers, though this is not guaranteed in general for the same reason.
\fi
Validation in \Cref{sec:validation-benchmarking} sets $\eta=80\%$ to match the RM3 design assumption of \cite{RM3}.

The full expansion of $p_{\text{avg},\text{elec}}$ and $p_{pk,\text{elec}}$ in terms of the optimal-control variables $B_l$, $K_l$, and the electrical-port current phasor $\hat{I}$ is given in \Cref{sec:appendix-power-elec}.



\subsubsection{Optimal Control}\label{sec:optimal-control}
This subsection presents the unconstrained and constrained optimal-control solutions used in the model.
\paragraph{Unconstrained}
In the absence of constraints, maximizing power at a given port requires impedance matching at that port due to the maximum power transfer theorem.
We apply matching at the generator electrical port, so the electrical load impedance $Z_l$ must be the complex conjugate of the Th\'{e}venin equivalent source impedance $Z_{s,th}$ seen from that port.
This is called reactive control, with subscript $reac$:
\begin{equation}\label{eq:matched-load}
    Z_{l,reac} = Z_{s,th}^*
\end{equation}

However, some PTOs lack reactive capability (the ability to function as a motor and add energy to the system at some times), such as rectifying hydraulic systems or winches where cable tension must be maintained.
Avoiding reactive power ($Q=0$) at the electrical port restricts the system to so-called damping control, with subscript $damp$, and requires $K_{l,\text{damp}}=0$.
The power-maximizing PTO damping $B_{l,\text{damp}}$ is then: 
\begin{equation}\label{eq:damping-control}
    B_{l,damp} = \sqrt{ (B_{l,reac})^2 + (K_{l,reac}/\omega)^2}
\end{equation}
\ifdefined\DISSERTATION
    When a port other than the electrical port must avoid reactive power, the constrained solution detailed later in this section can be used to enforce $Q=0$ at that port.

    Both of these controllers frequently result in excessively large forces and thus structural and powertrain requirements, especially in energetic sea states or in reactive control when $K_l$ is far from zero.
    $K_{l,\text{reac}}$ is often a large negative number because cost minimization favors a low mass, and the negative PTO stiffness reduces the natural frequency to better match the ocean wave frequency with a smaller device mass.
    Control laws \eqref{eq:matched-load} and \eqref{eq:damping-control} provide no mechanism to reduce this torque, so below we formulate a control law with the ability to obey constraints.
\else
    Both controllers frequently produce excessively large forces in energetic sea states or under reactive control, and provide no mechanism to obey limits.
    We therefore formulate a control law that handles constraints below.
\fi

\paragraph{Constrained}
We first compile expressions for all metrics $M$ which must be constrained.
\ifdefined\DISSERTATION
    For the time being, we keep these expressions generic so they apply to any port.
    To facilitate the calculations in transmission form, we write the metrics as the quadratic product of the port variable vector with a suitable matrix $\mathbf{A}_M$:
\else
    Each metric is written as a quadratic product of the port variable vector with a matrix $\mathbf{A}_M$:
\fi
\begin{equation}\label{eq:metrics}
    M = \frac{1}{4} \begin{bmatrix}
    \hat{e} \\ \hat{q}
    \end{bmatrix}^*
    \mathbf{A}_M
    \begin{bmatrix}
    \hat{e} \\ \hat{q}
    \end{bmatrix}
\end{equation}
The matrices for various metrics are given in \Cref{tab:power-metrics}.
\begin{table}[bhp]
\centering
\caption{Matrices for calculating various metrics at a port via \Cref{eq:metrics}}
\label{tab:power-metrics}
\begin{tabular}{>{\raggedright\arraybackslash}p{0.1\linewidth}|ccccc}
    Metric $M$ & $p_{\text{avg}}=P$ & $Q$ & $S$ & $|\hat{e}|^2$ & $|\hat{q}|^2$ \\ \hline
    Matrix $\mathbf{A}_M$ &
    $\begin{bmatrix}
        0 & 1 \\
        1 & 0
    \end{bmatrix}$
    &
    $\frac{1}{i}
    \begin{bmatrix}
        0 & -1 \\
        1 & 0
    \end{bmatrix}$
    &
    $\begin{bmatrix}
        0 & 0 \\
        2 & 0
    \end{bmatrix}$
    &
    $\begin{bmatrix}
        1 & 0 \\ 
        0 & 0
    \end{bmatrix}$
    & 
    $\begin{bmatrix}
        0 & 0 \\ 
        0 & 1
    \end{bmatrix}$
\end{tabular}
\end{table}

Note that the apparent power $|S|$, and therefore the peak power from \Cref{eq:power-avg-peak}, cannot be written in the quadratic form of \Cref{eq:metrics} and is instead fourth-order:
\begin{equation}
    |S|^2 = S S^* = \frac{1}{16} \begin{bmatrix} \hat{e} \\ \hat{q} \end{bmatrix}^*
    \mathbf{A}_S
    \begin{bmatrix} \hat{e} \\ \hat{q} \end{bmatrix}
    \begin{bmatrix} \hat{e} \\ \hat{q} \end{bmatrix}^*
    \mathbf{A}_S^T
    \begin{bmatrix} \hat{e} \\ \hat{q} \end{bmatrix}
\end{equation}
Peak power constraints use a more complicated approach discussed in \Cref{sec:appendix-mag-S-constraints} and are left out of the formulation below for simplicity.

To implement the constraints, we create an intentional linear impedance mismatch, sacrificing power to satisfy the limits (method M2 in \Cref{tab:constraint-approaches}).
For select constraints where a describing function is available (currently the torque limit), we additionally use nonlinear control to satisfy the limit with less of a power sacrifice (method M1.2 in \Cref{tab:constraint-approaches}).

For the linear impedance mismatch, the average electrical power from \Cref{eq:power-avg-peak} can be maximized subject to quadratic constraints on the effort and flow variables.
This is a quadratically-constrained quadratic program (QCQP):
\begin{equation}\label{eq:opt-problem}
\begin{aligned}
& \max_{\vec{x} \in \mathbb{C}^2} & p_{avg,VI} &= \frac{1}{4}\vec{x}^{\,*} \mathbf{A}_P \vec{x} \\[3pt]
& \text{s.t.}
& \vec{x}^{\,*}\mathbf{Q}_i\vec{x}
    + 2\Re\!\left\{\vec{a}_i^{\,*}\vec{x}\right\}
    &\le b_i,
    \quad \forall i \in [1,\cdots,N] \\[3pt]
& & \vec{c}^{\,*}\vec{x} &= d \\[3pt]
& \text{where} 
& \vec{x} &= \begin{bmatrix} \hat{V} & \hat{I} \end{bmatrix}^T \\[3pt]
& & \mathbf{Q}_i &\in \mathbb{C}^{2\times 2}, \quad \mathbf{Q}_i = \mathbf{Q}_i^{\,*} \\[3pt]
& & \vec{a}_i &\in \mathbb{C}^2 \\[3pt]
& & b_i &\in \mathbb{R} \\[3pt]
& & \vec{c}^{\,*} &= \begin{bmatrix} 1 & Z_{s,th} \end{bmatrix} \\[3pt]
& & d &= \hat{V}_{s,th}
\end{aligned}
\end{equation}

The linear equality constraint $\vec{c}^{\,*}\vec{x} = d$ enforces the WEC and PTO dynamics through a voltage-current relationship at the generator electrical port, while the quadratic inequality constraints encode limits on linear or quadratic combinations of effort and flow at any port.
\ifdefined\DISSERTATION
    The specific quantities constrained in this case are shown in \Cref{tab:qp-constraints}.
    More precisely, this simulation implements constraints on the maximum allowed generator torque and mechanical power, float and spar body amplitudes, and PTO amplitude, as well as a positivity constraint on the electrical power.
    In addition to the amplitude limits shown in the first three rows of \Cref{tab:qp-constraints}, a differently-structured quadratic amplitude constraint will be presented in \Cref{sec:design-load-cases} (see \Cref{eq:slam-constraint-quadratic}) representing a coupled amplitude-phase criterion for the bodies to not exit the water or be fully submerged.
\else
    Constraints implemented in MDOcean include generator torque, mechanical power, float and spar amplitudes, PTO amplitude, and electrical power positivity; quadratic coefficients $\{\mathbf{Q}_i, \vec{a}_i, b_i\}$ for each are given in \Cref{tab:qp-constraints} (\Cref{sec:appendix-qp-constraints-table}).
    A differently-structured quadratic amplitude constraint for float and spar slamming/submersion is presented in \Cref{sec:design-load-cases}.
\fi

\Cref{eq:opt-problem} has the following solution:
\begin{equation}\label{eq:constrained-qp-solution}
\begin{aligned}
    \vec{x}_{opt} &=
    \frac{\hat{V}_{s,th}}{2\Re(Z_{s,th})}
        \begin{bmatrix} Z_{s,th}^* (1+\Gamma_{\text{opt}}) \\ 1-\Gamma_{\text{opt}} \end{bmatrix} \\
    p_{avg,VI,opt} &= \frac{|\hat{V}_{s,th}|^2}{8\Re(Z_{s,th})}~ (1 - |\Gamma_{\text{opt}}|^2)
\end{aligned}
\end{equation}
where $\Gamma_{\text{opt}}$ is a complex scalar with magnitude between 0 and 1 called the optimal reflection coefficient, with formula given in \Cref{eq:gamma-opt-quadratic,eq:candidate-points,eq:circle-intersection} of \Cref{sec:appendix-qp-solution}.
The reflection coefficient represents the degree to which the constraints introduce an impedance mismatch.
$\Gamma_{\text{opt}}=0$ corresponds to the impedance-matched solution (no active inequality constraints).

\ifdefined\DISSERTATION
    As \Cref{sec:mod-freq-domain} describes, the solution for $\Gamma_{\text{opt}}$ can be found numerically or analytically, and we further distinguish that the numerical approach can be accomplished with either direct or indirect optimal control.
    Direct methods solve the optimization problem itself, while indirect methods first derive the optimality conditions and then solve those conditions.
    MDOcean currently implements both the analytical approach, which corresponds to the explicit architecture in \Cref{fig:control-arch}, and the numerical indirect approach, which corresponds to the implicit architecture in \Cref{fig:control-arch}.
    The analytical approach is described in \Cref{sec:appendix-qp-solution}, and the numerical indirect approach in \Cref{sec:appendix-qp-numerical}.

    The numerical direct approach can correspond to either the nested architecture or the simultaneous architecture of \Cref{fig:control-arch}.
    While the nested architecture solves the quadratic program \Cref{eq:opt-problem}, the simultaneous architecture integrates the QCQP into the nonlinear design optimization, which does not take advantage of the efficiency available with the QCQP formulation.
    It is unknown whether the numerical direct approach to solving the QCQP would be more efficient than the numerical indirect approach, but both are expected to be significantly less efficient than the analytical approach, so the analytical approach is used for all results and benchmarks presented here.
\else
    Both analytical and numerical-indirect solutions for $\Gamma_{\text{opt}}$ are implemented in MDOcean.
    This study focuses on the superior analytical approach described in \Cref{sec:appendix-qp-solution}, which has a geometric interpretation as a minimum-distance problem on the complex plane of $\Gamma_{\text{opt}}$ and is used for all results presented here.
    \citet[Appendix A.3.3]{mccabe_dissertation_2026} describes the numerical-indirect approach.
\fi

\begin{figure}[htbp]
\centering
\begin{subfigure}[t]{0.48\linewidth}
    \includegraphics[width=\linewidth]{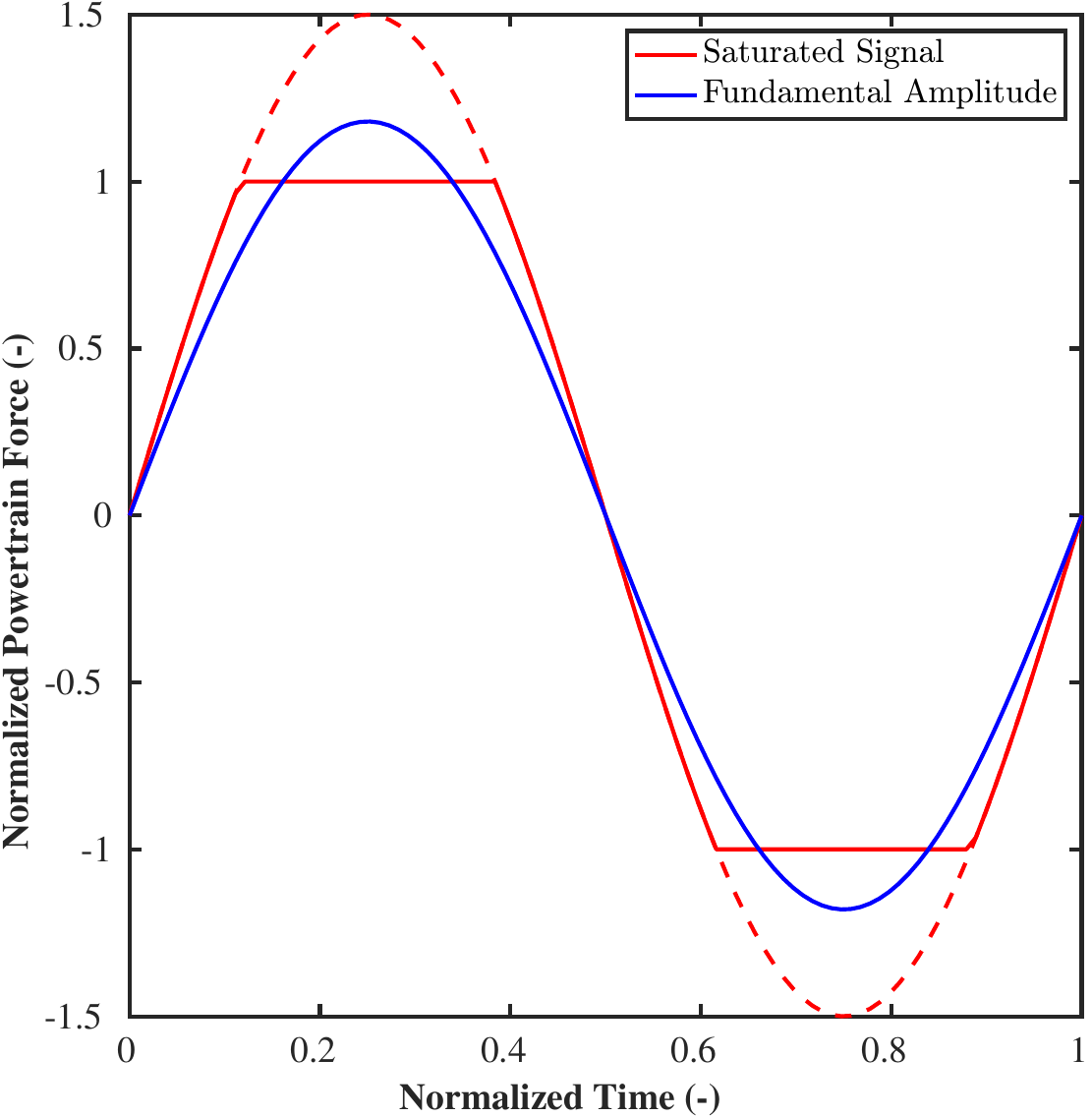}
    \caption{Generator torque saturation}\label{fig:sat}
\end{subfigure}
\hfill
\begin{subfigure}[t]{0.48\linewidth}
    \includegraphics[width=\linewidth]{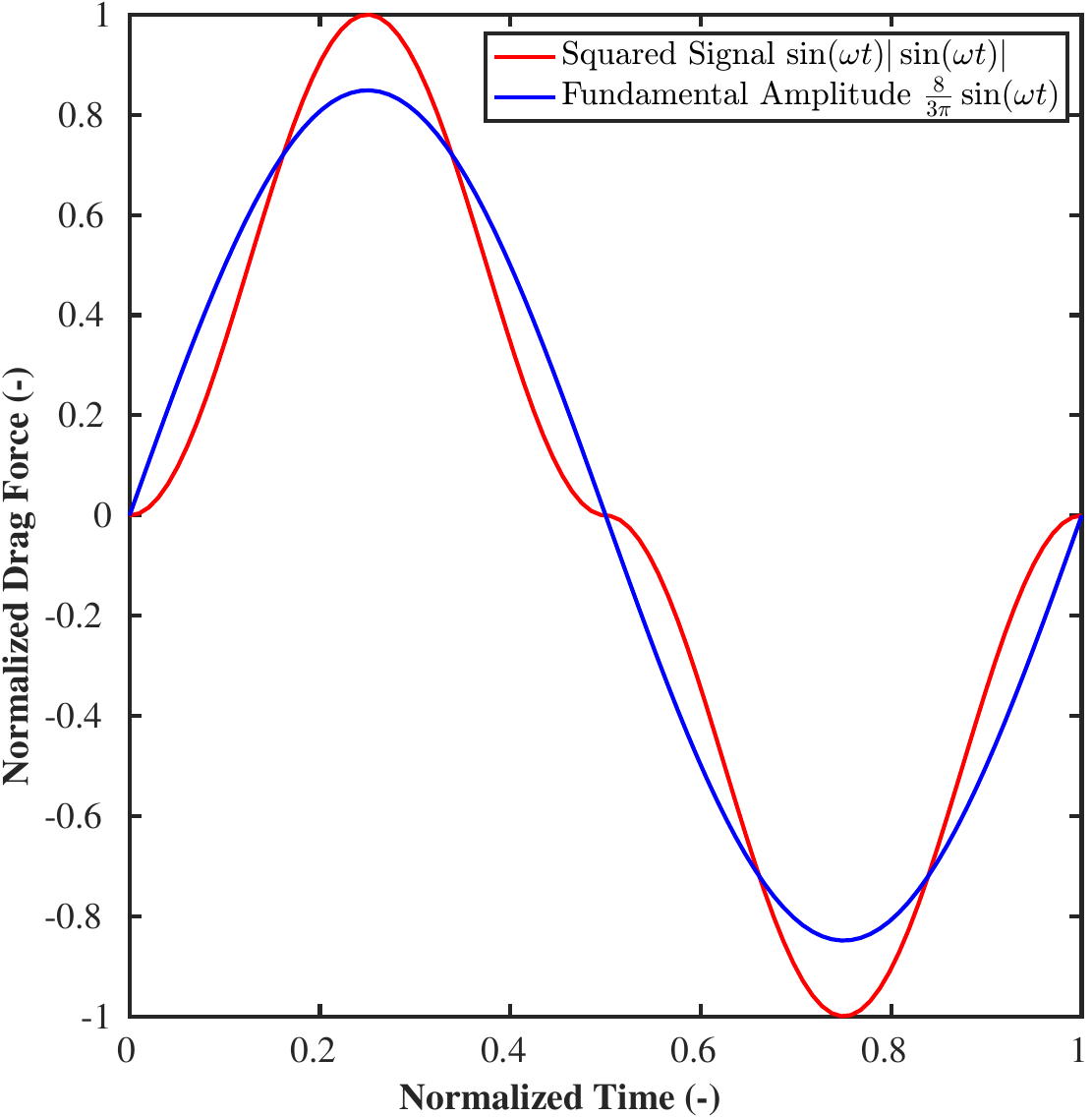}
    \caption{Quadratic drag}\label{fig:drag-df}
\end{subfigure}
\caption{Conceptual demonstration of describing function approximations. Plots show force versus time. Nonlinear signals in red; fundamental amplitudes in blue.}\label{fig:desc-fcns}
\end{figure}

\paragraph{Nonlinear controller}
\ifdefined\DISSERTATION
    For the nonlinear controller, the maximum generator torque constraint is considered separately from the others.
    This constraint changes the optimal torque signal from a pure sinusoid to a saturated waveform, approaching a square wave for sufficiently low torque limit.
    As is, this nonlinear signal could not be handled with a frequency-domain transfer function and would require time-domain simulation.
    Conveniently, through the describing function method, the saturated signal can be represented with Fourier analysis as a superposition of harmonics.
    Furthermore, because the plant is a second order oscillator and filters out high frequency inputs, harmonics other than the fundamental contain very little energy and can be safely neglected.
    \Cref{fig:desc-fcns}(a) shows this approximation graphically for a saturated-sine force signal.
\else
    The maximum generator torque constraint is handled separately.
    Saturation makes the optimal torque non-sinusoidal (approaching a square wave for low torque limits), which would normally require time-domain simulation.
    The describing function method instead represents the saturated signal by its fundamental Fourier component (\Cref{fig:desc-fcns}a), with higher harmonics filtered out by the second-order plant.
\fi

Details of the calculation are provided in \Cref{sec:appendix-force-sat}.
The result is updated effective values for control gains $B_l$ and $K_l$ that represent the nonlinear saturated controller with a torque that obeys the constraint and is used in place of \Cref{eq:constrained-qp-solution}.

\subsubsection{Drag}\label{sec:drag}
\ifdefined\DISSERTATION
    The hydrodynamics model of \Cref{sec:hydro} does not capture all fluid forces on the body due to its linear, irrotational, and inviscid assumptions.
    More specifically, viscous shear stresses (skin friction),
    higher-order nonlinear potential flow effects, and
    steady and unsteady viscous separation phenomena are all neglected.
    In this section, we model the form drag due to viscous flow separation, including steady wake losses, turbulence, and cycle-averaged vortex shedding.
    These effects have been found to be the dominant drag force for heaving WECs \cite{quartier_influence_2021} and are important to avoid unrealistically high resonant peaks in the response amplitude which would lead to overestimates of power production.
\else
    The linear, inviscid hydrodynamics model of \Cref{sec:hydro} omits viscous flow separation, which is the dominant drag mechanism for heaving WECs \citep{quartier_influence_2021}.
    Without modeling drag, the predicted resonant amplitude is unrealistically high and power is overestimated.
\fi

Drag force $\vec{F}_d$ is proportional to the product of the relative velocity of the WEC and incident wave, $\vec{v}_{\text{rel}}(y,t)=\vec{\dot{\xi}}(t)-\vec{v}_{\text{wave}}(y,t)$, with its absolute value.
The resulting quadratic form $|\vec{F}_d|\sim|\vec{v}_{\text{rel}}|^2$ implies a nonlinear input-output response that would normally require time-domain simulation.
To improve computational efficiency and allow solution in the frequency domain, the nonlinearity is quasi-linearized with a describing function, following \cite{quartier_influence_2021}.
The pressure along the bottom surface of each body due to drag $\vec{p}_d(y,t)$ is modeled with empirical drag coefficient vector $\vec{C}_d$:
\begin{equation}\label{eq:drag-pressure}
    \vec{p}_{d}(y,t) = \frac{1}{2}\rho_w \vec{C}_d ~\vec{v}_{rel}(y,t) |\vec{v}_{rel}(y,t)|
\end{equation}
\Cref{eq:drag-pressure-approx} expands the velocity term then applies the describing function approximation:
\begin{equation}\label{eq:drag-pressure-approx}
\begin{aligned}
    \left(\vec{v}_{rel}(y,t)~\cdot\right.\\
    \left.\left|\vec{v}_{rel}(y,t)\right|\right)
    &= |\vec{\hat{V}}_{rel}(y)|^2 \cos(\omega t+\angle \vec{\hat{V}}_{rel}(y)) 
    \left| \cos(\omega t+\angle \vec{\hat{V}}_{rel}(y))\right|\\
    &\approx~|\vec{\hat{V}}_{rel}(y)|^2\frac{8}{3\pi} \cos(\omega t+\angle \vec{\hat{V}}_{rel}(y))
\end{aligned}
\end{equation}
\ifdefined\DISSERTATION
    Computation of the relative velocity uses the following expression for the incident wave velocity phasor in finite depth water evaluated at the body draft, with $\vec{T}=[T_{f,2},T_s]$ collecting the bottom draft of the float and spar:
    \begin{equation}\label{eq:wave-velocity-phasor}
        \vec{\hat{V}}_{wave}(y) = \frac{H}{2} \frac{g k}{\omega} e^{-k\vec{T}} e^{-iky}
    \end{equation}
\else
    The relative velocity uses the standard linear-wave velocity phasor evaluated at body draft; the explicit expression is in \Cref{sec:appendix-drag} (\Cref{eq:wave-velocity-phasor}). 
\fi

The time-domain nonlinear waveform and describing function approximation for the drag pressure/force are compared in \Cref{fig:desc-fcns}(b).
The amplitude of the sinusoidal approximation equals $\frac{8}{3\pi}\approx0.85$ times the peak of the nonlinear waveform.

Integrating the drag pressure over the wetted surface via strip theory (\Cref{sec:appendix-drag}) yields a quasi-linearized drag force consisting of a damping term in phase with the body motion and an excitation term potentially out of phase:
\begin{equation}\label{eq:drag-damping-excitation}
    \vec{\hat{F}}_d = -\mathbf{B}_{d} \vec{\hat{\dot{\xi}}}
    + \vec{\gamma}_{d} \hat{\zeta}
\end{equation}
The excitation term arises from the use of the relative velocity rather than direct WEC velocity, which is often overlooked in WEC models and has important implications for the phase of the drag force.
The damping-excitation grouping of the drag terms, as well as the dependence of the relative velocity on the direction of wave propagation $y$, improves on the approach of \cite{quartier_influence_2021} and is discussed more in \Cref{sec:appendix-drag}.
Explicit expressions for $\mathbf{B}_d$ and $\vec{\gamma}_d$ in terms of nondimensional integrals over body geometry are given in \Cref{sec:appendix-drag} (\Cref{eq:drag-coeffs}).

\ifdefined\DISSERTATION
    The nondimensional integrals are computed numerically, and for applications like optimization where the simulation must be run many times, they are pre-computed for a range of nondimensional input values and accessed via a lookup table to reduce runtime.

    The resulting magnitude and phase of WEC motion are obtained by numerically iterating the state-dependent coefficients $\mathbf{B}_{d}$ and $\vec{\gamma}_{d}$.
    Further derivation, convergence details, and control implications are discussed in \Cref{sec:appendix-drag}.
\else
    These integrals are precomputed and stored as a lookup table to support fast optimization.
    The state-dependent coefficients $\mathbf{B}_{d}$ and $\vec{\gamma}_{d}$ are converged by fixed-point iteration, typically in 5-8 iterations; details are in \Cref{sec:appendix-drag}. 
\fi


\subsubsection{Energy Production}
The analysis thus far has either been sea-state agnostic (\Cref{sec:geom}) or applied separately to each individual sea state (\Cref{sec:hydro,sec:dynamics}).
We now combine the results across sea states to find the long-term energy statistics in a given wave environment.

\ifdefined\DISSERTATION
    While matrices in \Cref{sec:hydro,sec:dynamics} represent coupled relationships between vector output variables (i.e.\ $\vec{\hat{\xi}}$) for systems with multiple bodies or degrees of freedom, in the current section each output element is treated as a scalar, either representing a single element of an output vector (i.e.\ $\hat{\xi}_f$) or the sum across all degrees of freedom (i.e.\ power in systems with multiple actuated degrees of freedom).
    These scalars are assembled into ``matrices'' (i.e.\ $\mathbf{\hat{\xi}}^{H,T}_f$) indexed by sea state rather than by degree of freedom, indicated by the $H,T$ superscript.
    Such ``matrices'' are incompatible with matrix multiplication and any matrix operations in this section should be interpreted as elementwise operations across sea states.
    To distinguish from the average and maximum signal values over a wave period denoted with $avg$ and $pk$ subscripts in \Cref{sec:power}, the long-term average across sea states is denoted with an overbar $\overline{(\cdot)}$ and long-term maximum with subscript $max$.
    This intentionally matches the $max$ subscript used in \Cref{tab:qp-constraints} for upper limits within each sea state because MDOcean currently applies the same limit value for a given constraint across all sea states, though this need not be the case in general.
    An exception is the slamming and submergence constraints, which will be discussed separately in \Cref{sec:design-load-cases}.
\else
    Quantities indexed by sea state are written as ``power matrices'' $(\cdot)^{H,T}$ (e.g.\ $\mathbf{P}_{\text{elec}}^{H,T}$),
    with operations interpreted elementwise across sea states rather than as matrix multiplication.
    Long-term averages across sea states are denoted with an overbar $\overline{(\cdot)}$ and long-term maxima with subscript $max$,
    distinguishing them from the within-period averages and peaks ($avg$, $pk$) introduced in \Cref{sec:power}.
    A separate quadratic amplitude limit for slamming and submergence, which varies across sea states, is treated in \Cref{sec:design-load-cases}.
\fi

\begin{figure}[b!]
\centering
\includegraphics[width=\linewidth]{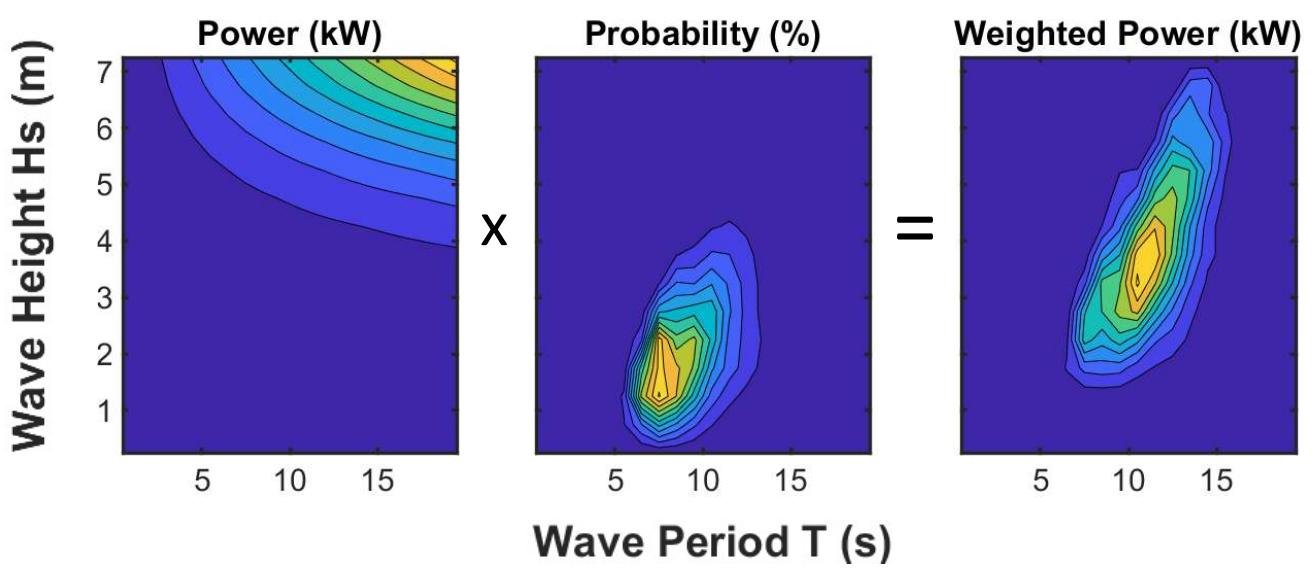}
\caption{Power matrix multiplication}
\label{fig:JPD-multiply}
\end{figure}

To find the long-term average power production $\overline{P}_{\text{elec}}$ in a location with a given distribution of sea states, the power matrix $\mathbf{P}^{H,T}_{\text{elec}}$ for each sea state is weighted by that sea state's probability using a Joint Probability Density (JPD) matrix and then summed.
This method is illustrated in \Cref{fig:JPD-multiply}.
The JPD used in this analysis represents wave conditions in Humboldt Bay, CA and is taken from the study by \cite{janzou_system_2020}.

The annual energy production $AEP$ in kWh/year for a farm of multiple devices is the product of the number of devices, the average power per device, an efficiency $\eta_{\text{array}}$ representing transmission losses and array downtime, and an appropriate unit conversion:
\begin{equation}\label{eq:AEP}
    AEP = N_{WEC} ~ \overline{P}_{elec} ~\eta_{array} \frac{8766}{1000}.
\end{equation}


\subsubsection{Dynamic Limits and Design Load Cases}
\label{sec:design-load-cases}
This study considers two design load cases: (1) cyclic operational loading in sea states where power is produced, and (2) storm loading in sea states where the device enters survival mode and stops producing power.
These correspond roughly with the 1-year and 50-year return periods specified in the wave energy design standard IEC TS 62600-2, and with the fatigue limit state (FLS) and ultimate limit state (ULS) in offshore wind standards.
When the WEC is in survival mode, an external brake locks the PTO, enforcing the float and spar to move together without loading the generator.
\ifdefined\DISSERTATION
    This section details the allowable amplitudes and forces in each case.
    Currently all sea states in the JPD with nonzero probability are considered operational, since the generator force and power limit serve to reduce loads in energetic sea states.
    Future studies could control the transition from operational to survival mode more finely, perhaps with an additional input variable representing an operational incident energy threshold.
\else
    All nonzero-probability sea states in the JPD are considered operational; the generator force and power limits reduce loads in the most energetic sea states.
\fi

\ifdefined\DISSERTATION
    The heave forces in equation \eqref{eq:eom} can be separated into fluid forces, which act on the vertically-oriented wetted surface of the body, the PTO force, which acts at the connection between the float and the spar, and the inertial force, which for dynamic purposes is assumed to act at the center of mass but for structural purposes actually acts distributed over the material.
    The distributed nature of the inertial force means that the float and spar structures experience internal loads that vary not only in time but also in space along their lengths.
    Due to force balance in dynamic equilibrium, the maximum magnitude of internal heave force in space at any given time, $F_{\text{heave},f}(t)$, is the greater of the absolute value of the PTO force and the fluid force.
    Expressing the latter as the difference between the inertial force and the PTO force, this gives:
\else
    By force balance, the peak internal heave force at any cross-section of each body is the greater of the PTO force and the fluid (inertial minus PTO) force:
\fi
\begin{equation}\label{eq:heave-force-decomp}
\begin{aligned}
    F_{heave,f}(t) &=\max\!\left(|F_p(t)|,~ |m_f\ddot{\xi}_f(t)-F_p(t)|\right) \\
    F_{heave,s}(t) &=\max\!\left(|F_p(t)|,~ |m_s\ddot{\xi}_s(t)-F_p(t)|\right)
\end{aligned}
\end{equation}
In operational seas, $F_p(t)$ is determined by the generator torque $\tau(t)$ cascaded through the PTO dynamics matrix $\mathbf{a}_{F_p\xi\leftarrow \tau\Omega}$.
In storm seas, $F_p(t)$ represents the brake force, and its fundamental amplitude is obtained by solving the 2-DOF equation of motion for the force necessary to enforce zero relative motion between the float and spar.
\ifdefined\DISSERTATION
    Maximizing over time requires more care because nonlinearity-induced harmonics cannot be neglected for peak force analysis like they can for power production.
    The peak of $\tau(t)$ is lower than its fundamental amplitude due to generator force saturation at $\tau_{\text{max}}$, so using $\tau_{\text{max}}$ rather than $|\hat{\tau}|$ is both less conservative and more realistic.
    Assuming conservatively that the shaft torque constructively interferes with the saturated generator torque, for the operational case we use:
\else
    Saturation caps the operational generator torque at $\tau_{\text{max}}$, which is a less conservative and more realistic peak than the fundamental amplitude $|\hat{\tau}|$.
    Assuming constructive interference with the shaft torque,
\fi
\begin{equation}\label{eq:peak-pto-force}
\max_t |F_{p,op}(t)| \approx \frac{1}{R} \left(\left| Z_{shaft} \hat{\Omega} \right| + \tau_{max} \right).
\end{equation}
The storm case lacks the control saturation nonlinearity and uses the fundamental amplitude $\hat{F}_{p,st}$ directly.

The fluid force nonlinearities $|m\ddot{\xi}(t)-F_p(t)|$ are harder to handle, since unlike the generator torque, the peak of the drag force exceeds its fundamental.
MDOcean currently uses the fundamental amplitude for both load cases:
\begin{equation}\label{eq:peak-fluid-force}
\max_t \left|m\ddot{\xi}(t)-F_p(t)\right| \approx \left|-m\omega^2 \hat{\xi} - \hat{F}_p\right|.
\end{equation}
This approximation can either under- or overestimate the true peak depending on whether harmonics interfere constructively or destructively with the fundamental; further discussion is in \Cref{sec:appendix-peak-fluid-force}.
\ifdefined\DISSERTATION
    The fundamental amplitudes are also used to calculate the peak float and spar displacements $\max_t |\xi_f(t)|$ and $\max_t |\xi_s(t)|$ in both cases.
    Using the fundamental for the response displacements is more accurate than for the forces because the system dynamics filter out high frequency inputs.
\else
    Peak float and spar displacements likewise use fundamental amplitudes; the second-order plant filters out higher harmonics, making the approximation more accurate for displacements than for forces.
\fi

\Cref{tab:DLCs} summarizes the peak forces and amplitudes based on the assumptions of each design load case.

\begin{table}[htbp]
    \centering
    \caption{Forces, amplitudes, and maximum allowable amplitudes in each design load case}
\label{tab:DLCs}
\begin{tabular}{>{\centering\arraybackslash}p{0.28\linewidth}>{\centering\arraybackslash}p{0.3\linewidth}>{\centering\arraybackslash}p{0.33\linewidth}}
    & \multicolumn{2}{c}{\textbf{Design Loadcase}}\\\cline{2-3}
    \textbf{Variable}& \textbf{1: Operational}&\textbf{2: Storm}\\ \hline
    \parbox[m]{\linewidth}{\centering \vspace{8pt}
    Peak powertrain force, $\max_t |F_p(t)|$ \vspace{8pt} }
        & \parbox[m]{\linewidth}{\centering 
        Given by \Cref{eq:peak-pto-force}} 
        & \parbox[m]{\linewidth}{\centering
        Solve \Cref{eq:eom-freq-domain} with \Cref{eq:op-hydro-coeffs} coefficients for $F_p$}\\ \hline
    \parbox[m]{\linewidth}{\centering 
    Peak fluid force, $\max_t |m\ddot{\xi}(t)-F_p(t)|$ }
        & \parbox[m]{\linewidth}{\centering 
        Given by \Cref{eq:peak-fluid-force} with \Cref{eq:op-hydro-coeffs} coefficients and $F_p$ from \Cref{eq:peak-pto-force}}
        & \parbox[m]{\linewidth}{\centering 
        Given by \Cref{eq:peak-fluid-force} with \Cref{eq:st-hydro-coeffs} coefficients and $F_p$ from storm case}\\ \hline
    \parbox[m]{\linewidth}{\centering 
    Peak float travel, $\max_t |\xi_f(t)|$ }
        & \parbox[m]{\linewidth}{\centering 
        Given by \Cref{eq:eom-freq-domain} with \Cref{eq:op-hydro-coeffs} coefficients}
        & \parbox[m]{\linewidth}{\centering 
        Given by \Cref{eq:eom-freq-domain} with \Cref{eq:st-hydro-coeffs} coefficients and $F_p=0$}\\ \hline
    \parbox[m]{\linewidth}{\centering 
    Peak spar travel, $\max_t |\xi_s(t)|$ }
        & \parbox[m]{\linewidth}{\centering 
        Given by \Cref{eq:eom-freq-domain} with \Cref{eq:op-hydro-coeffs} coefficients}
        &Equals $\xi_f$ due to brake\\ \hline
    \parbox[m]{\linewidth}{\centering \vspace{8pt}
    Maximum allowable relative amplitude between float and spar, $|\xi_f-\xi_s|_{\text{max}}$ \vspace{8pt}}
        & \shortstack{$\min(h_{fs,\text{clear}},h_{fs,\text{up}},$\\$h_{fs,\text{down}})$} 
        &N/A\\  \hline
    \parbox[m]{\linewidth}{\centering \vspace{8pt}
    Maximum allowable float amplitude, $|\xi_f|_{\text{max}}$ \vspace{8pt} }
        & $\min(\xi_{f,\text{linear}}, \xi_{f,\text{slam}})$
        & N/A \\ \hline 
    \parbox[m]{\linewidth}{\centering \vspace{8pt}
    Maximum allowable spar amplitude, $|\xi_s|_{\text{max}}$ }
        & $\min(\xi_{s,\text{linear}}, \xi_{s,\text{slam}})$
        & N/A 
\end{tabular}
\end{table}

The surge force for each component is calculated as \citep{newman_motions_1963}
\begin{equation}\label{eq:surge-force}
   F_{surge} = \frac{H\rho_w \omega^2 A_w}{k} (e^{-kz_{\text{top}}}-e^{-kT})
\end{equation}
using the respective waterplane areas $A_w$  and drafts $T$ for the float and spar.
The top vertical height $z_{\text{top}}$ is set to zero for the float and $T_{f,2}$ for the spar.
The design surge force is the maximum surge force over all nonzero-probability sea states.
Equation~\eqref{eq:surge-force} assumes a slender cylinder with diameter much smaller than the wavelength and may be less accurate for the float due to diffraction effects.
Incorporating semi-analytical hydrodynamic coefficients for surge is a possible area of future work. 
Maximum acceptable heave 
forces will be evaluated with structural factors of safety in \Cref{sec:structures}.
The structures model currently lacks the ability to account for surge forces, so the surge force calculation is not used in the simulation.

The table also indicates the maximum permissible amplitudes.
In operational sea states, the geometric clearance between the float and spar is enforced to prevent overtravel.
The permissible heights for upward and downward motion of the float relative to the spar are:
\begin{equation}\label{eq:h-fs-up-down}
\begin{aligned}
    h_{fs,\text{up}} &= h_s - T_s - (h_f- T_{f,2}) \\
    h_{fs,\text{down}} &= T_s - T_{f,2} - h_d
\end{aligned}
\end{equation}
Additionally, a limit of $h_{fs,\text{clear}}$ (see definition in \Cref{fig:dims}) is required to ensure clearance of the tubular structure connecting the float and spar.

The conditions for linear wave-body interaction are enforced to maintain compatibility with the linear potential flow theory approach of \Cref{sec:hydro}.
This maximum amplitude $\xi_{\text{linear}}$ for the float and spar respectively is
\begin{equation}
    \xi_{f,\text{linear}} = \frac{1}{10}(h-T_{f,2}), \quad \xi_{s,\text{linear}}  = \frac{1}{10}(h-T_s)
\end{equation}
\ifdefined\DISSERTATION
    which is derived by requiring that the height of the water column below the body (the relevant dimension determining the hydrodynamic coefficients in MEEM) does not change by more than 10\% from its equilibrium value when the body is at its peak displacement.
\else
    derived by requiring the water column height below the body (relevant for the MEEM hydrodynamic coefficients) to vary by no more than 10\% from equilibrium.
\fi

In both operational and storm states, the WEC surface should not transiently pierce the wave free surface, since this causes extremely high loads.
This includes slamming (the bottom surface rises out of the water and impacts on reentry) and submersion (the top of the WEC becomes fully submerged).
The condition to prevent slamming and submersion in regular waves of height $H$ is derived in \Cref{sec:appendix-slam}.
These limits depend on the phase of the response, not just the magnitude, so they do not use the quadratic coefficients of \Cref{tab:qp-constraints} but can nonetheless be expressed in a quadratic form compatible with \Cref{eq:opt-problem}:
\begin{equation}\label{eq:slam-constraint-quadratic}
    \hat{\xi}^*\hat{\xi} - H\Re(e^{iky}\hat{\xi}) < \Delta z^2-\left(\frac{H}{2}\right)^2
\end{equation}
for relevant vertical body dimension $\Delta z$ and horizontal coordinate $y$.
The values of $\Delta z$ used for float and spar in operational and storm cases and the rationale (avoiding slanted-surface slamming, damping plate surfacing, and related effects) are given in \Cref{sec:appendix-slam}.

\ifdefined\DISSERTATION
    The values $h_{fs,\text{clear}}$, $h_{fs,\text{up}}$, $h_{fs,\text{down}}$, $\xi_{f,\text{linear}}$, and $\xi_{s,\text{linear}}$ are all upper limits and are independent of sea state, while $\xi_{\text{slam}}$ depends on the sea state and can separately act as both an upper and lower limit on the oscillation magnitude, as shown in \Cref{sec:appendix-slam}.
    In early design studies that do not directly compute the hydrodynamic coefficients and associated power production, the amplitude limits proposed here could be plugged into Budal's upper bound, a common analytical expression that combines an upper bound on the excitation force with the maximum permissible body velocity to yield an upper bound on power production \cite{zou_practical_2023}.

    Note that the analysis of this section continues to use standard linear wave theory with equivalent regular waves, even for the storm condition.
    This is a questionable assumption, but it is necessary because optimization-suitable models for nonlinear wave forces do not currently exist.
    \Cref{sec:discussion} will discuss possible future alternatives to this modeling choice.
\else
    Both load cases use standard linear wave theory with equivalent regular waves, an approximation whose limitations and possible future alternatives are discussed in \Cref{sec:unmodeled-effects}.
\fi




\subsubsection{Irregular Waves}\label{sec:irregular-waves}
MDOcean uses a JPD (joint probability density matrix) to capture the long-term variation in wave conditions and accurately estimate annual average power and fatigue load.
Additionally, it represents each irregular wave condition within the JPD as an equivalent sinusoidal regular wave.
Regular wave height $H$ and period $T$ are selected to provide equal energy as the irregular sea state:
\begin{equation}\label{eq:regular-wave-equiv}
    H = \frac{H_s}{\sqrt{2}}, \qquad T = T_e \approx 0.857 ~T_p,
\end{equation}
where the irregular sea state has significant wave height $H_s$, energy period $T_e$, and (for a Pierson-Moskowitz spectrum) peak period $T_p$.
This regular wave assumption avoids the time-consuming convolution integral or state-space identification that is typically required to simulate nonlinear dynamics subject to irregular wave forcing.

In the absence of dynamic constraints, the regular wave assumption does not significantly affect power production if the ultimate design utilizes a high-order linear controller capable of impedance matching over the bandwidth of each sea state.
If the ultimate design instead utilizes a simple PI controller, the assumption slightly overestimates the power produced in a broadband wave environment.
\cite{coe_practical_2021} used a Bretschneider spectrum and found the irregular wave power generation for a point absorber with PI control to be around 92\% of the perfectly matched power, which is sufficiently close to justify the approximation for early design phases without worrying about the complexity tradeoff of controller order.

\ifdefined\DISSERTATION
    Nonetheless, irregular time-domain transient peaks and the variation of hydrodynamic coefficients over the spectral width of each individual sea state are not considered.
    Therefore, MDOcean likely underestimates the sensitivity to peak force, power, and amplitude constraints.
    \Cref{sec:discussion} will discuss possible future extensions to incorporate irregular wave effects into the analysis.
\else
    Transient peaks and spectral coefficient variation within sea states are not captured, so MDOcean likely underestimates sensitivity to peak force, power, and amplitude constraints.
    Stochastic linearization is a candidate for future extension; see \Cref{sec:unmodeled-effects}.
\fi

\subsection{Structures}\label{sec:structures}
The MDOcean structures module uses forces from the dynamics module as well as area and volume outputs from the geometry module to calculate the factors of safety for various structural failure modes.
The factors of safety represent the multiplicative increase in force before stress would surpass peak limits for storm loads and the endurance limit under operational loads.

\ifdefined\DISSERTATION
    Again aligning with the modeling philosophy of investing the up-front development time to create a simulation that runs extremely quickly with moderate accuracy, MDOcean integrates a combination of explicit algebraic and tabulated semi-analytical structural models that are well-established in the structures community yet rarely utilized for WEC design.
    The structures module therefore runs significantly faster than traditional finite element method (FEM) simulations, though it lacks the ability to easily generalize to other geometries, boundary conditions, and load cases as a FEM model could.

    More specifically, the peak limits analyzed include yield, global buckling, local buckling, and ultimate where appropriate, while the endurance limit assesses high-cycle fatigue.
    The latter is potentially conservative, and future work could implement Miner's law to better represent infrequent operational loads (low-cycle fatigue) \cite{ove_arup__partners_ltd_structural_2016}.
\else
    Peak-limit checks include yield, global buckling, local buckling, and ultimate where appropriate;
    the endurance limit assesses high-cycle fatigue.
\fi

The factor of safety is directly calculated for the most heavily loaded subcomponent in each of the three components: the bottom plate in the float, the cylindrical shell of the spar, and the annular disc of the damping plate.
Thicknesses of other subcomponents are either held constant or scaled with the dimensions of subcomponents that are directly assessed, as outlined in \Cref{tab:struct-calc-vs-scale}.
\ifdefined\DISSERTATION
    In the future, extending the structures module to directly calculate the factor of safety for each subcomponent would allow for more realistic structural design.
\fi

\begin{table}[t!]
    \centering
    \caption{Structural analysis methodology by subcomponent}
    \label{tab:struct-calc-vs-scale}
\begin{tabular}{M{0.15\linewidth} M{0.4\linewidth} M{0.3\linewidth}}
         \textbf{Component}&  \textbf{Structural Subcomponent}&  \textbf{Method of Analysis}\\ \hline
         \multirow[c]{5}{*}{\centering Float}&  \begin{itemize}\item Bottom trapezoidal plate\end{itemize}&  Directly calculated\\
         &  \begin{itemize}\item Top trapezoidal plate\end{itemize}&  Scaled from bottom plate\\
         &  \begin{itemize}\item Radial rectangular plate\end{itemize}&  Scaled from bottom plate\\
        & \begin{itemize}\item Circumferential rectangular plates\end{itemize}& Scaled from bottom plate\\
        & \begin{itemize}\item PTO connection tubes\end{itemize}& Held constant\\ \hline
        Spar& Cylindrical shell& Directly calculated\\ \hline
        \multirow[c]{2}{*}{\centering Damping plate}& \begin{itemize}\item Annular plate\end{itemize}& Directly calculated\\
        & \begin{itemize}\item Spar connection tubes\end{itemize}& Held constant\\
    \end{tabular}
    \end{table}

\begin{figure}[t!]
\centering
\includegraphics[width=.9\linewidth]{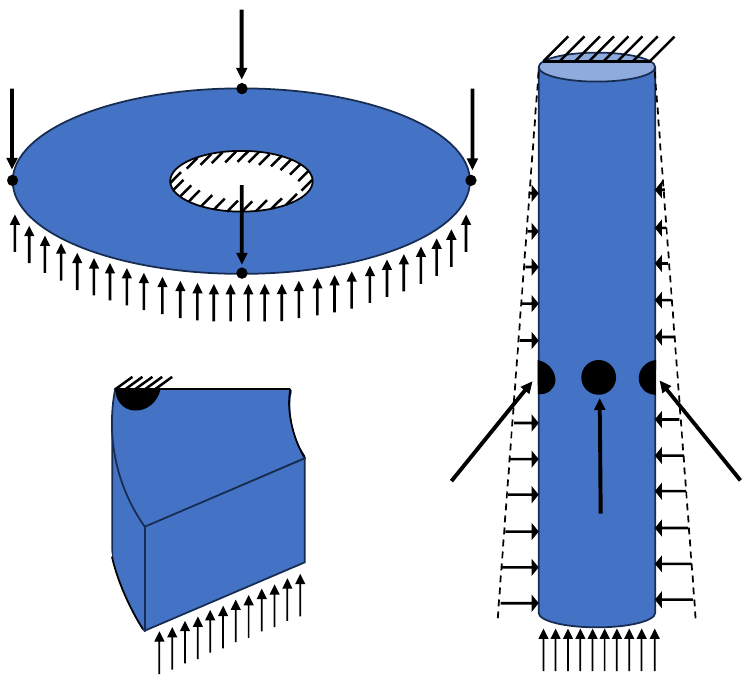}
\caption{Applied loads and fixed points of each structure}\label{fig:FBDs}
\end{figure}

\Cref{fig:FBDs} shows a free body diagram of the applied load on each of the three structural units and the locations where they are modeled as structurally fixed.

Each of the twelve float sections is modeled as a stiffened shell subject to distributed bottom loading and top fixity at the tube support weld joint.
\ifdefined\DISSERTATION
    To simplify the representation of load transfer between plate elements, the bottom 
    float plate is analyzed individually, assuming fixed edges and solving for the reaction loads at these edges.
    This means that the 
    plate is conservatively sized assuming the other plate elements are perfectly rigid. 
\else
    The bottom plate is analyzed individually with fixed-edge boundary conditions, conservatively sizing it as if the other plate elements were perfectly rigid.
\fi

The cylindrical shell of the spar is modeled as a short column under compression and hydrostatic hoop stress.
The column's short length requires an intermediate solution between pure compression and Euler buckling, a consideration that the original reference model design did not account for \citep{previsic_reference_2011}.

The damping plate is a thin stiffened annular plate subject to a vertical distributed force on its bottom surface, reacted by the welded connection to the column along the column's circumference and by four welded tubular supports.
\ifdefined\DISSERTATION
    Two analytical solutions for a thin annular (cylindrical with central hole) plate fixed at its inner radius are superimposed: a distributed pressure and four point loads at the edge, with appropriate modifications to account for the stiffeners.
\else
    Two analytical solutions for a fixed-inner-radius annular plate (distributed pressure and four point loads at the edge) are superimposed, with modifications for the stiffeners.
\fi

Further details of all structural calculations performed in the simulation are provided in \Cref{sec:appendix-structures}.
For illustrative purposes only, assuming a fixed maximum stress and considering only dominant terms for geometries similar to the nominal design, the scaling of the required structural thicknesses with bulk dimensions simplifies to the following:
\begin{equation}\label{eq:struct-scaling}
\begin{aligned}
    t_{f} h_{\text{stiff},f}^2 &\sim \frac{F_{\text{heave}} D_f}{\left(\frac{D_f}{D_s}-1\right)^2} \\
    t_{s}^4 &\sim F_{\text{heave}} h_s^2 \\
    t_{d} h_{\text{stiff},d}^2 &\sim \frac{F_{\text{heave}}D_d^2}{D_s} \\
\end{aligned}
\end{equation}

\ifdefined\DISSERTATION
    Since structural material scales linearly with both the thickness $t$ and the stiffener height $h_{\text{stiff}}$, the squaring of stiffener heights in \Cref{eq:struct-scaling} means that increasing the stiffener height is generally more structurally efficient than increasing the thickness.
    We therefore expect that an optimization would potentially drive the thicknesses to their minimum values, and thus it is also required to model the minimum value of $t/h_{\text{stiff}}$ for which the stiffened plate assumptions hold.
    These limits are also derived in \Cref{sec:appendix-structures}.
\else
    Because structural material volume scales linearly in both $t$ and $h_{\text{stiff}}$, the squared dependence on stiffener height in \Cref{eq:struct-scaling} implies that taller stiffeners are more material-efficient than thicker plates.
    Optimization therefore tends to drive thicknesses toward the minimum value of $t/h_{\text{stiff}}$ for which the stiffened-plate assumptions hold;
    these limits are derived in \Cref{sec:appendix-structures}.
\fi

\subsection{Economics}\label{sec:econ}
The common economic metric $LCOE$ is calculated as the ratio of annualized expenditures to the annual energy production $AEP$:
\begin{equation}\label{eq:lcoe}
    LCOE = \frac{CAPEX\cdot  FCR + OPEX}{AEP}
\end{equation}
where the numerator consists of the up-front capital expenditure $CAPEX$, annualized via a fixed charge rate $FCR$, plus an annual operational expenditure $OPEX$.
\ifdefined\DISSERTATION
    The fixed charge rate incorporates the effect of device lifetime as well as a variety of financial values affecting the time-value of money such as tax rate, interest rate, depreciation, and how much of the project is financed with debt (loans) compared to equity (investors).
\fi

\ifdefined\DISSERTATION
    The MDOcean economics model aims not to predict the precise cost breakdown of a given design but to accurately reflect the scaling of cost with design variables to facilitate design optimization and tradeoff analysis.
    With this in mind, the model lumps costs into categories that are either constant or scale with design, and the constant costs and scale factors are tuned to match more detailed cost breakdowns.
    The cost data chosen for tuning is the RM3 Cost Breakdown Structure (CBS) \cite{neary_reference_2014} (\Cref{tab:CBS} in \Cref{sec:appendix-econ}).
\else
    The MDOcean economics model captures the scaling of cost with design variables rather than predicting absolute cost.
    Constant and design-dependent cost categories are calibrated against the RM3 Cost Breakdown Structure (CBS) \citep{neary_reference_2014} (\Cref{tab:CBS} in \Cref{sec:appendix-econ}).
\fi

Two CBS categories scale with design: structural cost (category 1.4) scales linearly with the volume of structural material $V_{\text{struct}}$, and PTO cost (category 1.5) scales with the peak electrical power $P_{pk,\text{elec}}$ and the force limit $F_{\text{max}}$.
All other categories are held constant per device but decrease with the number of devices $N_{WEC}$ via a power law calibrated against the CBS estimates for $N_{WEC} = (1, 10, 50, 100)$ \citep{neary_reference_2014}.

\ifdefined\DISSERTATION
    Some cost components assumed constant in the model would scale with design in reality, and modifying the model to reflect this is an area for future work.
    For example, installation and decommissioning costs (categories 1.7 and 1.8) could scale with the device maximum dimension since a larger device requires a larger ship, but this is not included due to the difficulty of determining dimension thresholds.
    Likewise, the mooring and foundation costs (category 1.3) could scale with storm surge force, and the replacement parts and consumables costs (categories 2.5 and 2.6) could scale with the PTO power and force as the PTO capital costs do.
\else
    Several categories that would scale with design in reality (installation by device size, mooring by storm surge force, replacement parts and consumables by PTO ratings) are held constant here and flagged in \Cref{sec:unmodeled-effects} as future work.
\fi

The chosen scalings are a power law for the number of devices to reflect economies of scale, direct proportionality between structural cost and material volume, and constant plus linear for the PTO cost since the main components of the PTO (generator, drivetrain, and structural support) are generally sized linearly with power and torque (\Cref{sec:appendix-econ}).
This yields:
\begin{equation}\label{eq:unit-cost}
\begin{aligned}
    C_{unit} &=  C_{pto} + C_{\text{struct}} + C_{\text{fixed}} \\
    C_{pto} &= C_{pto,\text{constant}} + p_{P} P_{pk,\text{elec}} +p_{F} F_{\text{max}} \\
    C_{\text{struct}} &= p_{s} V_{\text{struct}}
\end{aligned}
\end{equation}
where $p_{P}$, $p_{F}$, and $p_{s}$ are the prices of power, force, and structural material in units of \$/kW, \$/N, and \$/m$^3$ respectively.

The per-WEC unit costs and prices $C_{\text{fixed}}$, $C_{pto,\text{constant}}$, $p_{P}$, $p_{F}$, and $p_{s}$ all decrease with the number of devices $N_{WEC}$ via a power law (\Cref{eq:cost-power-law} in \Cref{sec:appendix-econ}), with curve-fit parameters given in \Cref{tab:econ-model-values}.

The total $CAPEX$ and $OPEX$ are found by multiplying the unit costs by the number of devices $N_{WEC}$.
For consistency with prior references \cite{neary_reference_2014,RM3}, all costs are intentionally kept in units of 2012 USD without adjusting for inflation.
\section{Validation and Benchmarking}
\label{sec:validation-benchmarking}

\subsection{Dynamic Validation Using WEC-Sim}\label{sec:dynamic-validation}
The popular time-domain hydrodynamic simulation software WEC-Sim \citep{ruehl_wec-simwec-sim_2024} is used to validate the dynamics module.
The WEC-Sim RM3 example is run with regular waves and with the device constrained to oscillate only in heave.
\ifdefined\DISSERTATION
    Notably, the RM3 geometry provided in WEC-Sim differs slightly from the published dimensions in the RM3 report, so for validation the dimensions input to MDOcean are adjusted to match WEC-Sim.
\fi

\ifdefined\DISSERTATION
    WEC-Sim runs utilize hydrodynamic coefficients obtained with the WAMIT BEM for dynamics, and control coefficients calculated with MDOcean for consistency.
    MDOcean is run with MEEM as usual, and separately also run with the WAMIT coefficients to distinguish differences caused by disparate hydrodynamic coefficients from those caused by the underlying dynamics.
\else
    Two MDOcean configurations are compared against WEC-Sim:
    one using MEEM hydrodynamic coefficients (the default) and one using WAMIT BEM coefficients matching WEC-Sim's,
    to separate dynamics-model error from hydrodynamic-coefficient error.
\fi

\begin{figure*}[htbp]
    \centering
    \includegraphics[width=1\linewidth]{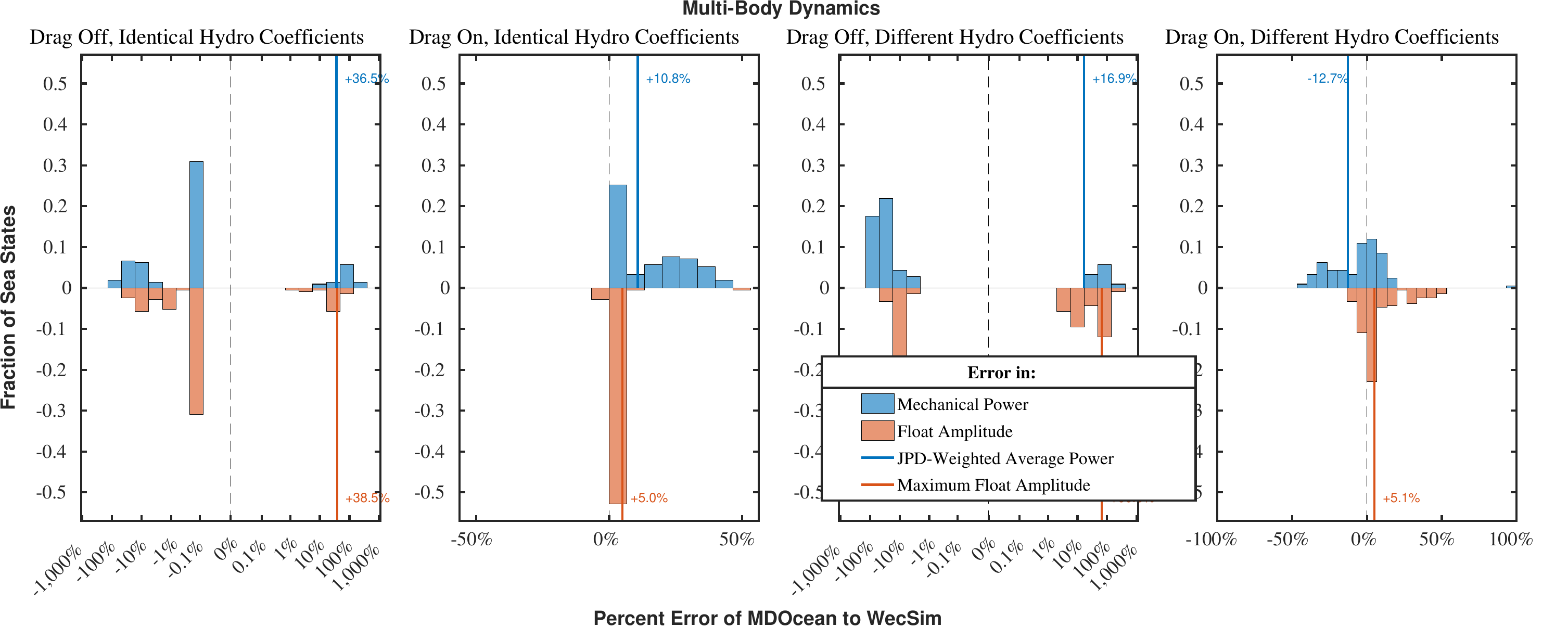}
    \caption{Error breakdown based on WEC-Sim Verification Runs}
    \label{fig:error-histogram}
\end{figure*}

The absolute error in average power compared to the WEC-Sim power is less than \resultsAOR[wecsimAvgPowerErrorBestCase] in the best case and \resultsAOR[wecsimAvgPowerErrorWorstCase] in the worst case, with an error breakdown for all simulation scenarios and sea states shown in \Cref{fig:error-histogram}.
The detailed error breakdown across drag-on/drag-off and MEEM/WAMIT coefficient configurations is provided in \Cref{sec:appendix-dynamic-validation}, revealing that the dominant error sources are interactions between drag, hydrodynamic-coefficient mismatch, and the inter-body phase relationship in the 2-DOF model.
\Cref{sec:appendix-dynamic-validation} also validates the describing-function approximation itself, showing total harmonic distortion below 1\% in the worst sea state and excellent agreement between the assumed and actual drag force waveforms at all four corners of the JPD.

These errors are deemed acceptable for the purposes of this study, since the goal is to demonstrate the value of simultaneously analyzing multiple disciplines and the ability to quickly evaluate a large number of design options.

\subsection{Static Validation}

\ifdefined\DISSERTATION
    The MEEM hydrodynamic coefficients for the nominal RM3 geometry match WAMIT BEM results closely, as previously shown in Fig.~5 of \citep{mccabe_open-source_2024}, except at frequencies below 0.3~rad/s.
    This is expected because the BEM solution assumes infinite depth, which requires a finite added mass at low frequency, whereas MEEM uses finite depth, where the added mass must grow logarithmically toward infinity at zero frequency \citep{mciver_added_1991}.
    For the site under consideration, the lowest-frequency sea state containing any energy is 0.4~rad/s, so the discrepancy is not a concern.
    After the publication of reference~\citet{mccabe_open-source_2024}, excitation phase was added to the MEEM implementation.
It also matches WAMIT results well.
\else
    The MEEM hydrodynamic coefficients for the nominal RM3 geometry match WAMIT BEM results closely for $\omega \geq 0.3$~rad/s (Fig.~5 of \citet{mccabe_open-source_2024});
    the divergence at lower frequencies arises from differing finite-depth versus infinite-depth assumptions \citep{mciver_added_1991} and is not a concern for the site considered (minimum sea-state frequency 0.4~rad/s).
\fi

The overall model is validated by comparing simulated structural forces, power, mass, and cost results to the nominal values in \citet{RM3}, as shown in \Cref{tab:validation}.
\ifdefined\DISSERTATION
    Volume and center of buoyancy results are also compared to the WEC-Sim RM3 tutorial geometry, which differs slightly from the nominal RM3 geometry.
    Structural calculations are additionally compared to \citet{previsic_reference_2011}, an unpublished report by an author of the reference model report providing more detail on the structural calculations.
\fi
The mass, power, and cost track well, but the structural force in storm waves has a significant discrepancy.
\ifdefined\DISSERTATION
    This is because load cases in \citet{RM3} are derived experimentally from wave tank tests rather than a linear model.
    For this study, a tuned scale factor on force is used to account for the discrepancy.
    Improving the force model to align with the wave tank data is an area of future work, with suggestions discussed in \Cref{sec:unmodeled-effects}.
\else
    The reference storm loads in \citet{RM3} are derived from wave-tank tests rather than a linear model, so a tuned scale factor on force is used here;
    improving this is identified in \Cref{sec:unmodeled-effects} as future work.
\fi

\ifdefined\DISSERTATION
    As with force, a tuned scale factor is also applied to the power when the geometry is set to the nominal RM3 geometry to obtain an equal average power to the reference model report (86~kW).
    Unlike the force, the power discrepancy is not attributed to a modeling error because the separate power validation against WEC-Sim tracks extremely well as described in \Cref{sec:dynamic-validation}.
    Rather, the discrepancy is attributed to the fact that the reference model report uses a different dynamic model that is not described in detail, and does not document the hydrodynamic coefficients, water depth, drag coefficient, and other necessary input parameters that would be required to obtain a close match.
    A final tuned scale factor is applied to the structural mass to account for the fact that not all structural components are modeled, including spar stiffeners, damping plate circumferential stiffeners, and miscellaneous features like flanges, enclosures, and access mounts.
\else
    A power scale factor is similarly applied to match the reference report's 86~kW average power;
    the discrepancy is attributed to the reference report's use of an undocumented dynamic model rather than to a modeling error in MDOcean
    (the WEC-Sim validation of \Cref{sec:dynamic-validation} tracks closely).
    A structural-mass scale factor accounts for unmodeled subcomponents (spar stiffeners, damping plate circumferential stiffeners, flanges, enclosures, access mounts).
\fi

After applying a structural force scale factor of \resultsAOR[forceScaleFactor],
a power scale factor of \resultsAOR[powerScaleFactor],
and a mass scale factor of \resultsAOR[massScaleFactor],
the scaled quantities match within 1\% of the reference model report.

    \begin{table*}[htbp]
        \centering
        \begin{tabular}{>{\centering\arraybackslash}p{0.26\linewidth}|c|c|r|c|c|r}
&\multicolumn{3}{c|}{DOE Report RM3 Design \cite{RM3}} & \multicolumn{3}{c}{WEC-Sim RM3 Design} \\ 
Variable & MDOcean & Actual & Error & MDOcean  & Actual  & Error \\ 
\hline 
Mass Float (kg) & $213 \cdot 10^{3}$ & $208 \cdot 10^{3}$ & $2.4\% $ & - & - & - \\ 
Mass Column (kg) & $209 \cdot 10^{3}$ & $224 \cdot 10^{3}$ & $6.6\% $ & - & - & - \\ 
Mass Plate (kg) & $257 \cdot 10^{3}$ & $245 \cdot 10^{3}$ & $4.7\% $ & - & - & - \\ 
Mass Total (kg) & $679 \cdot 10^{3}$ & $680 \cdot 10^{3}$ & $0.2\% $ & - & - & - \\ 
Power Average (W) & $88 \cdot 10^{3}$ & $85.9 \cdot 10^{3}$ & $2.5\% $ & $135 \cdot 10^{3}$ & - & - \\ 
Power Max (W) & $286 \cdot 10^{3}$ & $286 \cdot 10^{3}$ & $0.0\% $ & - & - & - \\ 
Force Heave (N) & $8.61 \cdot 10^{6}$ & $8.5 \cdot 10^{6}$ & $1.3\% $ & - & - & - \\ 
FOS Spar (-) & $11.7 $ & $11.1 $ & $5.8\% $ & - & - & - \\ 
C V (\%) & $86.9 $ & $71.1 $ & $22.4\% $ & $66.9 $ & - & - \\ 
Vol Float (m$^3$) & $702 $ & $702 $ & $0.0\% $ & $727 $ & $726 $ & $0.1\% $ \\ 
Vol Spar (m$^3$) & $1.01 \cdot 10^{3}$ & $1.01 \cdot 10^{3}$ & $0.1\% $ & $888 $ & $887 $ & $0.1\% $ \\ 
CB Float (m) & - & - & - & $1.29 $ & $1.29 $ & $0.0\% $ \\ 
CG Float (m) & - & - & - & $283 \cdot 10^{-3}$ & $283 \cdot 10^{-3}$ & $0.0\% $ \\ 
CAPEX avg error & - & - & $1.4\% $ & - & - & - \\ 
OPEX avg error & - & - & $1.4\% $ & - & - & - \\ 
LCOE avg error & - & - & $1.8\% $ & - & - & - \\ 
\end{tabular}
        \caption{Verification}
        \label{tab:validation}
    \end{table*}


The scaling behavior of economic outputs against the number of WECs is validated by sweeping $N_{WEC}$ from 1 to 100 and comparing the simulated CAPEX, OPEX, and LCOE against the values in the reference model report \citep{RM3}.
The average percent errors are included in \Cref{tab:validation}, with additional detail in \Cref{fig:econ-nwec-validate} of \Cref{sec:appendix-econ-validation}.

\subsection{Runtime Benchmarking}
\label{sec:sim-runtime}
Benchmarking the runtime of the MDOcean simulation is important to verify it achieves the desired speed to facilitate rapid design optimization.
An initial speed requirement order of magnitude of 100~ms for all modules was set to enable a 100-iteration finite difference optimization with 12 design variables to complete in approximately two minutes.
Ultimately, each simulation run takes around \resultsAOR[simRuntime], solidly meeting the goal.
The timings in this section are performed on an Ubuntu 20.04 LTS server with a 14-core Intel Core i9-10940X CPU (3.3~GHz base clock) and 256~GB of DDR4 RAM at 3200~MHz, running MATLAB R2024b.

\begin{figure}[b!]
\centering
\includegraphics[width=\linewidth]{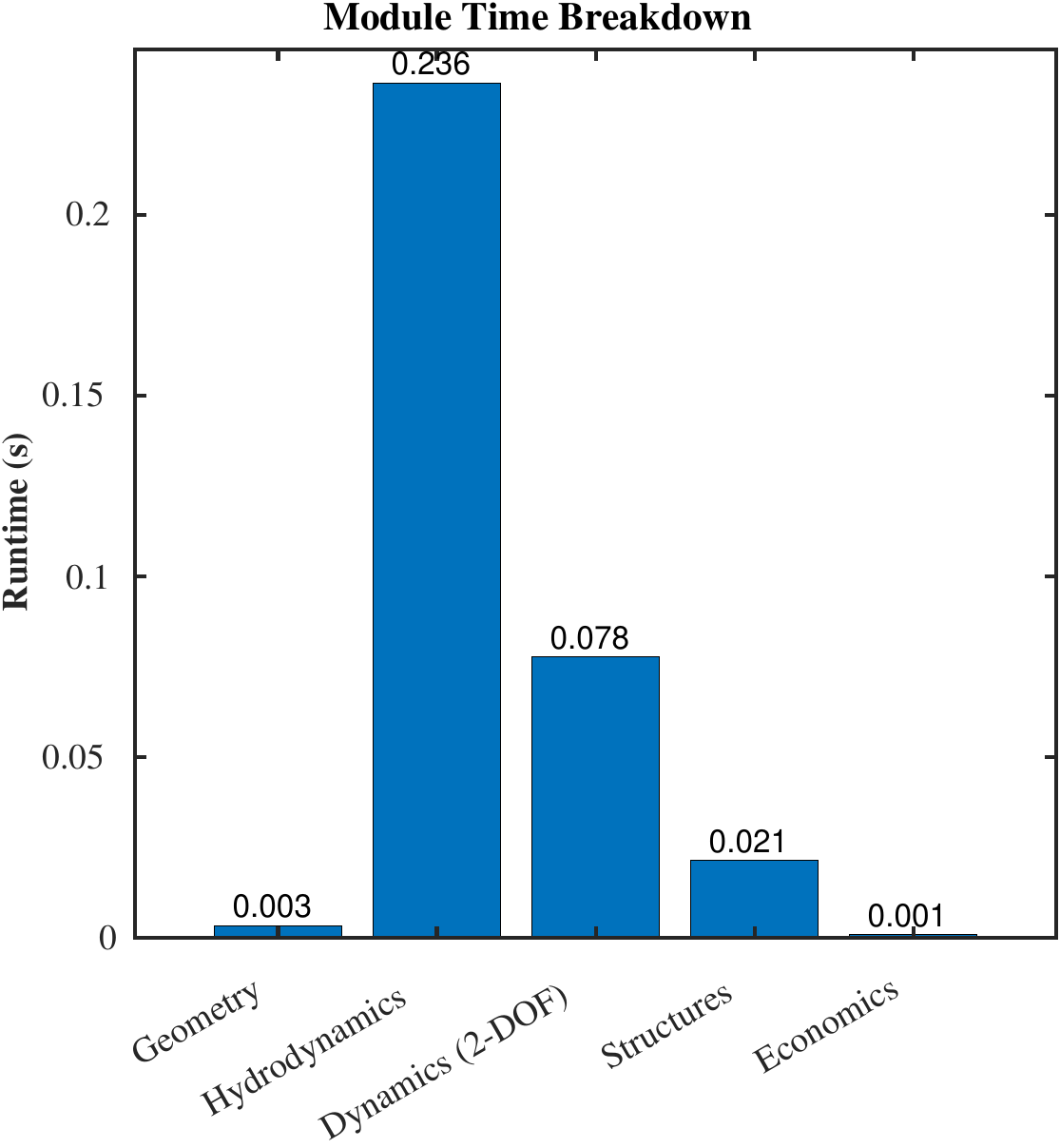}
\caption{Bar chart showing simulation runtime breakdown between modules}\label{fig:runtime-modules}
\end{figure}

\begin{figure}[t!]
\centering
\includegraphics[width=\linewidth]{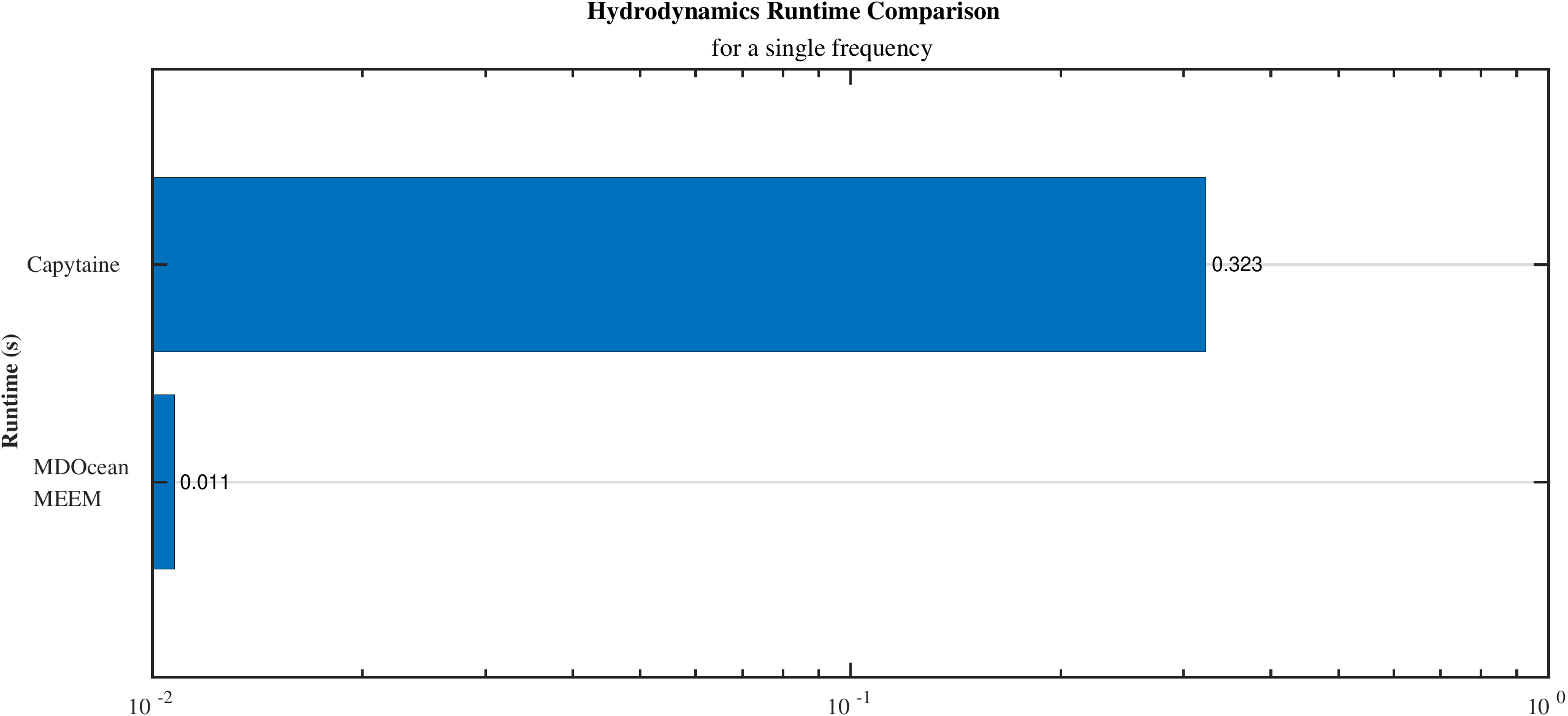}
\caption{Bar chart demonstrating the speed improvement of MDOcean's hydro module over baseline solver Capytaine}\label{fig:runtime-hydro}
\end{figure}

\Cref{fig:runtime-modules} visualizes the breakdown of runtime between modules.
The MEEM hydrodynamics module takes the majority (\resultsAOR[pctRuntimeMEEM]) of the time and is broken down in \Cref{fig:runtime-hydro}.
\ifdefined\DISSERTATION
    The biggest portion is dedicated to evaluating Bessel functions in the semi-analytical solution, another large portion to unpacking variables from the cell data structure, and a smaller portion to solving the imaginary modes of the dispersion relation and the linear matrix equation.
\fi
The simulation is an order of magnitude faster than the Capytaine BEM solver for similar convergence levels.

\begin{figure}[t!]
\centering
\includegraphics[width=\linewidth]{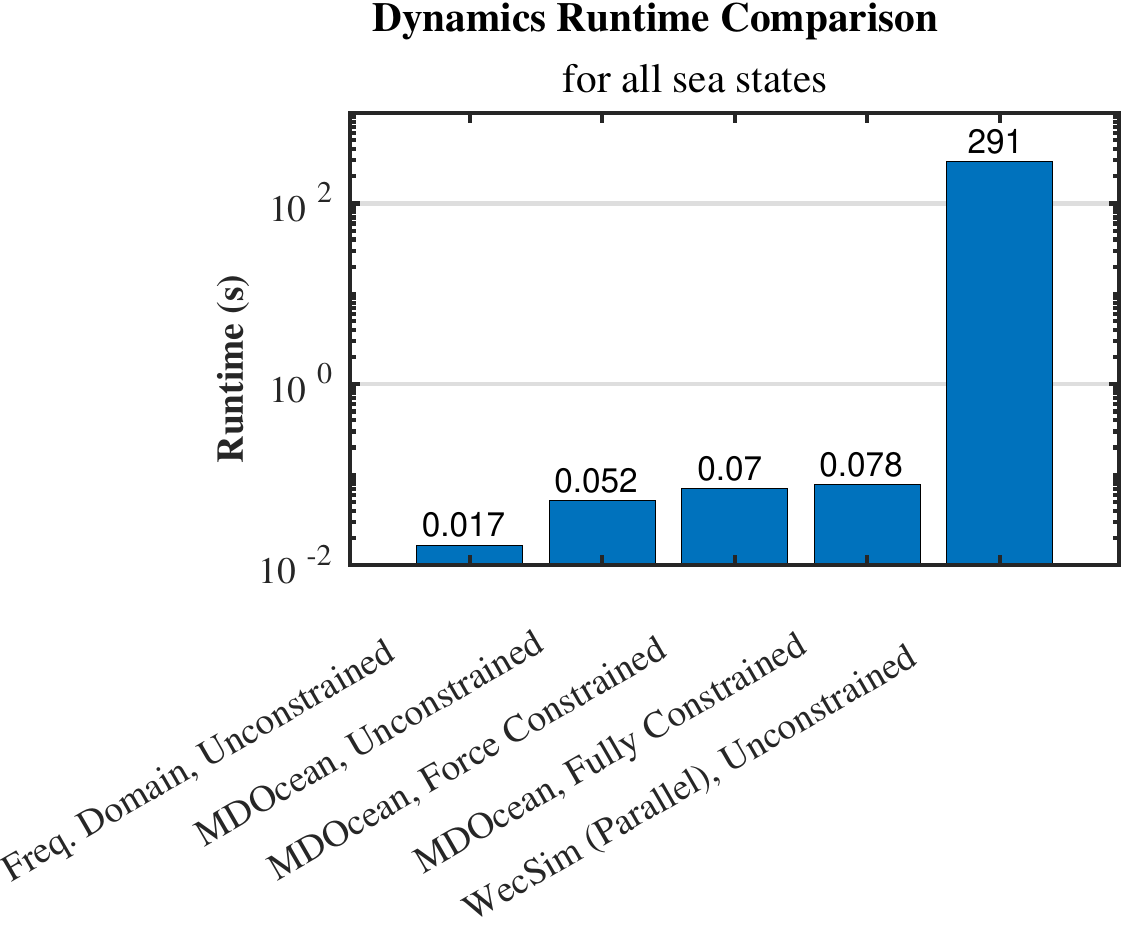}
\caption{Bar chart demonstrating the speed improvement of MDOcean's dynamics module over baseline solver WEC-Sim}\label{fig:runtime-dynamics}
\end{figure}

The dynamics and controls module takes the next longest (\resultsAOR[pctRuntimeDynamics], enlarged in \Cref{fig:runtime-dynamics}), with contributions from force saturation, spar analysis, drag linearization, and evaluation of the motion transfer function.
This represents a three-order-of-magnitude improvement over the equivalent regular-wave WEC-Sim simulation run in parallel.
\ifdefined\DISSERTATION
    Simplifying the dynamics to a single degree of freedom (DOF) achieves another order of magnitude speedup, although the optimization and benchmarking results presented here utilize the 2-DOF model.
\fi

The structures, geometry, and economics modules are not computationally expensive and together compose the remaining \resultsAOR[pctRuntimeOther]~of the total runtime.


\subsection{Discussion}
\ifdefined\DISSERTATION
    \Cref{sec:validation-benchmarking} demonstrates that MDOcean achieves accuracies generally within single-digit-percent JPD-weighted annual average power under realistic conditions, despite worst-case per-sea-state errors that can be larger due to drag and inter-body phase sensitivity in the 2-DOF model 
    Just three scale factors are used to tune results across disciplines and achieve an interdisciplinary model fully consistent with all relevant outputs in the reference model report, providing a valuable benchmark to compare technology improvements while retaining the flexibility to adjust scale factors as needed to represent other scenarios.
    The model is quite accurate for inputs near the nominal RM3 design and therefore its ability to capture multidisciplinary interactions in a single software package while running 1-3 orders of magnitude faster than established software is considered a major success.
    Still, the model is not perfect, and although only minor tuning was performed, the model's development centers around the Reference Model 3 and there may be unidentified discrepancies for geometries far from the reference design.
    We emphasize that the model is not intended to replace detailed models for final design and analysis, but rather to enable rapid design space exploration and optimization in the early stages of design.
    Since the values without scale factors are of the correct order of magnitude and the trends are reasonable, the use of scale factors to resolve discrepancies in the force, power, and mass models is deemed acceptable for the purposes of this study.
\else
    MDOcean reproduces the dynamics of an established time-domain solver within a few percent under matched modeling assumptions and within single-digit-percent JPD-weighted annual average power under realistic conditions, despite worst-case per-sea-state errors that can be larger due to drag and inter-body phase sensitivity in the 2-DOF model.
    Three scale factors are used to tune force, power, and mass outputs across disciplines for consistency with the reference model report.
    The \resultsAOR[simRuntime] runtime is 1-3 orders of magnitude faster than established software, enabling multidisciplinary optimization workflows that would otherwise be prohibitive.
    The model is intended to support rapid design-space exploration in early design, not to replace detailed FEM or BEM simulations for final analysis.
    Accuracy may degrade for geometries far from the reference design; users should validate against detailed models before drawing conclusions about previously-unexplored regions of the design space.
\fi
\section{Insights and Discussion}
\label{sec:discussion}

\subsection{Variable Sweeps}
This subsection summarizes insights from variable-sweep analyses before presenting specific trend breakdowns.

\subsubsection{Effect of Bulk Dimensions on Hydrodynamic Efficiency}
\ifdefined\DISSERTATION
    A brute-force sweep of nondimensionalized bulk geometry variables and the nondimensional wavenumber $m_0h$ allows for the identification of design trends.
    MDOcean performs a six-dimensional sweep with \resultsAOR[meemSweepSize] combinations in only \resultsAOR[meemSweepTime] seconds, an average of \resultsAOR[meemSweepTimePerCombo] per combination.
    Using only the hydrodynamics and dynamics modules for the present, we examine nondimensional outputs including radiation efficiency, capture width ratio, and nondimensionalized surface area.
    These serve respectively as rough proxies for power, power per unit cost, and cost, allowing for the visualization of tradeoffs between power and cost across the design space.
    Radiation efficiency was defined in \Cref{eq:power-matrix-decomposition} as the ratio of the capture width to the maximum radiation-limited capture width, and represents how well the device's geometry allows it to absorb power at a given frequency.
    Capture width ratio is the ratio of capture width to a characteristic length scale, often taken as the device diameter but taken here as the square root of the wetted surface area to better serve as a proxy for cost.
    (The cube root of submerged volume was also considered as a characteristic length scale; results were qualitatively similar to the square root of surface area and are not shown here.
    If wavelength were used as the characteristic length scale, the resulting capture width ratio would be a constant multiple of the radiation efficiency).
    For the final output, to nondimensionalize the wetted surface area, we divide it by the wavelength squared.
\else
    A brute-force sweep of nondimensionalized bulk geometry variables and the nondimensional wavenumber $m_0h$ identifies design trends.
    MDOcean performs the six-dimensional sweep with \resultsAOR[meemSweepSize] combinations in \resultsAOR[meemSweepTime].
    We examine three nondimensional outputs: radiation efficiency (proxy for power), capture width ratio (proxy for power per unit cost, normalized by $\sqrt{A_{wet}}$), and nondimensional wetted surface area (proxy for cost).
\fi

\begin{figure*}[htbp]
    \centering
    \begin{subfigure}[t]{0.48\linewidth}
        \includegraphics[width=\linewidth]{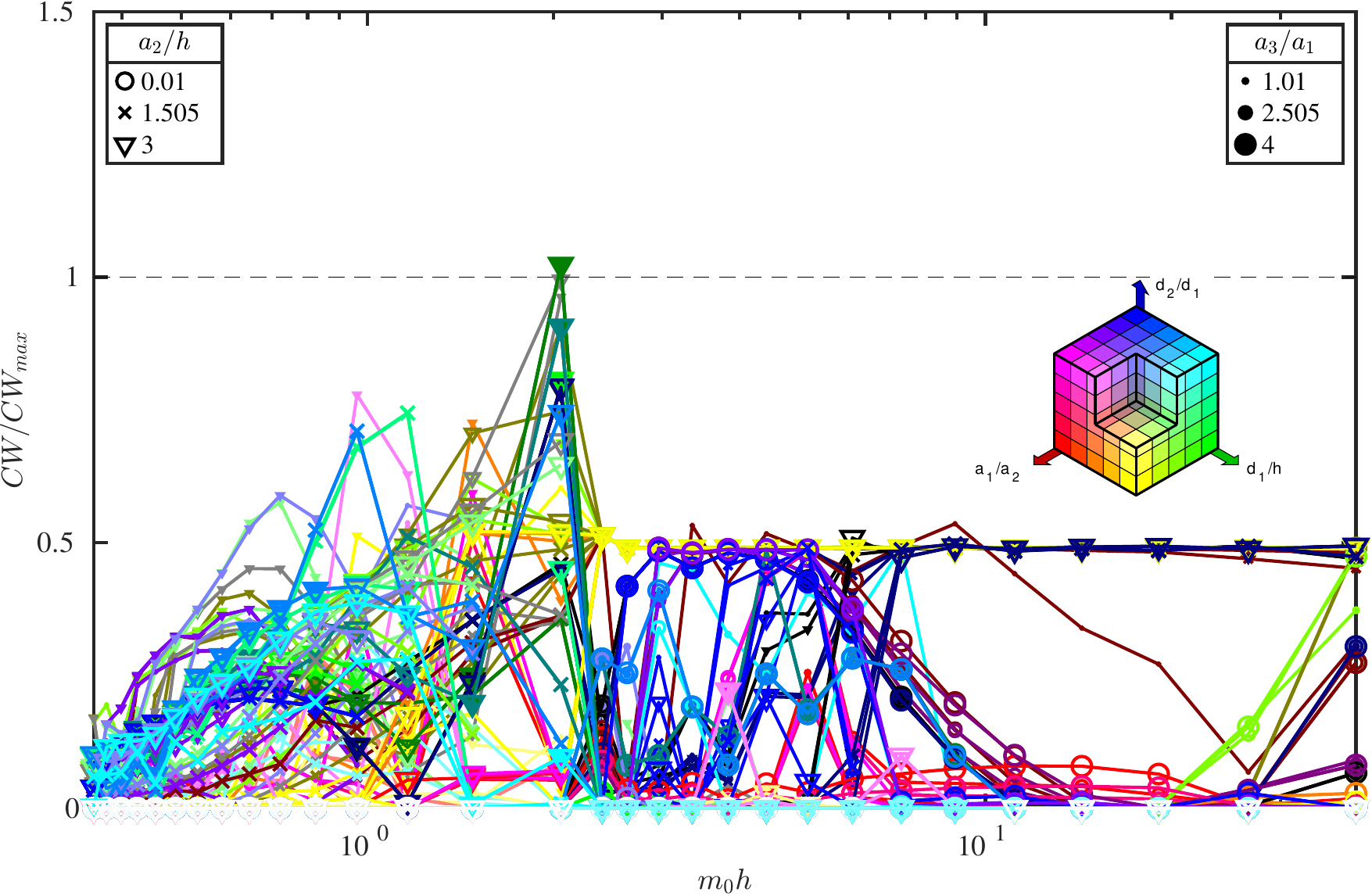}
        \caption{Radiation efficiency}
        \label{fig:meem-sweep-eff}
    \end{subfigure}
    \hfill
    \begin{subfigure}[t]{0.48\linewidth}
        \includegraphics[width=\linewidth]{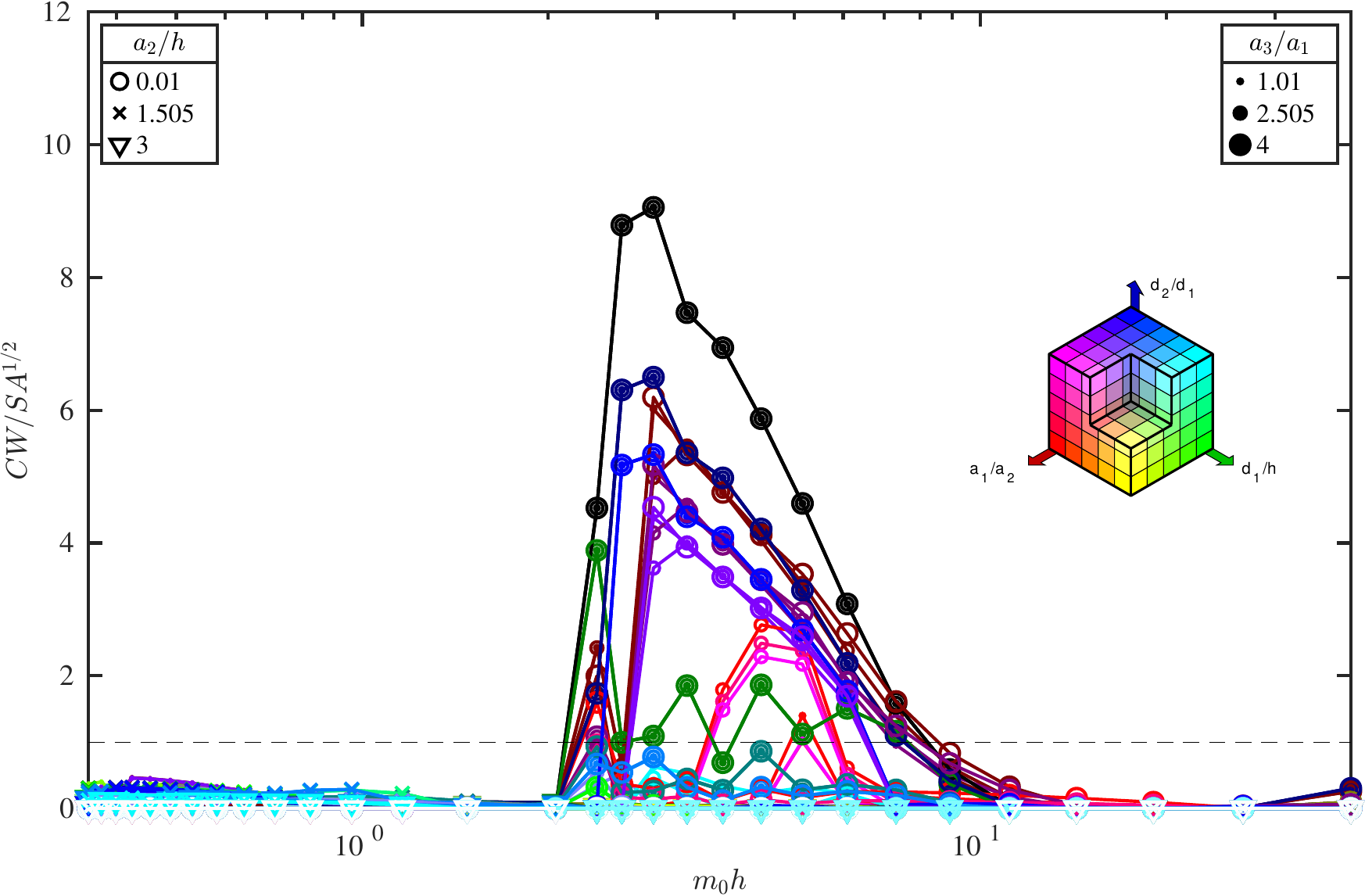}
        \caption{Capture width ratio}
        \label{fig:meem-sweep-cwr}
    \end{subfigure}
    \caption{Effect of wave environment and hydrodynamic design variables on (a) radiation efficiency and (b) capture width ratio}
    \label{fig:meem-sweep-m0h}
\end{figure*}

\begin{figure}[htbp]
    \centering
    \includegraphics[width=.9\linewidth]{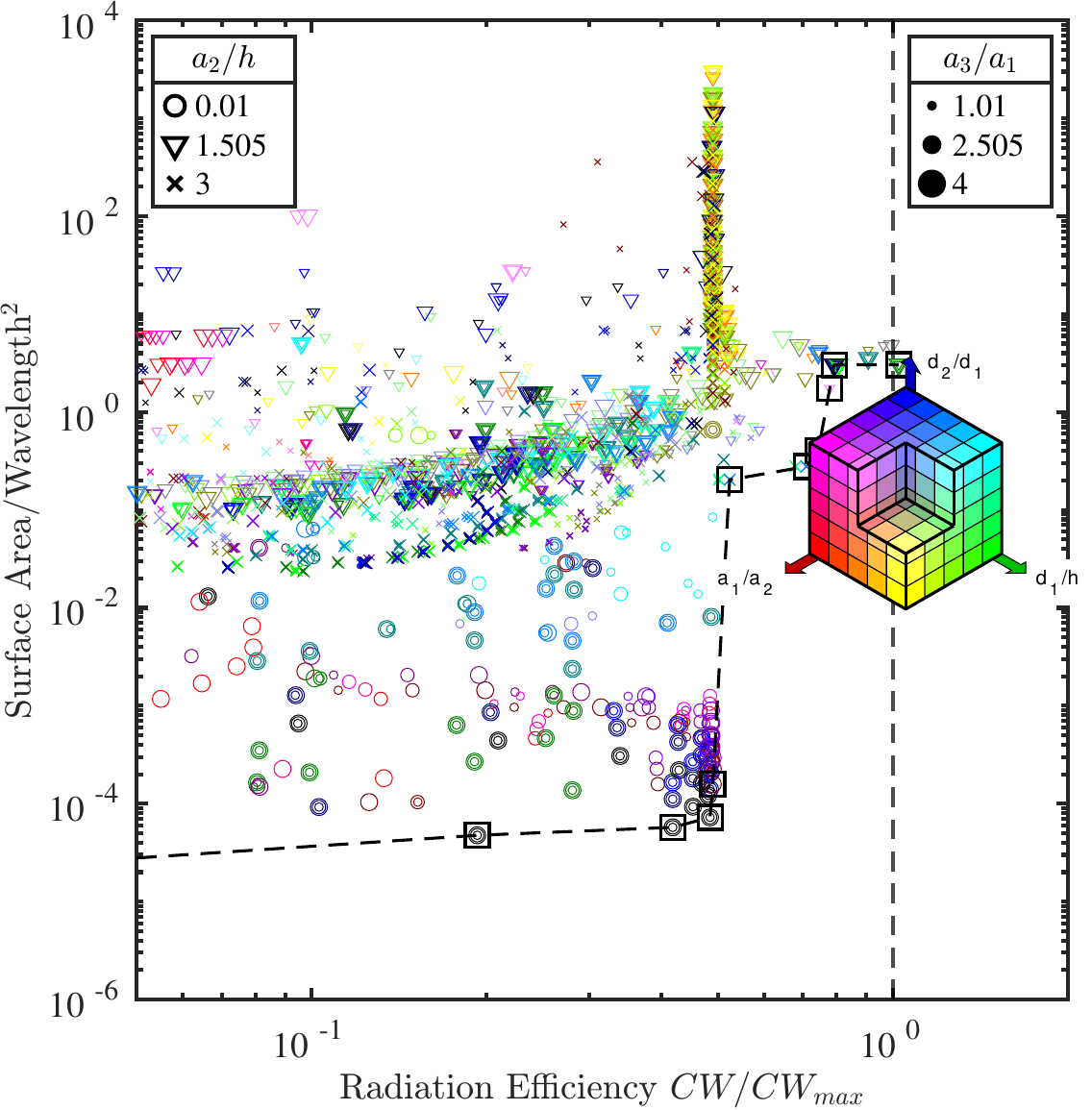}
    \caption{Effect of hydrodynamic design variables on radiation efficiency and power per unit surface area}
    \label{fig:meem-sweep-pareto}
\end{figure}

\ifdefined\DISSERTATION
    Visualizing this many-dimensional dataset (six inputs, three outputs) can pose a challenge.
    \Cref{fig:meem-sweep-m0h} will provide a view of all six input dimensions simultaneously, focusing centrally on the nondimensional wavenumber $m_0h$, and examine two outputs of radiation efficiency and capture width ratio.
    Meanwhile, \Cref{fig:meem-sweep-pareto} will show the tradeoff of radiation efficiency with nondimensionalized surface area and denote five of the six input dimensions, excluding only $m_0h$.
    In both figures, marker size and shape represent $a_2/h$ and $a_3/a_1$, which are swept over a range of 0-3 and 1-4 respectively.
    Ratios $a_1/a_2$, $d_1/h$, and $d_2/d_1$ must fall between zero and one and are represented as the red, blue, and green decimal values of the RGB color space, respectively.
\fi

\paragraph{Best Design as a function of Wave Environment}

\ifdefined\DISSERTATION
    Starting with \Cref{fig:meem-sweep-eff}, we observe that the most efficient design depends on the frequency-depth regime $m_0h$ of the wave environment, shown on the x-axis.
    At low $m_0h$ (shallow water, slow waves), the designs with the highest radiation efficiency are those with $a_1/a_2 \approx 1$, $a_2/h \ll 1$, and $a_3/a_1\approx 1$ with any value of $d_1/h$ (the blue/green/purple curves on the left).
    This represents a pure cylinder (float, spar, and damping plate all equal in diameter) with a small radius and any draft.
    
    The radiation limit is marked with a horizontal dashed line, and efficiencies above the limit are obtained around $m_0h\approx 1, a_1/a_2\approx 1, a_2/h\approx 3, a_3/a_1\approx 1$ (the pink/orange/yellow curves in the center of the figure).
    This indicates an issue with the simulation and data over the limit is disregarded.
    At this moderate $m_0h$ regime, the most efficient designs that stay within the radiation limit are green/blue, generally indicating a low $a_1/a_2$ (small spar).
    At high $m_0h$ (deep water, fast waves), the most efficient designs (pink/red) have
\else
    The most efficient design depends on the frequency-depth regime $m_0h$.
    Low $m_0h$ (shallow water, slow waves) favors a pure cylinder geometry with small radius.
    Moderate $m_0h$ favors low $a_1/a_2$ (a small spar relative to the float).
    High $m_0h$ favors qualitatively different geometries.
\fi

\paragraph{Design Cost-Power Tradeoffs}
\Cref{fig:meem-sweep-pareto} shows a Pareto front for the tradeoff of radiation efficiency (power proxy) and normalized surface area (cost proxy) across the six geometric design variables.
\ifdefined\DISSERTATION
    The normalized surface area does not depend on the seventh input variable $m_0h$, and the largest radiation efficiency for that geometry across all $m_0h$ is plotted, essentially showing the best possible radiation efficiency for each geometry across all wave environments.
    Black squares denote the non-dominated Pareto-optimal points: the geometries that achieve the highest radiation efficiency for a given normalized surface area.
\fi

\subsubsection{Damping Plate Size}
\begin{figure}[htbp]
    \centering
    \includegraphics[width=.85\linewidth]{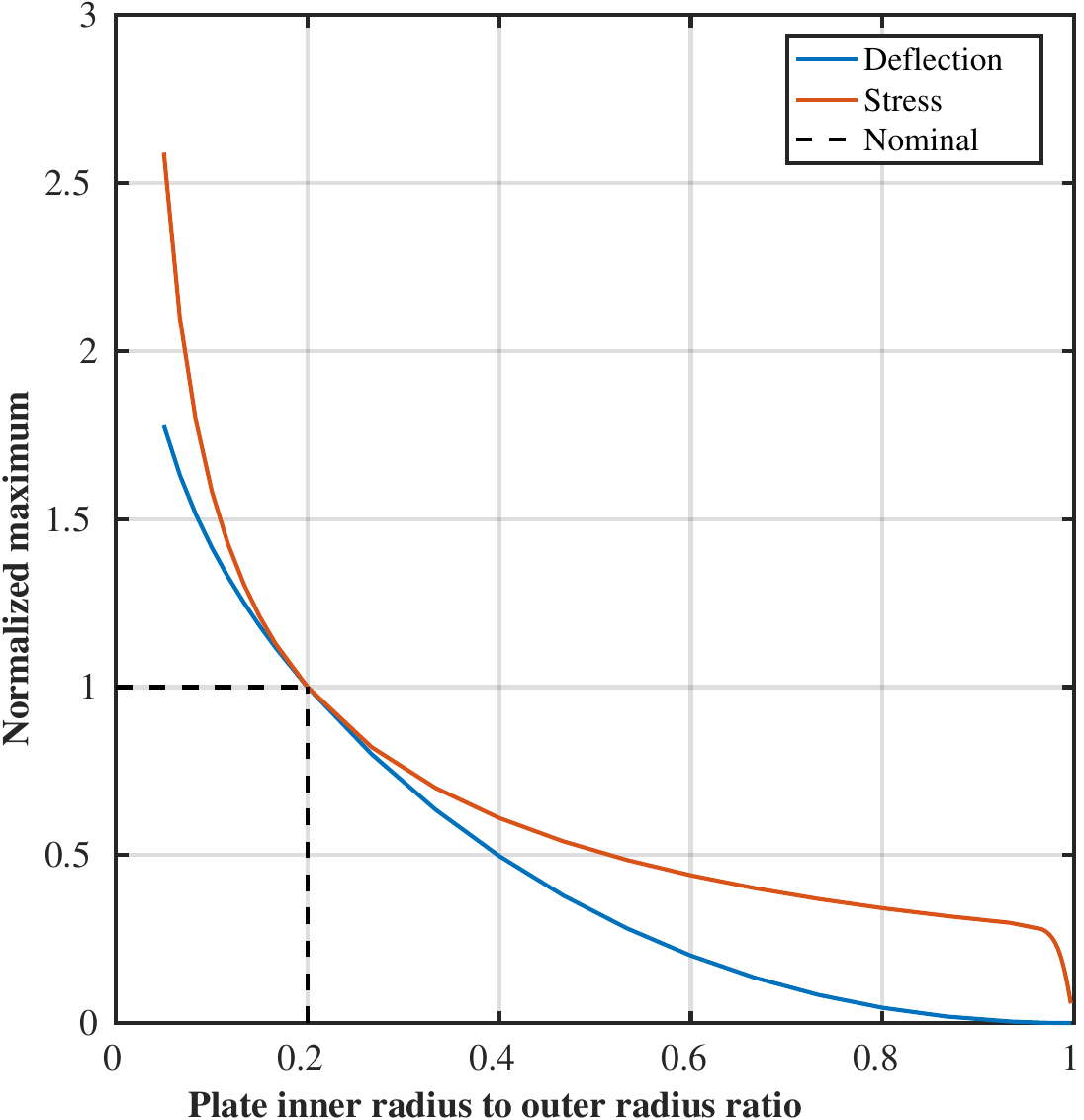}
    \caption{Normalized effect of damping plate aspect ratio on maximum stress and deflection.
The dashed lines indicate the nominal design point at $b/a = 0.2$.}
    \label{fig:damping-plate-maxs}
\end{figure}

\Cref{fig:damping-plate-maxs} shows that damping plate stress can be reduced substantially by increasing the plate inner radius.
This is because the same force has a lower lever arm to the column and therefore creates less bending moment.

\subsubsection{Effect of PTO Force and Power Limits}
\begin{figure}[htbp]
\centering
\includegraphics[width=.8\linewidth]{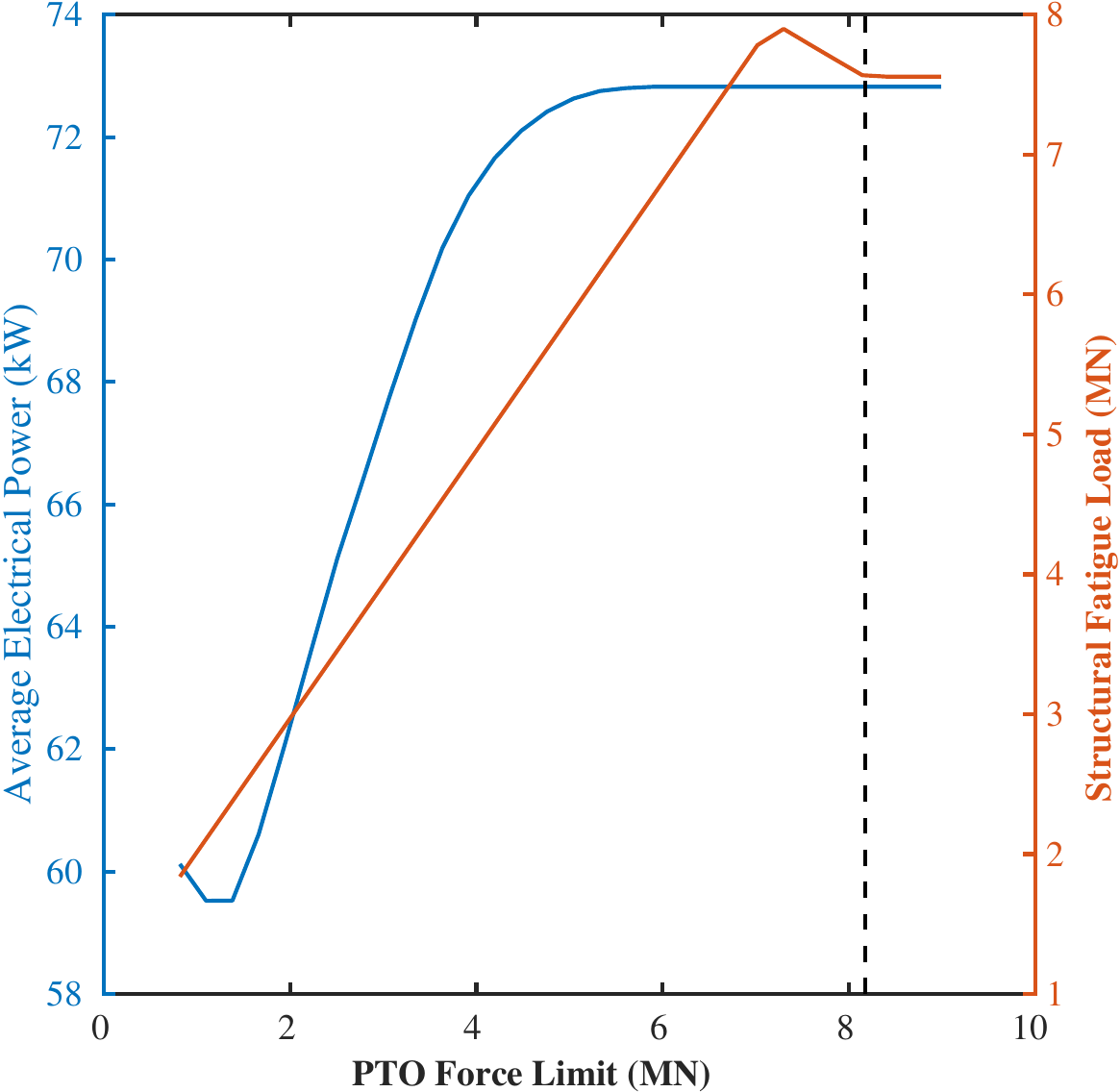}
\caption{Effect of Force Limit on Annual Average Power and Peak Structural Load}
\label{fig:force-limit}
\end{figure}

\ifdefined\DISSERTATION
    Previous work has shown that capping the PTO force can substantially reduce PTO size and cost with minimal decrease in power.
    \citet{devin_high-dimensional_2024} finds that impedance matching at resonance has a large effect on power and PTO sizing has little effect.
    This suggests that if cost were included, it may be preferable to have an undersized PTO that frequently saturates its control force.
    This agrees with the conceptual argument of \citep{coe_maybe_2021} and the numerical findings of \citep{mcgilton_optimal_2024}.
    The latter shows that reducing the maximum PTO force by 70\% leads only to a 3\% decrease in annual energy production, a trend consistent across two device archetypes (RM3 and RM5), four locations, and several dimensional scales.
    The recent RM3 optimization \citep{gaebele_tpl_2025} finds that the device should be scaled down by a factor of 0.55 to a new float diameter of 11~m in order to maximize the power per unit surface area, and that the continuous torque constraint is often active, illustrating the importance of including this constraint in the model.
    \citet{mccabe_force-limited_2024} shows analytically that for a given sea state, power decreases quadratically with decreasing force limit for the worst-case scenario of zero intrinsic reactance, and that highly reactive devices are even less sensitive to force limits.
    Intuitively, the sensitivity of annual average power to force limit should be even lower when considering a range of realistic sea states, because large sea states that require higher forces are comparatively rare.
\else
    Prior work consistently shows that capping the PTO force can substantially reduce PTO size and cost with minimal decrease in power \citep{devin_high-dimensional_2024,coe_maybe_2021,mcgilton_optimal_2024,gaebele_tpl_2025,mccabe_force-limited_2024}.
\fi
Using MDOcean, we investigate the effect of PTO force limit on average power and structural load in \Cref{fig:force-limit}.

Decreasing the force limit has almost no effect on power until the force limit is around 50\% of the nominal value, after which power falls off steeply.
\ifdefined\DISSERTATION
    From theory (see \Cref{sec:appendix-constraint-sensitivity}), we expect a smooth nonlinear relationship between power and force limit when considering a single sea state, and a piecewise continuous nonlinear relationship when considering multiple sea states as in \Cref{fig:force-limit}.
\fi
On the other hand, the structural load for the operational design load case scales nearly linearly with the force limit.
\ifdefined\DISSERTATION
    For structural load, we expect a linear relationship when considering a single sea state and a piecewise linear relationship when considering multiple sea states.
    The fact that the linear relationship holds when considering multiple sea states implies that both the worst-case sea state and its quasi-linearized drag coefficients hardly change as the force limit is swept.
    If this trend holds true for other designs, it could suggest a more efficient analysis strategy where the worst-case sea state is identified at the nominal design and then maintained (or only infrequently updated) throughout a design optimization, rather than performing a full multi-sea-state structural analysis for each design being considered.
\fi

\begin{figure}[htbp]
\centering
\includegraphics[width=.9\linewidth]{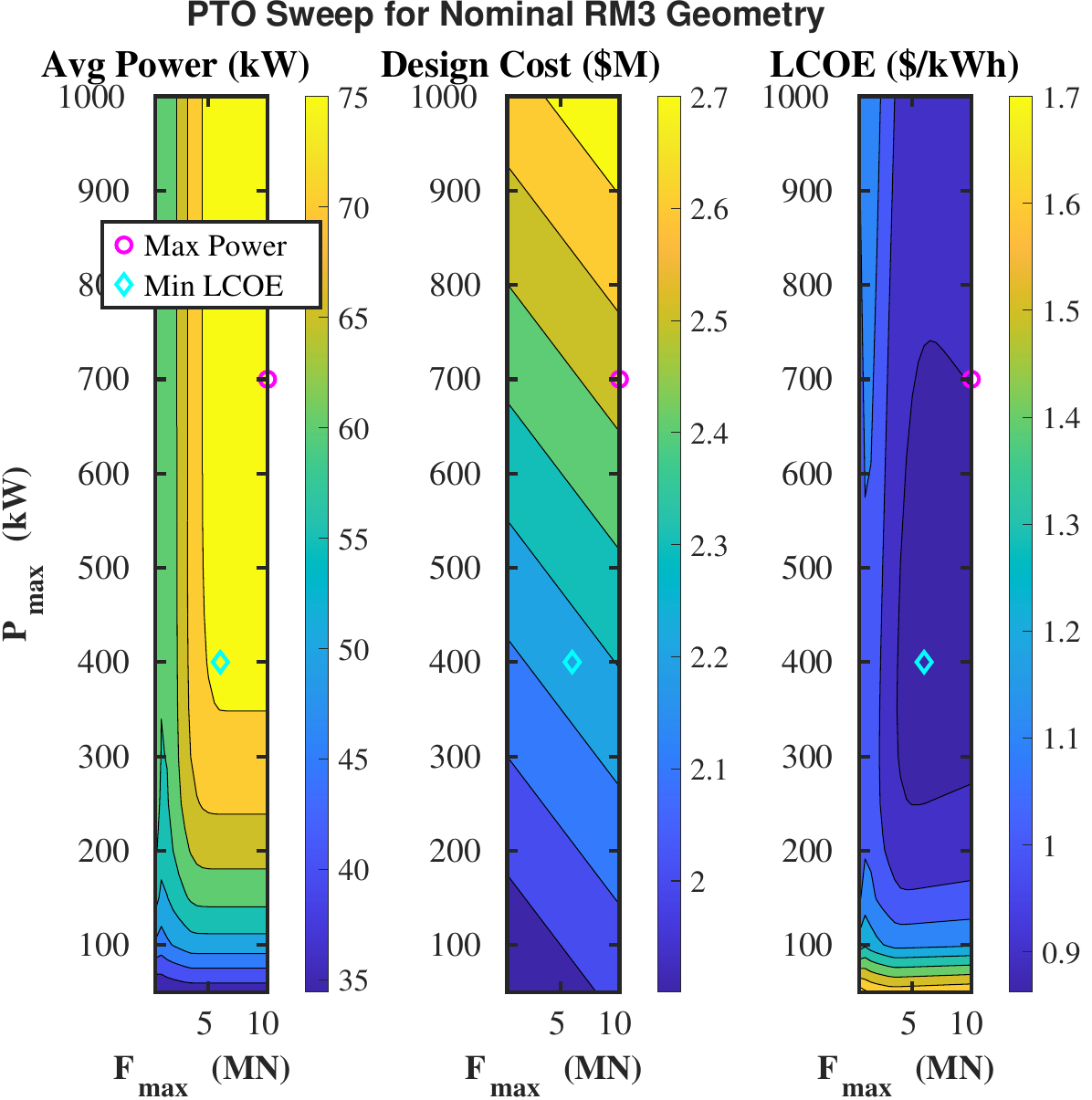}
\caption{Effect of Generator Force Limit and Power Limit on Annual Average Power and LCOE}
\label{fig:force-power-limit-sweep}
\end{figure}

We next move to a simultaneous sweep of both the force limit and the power limit, which is more relevant for PTO design optimization since generator selection dictates both limits.
\ifdefined\DISSERTATION
    Here we hold structural thicknesses constant, so any cost savings only reflect reduced direct cost of the PTO and not the cost savings from the reduced structural material that would be possible with lower loads.
\fi
\Cref{fig:force-power-limit-sweep} shows the results for both power and LCOE, with hatched shading to indicate regions that are infeasible.
\ifdefined\DISSERTATION
    The unsaturated solution is in the top right corner, past which power is constant because the unconstrained reactive controller satisfies the limit in all sea states.
\fi

Notably, the minimum LCOE point occurs at a force limit far below the unsaturated solution because the cost savings from a smaller PTO outweigh the loss in power.
\ifdefined\DISSERTATION
    If a full design optimization were performed to adjust the structural thicknesses to take advantage of the reduced structural loads, we expect an even lower force limit to be optimal.
\fi
This illustrates the importance of including the force-cost tradeoff in an optimization model rather than assuming an unconstrained solution.
\ifdefined\DISSERTATION
    We also observe that power and LCOE both appear to be convex with respect to the force and power limits, and that the feasible region is convex as well.
    The theoretical analysis in \Cref{sec:appendix-constraint-sensitivity} derives mathematical conditions for when this is the case.
    Convexity suggests that local optimization methods should be effective for finding the optimal PTO design.
\else
    Power and LCOE both appear to be convex with respect to the force and power limits across the feasible region, a structural property that future optimizations can exploit.
\fi

\subsubsection{Design Space Exploration}
\begin{figure}[htbp]
\centering
\includegraphics[width=.9\linewidth]{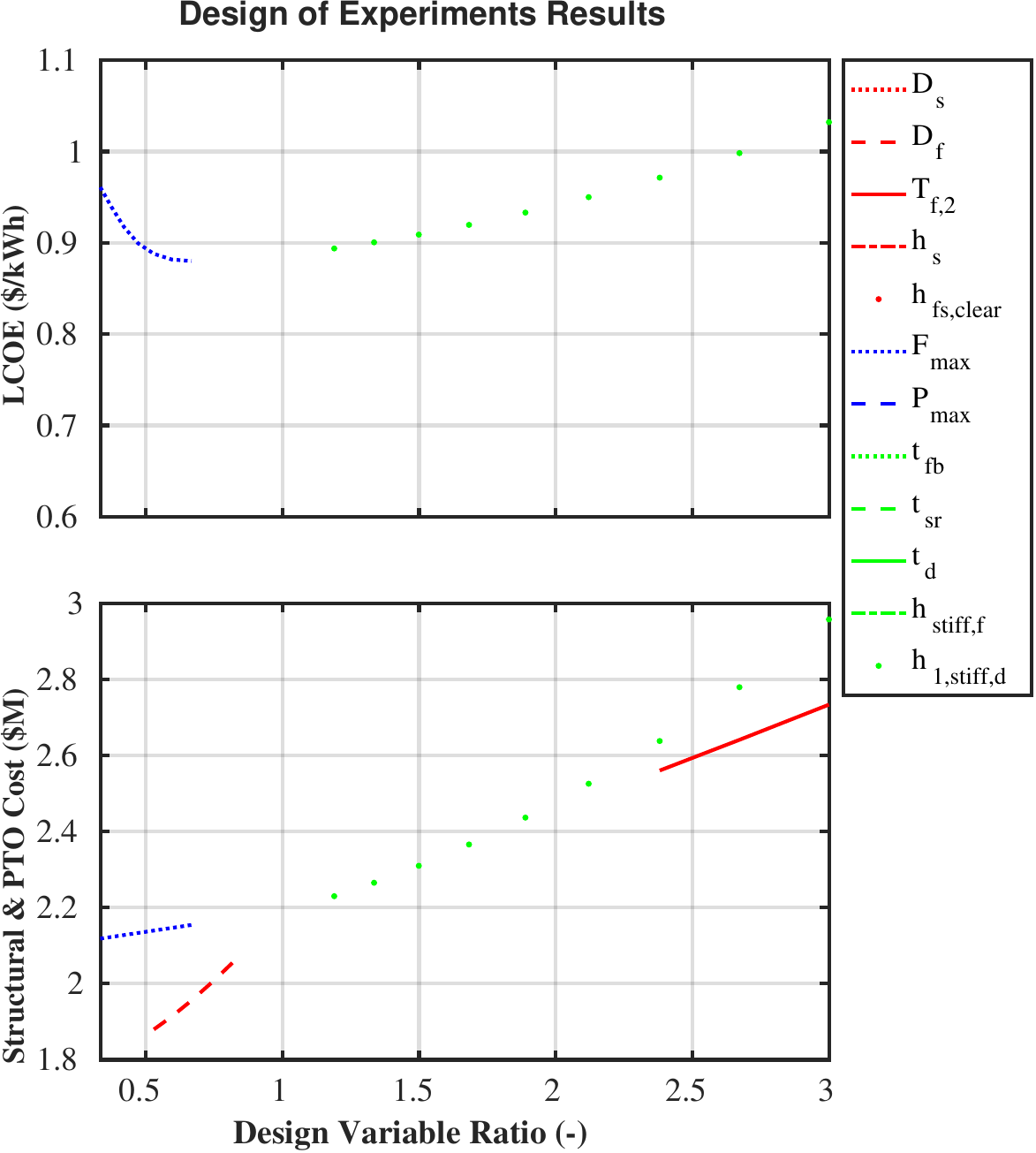}
\caption{Design of experiments}\label{fig:experiments}
\end{figure}

\ifdefined\DISSERTATION
    The OAT design sweep, shown in \Cref{fig:experiments}, immediately draws attention to the float diameter $D_f$, shown with a red dashed line, as the variable with the strongest effect on the objectives.
    A slight reduction in the diameter from its nominal value causes the cost to decrease close to linearly but the LCOE to skyrocket due to a substantial reduction in power generation.
    Meanwhile, slightly increasing the diameter causes the power production to grow significantly but increases loads, leading to fatigue failure of the damping plate.
    This is shown as blank on the diagram to indicate infeasibility.
    A larger $\sim30\%$ diameter increase restores feasibility of the damping plate and achieves a roughly 15\textcent/kWh reduction in LCOE, representing the best design of the OAT sweep.
    The results also point to the possibility of an even lower LCOE design with float diameter perhaps $\sim20\%$ larger than nominal, if the damping plate can be made feasible by increasing $t_d$ or $h_{1,\text{stiff},d}$ without a major cost penalty.
\else
    The one-at-a-time (OAT) design sweep (\Cref{fig:experiments}) reveals that float diameter $D_f$ has the strongest effect on objectives.
    The feasible design space is highly fragmented: an ``island'' of low-LCOE designs at $1.3 < D_f/D_{f,nom} < 2.1$ is separated from the nominal design by the damping plate fatigue constraint.
\fi

\ifdefined\DISSERTATION
    Besides $D_f$, the other design variable displaying a non-monotonic relationship with $LCOE$ is the force limit $F_{\text{max}}$, with an optimal at around half the nominal force limit.
    Together, these insights inform the selection of the starting design for optimization, $\vec{x}_o$.
    The OAT sweep also does not reveal any multimodality in the objectives, a promising sign that only a limited number of multi-starts may be necessary, although objective multimodality could still exist in other parts of the design space.
    Finally, the OAT design sweep shows the highly constrained nature of the problem, with large regions of the design space infeasible.
    In particular, the feasible region $1.3<D_f/D_{f,nom}<2.1$ is an ``island'' in the feasible design space, separated from the other feasible points at $D_f/D_{f,nom}<1 $ by the damping plate fatigue constraint.
    Thus even without objective multimodality, multimodality exists in the Lagrangian (objective plus weighted active constraints), indicating the need for a multi-start optimization procedure as planned.
\else
    Force limit $F_{\text{max}}$ also shows non-monotonic LCOE dependence, with an optimum near half the nominal limit.
    Although no objective multimodality is observed within the swept region, the fragmented feasible space introduces multimodality in the Lagrangian, indicating the need for a multi-start optimization procedure \citep{mccabe_leveraging_2026}.
\fi

\subsection{Multidisciplinary Insights}
This section leverages the analytical multidisciplinary nature of the model to draw intuitive insights on limit cases and tradeoffs, and observe nondimensional relationships and scaling laws that would not be readily apparent in a purely numerical or single-discipline model.

\begin{figure*}[htbp]
\centering
\includegraphics[width=.95\linewidth]{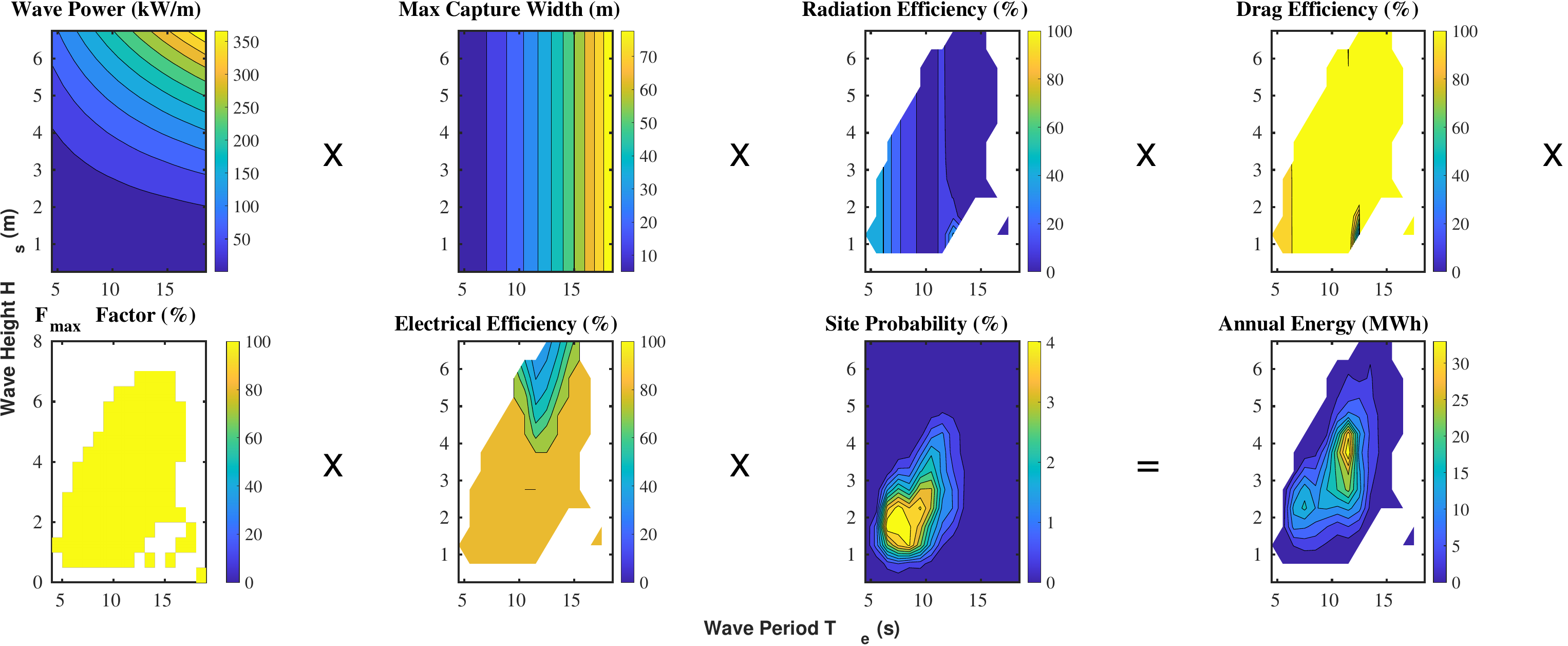}
\caption{Power matrix decomposition}
\label{fig:power-matrix-decomposition}
\end{figure*}

\paragraph{Power Matrix Drivers}
Similar to \Cref{fig:JPD-multiply} showing the multiplication of the power matrix by the JPD to obtain average power, \Cref{fig:power-matrix-decomposition} further decomposes the power matrix into component matrices showing the impact of shape, drag, maximum force, and PTO losses, with the following breakdown:
\begin{equation}\label{eq:power-matrix-decomposition}
\mathbf{P}^{H,T}_{elec} = \mathbf{J}^{H,T}_{wave} ~\mathbf{CW}^{H,T}_{max} ~\mathbf{\eta}^{H,T}_{rad} ~\mathbf{\eta}^{H,T}_{drag} ~\mathbf{\eta}^{H,T}_{limits} ~\mathbf{\eta}^{H,T}_{elec}
\end{equation}

Here $\mathbf{J}^{H,T}_{\text{wave}}$ is the wave power density (design-independent) and $\mathbf{CW}^{H,T}_{\text{max}}$ is the radiation-limit capture width (depends only on the mode of motion).
The four efficiency matrices isolate the design dependencies: $\mathbf{\eta}^{H,T}_{rad}$ for hull-shape radiation efficiency, $\mathbf{\eta}^{H,T}_{\text{drag}}$ for drag, $\mathbf{\eta}^{H,T}_{\text{limits}}$ for dynamic constraints, and $\mathbf{\eta}^{H,T}_{\text{elec}}$ for PTO losses and dynamics.
Explicit definitions are given in \Cref{sec:appendix-power-decomposition}.

\ifdefined\DISSERTATION
    Two features of this decomposition deserve emphasis.
    First, $\mathbf{\eta}^{H,T}_{\text{drag}}$ may exceed 1 in sea states where the drag-free system is so undamped that the unconstrained optimal controller would induce instability and a sub-optimal controller is used instead --- this reflects a denominator artifact, not an actual energy gain from drag.
    Second, the multiplicative structure depends on the order chosen: alternative factorizations (e.g., defining drag effect relative to the saturated rather than the ideal system) would represent independent sensitivities and could not be multiplied to recover the overall power matrix.
\fi

This multiplicative decomposition allows isolation of the effect of each design variable and parameter on power, intuitively identifying which design variables are most important for improving power.
\Cref{tab:power-matrix-dependence} maps the dependence explicitly.
{
\renewcommand{\arraystretch}{1.35}
\begin{table}[h]
\centering
\caption{Dependence of efficiencies on inputs}\label{tab:power-matrix-dependence}
\begin{tabular}{lcccccc}
\hline
 & $x_{\text{shape}}$ & $C_{d,f}$ & $C_{d,s}$ & $F_{\text{max}}$ & $P_{\text{max}}$ & $\eta$ \\
\hline
$\mathbf{\eta}^{H,T}_{rad}$   & $\checkmark$ &  &  &  &  &  \\
$\mathbf{\eta}^{H,T}_{\text{drag}}$  & $\checkmark$ & $\checkmark$ & $\checkmark$ &  &  &  \\
$\mathbf{\eta}^{H,T}_{\text{limits}}$ & $\checkmark$ & $\checkmark$ & $\checkmark$ & $\checkmark$ &  &  \\
$\mathbf{\eta}^{H,T}_{\text{elec}}$  & $\checkmark$ & $\checkmark$ & $\checkmark$ & $\checkmark$ & $\checkmark$ & $\checkmark$ \\
\hline
\end{tabular}
\end{table}
}

\ifdefined\DISSERTATION
    Importantly, the later matrices in the product depend on the design variables and parameters that affect the earlier matrices.
    This coupling means that the effect of the drag coefficients, for example, on power in a device with PTO limits and dynamics must be understood not only through $\mathbf{\eta}^{H,T}_{\text{drag}}$ but also through their effect on $\mathbf{\eta}^{H,T}_{\text{limits}}$ and $\mathbf{\eta}^{H,T}_{\text{elec}}$.
    While it would be possible to formulate variants such as $\mathbf{\eta}^{H,T}_{\text{drag},var} = \mathbf{P}_{IV}/\mathbf{P}_{IV, no \text{drag}}$ that capture the full effect of drag on the saturated system, such matrices would represent independent sensitivities and could not be multiplied together to obtain the overall power matrix, which is a key advantage of the current formulation.
\else
    The later matrices depend on the design variables affecting earlier matrices, so the total impact of (e.g.) drag must be assessed through its effects on $\mathbf{\eta}^{H,T}_{\text{drag}}$, $\mathbf{\eta}^{H,T}_{\text{limits}}$, and $\mathbf{\eta}^{H,T}_{\text{elec}}$ jointly.
\fi

\Cref{fig:power-matrix-decomposition} shows the power matrix decomposition for the nominal RM3 design.
\ifdefined\DISSERTATION
    Aligning with \citet{zou_practical_2023}, low-period sea states achieve mechanical powers closer to the radiation limit (higher radiation efficiency).
    High-period sea states have lower radiation efficiencies due to the amplitude limit yet still contribute significantly due to their higher energy content.
    Despite the most probable wave period being 6-8 seconds, 10 second waves contribute the most to energy production for this reason, though even longer waves contribute little due to their rarity.
\else
    Low-period sea states achieve higher radiation efficiency, while high-period sea states have lower radiation efficiency but contribute significantly to total production due to their higher energy content; despite the most probable period being 6--8~s, 10~s waves contribute the most to annual energy production \citep{zou_practical_2023}.
\fi

The effect of drag is most significant in the 11-12 second range, corresponding to the spar's natural frequency.
\ifdefined\DISSERTATION
    Interestingly, this is also where the force limit has the strongest effect, indicating that the large amplitudes at the natural frequency drive up the force more than the larger stiffnesses required for reactive control at frequencies far from the natural frequency.
\fi

\paragraph{Constraint Sensitivities}
The structure of the constrained optimal control problem allows for quantification of the sensitivity of optimal power to each constraint limit, formalizing the intuition behind $\eta^{H,T}_{\text{limits}}$.
From \Cref{eq:constrained-qp-solution,eq:power-elec}, the average electrical power (the LCOE denominator) is a sea-state-weighted sum of the impedance-matched power in each sea state, each scaled down by the factor $(1-|\Gamma_{\beta,opt}|^2)$ that captures the effect of active dynamic constraints:
\begin{equation}\label{eq:power-elec-sum}
\begin{aligned}
    \overline{P}_{elec} &= \eta \sum_{\beta=1}^{N_{sea}} JPD_\beta \cdot
    \frac{|\hat{V}_{s,th,\beta}|^2}{8\Re(Z_{s,th,\beta})}~ (1 - |\Gamma_{\beta,opt}|^2)
    \\ s.t.
\quad
    |\hat{F}_{p,\beta}| &\leq F_{max},\quad
    |p_{pk,elec,\beta}| \leq P_{max,elec},\\
    |\hat{\xi}_{f,\beta}| &\leq X_{f,max},\quad
    |\hat{\xi}_{s,\beta}| \leq X_{s,max}\\
\end{aligned}
\end{equation}

Power is maximized when $|\Gamma_{\beta,opt}|=0$ in every sea state, which corresponds to unconstrained impedance-matched control.
When constraints are active, the optimal $|\Gamma_{\beta,opt}|>0$ represents the impedance mismatch required to satisfy them, and power decreases quadratically with $|\Gamma_{\beta,opt}|$.

\Cref{sec:appendix-constraint-sensitivity} uses this relationship to derive a closed-form scaling law for the optimal annual average power $\overline{P}_{\text{elec}}$ as a function of the constraint coefficients $b_\mu$ (which are affine or square functions of the constraint limits $\tau_{\text{max}}$, $P_{\tau\Omega,\text{max}}$, $\xi_{f,\text{max}}$, $\xi_{s,\text{max}}$, and $X_{PTO,\text{max}}$ defined in \Cref{tab:qp-constraints}).

The result is:
\begin{equation}\label{eq:power-double-sum-main}
\begin{aligned}
    \overline{P}_{elec} =~&\overline{P}_{elec,0}
    - \eta \sum_{\mu\nu\beta}
    \left[ 
        \sqrt{\textrm{aff}_{\mu\nu\beta}(b_{\mu})}
        ~+ \right. \\
        & \left. \textrm{quad}_{\mu\nu\beta}(b_{\mu}, b_{\nu})
    \right.+
     \\
    &\left.
        \left(\textrm{aff}_{\mu\nu\beta}(b_{\mu}) + \textrm{aff}_{\mu\nu\beta}(b_{\nu})\right)
        \sqrt{\textrm{quad}_{\mu\nu\beta}(b_{\mu}, b_{\nu})}~
    \right]
\end{aligned}\end{equation}
where $\overline{P}_{\text{elec},0}$ is the unconstrained power and $\textrm{aff}_{\mu\nu\beta}$, $\textrm{quad}_{\mu\nu\beta}$ are affine and quadratic functions of $b_\mu$ and $b_\nu$, respectively.
$\mu$ and $\nu$ are constraint indices, expanded in \Cref{sec:appendix-constraint-sensitivity}.

Under conditions derived in \Cref{sec:appendix-constraint-sensitivity}, this scaling law is convex in the constraint coefficients $b_\mu$.
While satisfaction of these conditions is not guaranteed, convexity is observed to hold across the PTO sweep of \Cref{fig:force-power-limit-sweep}.

The scaling law also displays coupling in the constraint variables: the effect of one constraint limit depends on the values of the other constraint limits.
This illustrates the importance of a multidisciplinary optimization considering all interactions, rather than a sequential optimization that optimizes one subsystem at a time holding the others fixed.

\paragraph{LCOE Drivers}
To better see the dependence of LCOE on design for a fixed $N_{WEC}$ and $FCR$, one can divide out the constant factors from \Cref{eq:lcoe}, leaving normalized price constants $p^{'}_{(*)}$:
\begin{equation}\label{eq:LCOE-scale}
  LCOE = \frac{
    p^{'}_{F} F_{max} + p^{'}_{P}P_{max,elec} + p^{'}_{s} V_{struct} + p^{'}_0
    }{\overline{P}_{elec}}
\end{equation}
The numerator is affine in the design inputs $F_{\text{max}}$ and $P_{\text{max},\text{elec}}$, while $V_{\text{struct}}$ is the product of structural thicknesses with their corresponding areas, which depend quadratically on the bulk dimensions.
The fact that decreasing a constraint limit causes the numerator of the LCOE to monotonically increase and the denominator to monotonically decrease implies that at system optimality, each constraint will be active in at least one sea state.

\ifdefined\DISSERTATION
    The LCOE minimization with design variables of $F_{\text{max}}$, $P_{\text{max},\text{elec}}$, and thicknesses is a fractional quadratic program, a structure common in resource allocation problems.
    When the denominator is convex with respect to the PTO limits, this means that LCOE is ``quasi-convex'' in the PTO limits for fixed geometry \citep{agrawal_disciplined_2020}.
    This quasi-convexity can be leveraged by using a subgradient method or by solving a small sequence of convex problems, which simplifies the optimization process significantly compared to a general nonlinear solver.
    This motivates a three-level nested optimization strategy:
    at the innermost level, the optimal control problem for fixed geometry and PTO limits is solved as an analytical QP;
    at the mid-level, the optimal PTO limits are found by exploiting the quasi-convex fractional structure when possible;
    and at the outer level, the bulk geometry is optimized with a general nonlinear program.
\else
    The fractional-quadratic structure of \Cref{eq:LCOE-scale} admits an efficient nested optimization that exploits convexity properties; see the companion paper \citep{mccabe_leveraging_2026} for the optimization strategy.
\fi

\subsection{Limitations and Future Work}\label{sec:unmodeled-effects}
\ifdefined\DISSERTATION
    Broadly, future work must both address limitations of the model and apply it to answer design questions.
    As MDOcean is open source, the authors encourage the wave energy community to contribute to both areas, and are themselves actively pursuing the latter with a multi-objective optimization of the RM3 WEC \citep{mccabe_leveraging_2026}.
    \Cref{tab:future-work} summarizes potential future improvements to the model, distinguishing between model improvements that would enhance the accuracy or realism of studies that can be conducted with the present model and those that would unlock the ability to answer design questions that the current model cannot.
    This section describes the relevance and possible implementation paths for each development area.
\else
    \Cref{tab:future-work} summarizes principal limitations and future work, distinguishing model improvements to enhance the accuracy of currently-achievable studies from extensions to unlock new design questions.
    MDOcean is open-source \citep{mccabe_mdocean_2024}; community contributions are welcome, and the authors are pursuing the multi-objective RM3 optimization in the companion paper \citep{mccabe_leveraging_2026}.
\fi

\newcommand{\modelTrustBuilders}{
    \begin{enumerate}[leftmargin=*]
        \item Surge force and mooring cost
        \item Nonlinear storm wave forces
        \item Irregular waves
        \item Structures: thinner plates, better validation
        \item Additional PTO nonlinearities
    \end{enumerate}
}

\newcommand{\modelStudyEnablers}{
    \begin{enumerate}[leftmargin=*]
        \item Spectral power, load, amplitude
        \item Different WEC archetypes
        \item Lifetime and sea state contours
        \item MEEM extensions for damping plate and 3+ cylinders
        \item Generator magnetics and cost
        \item Other structural materials
    \end{enumerate}
}

\begin{table}[htbp]
\caption{Future model improvements}
    \label{tab:future-work}
\begin{tabular}{
        >{\raggedright\arraybackslash}m{0.35\linewidth}
        >{\raggedright\arraybackslash}m{0.5\linewidth}}
Enhance Trust in Achievable Studies  & Unlock New Studies  \\ \hline
\modelTrustBuilders                & \modelStudyEnablers \\
    \end{tabular}
    \end{table}

\ifdefined\DISSERTATION
    \paragraph{Neglect of Surge Force and Mooring Cost}
    \Cref{sec:design-load-cases} establishes that although the dynamics model estimates the surge force, the structures model uses only the heave force to calculate the factors of safety.
    The neglect of surge force means that geometries with significant lateral areas that would experience high surge force appear to perform better than they should.
    This is not only because of the impact on the structures module, but also because the surge force drives mooring cost, an effect which is not modeled in this study.
    The model predicts that the nominal RM3 design experiences a storm surge force of \resultsAOR[surgeForceFloatNominal] on the float and \resultsAOR[surgeForceSparNominal] on the spar, compared to the \resultsAOR[heaveForceNominal] heave force.
    With mooring and foundation comprising 12\% of the capital cost (see \Cref{tab:CBS}), this effect may be substantial.
    Incorporating surge force into the structures module and adding a mooring cost module are recommended for future work.
    If these effects are implemented and confirmed to be important, then it may also be desired to obtain a more accurate estimate of the surge force across frequencies by utilizing the surge hydrodynamic coefficients instead of the long-wavelength approximation in \Cref{eq:surge-force}. 
    
    \paragraph{Nonlinear Storm Loads}
    As \Cref{sec:irregular-waves} describes, MDOcean uses standard linear wave theory with equivalent regular waves, even for the storm condition.
    However, storm waves are nonlinear, and it is generally recommended to utilize higher order potential flow, CFD simulations, or wave tank tests to obtain storm loads \citep{coe_survey_2018}.
    Additionally, even when storm wave analysis uses linear or quasi-linear hydrodynamics, other design procedures typically utilize a time-domain ``design wave elevation'' signal to capture transient peaks.
    The regular wave analysis used in MDOcean cannot capture nonsinusoidal peaks.
    Therefore, determination of design loads from a given extreme sea state represents a major area of uncertainty in this model and a challenge for future development.
    Open-source tools for WEC extreme response, such as MHKiT\footnote{\url{https://mhkit-software.github.io/MHKiT/}} and the DLC Generator\footnote{\url{https://dlc.primre.org/DLCGenerator}}, apply extreme statistics to facilitate the creation of a design wave elevation but do not calculate loads from this elevation.
    Finding a modeling methodology that captures hydrodynamic nonlinearities and transient peaks without optimization-prohibitive computational expense remains an open problem.
    Second-order MEEM can potentially address the former, but nonlinearities require solving the radiation and diffraction problems separately, and corner discontinuities present new analytical challenges \citep{cong_novel_2020,mavrakos_second-order_2009}.
    Another option is a slender-body approximation for second order loads that was recently implemented in the frequency-domain quasi-linear hydrodynamics solver RAFT \citep{carmo_slender-body_2025}.
    Meanwhile, to address the transient dynamics of an irregular design wave elevation, it requires more investigation to determine whether the semi-analytical dynamics model used here can be adequately extended.
    Alternatives include integrating second-order hydrodynamics into pseudo-spectral methods \citep{coe_initial_2020}, or an effort to speed up time-domain models where second-order hydrodynamics have recently been integrated \citep{issoglio_second-order_2026}.
    
    \paragraph{Irregular Waves}
    Besides storm loads, the regular wave assumption also affects the operational power and load calculations.
    In the future, the interaction of irregular waves with dynamic constraints could be roughly approximated in MDOcean with minimal implementation effort by computing the probability density function for each signal and saturating any amplitude above the constraint threshold.
    To more fully capture nonlinearities in irregular waves, the describing function mentioned earlier (which quantifies the fundamental amplitude of the response to a deterministic sinusoidal input) could be replaced with its probabilistic counterpart, stochastic linearization (which quantifies the expected value of the fundamental over the spectral input).
    This technique has been explored in several WEC papers \citep{da_silva_statistical_2020,da_silva_stochastic_2023,kluger_synergistic_2017,folley_spectral-domain_2016,spanos_efficient_2016}, including one that performs multi-objective design optimization \citep{neshat_enhancing_2024}.
    
    \paragraph{Probabilistic Variation in Load, Amplitude, and Power}
    Once implemented, an irregular wave formulation would allow for the quantification of the short-term variability in power, load, and response amplitude within a sea state.
    Specifically, the Wiener-Khinchin theorem describes the spectrum of the output of a linear system in terms of the input spectrum and the transfer function.
    This opens the door for investigations in other disciplines, including the development of a spectral fatigue model, insight into broadband impedance matching and dynamic constraints, and analysis of grid integration.
    A broadband matching study could consider the complementary $Z_i(\omega)$ shapes of different PTO types, WEC shapes, and wave spectra.
    Use of probabilistic amplitude constraints rather than the regular (or equivalent wave-by-wave) analysis assumed in MDOcean would allow a designer to encode a desired level of risk tolerance for extreme events, which may be more appropriate than the current approach of using the maximum nonzero joint probability density for the operational amplitude constraints. 
    
    Additionally, while MDOcean accounts for the effect of the power limit on PTO cost and models long-term power variation across sea states, its failure to model short-term power variation within a sea state means that it cannot reliably be used for sizing energy storage or grid connection.
    In particular, the temporal alignment of power production with grid demand can affect the value of the produced electricity, and therefore the economic viability.
    The authors are actively pursuing this extension, with an initial modeling process described in the study \citep{mccabe_wec_2025} and ongoing refinements in future work.
    Going further, future studies could investigate whether the addition of local or grid-connected electrical energy storage or the modulation of hydrodynamic design to smooth the dynamic response is more cost-effective at reducing power variability.
    
    \paragraph{Multi-cylinder MEEM with damping plate}
    Recall from \Cref{sec:hydro-meem} that the present study utilizes a two-cylinder MEEM formulation that neglects the slanted float bottom and damping plate.
    The multi-cylinder MEEM has recently been implemented \citep{best_openflash_2026,bimali_matrix_2026}, but has not yet been integrated into MDOcean.
    Doing so would not only improve the accuracy of the hydrodynamic coefficients, allowing calculation of the spar coefficients without relying on the approximations of \Cref{sec:hydro-other}, but also expand the reachable design space to include qualitatively different frequency-dependence of the hydrodynamic coefficients.
    Implementing the damping plate in the MEEM model would further expand the reachable design space to include spar designs that deviate from the nominal aspect ratios imposed by the interpolation in \Cref{sec:hydro-other}.
    
    \paragraph{Additional PTO nonlinearities}
    The current model includes describing functions for a force saturation nonlinearity and a drag nonlinearity.
    Other relevant nonlinearities that could be added include drivetrain static friction, hydraulic check valve rectification, and power saturation.
    It would also be possible to utilize the higher harmonics of the existing describing functions if the fundamental approximation is found to be insufficient and a fully nonlinear model is desired.
    This would require a major overhaul to the dynamics model to transition from the quasi-linear frequency domain analysis to a spectral or pseudo-spectral analysis, and might best be implemented in an existing tool such as WecOptTool.
    
    \paragraph{Generator magnetics and cost model}
    Building off the PTO CCD study that is easily realizable with the current model, a generator model would unlock control co-design with the generator itself, for example determining the conditions under which an expensive high-torque generator outperforms a cheap low-torque generator.
    The torque limit would essentially impose a constraint on the $F_{\text{max}}$ design var, the core losses introduce nonlinear damping, and impose a relation between generator torque limit and max/min generator inertia.
    This analysis could leverage a simplified generator model as explored in the RM3 CCD study \citep{anderson_re-imagining_2024}, or a full generator model as in the offshore wind MDO study \citep{barter_beyond_2023}.
    
    \paragraph{Structural model extensions}
    The structures module could be improved for geometries with more prominent stiffeners and thinner skins by adding local buckling failure criteria and effective breadth considerations.
    Confidence in the semi-analytical structures model would be enhanced by performing a systematic validation study using FEA, rather than relying on minimally-documented FEA results in the reference model report.
    
    Extending the structures model to other types of materials would allow the comparison of novel WEC materials with traditional steel designs.
    The existing relations between force and stress should still hold, but the relation between stress and factor of safety require modification for materials with different limit states.
    This is of interest because steel accounts for a significant portion of device cost, and the assumption of steel construction may be holding the WEC industry back from economic viability.
    For example, \citet{roberts_bringing_2021} finds inflatable polyurethane coated nylon fabric to be more optimal than steel, reinforced concrete, fiberglass, and rubber based on basic density and cost considerations.
    However, unlike ductile metals, composites require anisotropic analysis, concrete requires brittle fracture analysis such as Mohr's circle, and inflatables require tensioned membrane modeling.
    
    \paragraph{Lifetime and storm contour modeling}
    MDOcean's conservative use of the endurance limit to model fatigue implies an infinite structural lifetime.
    A less conservative fatigue analysis that quantifies the fatigue lifetime, such as Miner's law, could in turn couple with the economics module and wave statistics to understand the tradeoff between capital cost and lifetime.
    Explicitly modeling the storm sea state contour as a function of location and device lifetime is possible by integrating with WDRT or a similar tool.
    This is of interest in design scenarios where the WEC lifetime is a design variable.
    This may prove useful to explore the tradeoffs of single-use (set-and-forget) WECs for far-offshore sensing applications where maintenance is particularly difficult or costly.
    
    \paragraph{Different WEC archetypes}
    Extending the model beyond 2-body heaving point absorbers to other WEC archetypes such as oscillating water columns, overtopping devices, or surge/pitch devices would allow systematic quantification of the comparative benefits and drawbacks of each WEC archetype in a consistent way.
    This capability is especially valuable in the early concept design phase to inform architectural decisions and could facilitate design convergence for the field as a whole, which is considered an industry bottleneck \citep{caio_tackling_2019}.
    Generalizing MDOcean's dynamics/controls and economics modules to other archetypes would require minor adaptations including alternate degrees of freedom and cost fits, but not structural modification of the core models.
    On the other hand, the hydrodynamics and structures models would require significant modification since separate semi-analytical continuum models would need to be implemented for each new geometry.
    Before initiating such an undertaking, it would be important to confirm that numerical alternatives remain too inefficient for optimization.
    Since development of MDOcean began in 2021, new numerical tools have emerged that could potentially reduce the computational cost of mesh-based geometry-agnostic computations.
    For example, a differentiable BEM solver in Julia \texttt{MarineHydro.jl} was released in 2025 and is in its infancy \citep{khanal_fully_2025}, while a lightweight FEA tool with plate/shell elements \texttt{pynite} was released in 2021 and has seen widespread adoption.
    Whether semi-analytical or numerical, extension to other archetypes would require significant development effort but unlock new comparative insights.
\else
    The most consequential current limitations are:
\begin{itemize}[leftmargin=*]
    \item \textbf{Neglect of surge force in the structures module and absence of a mooring cost model.} Surge forces on the nominal RM3 are \resultsAOR[surgeForceFloatNominal] (float) and \resultsAOR[surgeForceSparNominal] (spar), and mooring/foundation accounts for 12\% of CAPEX (\Cref{tab:CBS}); incorporating these would likely affect optimal designs.
    \item \textbf{Regular-wave assumption in storm load cases.} Storm waves are nonlinear, and the regular-wave equivalent cannot capture transient peaks; second-order MEEM \citep{cong_novel_2020,mavrakos_second-order_2009} or the slender-body approximation in RAFT \citep{carmo_slender-body_2025} are candidate extensions.
    \item \textbf{Regular-wave assumption in operational loads.} Stochastic linearization \citep{da_silva_statistical_2020,da_silva_stochastic_2023,kluger_synergistic_2017,folley_spectral-domain_2016,spanos_efficient_2016,neshat_enhancing_2024} could replace the describing function to handle spectral inputs and enable spectral fatigue, grid-integration, and storage-sizing analyses \citep{mccabe_wec_2025}.
    \item \textbf{Single WEC archetype.} Extending MEEM to handle the damping plate or 3+ concentric cylinders \citep{best_openflash_2026,bimali_matrix_2026} would broaden applicability, as would generalizing the framework to non-point-absorber archetypes.
\end{itemize}
    
    Additional model refinements that would build trust without unlocking new study types include local buckling and effective-breadth criteria in the structures module, FEA-based validation of the structural model, additional PTO nonlinearities (drivetrain friction, check-valve rectification, power saturation), and Miner's-law fatigue analysis coupled with storm-sea-state contour modeling for finite-lifetime designs.
    Per-item rationale and implementation paths are in \citet{mccabe_dissertation_2026}.
\fi

\section{Conclusion}
This paper develops MDOcean, an open-source semi-analytical modeling framework for two-body axisymmetric point-absorber wave energy converters that integrates hydrodynamics, dynamics, control, structures, and economics in a single fast simulation.

\ifdefined\DISSERTATION
    The geometry module calculates volumes, areas, and the pitch stability margin with simple algebraic relations, and the economics module calculates design-dependent cost from structural material volume and powertrain force and power ratings.
    The hydrodynamics module implements the matched eigenfunction expansion method (MEEM) for a dual concentric cylinder geometry to obtain the float radiation coefficients, and interpolates existing BEM solutions to obtain the spar radiation coefficients due to the lack of a damping plate in the current MEEM model.
    This achieves an order of magnitude speedup compared to the popular BEM solver Capytaine.
    The dynamics module employs describing functions to extend the typical frequency-domain formulation to capture nonlinearities and dynamic constraints, including viscous drag and generator force saturation, and includes a novel analytical quadratically-constrained quadratic program (QCQP) for the constrained linear controller.
    This achieves a three-order-of-magnitude computational speedup over the common time-domain solver WEC-Sim.
    The structures module uses an equivalent-thickness model of stiffened plates and a library of canonical solutions to structural boundary-value problems to assess the factor of safety of the float bottom plate, spar shell, and damping plate in cyclic fatigue and storm load cases.
    This represents, to the authors' knowledge, the only public WEC structural simulation regardless of modeling technique.
    Finally, the economics module calculates the design-dependent cost based on structural material volume and powertrain force and power rating, performing a straightforward LCOE calculation.
\else
    The geometry module handles hydrostatics.
    The hydrodynamics module uses the matched eigenfunction expansion method (MEEM) for the float, achieving an order-of-magnitude speedup over BEM.
    The dynamics module employs describing functions and a novel analytical QCQP for constrained linear control, achieving a three-order-of-magnitude speedup over time-domain solvers.
    The structures module uses semi-analytical stiffened-plate models with realistic failure criteria for fatigue and storm load cases---to the authors' knowledge, the only public WEC structural simulation.
    The economics module includes a calibrated cost model.
    Full MEEM derivation appears in the forthcoming paper by \citet{bimali_matrix_2026}.
\fi

Validation against WEC-Sim and the RM3 reference design demonstrates that MDOcean reproduces the dynamics of an established time-domain solver to within a few percent under matched modeling assumptions and within single-digit-percent JPD-weighted annual average power under realistic conditions, despite worst-case per-sea-state errors that can be larger due to drag and inter-body phase sensitivity in the 2-DOF model.
Three tuned scale factors, on storm force, average power, and structural mass, bring the simulated outputs within 1\% of the reference model report across all major modules.
Total simulation runtime is \resultsAOR[simRuntime], which is 1-3 orders of magnitude faster than established baselines and enables multidisciplinary optimization workflows that would otherwise be prohibitive.

The methodological contribution most relevant to the broader WEC modeling community is the linearized pseudo-spectral dynamics formulation, which unifies prior describing-function and constrained-control approaches and adds a closed-form QCQP solution with a geometric interpretation on the complex plane of the reflection coefficient.
This solves analytically a problem that previously required numerical optimization, and we believe it is the first application of describing functions for both controller synthesis and evaluation in wave energy.

More broadly, the results suggest that semi-analytical multidisciplinary modeling can provide a useful intermediate between simplified conceptual studies and computationally intensive high-fidelity simulation, enabling both rapid optimization and improved physical insight during early-stage WEC development.

\ifdefined\DISSERTATION
    A key feature of this study that other WEC models lack is the inclusion of more realistic effects such as shape-informed structural stress, force saturation, amplitude limits, and a nonlinear drag model.
    Nonetheless, the model has limitations: the structural loads and stresses have non-trivial uncertainty due to the regular-wave assumption in storm load cases; the impact of irregular waves on operational power and constraints is approximated rather than directly modeled; surge force is not propagated to the structures module and no mooring cost model is included; and only a single WEC archetype (two-body axisymmetric point absorber) is considered.
    Future modeling work should pursue these enhancements alongside extensions to additional WEC archetypes, generator magnetics, and probabilistic / spectral analyses for fatigue, grid integration, and storage sizing.
\else
    Principal limitations include the regular-wave assumption in storm load cases, the approximation of irregular-wave operational dynamics through equivalent regular waves, the absence of surge force in the structures module and of a mooring cost model, and restriction to a single WEC archetype.
Detailed future-work directions are given in \Cref{sec:unmodeled-effects}.
\fi

The companion paper \citep{mccabe_leveraging_2026} applies MDOcean to a multidisciplinary techno-economic optimization of the RM3 reference design.
MDOcean is open-source and is intended to support continued community development of fast multidisciplinary WEC modeling tools that integrate insights from hydrodynamics, controls, structures, and economics in a unified framework.

\section*{Acknowledgements}

The authors thank Kapil Khanal, Yinghui Bimali, En Lo, and John Fernandez for assistance with hydrodynamics;
Fabien Royer for guidance on structures;
Ryan Coe, Jacob Mays, Patrick Reed, Nate DeGeode, and Alaa Ahmed for technical feedback on a draft manuscript; and Nola McCabe for proofreading support.

R.M. acknowledges funding from the National Science Foundation Graduate Research Fellowship.
M.D. acknowledges funding from the Fund for Undergraduate Research on Solutions to Climate Change and the Bill Nye ’77 Award in Undergraduate Research.

This material is based on work supported by the National Science Foundation Graduate Research Fellowship under Grant No.~DGE--2139899.
Any opinions, findings, conclusions, or recommendations expressed in this material are those of the authors and do not necessarily reflect the views of the National Science Foundation.


\section*{Data availability statement}

The MATLAB code for all simulation, analysis, and visualization to fully reproduce this work is available open-source via the MDOcean project at
\url{https://github.com/symbiotic-engineering/MDOcean} \citep{mccabe_mdocean_2024}.
Questions and contributions via GitHub issues and pull requests are welcomed.
All computational environments, results, and \LaTeX~artifacts can be reproduced via a data version control pipeline built with the \texttt{Calkit} package
and accessed at \url{https://calkit.io/symbiotic-engineering/mdocean}.

\appendix
\singleColMacro{

\section{Matched Eigenfunction Expansion Method}
\label{sec:appendix-meem-details}
As introduced in \Cref{sec:hydro-meem}, the hydrodynamic coefficients are computed semi-analytically using MEEM.
This section explains the MEEM formulation and solution methodology for the radiation of a dual truncated concentric cylinder geometry, originally presented by \citet{mavrakos_hydrodynamic_2004,chau_inertia_2010,chau_inertia_2012} as an extension of the single-cylinder MEEM radiation solution in the study by \citet{yeung_added_1981}.
Further numeric and realization details of the authors' implementation may be found in references \citet{bimali_matrix_2026,best_openflash_2026,mccabe_open-source_2024}.
The computation involves splitting the fluid domain into regions, approximating an infinite series by truncation, and solving a matrix equation to enforce the continuity of potential and velocity across regions. 

\subsection{Linear Hydrodynamics and Eigenfunctions}
The dynamics of a floating body in water waves are well-described by linear potential flow theory, a simplification of the Navier-Stokes equation.
This theory states that the fluid velocity field is the gradient of some complex potential $\phi$, $\vec{v}=\nabla\phi$, and $\phi$ satisfies the Laplace equation, $\nabla^2\phi=0$.
Adding the free surface condition, far-field or incident waves, and body surface and sea-bed conditions detailed in \citet{chatjigeorgiou_analytical_2018} yields a boundary value problem.
When boundary conditions correspond to the heave radiation problem (body moving vertically, no incident waves), solving for the potential $\phi(r,\theta,z)$ determines the heave added mass and damping $A_h$ and $B_h$, hereafter ``hydro coefficients.”

For appropriate geometries, the partial differential equation is separable and $\phi$ can be expressed as the product of radial, vertical, and circumferential basis functions called eigenfunctions.
In cylindrically symmetric problems, the radial eigenfunctions are a family of transcendental functions called Bessel functions.
The fluid is then divided into cylindrical regions.
Arbitrarily many fluid regions can exist, so the method applies to any axisymmetric geometry, including multiple concentric bodies that oscillate independently.
Here, two concentric cylinders and thus three fluid regions are demonstrated.
Extension to many regions is discussed in section 3.4.
\Cref{fig:meem-regions} illustrates the regions and dimensions: two internal regions \textit{i1} and \textit{i2}, and an external region \textit{e} extending to infinity.

\begin{figure}[htb]
\centering
\includegraphics{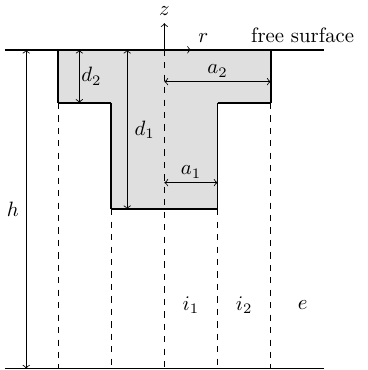}
\caption{Fluid regions and dimensions used in the dual concentric cylinder MEEM}
\label{fig:meem-regions}
\end{figure}

 The potential in each region is split into a homogeneous part for the unforced solution and a particular part due to body motion: $\phi=\phi_h+\phi_p$.
Boundary conditions dictate $\phi_p$ and the eigenfunctions for $\phi_h$ in each region, which the textbook \citet{chatjigeorgiou_analytical_2018} describes in detail.
\Cref{tab:MEEM-eigenfunctions} shows the equations originally presented in the studies by \citet{chau_inertia_2010,chau_inertia_2012} for the potential and eigenfunctions in each region, which include infinitely many unknown eigencoefficients $C_{1n}^{i1}$, $C_{1m}^{i2}$, $C_{2m}^{i2}$ and $B_{k}^{e}$.
By construction, this potential obeys all boundary conditions except for zero radial velocity on radial body surfaces.
The unknown coefficients must be computed to enforce this final condition as well as continuity across regions, which will be the subject of \Cref{sec:appendix-meem-matching}.

\begin{table}[htbp]
    \centering
    \begin{tabular}{|
    M{0.069\linewidth}|
    M{0.29\linewidth}|
    M{0.29\linewidth}|
    M{0.25\linewidth}|
    } \hline 
         Region&  $i1$&  $i2$& $e$\\ \hline 
         Homog. potential $\phi_h(r,z)$&  $\displaystyle\sum_n C_{1n}^{i1} R_{1n}^{i1}(r) Z_n^{i1}(z)$&  $\displaystyle\sum_m \left(C_{1m}^{i2} R_{1m}^{i2}(r) + C_{2m}^{i2} R_{2m}^{i2}(r) \right) Z_m^{i2}(z)$& $\displaystyle\sum_k B_k^e \Lambda_k(r) Z_k^e(z)$\\ \hline 
         Partic. potential $\phi_p(r,z)$&  $\displaystyle\frac{1}{2(h-d_1)}\left[ (z+h)^2 - \frac{r^2}{2}\right] $&  $\displaystyle\frac{1}{2(h-d_2)}\left[ (z+h)^2 - \frac{r^2}{2}\right]$& $0$\\ \hline 
         Radial eigen-function $R(r)$&  $R_{1n}^{i1}(r) = \begin{cases}
            \frac{1}{2} &  n=0 \\[1em]   
            \frac{\mathrm{I}_0(\lambda_{n}^{i1}r)}{\mathrm{I}_0(\lambda_{n}^{i1}a_{2})} & n \ge 1
        \end{cases} $&  \shortstack{$R_{1m}^{i2}(r) = \begin{cases}
            \frac{1}{2} &  m=0 \\[1em]   
            \frac{\mathrm{I}_0(\lambda_{m}^{i2}r)}{\mathrm{I}_0(\lambda_{m}^{i2}a_{2})} & m \ge 1
        \end{cases}$   \\ $R_{2m}^{i2}(r) = \begin{cases}
           \frac{1}{2}\ln(\frac{r}{a_2}) &  m = 0 \\
        \frac{\mathrm{K}_0(\lambda_{m}^{i2}r)}{\mathrm{K}_0(\lambda_{m}^{i2}a_{2})} & m \ge 1
        \end{cases}$}& $\Lambda_k(r) = \begin{cases}
           \frac{\mathrm{H}_0^{1}(m_0r)}{\mathrm{H}_0^{1}(m_0a_2)} & k = 0 \\[1em] 
          \frac{\mathrm{K}_0(m_kr)}{\mathrm{K}_0(m_ka_2)} &  k \ge 1
        \end{cases}$\\ \hline 
 Vertical eigen-function $Z(z)$& $Z_n^{i1}(z) = \begin{cases}
           1 & n=0 \\[1em]   
           \sqrt{2}\cos(\lambda_{n}^{i1}(z+h)) & n \ge 1
        \end{cases}$& $Z_m^{i2}(z) = \begin{cases}
           1 & m=0 \\[1em]   
           \sqrt{2}\cos(\lambda_{m}^{i2}(z+h)) & m \ge 1
        \end{cases}$&$    Z^{e}_{k}(z) = \begin{cases}
           N_0^{-\frac{1}{2}}\cosh( m_0(z+h)) &  k=0 \\[1em]   
           N_k^{-\frac{1}{2}}\cos( m_k(z+h)) &  k \ge 1
        \end{cases}$\\ \hline
 Eigen-value& $\displaystyle \lambda_{n}^{i1} = \frac{n\pi}{h-d_{1}},  n \geq 1$& $\displaystyle \lambda_{m}^{i2} = \frac{m\pi}{h-d_{2}}, m \geq 1$&
 $\displaystyle \begin{cases} m_0 \tanh(m_0h)= \omega^2/g, & k=0 \\ m_k \tan(m_kh) = -\omega^2/g, & k \geq 1\\ \end{cases} $\\\hline
    \end{tabular}
    \caption{Equations for potential (homogeneous and particular) and eigenfunctions (radial and vertical) for each region.}
    \label{tab:MEEM-eigenfunctions}
\end{table}

In \Cref{tab:MEEM-eigenfunctions},  $\textrm{I}_0$,  $\textrm{K}_0$, and $\textrm{H}_0^1$ are different Bessel functions of order zero; 1M and 2M mean body 1 and 2 (spar and float) are moving respectively, while 1S and 2S mean each is stationary; and the $N_k$ expression is defined as:

\begin{equation}
    N_k = \frac{1}{2}\left(1+\frac{f_k}{2m_kh}  \right)~
    \textrm{ where }
    f_k = 
    \begin{cases}
        \sinh(2m_0h), & k=0 \\ \sin(2m_kh), & k\geq1
    \end{cases}
\end{equation}

\newcommand{\RintOneDefn}{
    \shortstack{
        $\displaystyle
            \boldsymbol{\mathcal{R}}_{1j} = 
            \int\limits_{a_{in}}^{a_{out}} \vec{R}_{1j}(r)r dr$ 
        \\
            for $j=(n,m)
        $
    }
}

\newcommand{\RintTwoDefn}{
    $\displaystyle \boldsymbol{\mathcal{R}}_{2m} = \int\limits_{a_{in}}^{a_{out}} \vec{R}_{2m}(r)rdr$
}

\newcommand{\ZmnDefn}{
    \shortstack{ 
        $\displaystyle \boldsymbol{\mathcal{Z}}_{nm} = \boldsymbol{\mathcal{Z}}_{mn}^T =$
        \\
        $\displaystyle \int_{-h}^{-d_1} \vec{Z}_n^{i1 ~T}\vec{Z}_m^{i2} dz$
    }
}
\newcommand{\ZmkDefn}{
    \shortstack{
        $\displaystyle \boldsymbol{\mathcal{Z}}_{mk} = \boldsymbol{\mathcal{Z}}_{km}^T = $
        \\
        $\displaystyle \int_{-h}^{-d_2} \vec{Z}_m^{i2 ~T}\vec{Z}_k^{e} dz $
    }
}

\newcommand{\RintOneJzero}{
    $\displaystyle\frac{a_{out}^2-a_{in}^2}{4}$
}

\newcommand{\RintOneJOne}{
    $\displaystyle 
        \frac
            {a_{out} \mathrm{I}_1(\vec{\lambda}_j a_{out}) - a_{in}\mathrm{I}_1(\vec{\lambda}_j a_{in})}
            {\vec{\lambda}_j \mathrm{I}_0(\vec{\lambda}_j a_2)}
    $
}

\newcommand{\RintTwoJzero}{
    $\displaystyle
        \frac
        {
            f_{2m}a_{in}^2-a_{out}^2
        }
        {8}
    $
    where $
        f_{2m} = 1 + 2\ln\frac{a_{out}}{a_{in}}
    $
}

\newcommand{\RintTwoJOne}{
    $\displaystyle
    \frac
    {
        a_{in}\,\mathrm{K}_1 (
                \vec{\lambda}_m a_{in} )
        -a_{out}\,{\mathrm{K}}_1(
                \vec{\lambda}_m a_{out} )
    }
    {
        \vec{\lambda}_m \,{\mathrm{K}}_0 (
                \vec{\lambda}_m a_2 )
    }
    $
}

\newcommand{\ZnZeroMZero}{
    $h-d_1$
}

\newcommand{\ZnZeroMOne}{
    $\displaystyle
     \frac
                {\sqrt{2}\sin\left(
				\vec{\lambda}_m (h - d_1)
			\right)}
                { \vec{\lambda}_m}
    $
}

\newcommand{\ZnOneMZero}{
    $\displaystyle 0
    $
}

\newcommand{\ZnOneMOne}{
$\displaystyle
\begin{cases}
\frac{
            2 (-1)^{\vec{n}}
            }
            {
                    1 - \left(
                                    \frac{\vec{\lambda}_n^T}{  \vec{\lambda}_m}
                        \right)^2
            }
\frac{
            \sin\left(
                \vec{\lambda}_m(h-d_1)
         	\right)
            }
            {
            \vec{\lambda}_m
            }
,&  \lambda_n \neq \lambda_m\\
h -d_1, & \lambda_n = \lambda_m
\end{cases}
$
}

\newcommand{\ZmZeroKZero}{
    $\displaystyle \frac{\sinh(m_0(h-d_2))}{m_0 \sqrt{N_0}}$
}

\newcommand{\ZmOneKZero}{
    $\displaystyle \frac{\sqrt{2}~m_0 (-1)^{\vec{m}} \sinh(m_0(h-d_2))}{\sqrt{N_0} \left(m_0^2+{(\vec{\lambda}_m^T)}^2\right)}$
}

\newcommand{\ZkOneMZero}{
    $\displaystyle\frac{\sin(\vec{m}_k(h-d_2))}{\vec{m}_k \sqrt{\vec{N_k}}}$
}

\newcommand{\ZkOneMOne}{
    $\displaystyle
    \begin{cases}
            \frac{1}{\sqrt{2\vec{N_k}}}
            \left(
                \frac
                    {\sin\left((h-d_2)(\vec{m}_k+ \vec{\lambda}_m^T)\right)}
                    {\vec{m}_k+ \vec{\lambda}_m^T} + \right.
                 \\
                \left.~~~~~~~~~
                \frac
                    {\sin\left((h-d_2)(\vec{m}_k- \vec{\lambda}_m^T)\right)}
                    {\vec{m}_k- \vec{\lambda}_m^T} 
            \right),
        & |m_k| \neq \lambda_m
        \\
        \frac{h-d_2}{2},
        & |m_k| = \lambda_m 
    \end{cases}$
}


\Cref{tab:meem-integrals} lists several integrals of the radial and vertical eigenfunctions, $\boldsymbol{\mathcal{R}}$ and $\boldsymbol{\mathcal{Z}}$ respectively, that will be needed in the calculations to follow.

\begin{table}[htbp]
    \centering
    \caption{Eigenfunction integrals}
    \label{tab:meem-integrals}
    \begin{tabular}{|c|c|>{\centering\arraybackslash}p{0.25\linewidth}|c|} \hline 
        \multicolumn{2}{|c|}{Integral}          & $m,j=0$           & $m,j\geq1$            \\ \hline 
        \multicolumn{2}{|c|}{\RintOneDefn}      & \RintOneJzero     & \RintOneJOne          \\ \hline 
        \multicolumn{2}{|c|}{\RintTwoDefn}      & \RintTwoJzero     & \RintTwoJOne          \\ \hline 
        \multirow{2}{*}{\ZmnDefn}   & $n=0$     & \ZnZeroMZero      & \ZnZeroMOne           \\ \cline{2-4} 
                                    & $n\geq1$  & \ZnOneMZero       & \ZnOneMOne            \\ \hline
        \multirow{2}{*}{\ZmkDefn}   & $k=0$     & \ZmZeroKZero      & \ZmOneKZero           \\ \cline{2-4}
                                    & $k\geq1$  & \ZkOneMZero       & \ZkOneMOne            \\ \hline
    \end{tabular}
\end{table}

\subsection{Matching Across Fluid Boundaries}
\label{sec:appendix-meem-matching}

The eigencoefficients must be selected to enforce the radial velocity body boundary condition and the matching of the potentials and radial velocities at the edges of each region, earning this technique the name Matched Eigenfunction Expansion Method (MEEM).
The radiation problem was first solved this way for a floating cylinder in 1980 \citep{yeung_added_1981}.

First, the infinite sums in $\phi_h$ must be truncated.
Assuming truncation to $N$ terms in \textit{i1}, $M$ terms in \textit{i2}, and $K$ terms in \textit{e}, the total number of eigencoefficients to solve for is $N+2M+K$.
For a 3-region problem, there are 2 boundaries.
Thus there are four matching equations: (1) potential at $a_1$, (2) potential at $a_2$, (3) velocity at $a_1$, and (4) velocity at $a_2$.
As-is, this is not enough equations ($4 < N+2M+K$).
We must leverage eigenfunction orthogonality to get enough equations.
The first equation will turn into $N$ equations; the second and third each give $M$; the fourth $K$.
The transformation uses the following property of orthogonality.
Consider a generic function $Y(x)$ expressed as a series with coefficients $\alpha$ and basis functions $e(x)$: $Y(x)=\sum_i \alpha_i e_i(x)$.
If $e_j(x)$ is orthogonal to $e_i(x)$ from $x = a$ to $b$, then:

\begin{equation}
\begin{aligned}
       & \int\limits_a^b Y(x)e_j(x)dx = (b-a) <Y,e_j>=(b-a)<\sum_i \alpha_i e_i, e_j> \\ &= (b-a) \sum_i \alpha_i <e_i,e_j> = (b-a) \sum_i \alpha_i \delta_{ij} = (b-a) \alpha_j
\end{aligned}
\end{equation}

where $<\cdot,\cdot>$ is the inner product and $\delta_{ij}$ is Kronecker’s delta.
In the current hydrodynamics problem, the basis functions are the vertical eigenfunctions $Z_n^{i1}$, $Z_m^{i2}$, and $Z_k^{e}$.
Orthogonality of each eigenfunction can be verified with the inner product.
In the first region, for example, $<Z_{n_1}^{i1},Z_{n_2}^{i1}>=\delta_{n_1n_2}$.
Note that eigenfunctions of different domains are not orthogonal, and their inner products will be expressed as coupling integrals in \Cref{tab:meem-integrals}.

For each of the four matching equations, the property of orthogonality applies only after multiplying by the appropriate eigenfunction and integrating over appropriate bounds.
For the potential matching equations, multiply both sides by the eigenfunction of the region with smaller fluid height (so $Z_n^{i1}$ at $a_1$ and $Z_m^{i2}$ at $a_2$).
Then integrate over that fluid height ($z=-h$ to $-d_1$ at $a_1$, and $-h$ to $-d_2$ at $a_2$).
For velocity matching, multiply instead by the eigenfunction corresponding to the larger region, while still integrating over the smaller region.
In velocity matching, an extra step is required to incorporate the boundary condition of zero radial velocity along the radial surface of the body.
Since it is zero-valued, the integral of this velocity may be added to one side of the equation (the one corresponding to the velocity of the larger region) to change the integration bounds only on that side.
This manifests in the bounds of the coupling integrals to be presented in \Cref{tab:meem-integrals}.
Other combinations of eigenfunction multiplication or integration besides those described above are not useful since they result in integrating a quantity on a region where it is undefined, or a form unsuitable for the application of the orthogonality property.

\subsection{Block Matrix Structure}

Once orthogonality is applied, the matching equations create a linear system $A\vec{x}=\vec{b}$ where $A$ is a complex sparse $(N+2M+K)$x$(N+2M+K)$ square matrix corresponding to the homogeneous case, $\vec{x}=[\vec{C_{1n}^{i1}}, \vec{C_{1m}^{i2}}, \vec{C_{2m}^{i2}}, \vec{B_{k}^{e}}]$ is the complex eigencoefficient vector, and $\vec{b}$ is the real boundary condition vector corresponding to the particular case.
We elaborate on the block structure of the A-matrix and b-vector, an implementation detail that prior discussion of MEEM overlooks.
The A-matrix and b-vector block structures are shown in \Cref{tab:MEEM-A-matrix,tab:MEEM-b-vector} respectively.
They are written in compact notation using row vectors of basis functions, so $\vec{R_{1n}^{i1}}=[R_{10}^{i1}, R_{11}^{i1}, ..., R_{1(N-1)}^{i1}]$ and so on.
Each basis function is evaluated at the radius described to the left of its row in the table. $0_{ij}$ and $1_{ij}$ are the $i$ x $j$ matrices of zeros and ones respectively; diag($\cdot$) constructs a diagonal matrix from a vector; and $\odot$ is the Hadamard (element-wise) product.

\begin{table}[htbp]
    \centering
    \caption{MEEM A-matrix}
    \label{tab:MEEM-A-matrix}
    \begin{tabular}{|M{0.12\linewidth}|c||c|c|c|c|}\hline
 & & $\vec{C}_{1n}^{i1}$& $\vec{C}_{1m}^{i2}$& $\vec{C}_{2m}^{i2}$&$\vec{B}_k^e$\\\hline 
          &size&  N&  M&  M& K\\ \hline \hline 
          \shortstack{$\phi^{i1}=\phi^{i2}$ \\ at $r=a_1$}&N&  $(h-d_1)~\mathrm{diag}(\vec{R}_{1n}^{i1})$&  $-\boldsymbol{\mathcal{Z}}_{nm}\odot 1_{N1}\vec{R}_{1m}^{i2}$&  $-\boldsymbol{\mathcal{Z}}_{nm}\odot 1_{N1}\vec{R}_{2m}^{i2}$& $0_{NK}$\\ \hline 
          \shortstack{$\phi^{i2}=\phi^{e}$ \\ at $r=a_2$}&M&  $0_{MN}$&  $(h-d_2)~\mathrm{diag}(\vec{R}_{1m}^{i2})$&  $(h-d_2)~\mathrm{diag}(\vec{R}_{2m}^{i2})$& $-\boldsymbol{\mathcal{Z}}_{mk}\odot 1_{M1}\vec{\Lambda}_{k}$\\ \hline 
          \shortstack{$\frac{\partial}{\partial r}\phi^{i1}=\frac{\partial}{\partial r}\phi^{i2}$ \\ at $r=a_1$}&M&  $- \boldsymbol{\mathcal{Z}}_{mn} \odot 1_{M1} \frac{\partial}{\partial r}\vec{R}_{1n}^{i1}$&  $(h-d_2)~\mathrm{diag}(\frac{\partial}{\partial r}\vec{R}_{1m}^{i2})$&  $(h-d_2)~\mathrm{diag}(\frac{\partial}{\partial r}\vec{R}_{2m}^{i2})$& $0_{MK}$\\ \hline 
          \shortstack{$\frac{\partial}{\partial r}\phi^{i2}=\frac{\partial}{\partial r}\phi^{e}$ \\ at $r=a_2$}&K&  $0_{KN}$&  $-\boldsymbol{\mathcal{Z}}_{km} \odot 1_{K1}\frac{\partial}{\partial r}\vec{R}_{1m}^{i2}$&  $-\boldsymbol{\mathcal{Z}}_{km}\odot 1_{K1}\frac{\partial}{\partial r}\vec{R}_{2m}^{i2}$& $h~\mathrm{diag}(\frac{\partial}{\partial r}\vec{\Lambda}_k)$\\ \hline
    \end{tabular}
\end{table}

\begin{table}[htbp]
    \centering
    \caption{MEEM b-vector}
    \label{tab:MEEM-b-vector}
    \begin{tabular}{|c|c|} \hline 
         N& $\displaystyle
\int_{-h}^{-d_1} (\phi_p^{i2} - \phi_p^{i1})\vec{Z}_n^{i1~T} dz
$ \\ \hline 
         M& $\displaystyle-
\int_{-h}^{-d_2} \phi_p^{i2} \vec{Z}_m^{i2~T} dz
$\\ \hline 
         M& $\displaystyle
\int_{-h}^{-d_1} \frac{\partial}{\partial r}\phi_p^{i1}\vec{Z}_m^{i2~T} dz - \int_{-h}^{-d_2}\frac{\partial}{\partial r}\phi_p^{i2}\vec{Z}_m^{i2~T} dz
$\\ \hline 
         K& $\displaystyle
\int_{-h}^{-d_2} \frac{\partial}{\partial r}\phi_p^{i2} \vec{Z}_k^{e~T} dz
$\\ \hline
    \end{tabular}
\end{table}

The dense blocks contain coupling integrals $\boldsymbol{\mathcal{Z}}$ of the vertical eigenfunctions.

Of the sixteen blocks that make up the matrix, six are diagonal, four are zero, and six are dense, resulting in the sparsity pattern shown in \Cref{fig:sparsity}.
An even sparser matrix could be obtained with the alternate eigenfunction scaling for the second region described in \citet{chau_inertia_2012}. 
\begin{figure}[htbp]
    \centering
    \includegraphics[width=0.45\linewidth]{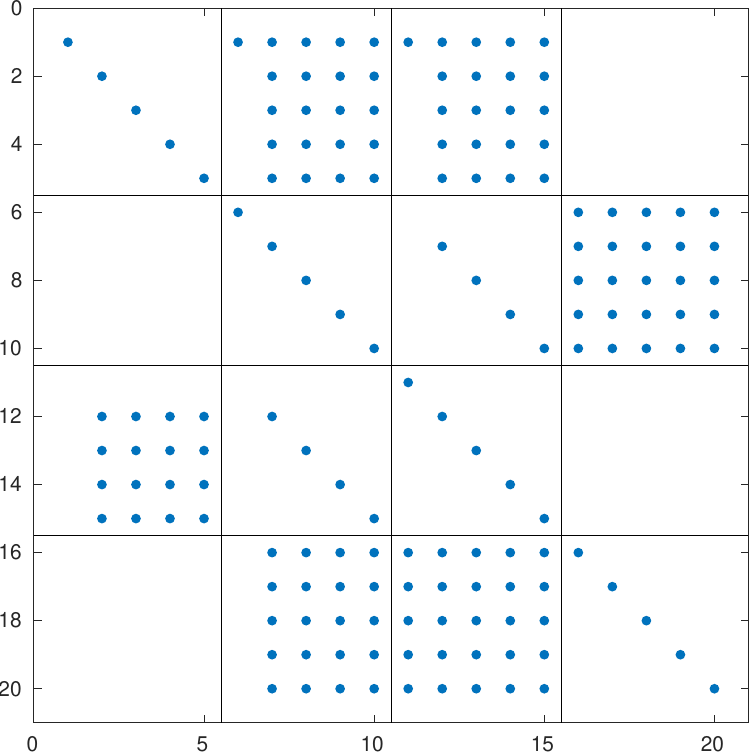}
    \caption{A-matrix sparsity pattern, shown for $N=M=K=4$}
    \label{fig:sparsity}
\end{figure}

\subsection{Calculation of Outputs}

Once the eigencoefficients have been calculated by solving the linear system for $\vec{x}$, the hydrodynamic radiation coefficients $A_h$ and $B_h$ are found by integrating the potentials and velocities over the body surface:
\begin{alignat}{5}\label{eq:meem-c-derivation}
    A_h + \frac{iB_h}{\omega}&=\rho h^3  \int_{\theta=0}^{2\pi} 
    & \int_{r=a_{in}}^{a_{out}}
    &\phi(r,z) \frac{\partial \phi(r,z)}{\partial z}
    r dr d\theta  \nonumber \\
    &= 2\pi \rho h^3 
    & \int_{r=a_{in}}^{a_{out}}&
   \left(\phi_p(r,z) + 
         \sum_{j_1} C_{j_1} R_{j_1}(r)Z_{j_1}(z)
   \right) \nonumber \\
  & & &  \left(\frac{\partial \phi_p(r,z)}{\partial z} 
               + \sum_{j_2} C_{j_2} R_{j_2}(r)
               \frac{\partial Z_{j_2}(z)} {\partial z} 
        \right)r dr \nonumber \\
    &= 2\pi \rho h^3 & \left[
    \int_{r=a_{in}}^{a_{out}} \right. &
    \left. \phi_p(r,z)
        \frac{\partial \phi_p(r,z)}{\partial z} rdr
    \right.
    \\
    & &  + \sum_{j_1} &C_{j_1} Z_{j_1}(z)
    \int_{r=a_{in}}^{a_{out}}R_{j_1}(r)
    \frac{\partial \phi_p(r,z)}{\partial z}rdr \nonumber
    \\
    &&   + \sum_{j_2}&
    \frac{\partial Z_{j_2}(z)}{\partial z} C_{j_2}
    \int_{r=a_{in}}^{a_{out}}
    \phi_p(r,z)  R_{j_2}(r)rdr \nonumber
    \\
    & & +  \sum_{j_1}& 
    \left.
        C_{j_1}Z_{j_1}(z) \sum_{j_2} C_{j_2} 
        \frac{\partial Z_{j_2}(z)}{\partial z}
        \int_{r=a_{in}}^{a_{out}} 
        R_{j_1}(r)R_{j_2}(r)rdr
    \right] \nonumber
\end{alignat}
where the generic indices $j_1$ and $j_2$ are used to represent $n$ or $m$ depending on the region.
For regions with multiple radial eigenfunctions or surfaces covering multiple fluid regions, summation over each eigenfunction and region respectively is implied.
After substituting the eigenfunctions from \Cref{tab:MEEM-eigenfunctions} into \Cref{eq:meem-c-derivation}, all $z$-dependent quantities are evaluated at the body draft on the integration surface, $z=-d$.
This means $\frac{\partial \phi_p}{\partial z}=1$, and $Z_j(z)$ simplifies to
\begin{equation}
    Z_j(z=-d) = \begin{cases}1,& j=0\\ \sqrt2(-1)^j,& j \geq 1\end{cases}
\end{equation}
The third and fourth integrals of \Cref{eq:meem-c-derivation} vanish, and the second integral is the radial eigenfunction integral $\boldsymbol{\mathcal{R}}$ expressed previously in \Cref{tab:meem-integrals}. 

The first and second term are independent of and scale linearly with the eigencoefficients $\vec{x}$, respectively.
The radiation coefficients are thus computed as
\begin{equation}\label{eq:hydro}
    A_h + \frac{iB_h}{\omega}=2\pi \rho h^3(c_0 + \vec{c}\cdot\vec{x}) =2\pi \rho h^3(c_0 + \vec{c} \cdot A^{-1} \vec{b})
\end{equation}
with output constant $c_0$ and output row vector $\vec{c}$ defined in \Cref{eq:c0,tab:MEEM-c-vector} respectively:
\begin{equation}\label{eq:c0}
    c_0 =  \frac{\left({a^2_{out}}-{a^2_{in}}\right)\,\left(-{a^2_{in}}-{a^2_{out}}+4(h-d)^2\right)}{16\,\left(h - d\right)}
\end{equation}
\begin{table}[h]
    \centering
    \caption{Output vector $\vec{c}$}
    \label{tab:MEEM-c-vector}
\begin{tabular}{|l|c|c|c|c|} \hline 
          Body&N&  M&  M& K\\ \hline 
          Float&$0_{1N}$&  $\vec{Z}_m(z=-d) \odot\boldsymbol{\mathcal{R}}_{1m}$&  $\vec{Z}_m(z=-d) \odot\boldsymbol{\mathcal{R}}_{2m}$& $0_{1K}$\\ \hline
 Spar& $\vec{Z}_n(z=-d) \odot\boldsymbol{\mathcal{R}}_{1n}$& $0_{1M}$& $0_{1M}$&$0_{1K}$\\\hline
    \end{tabular}
    \end{table}
The appropriate dimensions are
\begin{equation}
    a_{out} = \begin{cases}
        a_2, & \text{float} \\
        a_1, & \text{spar}
    \end{cases} 
    \qquad
    a_{in} = \begin{cases}
        a_1, & \text{float} \\
        0, & \text{spar}
    \end{cases} 
    \qquad
   d = \begin{cases}
        -d_2, & \text{float} \\
        -d_1, & \text{spar}
    \end{cases} 
\end{equation}
to obtain the hydrodynamic coefficients of the float or the spar respectively.

Using the A-matrix and b- and c-vectors, the MEEM solution directly calculates radiation coefficients $B_h$ and $A_h$.
The hydrostatic stiffness $K_h$ and the excitation coefficient $\gamma$ are calculated as follows:
\begin{equation}\label{eq:gamma-K}
    |\gamma|  = \sqrt{\frac{ 4 \rho_w g V_g  B_h} {m_0}}, \quad 
   \angle \gamma = -\frac{\pi}{2} + \angle\frac{ B_{k=0}^e}{\textrm{H}_0^{1}(m_0 a_2)},\quad
    K_h       = \rho_w g \underbrace{\frac{\pi}{4} (D_f^2 - D_s^2)}_{A_{w,f}},\quad  
    K_s = \rho_w g \underbrace{\frac{\pi}{4} D_s^2}_{A_{w,s}}
\end{equation}
where $g$ is the acceleration due to gravity, $\rho_w$ is the density of water, $m_0$ is the wavenumber, $V_g$ is the finite depth group velocity, $\textrm{H}_0^1$ is the zeroth-order Hankel function of the first kind, and $A_w$ is the waterplane area \citep{newman}.
This method of calculating excitation from damping is the well-known Haskind relation.
Note that while the excitation magnitude $|\gamma|$ depends on the radiation damping $B_h$ which in turn depends on all the inner region eigencoefficients ($\vec{C}_{m}^{i2}$ for float excitation and $\vec{C}_{1n}^{i1}$ for the spar excitation), the excitation phase $\angle\gamma$ depends only on the first exterior eigencoefficient, $B_{k=0}^e$. 

\subsection{Verification and Validation}

Hydro coefficient results are verified by comparing to a benchmark shallow-water concentric-cylinder MEEM solution in \citet{chau_inertia_2012}.
Excellent agreement is observed, shown in \Cref{fig:meem-yeung-validation}.
The results are also experimentally validated in the studies \citet{chau_inertia_2012,son_performance_2016}.
Previously in \Cref{sec:dynamic-validation}, the N=M=K=11 case was compared to WAMIT results for RM3 in deep water.

\begin{figure}[htbp]
    \centering
    \includegraphics[width=0.5\linewidth]{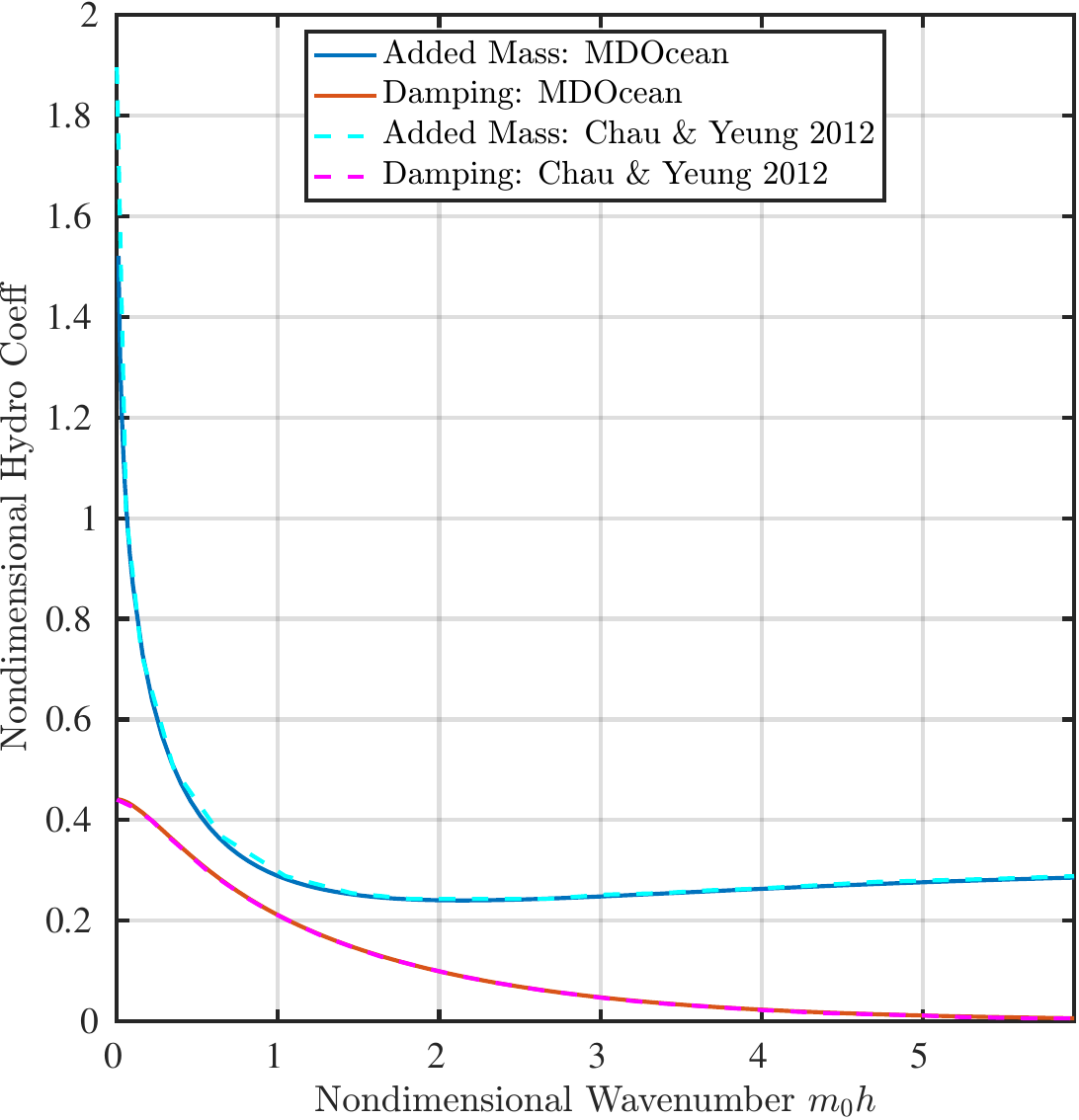}
    \caption{Nondimensional added mass and damping coefficient validation against \citep{chau_inertia_2012}}
    \label{fig:meem-yeung-validation}
\end{figure}

\subsection{Convergence}

As $N,M,K\rightarrow\infty$, matching quality improves, and hydro coefficients converge toward their true values.
Previous MEEM papers use $N=M=K=50$ to obtain 4-digit matching accuracy without elaborating on convergence properties \citep{chau_inertia_2012}.
We observe that potential matching converges faster than velocity matching.
\Cref{fig:meem-matching} shows the matching behavior for $N=M=K=11$, where potential matches well but velocity still has noticeable mismatch. 
\begin{figure}[htbp]
    \centering
    \includegraphics[width=0.5\linewidth]{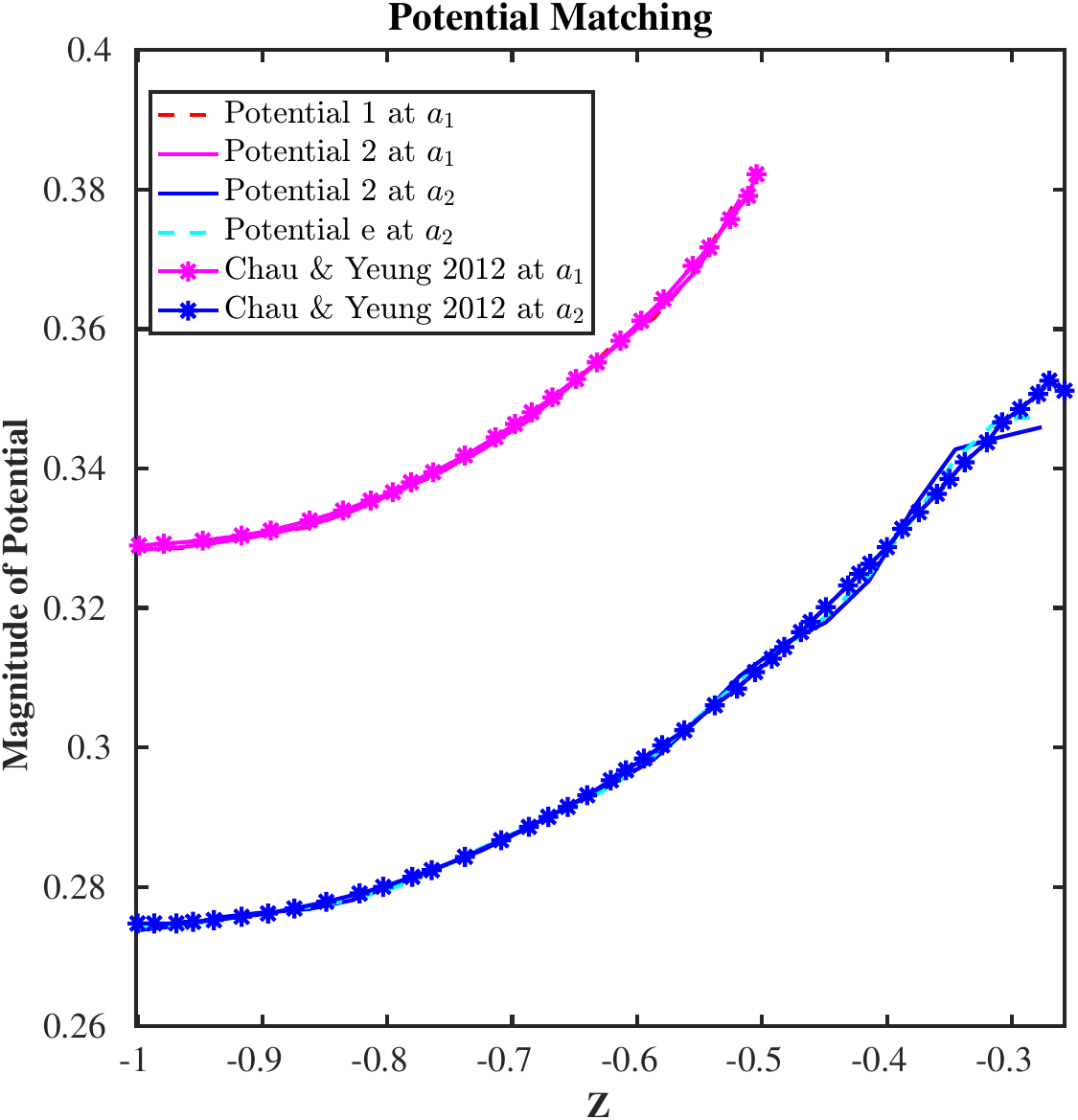}
    \caption{Matching for $N=M=K=11$ for benchmark geometry}
    \label{fig:meem-matching}
\end{figure}

Hydro coefficient convergence depends on the geometry: the benchmark shallow-water geometry of \citep{chau_inertia_2012} converges to within 0.25\% with only $N=M=K=4$, but RM3 requires $N=M=K>10$, shown in \Cref{fig:meem-convergence}.
There, damping converges well at low frequencies but requires more harmonics at higher frequencies, while added mass has similar convergence across frequencies. 
\begin{figure}[htbp]
    \centering
    \includegraphics[width=0.5\linewidth]{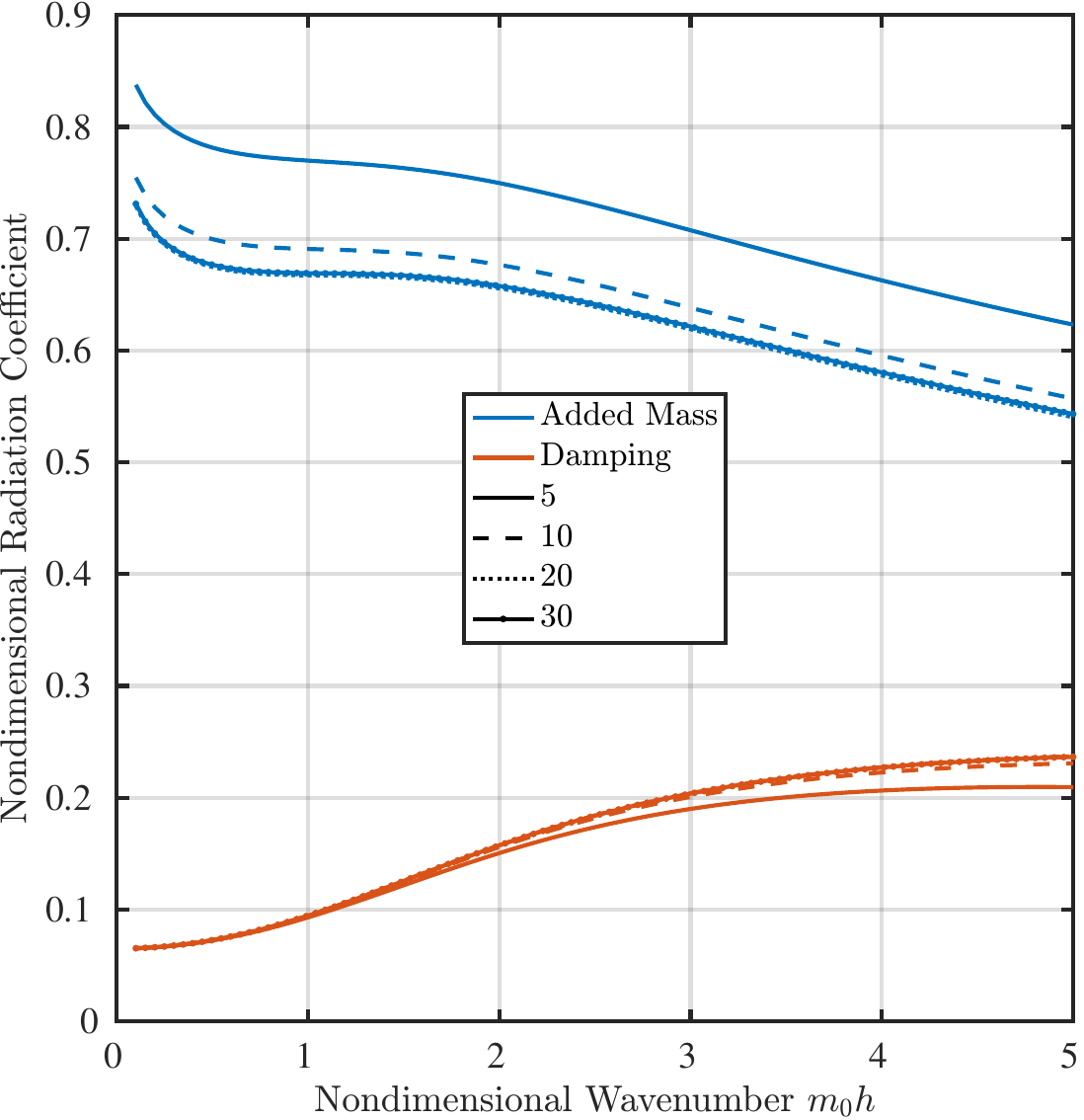}
    \caption{Convergence for $N=M=K=(5,10,20,30)$ for RM3}
    \label{fig:meem-convergence}
\end{figure}

\subsection{Numerical Notes}\label{sec:appendix-meem-numerics}
Note that radial eigenfunctions $R_{1n}^{i1}(r)$ and $R_{1m}^{i2}(r)$ contain the modified Bessel function of the first kind $\mathrm{I}_0(\chi)$, which diverges for $\chi\rightarrow \infty$, resulting in numeric overflow. 
To prevent overflow, each fluid region must exceed a minimum height $ \Delta z_{\text{min}}$ that is proportional to the region diameter $D$ for both the spar ($\Delta z_s, D_s$) and the float ($\Delta z_f, D_f$), with variables shown in \Cref{fig:meem-geom}.
The constant of proportionality between $\Delta z_{\text{min}}$ and $D$ is a function of the number of harmonics $N_{\text{harmonics}}$ used in that region ($N$ for spar, $M$ for float), as well as the maximum argument to the \texttt{besseli} function in MATLAB before overflow, $\chi_{\text{max}}$:
\begin{equation}\label{eq:delta-z-min}
    \frac{\Delta z_{min}}{D} = \frac{\pi N_{harmonics}}{2\chi_{max}} \approx \frac{N_{harmonics}}{446}.
\end{equation}
By trial, $\chi_{\text{max}}$ is found to be $\approx 700.5$, close to the theoretical value of \texttt{log(realmax)} $\approx 709.8$ for exponential scaling.
Since $N_{\text{harmonics}}=10$ gives adequate convergence for most geometries, this condition is trivially satisfied for nearly all floating bodies of practical relevance, but must still be added as a constraint to prevent the optimizer from exploiting the numerical divergence.
In future work, if high-accuracy solutions are desired for large bodies close to the sea floor, exponentially scaled Bessel functions could be used instead, with symbolic cancellation of the exponential scaler from the eigenfunction numerator and denominator.

Likewise, the vertical eigenfunction $Z_k^e$ for $k=0$ contains the $\cosh$ and $\sinh$ functions, which diverge for large values of $m_0h$ (high frequencies or deep water).
Since the largest relevant value of $m_0h$ depends on the site rather than on the WEC design, it is not possible to add geometric constraints to prevent overflow as it was above.
Therefore, the limit is derived analytically. 
\begin{equation}
    \lim_{m_0h\rightarrow\infty} Z_0^e(z)= \frac{\cosh^2(m_0h)}{\sqrt{2m_0h}}\exp\left(1+\frac{z}{h}\right)
\end{equation}
Plugging this into the first element of the bottom block of the b-vector (position $N+2M+1$) results in 
\begin{equation}
    \lim_{m_0h\rightarrow\infty}b_{N+2M+1}=\frac{-a_2}{h-d_2} \sqrt{\frac{h}{2m_0}}\exp(-d_2m_0)
\end{equation}
\begin{figure}[htbp]
    \centering
    \includegraphics[width=.5\linewidth]{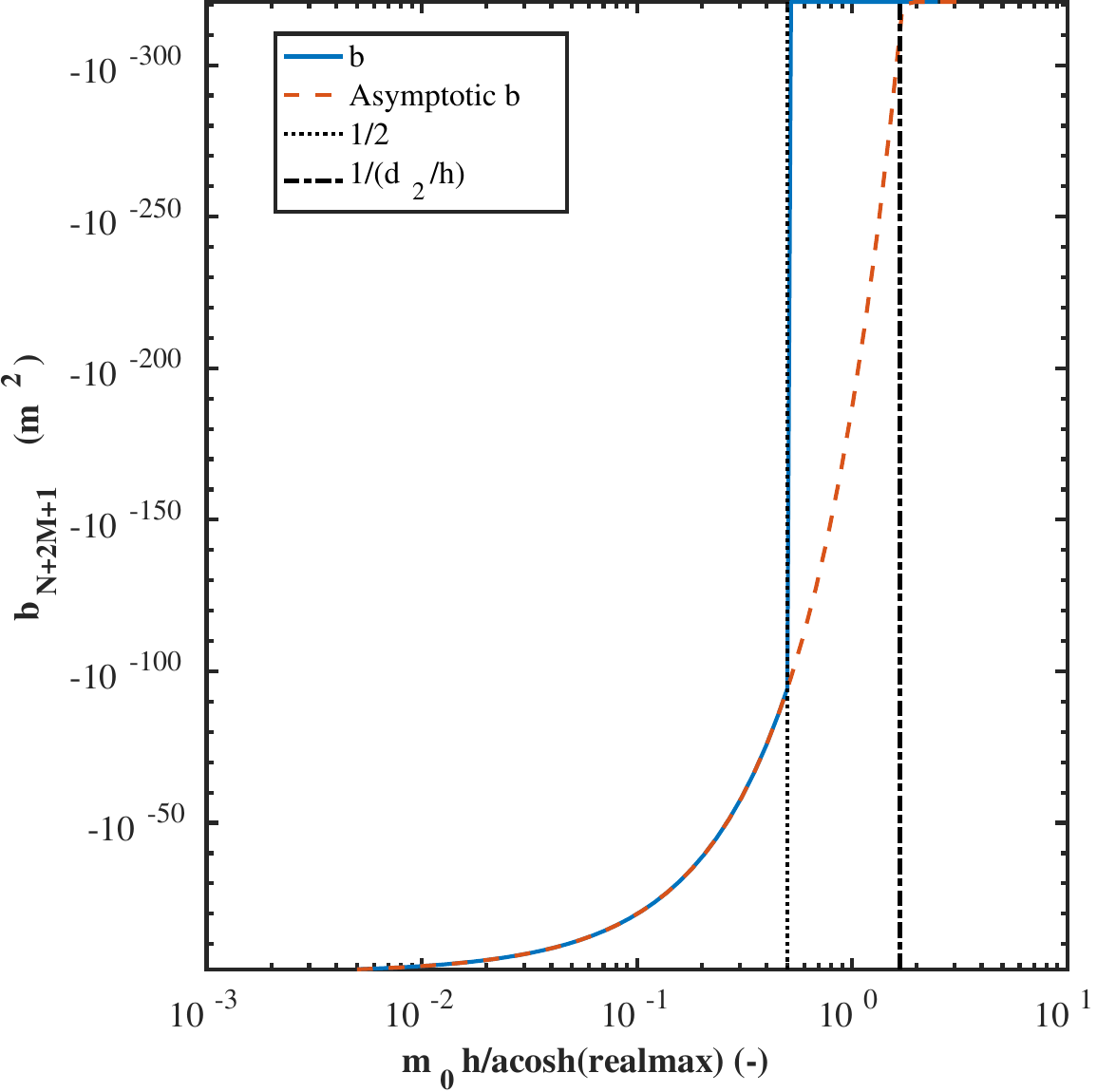}
    \caption{Asymptotic b-vector for large $m_0h$}
    \label{fig:meem-b-limit}
\end{figure}

\Cref{fig:meem-b-limit} shows the asymptotic approximation as well as the standard expression for the relevant b-vector entry as a function of nondimensional ratio $m_0h/\textrm{acosh(realmax})$.
With the standard expression, values for inputs above $\frac{1}{2}$ become \texttt{NaN}, which poses problems in deep water.
With the asymptotic approximation, we are able to avoid \texttt{NaN}s for the zone $\frac{1}{2}<m_0h/\textrm{acosh(realmax})<\frac{h}{d_2}$.
Above this threshold, \texttt{NaN}s occur even with the approximation, and the value in the b-vector is set to zero.
This introduces negligible error because in this range, the magnitude of the true value is less than $10^{-300}$ for the nominal RM3 geometry.
The asymptotic approximation is accurate for values of $m_0h/\textrm{acosh(realmax})$ as low as $10^{-2}$, though it is only applied for values above $\frac{1}{2}$.

The corresponding limit for the vertical coupling integral is:
\begin{equation}
\begin{aligned}
    \lim_{m_0h\rightarrow\infty}\boldsymbol{\mathcal{Z}}_{m,k=0} &
    = h~\frac{\cosh^2(m_0h)}{\sqrt{2m_0h}}\cdot \frac{-1+(-1)^m\exp(1-\frac{d_2}{h})}{f_{m}} \\
    \text{where}~~f_m&= \begin{cases}
        1, & m=0 \\
        h^2\lambda_m^2+1, & m \geq 1
    \end{cases}
    \end{aligned}
\end{equation}

A final numerical subtlety worth discussing is finite precision effects in calculating $m_k$.
Bounds of $180^\textrm{o}\cdot[k-\frac{1}{2}, k]$ are placed on $m_kh$ in a root-finding algorithm to ensure the $k$th root is identified.
Without these bounds, sometimes the \texttt{fsolve} solver would wrongly converge to values where the residual approaches $+\infty$ from one side and $-\infty$ from the other (asymptotes of $\tan()$) rather than true zeroes.
Degrees are used instead of radians so asymptotes occur at rational values.

\subsection{Runtime and Computational Cost Scaling}

The runtime of the MEEM method is the time required to find the eigencoefficients, then obtain the hydrodynamic coefficients from eigencoefficients.
First, a nonlinear root-finding algorithm runs $K-1$ times to generate the $m_k$ inputs used in the A-matrix and b-vector.
Then $3N+8M+2K-11$ Bessel functions must be evaluated for the radial terms of the A-matrix. 
(Note that this would reduce to $2N+8M+2K-10$ evaluations if the definition of the $R^{i1}_{1n}(r)$ eigenfunction were modified to use a scale factor based on $a_1$ rather than $a_2$, although this enhancement is not pursued here to maintain consistency with previous work).
The cost of evaluating vertical coupling integrals (\Cref{tab:meem-integrals}) is negligible since they are trigonometric.
Linear solves scale almost cubically with matrix size, so this step scales with $(N+2M+K)^3$.
The radial integrals for the c-vector do not require evaluating Bessel functions with any arguments that were not already evaluated for the A-matrix.
For $N=M=K=10$, the simulation of a single frequency averages \resultsAOR[MEEMRuntime] on the hardware described in \Cref{sec:sim-runtime}.
\Cref{fig:runtime-hydro} shows the time breakdown.
Most of the time is spent evaluating Bessel functions, so future code optimization should focus on speeding up Bessel evaluations, such as with lookup tables.
\citet{chau_inertia_2012} proposes using the sparsity pattern to reduce matrix size from N+2M+K to 2M, but this seems low impact since the linear solve only takes a few percent of compute time and the reduced matrix would require an equal number of Bessel evaluations.
On the other hand, matrix size in a boundary element method solver is much larger (meshes may have 1000s of panels) and the linear solve can drive computation cost.
On the same machine, Capytaine boundary element method for the same geometry takes an average of 323 ms for a 710 panel mesh (1\% convergence). 
Thus, MEEM achieves a 10x time reduction over Capytaine.

\section{Additional Hydrodynamics Details}
\label{sec:appendix-additional-hydro}
\subsection{Spar Coefficients}
As \Cref{sec:hydro-other} introduces, because of the importance of the damping plate, the spar hydrodynamic coefficients do utilize other methods besides MEEM.
We approximate the spar added mass coefficient $A_s$ as a frequency-independent quantity using a formula proposed by \cite{philip_damping_2012} based on the displaced water volume in the spherical projection of the damping plate:
\begin{equation}\label{eq:As}
 A_s =\frac{1}{3}\rho_w D_d^3\left(1 - \frac{3}{4}r^2\sqrt{1-r^2} - \frac{1}{4}(1-\sqrt{1-r^2})^2(2+\sqrt{1-r^2})\right),
\end{equation}
where $r = D_s/D_d$.
This expression smoothly transitions from a pure plate limit ($\frac{1}{3}\rho_w D_d^3$ at $r=0$) to a pure column limit ($\frac{1}{6}\rho_w D_d^3$ at $r=1$).

The other spar coefficients are calculated by fitting and interpolating data from WAMIT for the nominal design, then adjusting for known dimensional scalings based on differences from the nominal design.
To derive the dimensional scalings, we use the numerical results that \cite{olaya_hydrodynamic_2015} present for a variant of MEEM that incorporates the damping plate.
By plotting the data for a variety of geometries on a single plot and scaling the axes by dimensionless geometric ratios until the points begin to overlap as a single trend,
we empirically arrive at the following form for the excitation coefficient magnitude:
\begin{equation}\label{eq:scale-wamit}
|\gamma_s| = \rho g \pi D_s^2/4 |H_0(kR_x)| \exp(-k e_1) kh \alpha \max(1,\alpha) 
\sqrt{\frac{\Upsilon\left(\frac{kh R_x \alpha^3}{\beta R_p}\right)}
           {\beta\left(1+\frac{\beta}{\alpha}\right)}}
\end{equation}
where $\Upsilon(\cdot)$ indicates a nonlinear relationship to be interpolated from data, and $\alpha$, $\beta$,  $e_1$, $R_p$, and $R_x$ are dimensions and nondimensional ratios defined in \citet{olaya_hydrodynamic_2015}.
\Cref{fig:olaya-hydro-data} shows this nonlinear relationship, which has collapsed remarkably well for the variety of $\alpha$ and $\beta$ values.
Case numbers in the legend refer to the numerical parameters used by \citet[Section~V.]{olaya_hydrodynamic_2015}
\begin{figure}[htbp]
\centering
    \includegraphics[width=.8\linewidth]{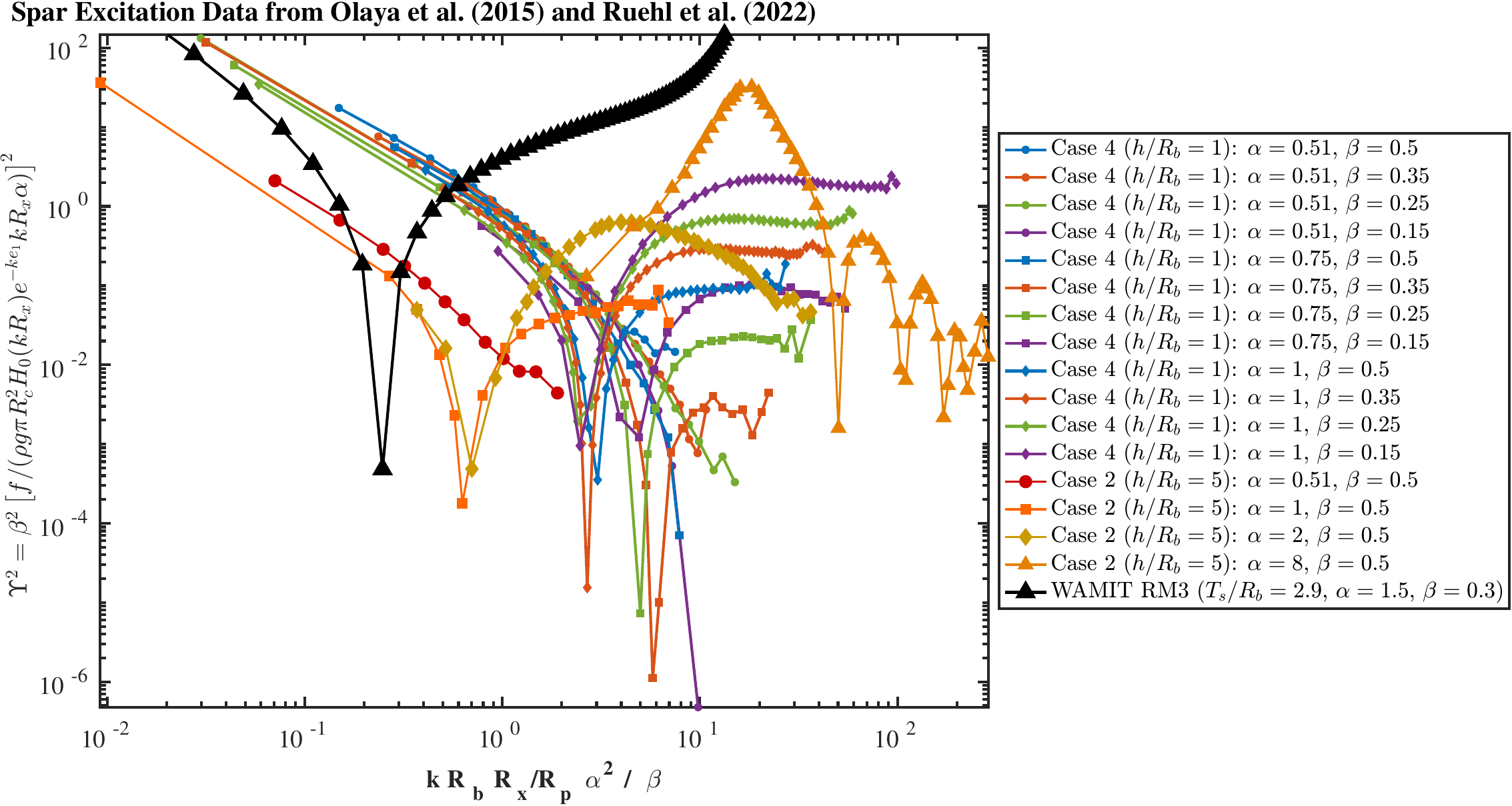}
\caption{Nondimensionalized hydrodynamic data used to interpolate spar coefficients, showing the rebound effect.}
\label{fig:olaya-hydro-data}
\end{figure}
The dip at mid frequencies shows the excitation and damping coefficients experiencing ``rebound'' effects where rather than slowly asymptoting to zero like those of a single cylinder, they go quickly to zero, increase to a peak, then slowly asymptote down to zero again.

\subsection{Positive-definiteness}
Computing the float coefficients in a different way than the spar and coupling coefficients introduces the possibility of violating physical requirements of the solution.
In particular, the damping matrix must be positive definite because it represents a dissipative force where outgoing waves remove energy from the system, and the added mass matrix must be positive definite because it represents the kinetic energy of the fluid which must be non-negative.
\begin{equation}\label{eq:open-loop-stabilty}
    \det(\mathbf{A}_{h,op})=A_f A_s - A_c^2>0 \,, \quad
    \det(\mathbf{B}_{h,op})=B_f B_s - B_c^2>0
\end{equation}
This positive-definiteness may be violated for certain combinations of coefficients and can cause the system dynamics to become unstable.
Even when other forces such as body inertia, drag, friction, and control stabilize the system, violation of the conditions of \Cref{eq:open-loop-stabilty} can artificially inflate power production.
These conditions must be checked for each sea state and the coefficients adjusted if needed.
MDOcean chooses to decrease the magnitude of the coupling coefficients $|A_c|$ and $|B_c|$ if a violation occurs because the coupling coefficients are hypothesized to correlate less strongly than the float and spar coefficients with the float and spar dimensions respectively.
This intends to lower the likelihood of creating saturation behavior (where a dimension can be perturbed without influencing power production) that could interfere with a gradient-based optimization routine.

\section{Dynamics and Control Module Details}
This section provides supplementary derivations and implementation details for the dynamics and control module.

\subsection{PTO Model Details}
\label{sec:appendix-pto-model}
We next turn our attention to the PTO kinematics and dynamics, to specify the generic impedance matrix $\mathbf{Z}_p$ in \Cref{eq:eom-freq-domain}.
A PTO model must capture the dynamics of any elements that transmit, transduce, or condition power between the rigid-body mechanical energy and the end energy product, in this case electricity.
Relevant dynamics include the inertia, stiffness, static friction, mechanical advantage, losses, and electromagnetic coupling of flywheels, shafts, springs, gears, hydraulic circuits, generators, and power electronics.
Reference \cite{penalba_review_2016} reviews WEC PTO models and reference~\cite{coe_co-design_2025} synthesizes these into a multiport impedance matching network framework to facilitate control co-design.
Reference \cite{coe_co-design_2025} presents a PTO model in cascade matrix form for a single degree of freedom system. 
This section provides a general structure in effort-flow form applicable across multiple energy domains for any number of degrees of freedom, as well as specific equations for a common PTO structure consisting of a mechanical drivetrain and a permanent magnet generator.

\subsubsection{Multiport Circuit Conventions}
\label{sec:appendix-multiport}

The intrinsic, powertrain, and Th\'evenin-equivalent representations used throughout \Cref{sec:dynamics} are all instances of multiport circuit models, with effort variables $\vec{e}$ and flow variables $\vec{q}$ at each port.
Following \cite{reveyrand_multiport_2018}, three matrix forms describe the relationships between port variables: impedance ($\mathbf{Z}$), cascade ($\mathbf{a}$, also called ABCD), and hybrid ($\mathbf{h}$):
\begin{equation}\label{eq:multiport-matrix-definitions}
\vec{e} = \mathbf{Z}\,\vec{q},
\qquad
\begin{bmatrix} \vec{e}_2 \\ \vec{q}_2 \end{bmatrix}
= [\mathbf{a}]_{2\leftarrow 1}
\begin{bmatrix} \vec{e}_1 \\ \vec{q}_1 \end{bmatrix},
\qquad
\begin{bmatrix} \vec{e}_1 \\ \vec{q}_2 \end{bmatrix}
= [\mathbf{h}]_{1,2}
\begin{bmatrix} \vec{q}_1 \\ \vec{e}_2 \end{bmatrix}.
\end{equation}
These equations apply to both time-domain and phasor representations of each port.
The three forms differ in how the ports are grouped:
\begin{itemize}
    \item \textbf{Impedance} ($\mathbf{Z}$) groups all ports together, describing the effort required to produce a unit of flow.
The intrinsic dynamics matrix $\mathbf{Z}_i$ and the PTO impedance $\mathbf{Z}_p$ in \Cref{eq:eom-freq-domain,eq:intrinsic-impedance} both use this form.
    \item \textbf{Cascade} ($\mathbf{a}$) divides the ports into two groups and describes how effort and flow transmit from one group to the other.
The arrow subscript indicates direction: $[\mathbf{a}]_{2\leftarrow 1}$ maps port group 1 to port group 2.
Reversing direction inverts the matrix,
    \begin{equation}\label{eq:cascade-inverse}
        [\mathbf{a}]_{1\leftarrow 2} = [\mathbf{a}]_{2\leftarrow 1}^{-1},
    \end{equation}
    and sequential multiplication represents serial composition of two-port elements.
This form is used for the PTO dynamics (drivetrain followed by generator).
    \item \textbf{Hybrid} ($\mathbf{h}$) mixes the two conventions, expressing one effort and one flow at the two port groups as functions of the complementary variables.
This form is used for the PTO kinematics, where it naturally accommodates underactuation.
\end{itemize}

Conversion between the three forms is accomplished using the formulas in reference \cite{reveyrand_multiport_2018}.
Some port relationships are physically well-defined in one convention but degenerate in another (for example, an ideal current source has no impedance representation but a well-defined hybrid representation) so the convention is chosen to match the physics of each subsystem.

The sign convention adopted here treats both flows as positive when entering the multiport, matching \cite{coe_co-design_2025}.

\subsubsection{Time-Domain Expansion of Instantaneous Power}
\label{sec:appendix-power-time}

Equation~\eqref{eq:power-PQS} can be used to rewrite the instantaneous power $p(t)$ at any port as \citep{saadat_power_1999}
\begin{equation}\label{eq:power-time-PQS}
    p(t) = P + |S|\cos\!\left(2(\omega t + \angle\hat{e}) - \tan^{-1}\!\left(\frac{Q}{P}\right)\right),
\end{equation}
a DC offset of $P$ plus an oscillation at $2\omega$ with amplitude $|S|$.
The average and peak powers in \Cref{eq:power-avg-peak} follow directly: $p_{\text{avg}} = P$ since the oscillation averages to zero, and $p_{pk} = P + |S|$ at the cosine maximum.
The minimum $p_{\text{min}} = P - |S|$ is negative when $|S|>P$, corresponding to instants when energy flows from the load back into the source under reactive control.

\subsubsection{Electrical-Port Power in Optimal-Control Variables}
\label{sec:appendix-power-elec}

Expressing the average and peak powers at the electrical port in terms of the constrained-control parameters introduced in \Cref{sec:optimal-control} (control damping $B_l$, control stiffness $K_l$, and current phasor $\hat{I}$), and applying the constant electrical efficiency $\eta$,
\begin{equation}\label{eq:power-elec}
\begin{bmatrix}
p_{avg,elec} \\
p_{pk,elec}
\end{bmatrix}
=
\eta
\begin{bmatrix}
p_{avg,VI} \\
p_{pk,VI}
\end{bmatrix}
=
\eta \tfrac{1}{2} B_l |\hat{I}|^2
\begin{bmatrix}
1 \\
1 + \sqrt{1 + \left(\tfrac{K_l}{B_l\omega}\right)^2}
\end{bmatrix}.
\end{equation}
The peak-power form is used as the apparent-power constraint in the QCQP of \Cref{sec:optimal-control}.

\subsubsection{PTO Kinematics and Underactuation}
The PTO kinematics describe the relationship between WEC body displacements $\vec{\hat{\xi}}$ to PTO displacements $\vec{\hat{X}}_{PTO}$, determined by linkages or similar mechanisms and not including subsequent PTO elements such as a mechanical drivetrain or generator.
In general form, the PTO kinematics can be expressed as:
\begin{equation}\label{eq:pto-kinematics-motion}
\vec{\hat{X}}_{PTO} = \mathbf{T}_{kin}~ \vec{\hat{\xi}}
\end{equation}
with kinematic transformation matrix $\mathbf{T}_{kin}$ that may depend on the body displacements $\vec{\hat{\xi}}$ if the linkages are nonlinear, but hereafter is assumed constant for simplicity.
This assumption means that \Cref{eq:pto-kinematics-motion} can be differentiated trivially to yield the same matrix relationship for velocities.
Assuming power conservation across the linkages, the kinematics describe the relationship between not only motions but also forces:
\begin{equation}\label{eq:pto-kinematics-force}
\vec{\hat{F}}_p = [\mathbf{T}_{kin}]^T  \vec{\hat{F}}_{PTO}
\end{equation}
To avoid confusion, subscript $PTO$ is used for forces in the PTO degrees of freedom (post-kinematics), while subscript $p$ denotes forces in the hydrodynamic body degrees of freedom (pre-kinematics).
\Cref{eq:pto-kinematics-motion,eq:pto-kinematics-force} imply rigid lossless massless linkages, and any linkage deflection, dissipation, and inertia would be included in the PTO dynamics rather than kinematics.

Multibody WECs are frequently underactuated, meaning that the PTO has fewer controlled degrees of freedom than the hydrodynamic bodies, and $\mathbf{T}_{kin}$ has more columns than rows.
Underactuated systems receive considerable attention in robotics \cite{tedrake_underactuated_2024} but have only recently been explored in the context of wave energy conversion \cite{faedo_principle_2022}.
Cascade matrices require an identical number of input and output ports and thus cannot be used to describe the unbalanced ports that underactuated kinematics create \cite{reveyrand_multiport_2018}.
Non-square impedance matrices may be used to capture unbalanced port relations in general, but the impedance is undefined for the specific relations of \Cref{eq:pto-kinematics-motion,eq:pto-kinematics-force}.
Instead, the kinematics can be packaged into a square hybrid matrix $[\mathbf{h}]_{kin}$ which accurately represents the underactuated system:
\begin{equation}\label{eq:h-matrix-kinematics}
    \begin{bmatrix}
        \vec{\hat{F}}_p \\
        \vec{\hat{\dot{X}}}_{PTO}
    \end{bmatrix}
    = \underbrace{
        \begin{bmatrix}
            \mathbf{0} & \mathbf{T}_{kin}^T \\
            \mathbf{T}_{kin} & \mathbf{0}
        \end{bmatrix}}
        _{[\mathbf{h}]_{kin}}
    \begin{bmatrix}
        \vec{\hat{\dot{\xi}}} \\ \vec{\hat{F}}_{PTO}
    \end{bmatrix}
\end{equation}

Here, we model two hydrodynamic degrees of freedom (heave of the float and spar respectively) and a single PTO degree of freedom corresponding to the relative heave motion between the float and spar.
The kinematics matrix is therefore:
\begin{equation}\label{eq:T-kin-specific}
\mathbf{T}_{kin} = \begin{bmatrix}1 & -1\end{bmatrix}
\end{equation}
Thus, even absent any mechanism with ``kinematics'' in the traditional sense, the kinematics matrix is filled with zeroes, ones, and negative ones to represent which degrees of freedom are controlled.

In the circuit representation with two hydrodynamic and one PTO degree of freedom, the hybrid matrix \Cref{eq:h-matrix-kinematics} can be represented as a three-port element, as shown in \Cref{fig:multiport-circuit-kinematics}.
\begin{figure}[htbp]
\centering
    \includegraphics[width=.7\linewidth]{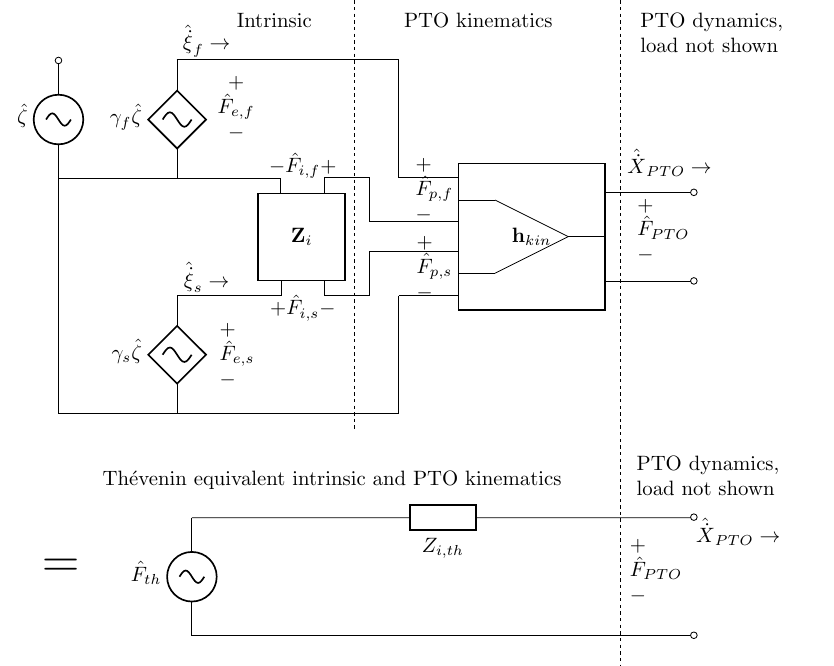}
\caption{Multiport circuit with powertrain kinematics represented as a hybrid matrix, followed by the Th\'{e}venin equivalent of the intrinsic dynamics and kinematics}
\label{fig:multiport-circuit-kinematics}
\end{figure}

\subsubsection{Combining Intrinsic Dynamics with PTO Kinematics}
Taking the Th\'{e}venin equivalent of the intrinsic dynamics and kinematics at the PTO port yields a scalar equivalent circuit for each PTO degree of freedom, with force sources $\vec{\hat{F}}_{th}$ and mechanical impedances $\vec{Z}_{i,th}$, shown in the bottom portion of \Cref{fig:multiport-circuit-kinematics} and computed with the following expressions:

\begin{equation}
\begin{aligned}
    \vec{\hat{F}}_{th} &= (\mathbf{T}_{kin}\mathbf{T}_{kin}^T)^{-1} \mathbf{T}_{kin} \mathbf{Z}_{i}^{-1} \vec{\gamma}\hat{\zeta}
    = \frac{1}{2} [1,~-1] \mathbf{Z}_{i}^{-1} \vec{\gamma}\hat{\zeta} \\
    \vec{Z}_{i,th} &= (-\mathbf{T}_{kin}\mathbf{Z}_{i}^{-1}\mathbf{T}_{kin}^T)^{-1} = -\frac{\det(\mathbf{Z}_{i})}{\Sigma \mathbf{Z}_{i}} = -\frac{Z_{i,f}Z_{i,s}-Z_{i,c}^2}{Z_{i,f}+Z_{i,s}+2Z_{i,c}}
\end{aligned}
\end{equation}
where the vector expressions for the general case are shown first, followed by the scalar result for the $\mathbf{T}_{kin}$ in \Cref{eq:T-kin-specific} with 1 controlled degree of freedom.

Although, as previously mentioned, a true cascade matrix must have the same number of input and output ports, it is possible to formulate a cascade-like matrix that can transform from the PTO port to the body ports.
However, it cannot transform in the other direction due to underactuation.

Even in the favorable direction, this matrix cannot transform to the body ports directly.
Due to the wave excitation, body velocities can be nonzero even when the PTO force and velocity are both zero.
Hence, there is no way to express $\hat{\dot{\xi}}$ as a weighted sum of $\hat{F}_{PTO}$ and $\dot{\hat{X}}_{PTO}$.
$\hat{\dot{\xi}}$ could be expressed as a function of the PTO variables by substituting $\mathbf{Z}_p=\mathbf{T}_{kin}^T (\vec{\hat{F}}_{PTO}/\vec{\hat{\dot{X}}}_{PTO}) \mathbf{T}_{kin}$ into \Cref{eq:eom-freq-domain}, but this yields a nonlinear expression of limited utility.
Instead, we can use cascade-like matrices to solve for a modified variable $\vec{\dot{\xi}}_\Delta$ defined as the change in the body velocities caused by the PTO, in other words the difference between the actual body velocities $\vec{\dot{\xi}}$ and their values if the PTO were disconnected with $F_{PTO}=0$, $\vec{\dot{\xi}}_{0}$:
\begin{equation}\label{eq:modified-body-velocity}
    \vec{\dot{\xi}}_\Delta = \vec{\dot{\xi}} - \vec{\dot{\xi}}_{0}
    \quad \textrm{ where } \quad
    \vec{\dot{\xi}}_{0} = \mathbf{Z}_i^{-1}~\vec{\gamma}~\frac{H}{2}
\end{equation}
The cascade-like matrix from the PTO port $\{F_{PTO}, \dot{X}_{PTO}\}$ to the modified body port $\{F_p, \dot{\xi}_\Delta\}$ is then:
\begin{equation}\label{eq:body-transmission-matrix}
[\mathbf{a}]_{F_p\xi_\Delta\leftarrow F\dot{X}} =
\begin{bmatrix}
\mathbf{I}_2 \\
-\mathbf{Z}_i^{-1}
\end{bmatrix}
\mathbf{T}_{kin}^T
\begin{bmatrix}
1 & 0
\end{bmatrix}
\end{equation}
noting that this matrix is non-square ($\in \mathbb{C}^{4\times2}$) since there are more hydrodynamic than PTO degrees of freedom, and that the matrix label drops the $PTO$ subscripts for compactness.
\Cref{fig:multiport-circuit-kinematics} represents this in circuit form, noting that the excitation source must be duplicated to create $\vec{\hat{\dot{\xi}}}_0$.

\subsubsection{PTO Dynamics}
\label{sec:appendix-pto-dynamics}
In addition to the PTO port already discussed, we define two additional ports of relevance: the generator mechanical port (torque $\tau$ and rotation speed $\Omega$) and the generator electrical port (voltage $V$ and current $I$).
Assuming a drivetrain consisting of a linear to rotational gear ratio with equivalent radius $R$ followed by a driveshaft with impedance $Z_{\text{shaft}}$, the PTO mechanical dynamics are expressed with cascade matrix $[\mathbf{a}]_{F\dot{X}\leftarrow \tau \Omega}$:
\begin{equation}\label{eq:mech-transmission-matrix}
\begin{bmatrix}
F_{PTO} \\ \dot{X}_{PTO}
\end{bmatrix}
=
\underbrace{\begin{bmatrix}
Z_{shaft}/R & 1/R \\
R & 0
\end{bmatrix}}
_{[\mathbf{a}]_{F\dot{X}\leftarrow \tau \Omega}}
\begin{bmatrix}
 \tau\\ \Omega
\end{bmatrix}
\end{equation}

Meanwhile, a synchronous permanent magnet generator modeled as an ideal gyrator with torque constant $k_\tau$ plus a winding impedance $Z_{w}$ has cascade matrix $[\mathbf{a}]_{\tau \Omega \leftarrow V I}$:
\begin{equation}\label{eq:elec-transmission-matrix}
\begin{bmatrix}
 \tau\\ \Omega
\end{bmatrix}
=
\underbrace{\begin{bmatrix}
0 & k_\tau \\
1/k_\tau & Z_{w}/k_\tau
\end{bmatrix}}
_{[\mathbf{a}]_{\tau \Omega \leftarrow V I}}
\begin{bmatrix}
 V \\ I
\end{bmatrix}
\end{equation}
with Park-transformed quadrature electrical voltage and current $V$ and $I$.

The system operating point is determined by a controller that applies some electrical load to the generator.
The controller can be formulated as an impedance at either the generator mechanical or electrical port.
The former is more convenient for hardware implementation, because modern drives typically expect torque commands, while the latter is more convenient for electrical power calculations and optimal control.
The mechanical control impedance $Z_u$ and electrical load impedance $Z_l$ are defined as follows:
\begin{equation}\label{eq:controller-impedances}
    Z_l = \frac{V}{I} = B_l + \frac{K_l}{i\omega}
    ,  \qquad
    Z_u = \frac{\tau}{\Omega}
\end{equation}
Combining \Cref{eq:controller-impedances,eq:elec-transmission-matrix} gives the following expression for $Z_u$ as a function of $Z_l$:
\begin{equation}\label{eq:control-impedance}
    Z_u = 
    \frac{
    \begin{bmatrix}
    1 & 0
    \end{bmatrix}
    [\mathbf{a}]_{\tau \Omega \leftarrow V I}
    \begin{bmatrix}
    Z_l \\ 1
    \end{bmatrix}
}{
    \begin{bmatrix}
    0 & 1
    \end{bmatrix}
    [\mathbf{a}]_{\tau \Omega \leftarrow V I}
    \begin{bmatrix}
    Z_l \\ 1
    \end{bmatrix}
}
\end{equation}
This equation allows the $Z_l$ identified with optimal control to be implemented as a corresponding $Z_u$ in hardware.

The PTO dynamics can be represented as a multiport circuit, as shown in \Cref{fig:multiport-circuit-pto}.
\begin{figure}[htbp]
\centering
    \includegraphics{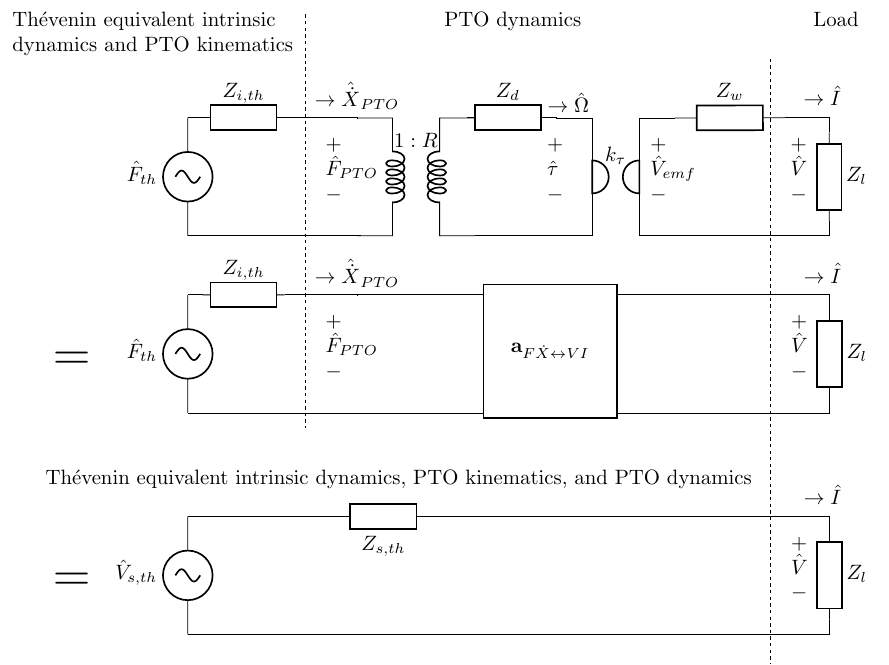}
\caption{PTO dynamics represented as a multiport circuit and electrical Th\'{e}venin equivalent}
\label{fig:multiport-circuit-pto}
\end{figure}

The intrinsic dynamics, PTO kinematics, and PTO dynamics can be combined into a Th\'{e}venin equivalent source voltage and impedance $\hat{V}_{s,th}$ and $Z_{s,th}$ as follows:
\begin{equation}\label{eq:thevenin-electrical}
    Z_{s,th} = \frac{
    \begin{bmatrix}1 & 0\end{bmatrix}
    \mathbf{a}_{VI \leftarrow F\dot{X}}
    \begin{bmatrix}-Z_{i,th} \\ 1\end{bmatrix}}
    {\begin{bmatrix}0 & 1\end{bmatrix}
    \mathbf{a}_{VI \leftarrow F\dot{X}}
    \begin{bmatrix}-Z_{i,th} \\ 1\end{bmatrix}
    } = \frac{B-AZ_{i,th}}{D-CZ_{i,th}}
    \qquad \hat{V}_{s,th} = \hat{F}_{th}(A-CZ_{s,th})
\end{equation}
where $A,B,C,D$ are the elements of $\mathbf{a}_{VI \leftarrow F\dot{X}}$.
This is shown in the bottom portion of \Cref{fig:multiport-circuit-pto}.
The load voltage and current can then be found as:
\begin{equation}\label{eq:voltage-current}
    \hat{V} = \hat{V}_{s,th}\frac{Z_l}{Z_{s,th}+Z_l}, \qquad \hat{I} = \frac{\hat{V}_{s,th}}{Z_{s,th}+Z_l}
\end{equation}

\subsection{Linear Solution Procedure}
\label{sec:appendix-dynamics-solution}
\Cref{sec:eom,sec:pto-dynamics-overview} introduce linear models for the rigid body dynamics, PTO kinematics, and PTO dynamics using impedance, hybrid, and cascade matrices respectively.
Here we unify the formulation and present a clear system-level solution procedure for all port variables:
\begin{enumerate}
\item Use \Cref{eq:thevenin-electrical} to assemble the Th\'{e}venin source voltage and impedance $\hat{V}_{s,th}$ and $Z_{s,th}$ as seen from the port on which power should be maximized (in this case, the electrical generator port $\{V,I\}$).
\item Set the load impedance $Z_l$.
Any load impedance can be simulated, but typically it is selected as a function of $Z_s$ to maximize the power at this port, possibly subject to constraints on various ports.
\Cref{sec:optimal-control} frames the load impedance selection as a constrained optimization problem.
\item From  $\hat{V}_{s,th}$, $Z_{s,th}$, and $Z_l$, calculate the $\{V,I\}$ port variables from \Cref{eq:voltage-current}.
\item Use cascade matrices \Cref{eq:mech-transmission-matrix,eq:elec-transmission-matrix,eq:body-transmission-matrix} to propagate the variables back toward the rigid body, yielding the $\{\tau,\Omega\}$, $\{F_{PTO},\dot{X}_{PTO}\}$, and $\{F_p, \dot{\xi}_\Delta\}$ port variables.
\item Transform from $\dot{\xi}_\Delta$ to $\dot{\xi}$ via \Cref{eq:modified-body-velocity}.
\end{enumerate}
Once all the port variables have been calculated, power at each port can be calculated according to \Cref{sec:power}.
This serves as the solution to a linear subproblem, which is then re-solved using updated quasi-linear coefficients until drag and force saturation nonlinearities have converged.

\ifdefined\DISSERTATION
    \subsection{Indirect Optimal Control Solution Procedure}
    \label{sec:appendix-qp-numerical}
    The indirect optimal control approach applies optimality conditions and then numerically solves the resulting equations.
    The optimality conditions depend on the active set of constraints.
    In the most straightforward case where simply one constraint exists, the constraint is active if and only if the unconstrained solution violates that constraint. 
    For example, if only a maximum force constraint exists, the constrained solution has a force $\hat{F}_{p,\text{constr}}$ that is the minimum of the unconstrained force $\hat{F}_{p,\text{unconstrained}}$ and the maximum allowed force fundamental $F_{\text{max}}$:
    \begin{equation}
    \hat{F}_{p,constr} = \min\left(\hat{F}_{p,unconstrained}, F_{max}\right)
    \end{equation}
    The solver guesses a value of $\Gamma$, calculates the resulting force $\hat{F}_{p,\text{guess}}$, and then uses $\hat{F}_{p,\text{guess}}-\hat{F}_{p,\text{constr}}$ as an error signal to adjust its next guess of $\Gamma$, iterating until the error converges to zero.
    This solver-based optimal control performs numerical root-finding, not numerical optimization, and can happen in an inner loop of the drag-linearization solver iteration (where $\hat{F}_{p,\text{unconstrained}}$ stays constant each control iteration) or simultaneously with the drag-linearization iteration (where $\hat{F}_{p,\text{unconstrained}}$ and the system linearization are updated each control iteration).
    MDOcean implements the latter, which simplifies the setup by requiring only one solver.
    Comparing the efficiency of the two approaches is left for future work.

    When multiple constraints exist, the situation is more complex, since it is possible for a given constraint to be violated by the unconstrained solution but not active in the constrained solution, or for multiple constraints to be active simultaneously.
    In the special case where the unconstrained solution violates no more than one constraint, that constraint is active and the solver can proceed as described above.
    When the unconstrained solution violates multiple constraints, the solver must try all subsets of 1-2 of those constraint(s) as the active set, find which of these active sets produces the largest power while satisfying all constraints, and then set the error as the deviation of the guessed solution from the constrained solution for that active set.
    Since the active set cannot be calculated a-priori as in the simpler case, this approach ends up looking more like a numerical optimization than a numerical root-finding, although the optimization is over a finite set of active set possibilities instead of a continuous control domain as it would be in a direct optimal control method. 
    The main advantage of the indirect approach is its ability to handle arbitrary nonlinear\footnote{Nonlinear meaning greater than second order in an optimization sense, not in a dynamic sense, because the dynamics are already quasi-linearized here} constraints, although for the quadratic constraints considered here, the analytical approach described in \Cref{sec:appendix-qp-solution} is more efficient.
\fi
\subsection{Analytical Quadratic Program for Constrained Power Maximization}
\label{sec:appendix-qp-solution}
This section describes the analytical solution process for the constrained optimal control problem \Cref{eq:opt-problem} introduced in \Cref{sec:optimal-control}.

\subsubsection{Quadratic Constraint Coefficients}
\label{sec:appendix-qp-constraints-table}

The quadratic coefficients $\{\mathbf{Q}_i, \vec{a}_i, b_i\}$ in \Cref{eq:opt-problem} for each constraint implemented in MDOcean are given in \Cref{tab:qp-constraints}.

\begin{table}[b!]
\centering
\caption{Quadratic constraint coefficients}
\label{tab:qp-constraints}
\begin{tabular}{lccc}
    Constraint & $\mathbf{Q}$ & $\vec{a}$ & $b$ \\ \hline
    $|\hat{\xi}_f| \leq \hat{\xi}_{f,\text{max}}$ &
        $\frac{1}{\omega^2}
        [\mathbf{a}_{(F_{p}\dot{\xi})_f\leftarrow VI}]^* 
        \mathbf{A}_{|\hat{q}|^2}
        [\mathbf{a}_{(F_{p}\dot{\xi})_f\leftarrow VI}]$ & 
        $[0,~ \hat{\xi}_{f,0}]^T$ & $\hat{\xi}_{f,\text{max}} - |\hat{\xi}_{f,0}|$ \\
    $|\hat{\xi}_s| \leq \hat{\xi}_{s,\text{max}}$ & 
        $\frac{1}{\omega^2}
        [\mathbf{a}_{(F_{p}\dot{\xi})_s\leftarrow VI}]^* 
        \mathbf{A}_{|\hat{q}|^2}
        [\mathbf{a}_{(F_{p}\dot{\xi})_s\leftarrow VI}]$ & 
        $[0,~ \hat{\xi}_{s,0}]^T$ & $\hat{\xi}_{s,\text{max}} - |\hat{\xi}_{s,0}|$ \\
    $|\hat{X}_{PTO}| \leq \hat{X}_{PTO,\text{max}}$ & 
        $\frac{1}{\omega^2}
        [\mathbf{a}_{F\dot{X}\leftarrow VI}]^* 
        \mathbf{A}_{|\hat{q}|^2}
        [\mathbf{a}_{F\dot{X}\leftarrow VI}]$ & 
        $\vec{0}$ & $\hat{X}_{PTO,\text{max}}^2$ \\
    $|\hat{\tau}| \leq \hat{\tau}_{\text{max}}$ & 
        $[\mathbf{a}_{\tau\Omega \leftarrow VI}]^* 
        \mathbf{A}_{|\hat{e}|^2}
        [\mathbf{a}_{\tau\Omega\leftarrow VI}]$ & 
        $\vec{0}$ & $\hat{\tau}_{\text{max}}^2$ \\
    $P_{\tau\Omega} \leq P_{\tau\Omega,\text{max}}$ & 
        $[\mathbf{a}_{\tau\Omega \leftarrow VI}]^* 
        \mathbf{A}_{P}
        [\mathbf{a}_{\tau\Omega\leftarrow VI}]$ & 
        $\vec{0}$ & $P_{\tau\Omega,\text{max}}$ \\
    $P_{VI} \geq 0$ & 
        $-\mathbf{A}_{P}$ & 
        $\vec{0}$ & $0$ \\
\end{tabular}
\end{table}

\subsubsection{Reduced Dimension Formulation}
The linear equality constraint $\vec{c}^{\,*}\vec{x} = d$ reduces the optimization problem dimension from $\vec{x} \in \mathbb{C}^2$ to $\mathbb{C}^1$.
Specifically, the equality constraint requires that the solution be a particular solution plus a scalar multiple of $\vec{n}$, the null space vector of $\vec{c}^{\,*}$:
\begin{equation}\label{eq:x-p-plus-nullspace}
    \vec{x} = \underbrace{
        \frac{\hat{V}_{s,th}}{2\Re( Z_{s,th})}
        \begin{bmatrix} \ Z_{s,th}^* \\ 1 \end{bmatrix} 
    }_{
        \textrm{particular solution } \vec{x}_p} 
    -~ \Gamma \cdot \frac{\hat{V}_{s,th}}{2\Re( Z_{s,th})}
    \underbrace{
        \begin{bmatrix} -\ Z_{s,th}^* \\ 1 \end{bmatrix}
    }_{
        \textrm{null space } \vec{n}}
\end{equation}
where $\Gamma$ is a nondimensional complex scalar representing the remaining degree of freedom in the optimization after enforcing the equality constraint.
For the particular solution, we have chosen to use the optimal solution in the absence of inequality constraints, which is identical to the solution implied by the impedance-matching condition \Cref{eq:matched-load} and can be derived from the quadratic program as follows:
\begin{equation}\label{eq:x-p}
    \vec{x}_{p} = \frac{d \mathbf{A}_P\vec{c}}{\vec{c}^{\,*}\mathbf{A}_P\vec{c}}
\end{equation}
This selection for $\vec{x}_p$ means that the degree of freedom $\Gamma$ represents the deviation of the solution from the impedance-matched solution.
Using $\hat{I}_p$ for brevity to denote the current in the particular solution, \Cref{eq:x-p-plus-nullspace} can be rewritten as:
\begin{equation}\label{eq:x-p-plus-nullspace-simplified}
\vec{x} = \vec{x}_p - \Gamma \hat{I}_p \vec{n}
\end{equation}

Inequality constraints 
$\vec{x}^{\,*}\mathbf{Q}_\mu\vec{x} + 2\Re\!\left\{\vec{a}_\mu^{\,*}\vec{x}\right\} \le b_\mu$
further reduce the feasible design space, and the presence of multiple inequality constraints introduces the possibility of \Cref{eq:opt-problem} having no solutions due to infeasibility.
While constrained quadratic programs like \Cref{eq:opt-problem} generally require numerical solutions, the low dimensionality allows an analytical solution with geometric intuition, even if the problem is non-convex (see \Cref{sec:appendix-qp-convexity} for a discussion on convexity). 
First, we rewrite the objective and inequality constraints in terms of $\Gamma$ by plugging \Cref{eq:x-p-plus-nullspace-simplified} into \Cref{eq:opt-problem}:
\begin{subequations}\label{eq:obj-inequality-gamma}
\begin{align}
P_{avg,VI}
&= \frac{|\hat{V}_{s,th}|^2}{8\Re(Z_{s,th})}~ (1 - |\Gamma|^2)
\label{eq:obj-gamma}
\\[6pt]
A_{q,\mu}|\Gamma|^2 +&
2\Re\!\left\{
    B_{q,\mu}\Gamma
\right\}+C_{q,\mu}
\le
0
\quad \forall \mu
\label{eq:ineq-gamma}
\end{align}
\end{subequations}
where quadratic coefficients $A_{q,\mu}$, $B_{q,\mu}$, and $C_{q,\mu}$ from \Cref{eq:ineq-gamma} are defined as:
\begin{equation}
\begin{aligned}
    A_{q,\mu} &= |\hat{I}_p|^{2},
    \vec{n}^{\,*}\mathbf{Q}_{\mu}\vec{n}& &\in \mathbb{R} \\
    B_{q,\mu} &= -\hat{I}_p,
    \left(\vec{x}_p^{\,*}\mathbf{Q}_{\mu} + \vec{a}_{\mu}^{\,*}\right)\vec{n}& &\in \mathbb{C} \\
    C_{q,\mu} &= 
        \vec{x}_p^{\,*}\mathbf{Q}_{\mu}\vec{x}_p
        +
        2\Re\!\left\{\vec{a}_{\mu}^{\,*}\vec{x}_p\right\}
        -b_{\mu}&
    &\in \mathbb{R}
\end{aligned}
\end{equation}
Since \Cref{eq:obj-gamma} shows that power decreases with increasing $|\Gamma|$, the power maximization problem \Cref{eq:opt-problem} can be recast as minimizing the norm of $\Gamma$:
\begin{subequations}\label{eq:opt-problem-gamma}
\begin{alignat}{2}
& \min_{\Gamma \in \mathbb{C}} &  |\Gamma|^2& \\[3pt]
& \text{s.t.} &
S_{\mu}|\Gamma - \Gamma_{c,\mu}|^2 &\le S_{\mu} r_{\mu}^2,
\quad \forall \mu \label{eq:disc-constraint}\\
& \text{where} &
\Gamma_{c,\mu}  &\in \mathbb{C} \nonumber\\
& & r_{\mu}^2 &\in \mathbb{R} \nonumber \\
& & S_{\mu} &\in \left\{-1,1\right\} \nonumber
\end{alignat}
\end{subequations}
where the constraint \Cref{eq:disc-constraint} is derived from \Cref{eq:ineq-gamma} by completing the square.
More precisely, when $A_{q,\mu} \neq 0$, the $\mu$th inequality constraint defines a circle in the complex plane of $\Gamma$ centered at $\Gamma_{c,\mu}$ with radius $r_{\mu}$.
The constraint requires that $\Gamma$ lie within this circle when $A_{q,\mu} > 0$ ($S_{\mu}=1$) or outside this circle when $A_{q,\mu} < 0$ ($S_{\mu}=-1$).
Parameters $\Gamma_{c,\mu}$, $r_{\mu}$, and $S_{\mu}$ are defined as:
\begin{subequations}\label{eq:gamma-disc-params}
\begin{align}
\Gamma_{c,\mu}
&= \frac{-B_{q,\mu}^{\,*}}{A_{q,\mu}} = 
\frac{
    \hat{I}_p^{\,*}\,
    \vec{n}^{\,*}
    \left(\mathbf{Q}_{\mu}\vec{x}_p + \vec{a}_{\mu}\right)
}{
    |\hat{I}_p|^{2}\,
    \vec{n}^{\,*}\mathbf{Q}_{\mu}\vec{n}
}
\label{eq:gamma-disc-center}
\\[6pt]
r_{\mu}^2
&=
\frac{|B_{q,\mu}|^2-A_{q,\mu}C_{q,\mu}}{A_{q,\mu}^2} = 
\frac{
b_{\mu}
-
\vec{x}_p^{\,*}\mathbf{Q}_{\mu}\vec{x}_p
-
2\Re\!\left\{\vec{a}_{\mu}^{\,*}\vec{x}_p\right\}
}{
|\hat{I}_p|^{2}\,
\vec{n}^{\,*}\mathbf{Q}_{\mu}\vec{n}
}
+
|\Gamma_{c,\mu}|^2 
\label{eq:gamma-disc-radius}
\\
S_{\mu} &= \textrm{sign}(A_{q,\mu}) = \textrm{sign}\left(\vec{n}^{\,*}\mathbf{Q}_{\mu}\vec{n}\right)
\label{eq:gamma-disc-sign}
\end{align}
\end{subequations}

\subsubsection{Solution}
\Cref{eq:opt-problem-gamma} can be interpreted graphically as finding the point closest to the origin that is interior to all constraint circles for which $S_{\mu}=1$ and exterior to all circles for which $S_{\mu}=-1$.
The unconstrained optimal solution of \Cref{eq:opt-problem-gamma} is the origin ($\Gamma=0$), and the constrained optimal solution will also be zero in the case where the origin is feasible (no active constraints), which occurs when $ S_{\mu} C_{q,\mu} \leq 0 \quad \forall \mu$.
If the origin is infeasible, the constrained optimal solution will lie on the boundary of one or more of the circles, specifically either at an intersection of two or more circles or at any circle's point closest to the origin.
We analytically compute all such candidate points and select the one with the lowest $|\Gamma|$ that satisfies all constraints.
For $N$ constraints, there are up to $N^2$ candidate points ($N^2-N$ at intersections and $N$ at closest points).
This leads to the following expression for the constrained optimal solution (valid when $A_{q,\mu} \neq 0 ~\forall \mu$):
\begin{equation}\label{eq:gamma-opt-quadratic}
\Gamma_{opt} =
\begin{cases}
 0 & \textrm{if } S_{\mu} C_{q,\mu} \leq 0 \quad \forall \mu \\[3pt]
 \arg\min\limits_{\Gamma \in \mathcal{F} \bigcap \mathcal{C}} |\Gamma| 
 & \textrm{otherwise} \\[3pt]
 \textrm{infeasible} & \textrm{if } \mathcal{F} \bigcap \mathcal{C} = \emptyset
\end{cases}
\end{equation}
where $\mathcal{F} = \bigcap_{\mu} \left\{\Gamma : S_{\mu}|\Gamma - \Gamma_{c,\mu}| \le S_{\mu} r_{\mu} \right\}$ is the set of all feasible points and $\mathcal{C}$ is the set of candidate points. $\mathcal{C}$ is the union of $\mathcal{C}_{int}$, the candidates at circle intersections, and $\mathcal{C}_{cl}$, the candidates at the closest point on each circle to the origin:
\begin{equation}\label{eq:candidate-points}
\begin{aligned}
\mathcal{C}_{int} &= \left\{
    \Gamma_{\mu,\nu}, \quad \forall \mu \neq \nu
\right\}.
\\
\mathcal{C}_{cl} &= \left\{
    (|\Gamma_{c,\mu}| - r_{\mu})e^{i\angle\,\Gamma_{c,\mu}}, \quad \forall \mu
\right\}.
\end{aligned}
\end{equation}
The intersection points of the $\mu$th and $\nu$th circles are given by:
\begin{equation}\label{eq:circle-intersection}
    \Gamma_{\mu,\nu} = 
    \Gamma_{c,\mu} + \frac{\Gamma_{c,\nu} - \Gamma_{c,\mu}}{d_{\mu\nu}}
    \left(
        a_{\mu\nu}
        \pm
        1i \sqrt{r_{\mu}^2-a_{\mu\nu}^2}
    \right)
\end{equation}
where $d_{\mu\nu} = |\Gamma_{c,\nu} - \Gamma_{c,\mu}|$ is the distance between the circle centers, 
$a_{\mu\nu} = \frac{1}{2}(r_{\mu}^2-r_{\nu}^2+d_{\mu\nu}^2)/d_{\mu\nu}$ is the distance from 
the center of the $\mu$th circle to the midpoint of the two intersection points, 
and $1i$ is the imaginary unit.
The closest-point candidates $\mathcal{C}_{cl}$ match \equationautorefname~4 of \cite{mccabe_force-limited_2024},
which provides the solution for problems with only one constraint.

Importantly, we need not compute the entire feasible space $\mathcal{F}$, only compute $\mathcal{C}$ and then check whether each candidate point is feasible.
\Cref{fig:qp-solution-geometry} illustrates the various cases for feasibility and optimality for a toy problem with $r_{\mu}=4$, $S_{\mu}=1$, and various $\Gamma_{c,\mu}$.
\begin{figure}[htbp]
    \centering
    \includegraphics[width=.8\textwidth]{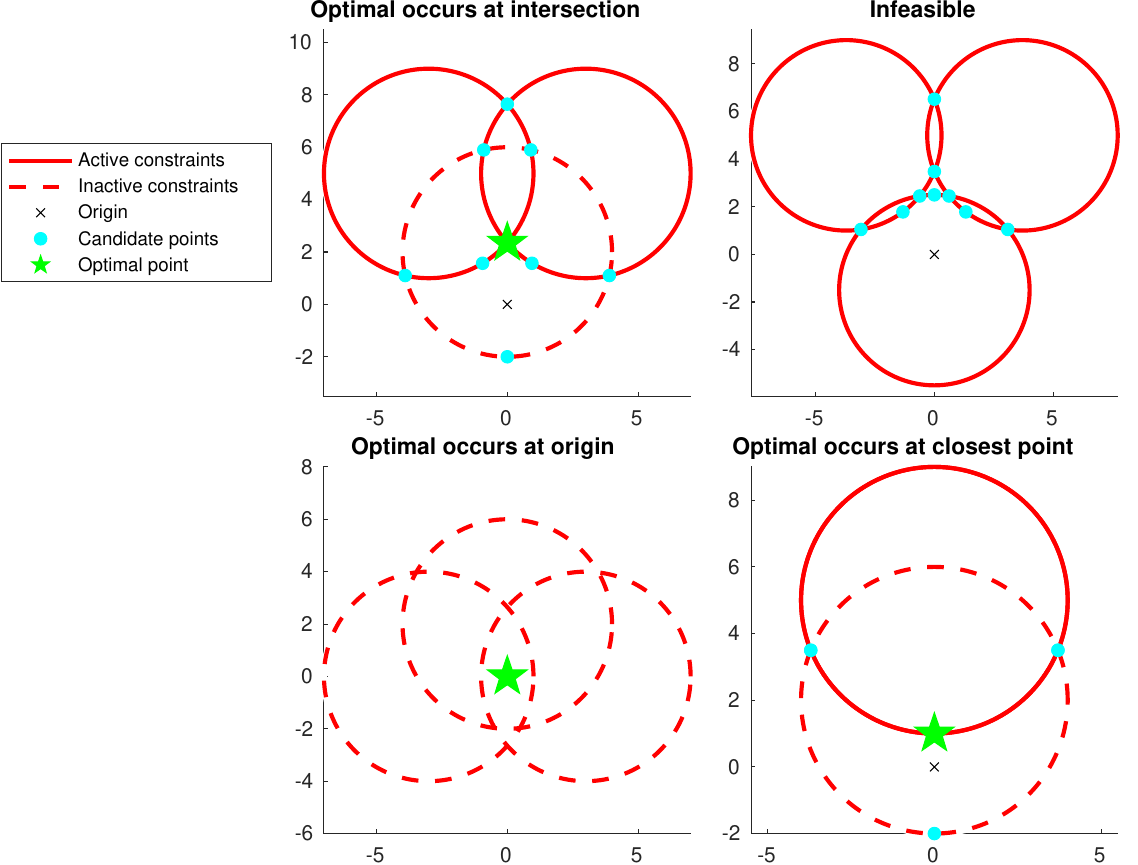}
    \caption{Visualization of the optimization problem \Cref{eq:opt-problem-gamma} in the complex plane of $\Gamma$.
Toy problems with various constraint circle centers $\Gamma_{c,\mu}$, all with radius $r_{\mu}=4$ and $S_{\mu}=1$, demonstrate the various solution cases in \Cref{eq:gamma-opt-quadratic}.}
    \label{fig:qp-solution-geometry}
\end{figure}

If $A_{q,\mu} = 0$ for any $\mu$, the $\mu$th inequality constraint defines a line in the complex plane of $\Gamma$ rather than a circle.
This can occur when $\mathbf{Q}_{\mu} = 0$ (the constraint on $\vec{x}$ is linear rather than quadratic, which is not the case for any of the constraints in \Cref{tab:qp-constraints}) or when $\vec{n}^{\, *}\mathbf{Q}_{\mu}\vec{n} = 0$ ($\mathbf{Q}_{\mu}$ is rank-deficient and $\vec{n}$ is in its nullspace).
The same types of critical points (now consisting of line-line, circle-line, and circle-circle intersections, as well as the closest point on each circle and line) apply, but for implementation simplicity, we replace $A_{q,\mu}$ with a small positive $\epsilon$ in this case and use the quadratic implementation.

\subsubsection{Quartic Apparent Power Constraints}
\label{sec:appendix-mag-S-constraints}
While constraints on effort and flow magnitudes and real and reactive power are quadratic in $\vec{x}$, constraints on the magnitude of apparent power $|S|<|S_{\text{max}}|$, and therefore the maximum and minimum instantaneous power, are fourth-order in $\vec{x}$ and thus $\Gamma$.
This results in a ``Cassini oval'' curve \citep{weisstein_cassini_2026} on the complex plane of $\Gamma$, which consists of two closed curves when $|S_{\text{max}}|<|S_0|$ and one curve when $|S_{\text{max}}|\geq|S_0|$, where $|S_0|$ is the apparent power of the unconstrained optimum.
Though the optimization is no longer a quadratically constrained quadratic program, the same geometric solution procedure applies, adding circle-oval intersections $\mathcal{C}_{int,\text{oval}-\text{circ}}$, oval-oval intersections, and oval closest points $\mathcal{C}_{cl,\text{oval}}$ to the set of candidate points $\mathcal{C}$.
The new candidate points are:
\begin{equation}
\begin{aligned}
\mathcal{C}_{int,oval-circ} &= \left\{
    \Gamma : 
    |\Gamma - \Gamma_{c,\mu}| = r_{\mu}, ~(1-|\Gamma|^2)^2 + 4\Im(\Gamma) = \frac{|S_{max,k}|}{|S_0|}, \quad \forall \mu,k
\right\}.
\\
\mathcal{C}_{cl,oval} &= \left\{
    \Gamma : 
    \Gamma = \begin{cases}
        \pm \sqrt{1-\frac{|S_{max,k}|}{|S_0|}}, \quad \forall k~ |S_{max,k}|<|S_0| \\
        \pm \sqrt{1+\frac{|S_{max,k}|}{|S_0|}}, \quad \forall k~ |S_{max,k}|>|S_0|
    \end{cases}
\right\}.
\end{aligned}
\end{equation}
where $\mu$ indexes circle constraints and $k$ indexes oval constraints.
While the earlier circle-circle intersections of \Cref{eq:candidate-points} result in a quadratic equation that is solved analytically for the coordinates of $\Gamma$, oval-circle intersections lead to quartic equations that are solved numerically with standard polynomial root-finding routines.
Typically there is only one apparent power constraint, so oval-oval intersections are not derived, but if present would likewise yield a fourth-order polynomial in $\Gamma$.
Therefore, the presence of apparent power constraints makes the optimal control method semi-analytical rather than analytical.

\subsubsection{Extension to Higher Dimensions and Convexity}
\label{sec:appendix-qp-convexity}
Future work may wish to extend the constrained power maximization to the full nonlinear dynamics, or to linear dynamics with irregular wave forcing as in the study \cite{bacelli_numerical_2015}.
In the terminology of \Cref{fig:venn-diagram}, these would be known as nonlinear and linear spectral optimal control respectively.
Both of these cases solve the full harmonic-balance equation, producing a higher-dimensional analog of \Cref{eq:opt-problem-gamma}.
The problem dimension equals twice the number of frequencies retained in the spectral expansion, preventing the use of the geometric analytical solution used in the monochromatic case.
Instead, a numerical solver must be used, and the computational scaling of that solver depends critically on whether the problem is \emph{convex}.

A convex optimization problem is one in which both the feasible set and the objective function are convex---informally, the feasible region has no holes or concavities, and the objective is bowl-shaped with no local valleys.
The key practical implication is that any locally optimal solution is also globally optimal, so gradient-based solvers converge to the true optimum without exhaustive search, and solution time scales polynomially with the number of decision variables rather than exponentially.

After reducing to the $\Gamma$ parameterization, the power objective $|\Gamma|^2$ is a sum of squares and therefore convex.
Even though the original power matrix $\mathbf{A}_P$ (see \Cref{tab:power-metrics}) has eigenvalues $\{-1,1\}$ and is indefinite, \citet{bacelli_numerical_2015} show that substituting the linear dynamics equality constraint into the objective---which is exactly what the reduction to $\Gamma$ in \Cref{eq:obj-gamma} accomplishes---renders the objective convex for both regular and irregular wave forcing.
Nonlinear dynamics can reintroduce non-convexity and should be checked on a case-by-case basis.

The inequality constraints are more problematic.
A constraint is convex if the set of points satisfying it forms a convex region.
In the $\Gamma$ plane, a convex constraint corresponds to a circle whose feasible region is the \emph{interior} ($S_\mu=1$ in \Cref{eq:disc-constraint}), while a non-convex constraint corresponds to a circle whose feasible region is the \emph{exterior} ($S_\mu=-1$).
Constraints on the effort and flow magnitude (e.g., force limits $|\hat{F}|\leq F_{\text{max}}$) can be convex or non-convex, depending on the complex angle of the Th\'{e}venin equivalent of the relevant source impedance and on the ratio of the constraint limit to its unconstrained value.
\citet{mccabe_force-limited_2024} visualize the feasible regions for families of such constraints on the complex $\Gamma$ plane. 
They show that the constraints are always convex for purely real impedances; that for reactive plants, nonconvexities occur when the value of the constraint limit exceeds some threshold; and that this threshold lowers as the reactivity increases.
The quadratic matrices $\mathbf{Q}$ in \Cref{tab:qp-constraints} confirm this: $\mathbf{A}_{|\hat{q}|^2}$ and $\mathbf{A}_{|\hat{e}|^2}$ are only positive \emph{semi}definite, indicating constraints that define half-spaces rather than bounded ellipsoids.
For the 2D problem studied here, non-convexity is handled by the exhaustive enumeration of candidate points in \Cref{eq:gamma-opt-quadratic}, which is efficient in two dimensions but does not scale to higher dimensions.
Future work should investigate semidefinite relaxations of these non-convex constraints using the S-procedure \cite{vanantwerp_tutorial_2000}, which replaces each non-convex constraint with a conservative but convex outer approximation.
It is expected that the optimal design will have low reactivity because reactivity increases signal peak-to-average ratios, implying that a semi-definite relaxation would likely improve in accuracy as it approaches the optimal.

\subsubsection{Comparison to Other WEC Constrained Control Formulations}
Notably, $\Gamma$ is equal to the reflection coefficient at the generator electrical port, which is commonly used in microwave circuit design to quantify impedance mismatch.
\citet{mccabe_force-limited_2024,coe_co-design_2025} further discuss the reflection coefficient for WECs.

Similar graphical interpretations of constraints for wave energy converters are presented in the studies \cite{mccabe_force-limited_2024,merigaud_geometrical_2023,bacelli_geometric_2013}, with all three visualizing quadratic constraints as circles on the 2D plane.
\citet{mccabe_force-limited_2024} focus on force/current limits but discuss amplitude, phase voltage, and apparent power constraints as well.
\citet{merigaud_geometrical_2023} consider exclusively amplitude limits while \citet{bacelli_geometric_2013} cover amplitude and force limits.
\cite{mccabe_force-limited_2024} use Smith charts to show constraints directly in $\Gamma$ space as we do here, while \citet{merigaud_geometrical_2023} work in the complex plane of the hydrodynamic transmission coefficient and \citet{bacelli_geometric_2013} in that of the PTO force.
\citet{bacelli_geometric_2013} also extend the formulation to capture higher harmonics (nonsinusoidal forces), where the exact constraint cannot be easily represented geometrically but sufficient conditions for constraint violation and satisfaction in the higher dimensional space can be represented as hyperspheres and hyperellipsoids.


\subsection{Constraint Sensitivity Analysis via Parametric Programming}
\label{sec:appendix-constraint-sensitivity}
To support the discussion in \Cref{sec:discussion}, this section derives the scaling of optimal average power with the constraint limits and demonstrates its conditions for convexity.

The constraint limits are 
$\tau_{\text{max}}$, $P_{\tau\Omega,\text{max}}$, $\xi_{f,\text{max}}$, $\xi_{s,\text{max}}$, and $X_{PTO,\text{max}}$, which we collectively denote as $L_\mu$.
The dependence of power on the first two limits (PTO sizing variables) is of more immediate relevance than the last three (amplitude limits) because the former can be independently manipulated through PTO design, while the latter can only be manipulated by changing the bulk dimensions.
The sensitivity of power to bulk dimensions is substantially more complex than the sensitivity to constraint limits for reasons discussed in \Cref{sec:appendix-dimension-sensitivity}, so we focus on the PTO sizing sensitivities here.

In a given sea state, the global sensitivity of optimal power to a constraint limit can be quantified using parametric programming theory, where a local sensitivity is found parametrically and then integrated over the parameter space to find the global parameter relationship.
The local sensitivity of power to a constraint parameter 
$\frac{\partial p_{\text{avg},VI,opt}}{\partial L_\mu}$
equals the product of the Lagrange multiplier for that constraint, $\lambda_\mu$,
and the partial derivative of the constraint function with respect to the parameter,
$\frac{\partial g_\mu}{\partial L_\mu}$ \cite[Section~5.9.3]{boyd_convex_2004}.
Since the constraint function is defined as 
$g_\mu(\vec{x}) = \vec{x}^{\,*}\mathbf{Q}_\mu\vec{x} 
                 + 2\Re\!\left\{\vec{a}_\mu^{\,*}\vec{x}\right\} - b_\mu\leq 0$,
and \Cref{tab:qp-constraints} shows that each constraint limit $L_\mu$ enters $g_\mu$ only through $b_\mu$, we have
$\frac{\partial g_\mu}{\partial L_\mu} = -\frac{\partial b_\mu}{\partial L_\mu}$.
This derivative can be trivially computed for all constraints in \Cref{tab:qp-constraints}.
The power sensitivity simplifies to:
\begin{equation}
    \frac{\partial p_{avg,VI,opt}}{\partial L_\mu} = -\lambda_\mu \frac{\partial b_\mu}{\partial L_\mu}
\end{equation}
If the optimization problem were solved with a numerical solver,
the Lagrange multipliers would be directly available as part of the solution.
In our case, we have a semi-analytical solution for the optimal reflection coefficient at each sea state in terms of the constraint centers and radii at that sea state,
$\Gamma_{opt}(r_{\mu}, \Gamma_{c,\mu})$,
so we can compute the Lagrange multipliers as follows:
\begin{equation}\label{eq:lagrange}
\begin{aligned}
   \lambda_\mu 
    &\equiv 
    \frac{\partial p_{avg,VI,opt}}{\partial g_\mu}
    \\
    &= - \frac{\partial p_{avg,VI,opt}}{\partial b_\mu}
    \\
    &= -
    \frac{\partial p_{avg,VI,opt}}{\partial |\Gamma_{opt}|}
    \frac{\partial |\Gamma_{opt}|}{\partial b_\mu}
    \\
    &= -
    \frac{\partial p_{avg,VI,opt}}{\partial |\Gamma_{opt}|}
    \left(
        \frac{\partial |\Gamma_{opt}|}{\partial r_{\mu}}
        \frac{\partial r_{\mu}}{\partial b_\mu}
        +
        \frac{\partial |\Gamma_{opt}|}{\partial \Gamma_{c,\mu}}
        \frac{\partial \Gamma_{c,\mu}}{\partial b_\mu}
    \right) 
\end{aligned}
\end{equation}
The computation of each of the five partials in the final expression of \Cref{eq:lagrange} is discussed next.

\subsubsection{Effect of Constraint Limits on Circle Radii}
Recall from \Cref{eq:gamma-disc-radius} that each inequality constraint maps to a circle in the complex $\Gamma$ plane 
with center $\Gamma_{c,\mu}$ and radius $r_\mu$ given by:
\begin{equation}\label{eq:radius-full}
    r_\mu^2
    = \frac{b_{\mu} - \vec{x}_p^{\,*}\mathbf{Q}_{\mu}\vec{x}_p
            - 2\Re\!\left\{\vec{a}_{\mu}^{\,*}\vec{x}_p\right\}}
           {|\hat{I}_p|^{2}\,\vec{n}^{\,*}\mathbf{Q}_{\mu}\vec{n}}
    + |\Gamma_{c,\mu}|^2
\end{equation}
The constraint limit $L_\mu$ enters $r_\mu$ only through $b_\mu$ 
(the right-hand side of the quadratic inequality in \Cref{tab:qp-constraints}),
while $\vec{x}_p$, $\vec{n}$, $\hat{I}_p$, $\mathbf{Q}_\mu$, $\vec{a}_\mu$, and $\Gamma_{c,\mu}$ depend only on the Th\'{e}venin impedance $Z_{s,th}$ and excitation $\hat{V}_{s,th}$ for the sea state,
which are independent of the constraint limits.
Defining $d_\mu = |\hat{I}_p|^{2}\,\vec{n}^{\,*}\mathbf{Q}_{\mu}\vec{n}$ 
and $c_\mu = d_\mu|\Gamma_{c,\mu}|^2 - \vec{x}_p^{\,*}\mathbf{Q}_{\mu}\vec{x}_p - 2\Re\!\left\{\vec{a}_{\mu}^{\,*}\vec{x}_p\right\}$, 
both independent of the constraint limits, \Cref{eq:radius-full} and its derivative simplify to:
\begin{equation}\label{eq:radius-simplified}
    r_\mu^2 = \frac{b_\mu + c_\mu}{d_\mu}
    , \quad
    \frac{\partial r_\mu}{\partial b_\mu} = \frac{1}{2\sqrt{d_\mu(b_\mu + c_\mu)}}
\end{equation}
Since $r_\mu\in\mathbb{R}$ (circle radii cannot have an imaginary part), then $b_\mu + c_\mu$ and $d_\mu$ must have the same sign, so $\frac{\partial r_\mu}{\partial b_\mu}$ is always positive.
In other words, increasing $b_\mu$ always enlarges the circle radius, regardless of whether the constraint represents the interior or exterior of the circle.
Meanwhile, \Cref{eq:gamma-disc-center} shows that the center $\Gamma_{c,\mu}$ does not depend on $b_\mu$, so $\frac{\partial \Gamma_{c,\mu}}{\partial b_\mu}=0$.
Thus, the constraint limit $L_\mu$ does not affect the circle center.

\subsubsection{Optimal Reflection Coefficient as a Function of Constraint Limits}
Now we combine the relationship between circle radius and optimal reflection coefficient $\Gamma_{opt}(r_\mu)$, with the relationship between circle radius and constraint limit $r_\mu(b_\mu)$ derived above
to obtain a relationship between reflection coefficient magnitude and constraint limit in a given sea state, $|\Gamma_{opt}|(b_\mu)$.
Substituting \Cref{eq:radius-simplified} into \Cref{eq:candidate-points} shows that
the squared optimal reflection coefficient magnitude in each sea state takes one of the following values depending on the number of active constraints:
\begin{equation}\label{eq:gamma-opt-summary}
    |\Gamma_{opt}|^2 =
    \begin{cases}
        0 
    & \text{none active} \\
        |\Gamma_{c,\mu^*}|^2 
        - 2|\Gamma_{c,\mu^*}|\sqrt{\frac{b_{\mu^*}+c_{\mu^*}}{d_{\mu^*}}} 
        + \frac{b_{\mu^*}+c_{\mu^*}}{d_{\mu^*}} 
    & \text{only $\mu^*$ active} \\
        \left|
            \Gamma_{c,\mu^*} + \frac{\Gamma_{c,\nu^*} - \Gamma_{c,\mu^*}}{d_{\mu^*\nu^*}}
            \left(
                a_{\mu^*\nu^*}
                \pm
                1i \sqrt{\frac{b_{\mu^*} + c_{\mu^*}}{d_{\mu^*}}-a_{\mu^*\nu^*}^2}
            \right)
        \right|^2 
    & \text{$\mu^*$ \& $\nu^*$ active}
    \end{cases}
\end{equation}
where $\mu^*,\nu^*$ refer to the indices of the active constraint(s) for a given sea state,
and $a_{\mu^*,\nu^*}$ and $d_{\mu^*,\nu^*}$ are real numbers defined in the discussion following \Cref{eq:circle-intersection}.
$d_{\mu^*,\nu^*}$ is independent of the constraint limits, while $a_{\mu^*,\nu^*}$ is affine in $b_{\mu^*}$ and $b_{\nu^*}$.
Thus, using \textrm{aff}() and \textrm{quad}() to denote affine and quadratic functions of the arguments, we have the following scale behavior:
\begin{equation}\label{eq:gamma-opt-scale}
    |\Gamma_{opt}|^2 =
    \begin{cases}
        0 
    & \text{none active} \\
        \textrm{aff}(b_{\mu^*}) + \sqrt{\textrm{aff}(b_{\mu^*})} 
    & \text{only $\mu^*$ active} \\
        \left|
            \textrm{aff}(b_{\mu^*})+\textrm{aff}(b_{\nu^*})
            \pm
            1i \sqrt{\textrm{quad}(b_{\mu^*}, b_{\nu^*})}
        \right|^2 
    & \text{$\mu^*$ \& $\nu^*$ active} \\
    = \textrm{quad}(b_{\mu^*}, b_{\nu^*}) + 
    (\textrm{aff}(b_{\mu^*}) + \textrm{aff}(b_{\nu^*}))
    \sqrt{\textrm{quad}(b_{\mu^*}, b_{\nu^*})} & 
    \end{cases}
\end{equation}
Constraint activity indices $\mu^*,\nu^*$ depend on the constraint limits,
and this dependence must be accounted for before the curvature of $|\Gamma_{opt}|^2$ can be analyzed.
From \Cref{eq:gamma-opt-quadratic}, we can define $\mu^*$ and $\nu^*$ more explicitly as functions of $b_\mu$ by defining a penalty function $f(\Gamma)$ that returns $\Gamma$ if $\Gamma$ is feasible and $\infty$ otherwise.
This function can be constructed as $f(\Gamma) = \Gamma + \sum_\mu \mathbb{I}_\mu(\Gamma)$,
where $\mathbb{I}_\mu(\Gamma)$ is an indicator function that returns $0$ if $\Gamma$ satisfies the $\mu$th constraint and $\infty$ otherwise.
The penalty function is convex in $\Gamma$ if and only if all constraints are convex, i.e., if $S_\mu=1 ~\forall \mu$.
Then, the optimal reflection coefficient of each candidate type can be written as the minimum of the candidate points passed through the penalty function.
\begin{equation}\label{eq:gamma-opt-min-per-candidate}
\begin{aligned}
   |\Gamma_{cl,\mu^*}| &= 
    \min_\mu \left\{
        | f\left(\Gamma_{cl,\mu}(b_\mu)\right) |
    \right\}, \quad
    |\Gamma_{int,\mu^*,\nu^*}| =
    \min_{\mu,\nu} \left\{
        | f\left(\Gamma_{\mu,\nu}(b_\mu, b_\nu)\right) |
    \right\}
\end{aligned}
\end{equation}
Finally, the overall optimal reflection coefficient $|\Gamma_{opt}|$ is the minimum of all the per-candidate-type minima:
\begin{equation}\label{eq:gamma-opt-min}
\begin{aligned}
    |\Gamma_{opt}| &=
    \min \left\{
        |\Gamma_{cl,\mu^*}(b_\mu)|,  |\Gamma_{int,\mu^*,\nu^*}(b_\mu, b_\nu)|
    \right\} 
\end{aligned}
\end{equation}

Working from the inside out, we now analyze the convexity of $|\Gamma_{opt}|^2$ with respect to $b_\mu$.
Starting with the candidates at closest points, the square root is a concave increasing function, 
and a concave increasing function of an affine function is concave, 
so $\sqrt{\textrm{aff}(b_{\mu})}$ is concave.
The sum of a concave function and an affine function is concave, 
so $|\Gamma_{cl}|=\textrm{aff}(b_{\mu}) + \sqrt{\textrm{aff}(b_{\mu})}$ is concave.
The penalty function is convex if and only if all constraints are convex, 
so $-|f(\Gamma_{cl})|$ is concave if and only if $S_\mu=1 ~\forall \mu$.
The minimum of a set of concave functions is a concave increasing function,
and $|\Gamma_{cl,\mu^*}|$ is the minimum of a set of concave functions if and only if all constraints are concave, 
so $|\Gamma_{cl,\mu^*}|$ is concave if and only if $S_\mu=1 ~\forall \mu$.

Moving onto the candidates at intersection points, a quadratic function is concave if its leading order term is negative, and the leading order term of $\textrm{quad}(b_{\mu}, b_{\nu})$ is negative (arising from the subtraction of squared-affine term $a_{\mu\nu}^2$).
The geometric mean of two positive values is a concave function, so the geometric mean of two positive concave functions is concave.
The term $\left(\textrm{aff}(b_{\mu}) + \textrm{aff}(b_{\nu})\right) \sqrt{\textrm{quad}(b_{\mu}, b_{\nu})}$
is the geometric mean of $(\textrm{aff}(b_{\mu})+\textrm{aff}(b_{\nu}))^2$ and $\textrm{quad}(b_{\mu}, b_{\nu})$.
The $(\textrm{aff}(b_{\mu})+\textrm{aff}(b_{\nu}))^2$ is positive but is convex rather than concave.
This means the curvature of $\left(\textrm{aff}(b_{\mu}) + \textrm{aff}(b_{\nu})\right) \sqrt{\textrm{quad}(b_{\mu}, b_{\nu})}$ cannot be established with the present analysis.
If the curvature of $\left(\textrm{aff}(b_{\mu}) + \textrm{aff}(b_{\nu})\right) \sqrt{\textrm{quad}(b_{\mu}, b_{\nu})}$ could be established as concave, 
then the sum of two concave functions is concave, so the candidate points $|\Gamma_{int}|$ would be concave.
If that were the case, then by the same argument as for the closest points, the minimum of feasible $|\Gamma_{int}|$ would be concave if and only if all constraints are concave.
The optimum $|\Gamma_{opt}|$ is the minimum of the feasible $|\Gamma_{cl}|$ and $|\Gamma_{int}|$, so if both of those are concave, then it is also concave.
Therefore, the criteria for the concavity of $|\Gamma_{opt}|$ with respect to $b_\mu$ is that all constraints are convex (interior of circles), and the curvature of $\left(\textrm{aff}(b_{\mu}) + \textrm{aff}(b_{\nu})\right) \sqrt{\textrm{quad}(b_{\mu}, b_{\nu})}$ is concave.
As future work, the above curvature analysis should be verified with a disciplined convex programming package such as CVXPY.

To facilitate the analysis of the next section, we observe that although the squared optimal reflection coefficient magnitude, $|\Gamma_{opt}|^2$,
has so far been expressed as a function only of the active constraints $(\mu^*,\nu^*)$,
it can be written as a sum over all constraint pairs $(\mu,\nu)$ multiplied by an appropriate Kronecker delta function:
\begin{equation} 
\label{eq:gamma-opt-scale-sum}
\begin{aligned}
    |\Gamma_{opt}|^2 
    = &\sum_\mu \sum_{\nu > \mu}
    \left[
        \delta_{\mu,\mu^*}(1-\delta_{\nu,\nu^*})
        \left(
            \textrm{aff}(b_{\mu})
            + \sqrt{\textrm{aff}(b_{\mu})}
        \right)
     + \delta_{\mu,\mu^*}\delta_{\nu,\nu^*}
    \right.
\\
    &\left.
        \left(
            \textrm{quad}(b_{\mu}, b_{\nu}) + 
            \left(\textrm{aff}(b_{\mu}) + \textrm{aff}(b_{\nu})\right)
            \sqrt{\textrm{quad}(b_{\mu}, b_{\nu})}
        \right)
    \right]
\end{aligned}
\end{equation}
where $\delta_{\mu,\mu^*}\delta_{\nu,\nu^*}$ is non-zero when both $\mu$ and $\nu$ are active in that sea state, 
and $\delta_{\mu,\mu^*}(1-\delta_{\nu,\nu^*})$ is non-zero when $\mu$ is active and $\nu$ is inactive in the sea state.
Both Kronecker delta functions are implicitly functions of the constraint limit $b_\mu$, as seen in \Cref{eq:gamma-opt-min}.  

\subsubsection{Average Power as a Function of Constraint Limits}
The results just obtained establish conditions for the convexity of power in a single sea state to the constraint limit in that sea state, which we now make explicit by adding a $\beta$ subscript to all sea state dependent quantities.
For LCOE, we are interested in the weighted average power across sea states as a function of the constraint limits.
All constraint limits are independent of the sea state ($L_{\mu\beta}=L_\mu$), with the exception of the amplitude limits due to slamming.
\Cref{sec:appendix-slam} will show that the slamming limits are a function of $\Delta z_{\text{slam}}/H$, which depends explicitly on the wave height, and $\theta$, which depends implicitly on both the wave period and wave height through the guessed phase $\angle\xi$.
Since the PTO sizing limits are of primary relevance, the following derivation assumes that the constraint limits are independent of the sea state to simplify the expression, although sea-state-dependent limits can easily be accommodated by adding a $\beta$ index on $b_\mu$ and carrying it through the derivation.

From \Cref{eq:power-elec-sum,eq:gamma-opt-summary}, the average power can be written as:
\begin{equation}\label{eq:power-elec-sensitivity}
    \overline{P}_{elec} = \overline{P}_{elec,0}
    - \eta \sum_{\beta=1}^{N_{sea}} w_\beta
    \cdot |\Gamma_{\beta,opt}|^2
\end{equation}
where $\overline{P}_{\text{elec},0}$ is the power available if all sea states were unconstrained (impedance-matched) and 
$w_\beta=JPD_\beta \cdot
    \frac{|\hat{V}_{s,th,\beta}|^2}
         {8\Re(Z_{s,th,\beta})}$ is a constant weight for each sea state.
The second term represents power lost to constraint-induced impedance mismatch and depends on the constraint limits through \Cref{eq:gamma-opt-summary}.

Substituting \Cref{eq:gamma-opt-scale-sum} into \Cref{eq:power-elec-sensitivity} requires a $\beta$ subscript on the starred index to the Kronecker delta functions since constraint activity varies by sea state, as well as a $\beta$ subscript on the affine and quadratic functions to capture sea-state dependence of the Th\'{e}venin parameters:
\begin{equation}
\begin{aligned}
    \overline{P}_{elec} = &\overline{P}_{elec,0}
    - \eta \sum_\mu \sum_{\nu > \mu}
    \sum_{\beta=1}^{N_{sea}} w_\beta
    \left[
        \delta_{\mu,\mu^*_\beta}(1-\delta_{\nu,\nu^*_\beta})
        \left(
            \textrm{aff}_\beta(b_{\mu})
            + \sqrt{\textrm{aff}_\beta(b_{\mu})}
        \right)
    \right.
\\
    &\left.
        + \delta_{\mu,\mu^*_\beta}\delta_{\nu,\nu^*_\beta}
        \left(
            \textrm{quad}_\beta(b_{\mu}, b_{\nu}) + 
            \left(\textrm{aff}_\beta(b_{\mu}) + \textrm{aff}_\beta(b_{\nu})\right)
            \sqrt{\textrm{quad}_\beta(b_{\mu}, b_{\nu})}
        \right)
    \right]
\end{aligned}
\end{equation} 
Expanding under the assumption that each constraint is active by itself in some sea states and active with another constraint in others, 
we find that $\overline{P}_{\text{elec}}$ is a piecewise function of each constraint limit $b_\mu$, with a quadratic leading order term as well as affine, square root of affine, and affine times square root of quadratic terms.
The two Kronecker delta terms are combined with the weighting term $w_\beta$ to create revised weighting terms $\tilde{w}_{1\mu\nu\beta}$ and $\tilde{w}_{2\mu\nu\beta}$ that are non-zero only when the corresponding constraint(s) are active in that sea state:
\begin{equation}\label{eq:power-triple-sum}
\begin{aligned}
    \overline{P}_{elec} &= \overline{P}_{elec,0}
    - \eta \sum_\mu \sum_{\nu > \mu}
    \left[
        \sum_{\beta=1}^{N_{sea}} \tilde{w}_{1\mu\nu\beta}
        \left(
            \textrm{aff}_\beta(b_{\mu})
            + \sqrt{\textrm{aff}_\beta(b_{\mu})}
        \right)
    \right. + \\
    &\left.
     \sum_{\beta=1}^{N_{sea}} \tilde{w}_{2\mu\nu\beta}
        \left(
            \textrm{quad}_\beta(b_{\mu}, b_{\nu}) + 
            \left(\textrm{aff}_\beta(b_{\mu}) + \textrm{aff}_\beta(b_{\nu})\right)
            \sqrt{\textrm{quad}_\beta(b_{\mu}, b_{\nu})}
        \right)
    \right]
\end{aligned}
\end{equation}
so the power-constraint relationship is a sum over sea states of piecewise functions of the constraint limit $b_\mu$.
Recognizing that the sea state sum in \eqref{eq:power-triple-sum} essentially just scales the individual piecewise functions, the expression can be simplified by absorbing the weights into the subscripts of the affine and quadratic terms, and combining the sum of an affine and quadratic term into a single quadratic term:
\begin{equation}\label{eq:power-double-sum}
\begin{aligned}
    \overline{P}_{elec} = \overline{P}_{elec,0}
    - \eta \sum_{\mu\nu\beta}
    &\left[
        \sqrt{\textrm{aff}_{\mu\nu\beta}(b_{\mu})}
        + \textrm{quad}_{\mu\nu\beta}(b_{\mu}, b_{\nu}) 
    +
        \left(\textrm{aff}_{\mu\nu\beta}(b_{\mu}) + \textrm{aff}_{\mu\nu\beta}(b_{\nu})\right)
        \sqrt{\textrm{quad}_{\mu\nu\beta}(b_{\mu}, b_{\nu})}
    \right]
\end{aligned}\end{equation}
We can then use a similar analysis as before to obtain conditions for the convexity of the overall power-constraint relationship.

\subsubsection{Extension to Dimension Sensitivity}
\label{sec:appendix-dimension-sensitivity}
In principle, the same parametric programming approach can be used to semi-analytically compute the derivatives of the constrained optimal power with respect to bulk dimensions, 
though the implementation remains out of reach.
These derivatives would be highly valuable, and indeed would enable the full wave-to-wire control co-design optimization problem to be solved semi-analytically, rather than the present approach which requires a numerical bulk-dimension optimization to be wrapped around the semi-analytical control optimization and simulation.
Computing bulk-dimension derivatives would require not only $\partial \Gamma_{opt}/\partial b_\mu$ as we derive here,
but also the derivatives of Th\'{e}venin parameters with respect to bulk dimensions, 
because bulk dimensions affect $\mathbf{Q}_\mu, \vec{a}_\mu$, and the unconstrained power through the Th\'{e}venin parameters.
The sensitivity formula would become significantly more complex and less intuitive with these additional terms, 
and more importantly, the Th\'{e}venin-to-dimension derivatives are not currently available in MDOcean because they depend on the hydrodynamic coefficient-to-dimension derivatives.
Application of the adjoint method has recently enabled efficient computation of hydrodynamic coefficient derivatives in the context of a numerical boundary element method solver \cite{khanal_fully_2025}, 
but the adjoint method has not yet been implemented for the semi-analytical MEEM model used here.
Implementing the adjoint for a linear solve is relatively straightforward in a multidisciplinary design optimization package like OpenMDAO, 
and would unlock a fully semi-analytical differentiation workflow.

\subsection{Force Saturation}
\label{sec:appendix-force-sat}
\Cref{sec:dynamics} introduces the use of describing functions to quantify the fundamental amplitude of non-sinusoidal force waveforms, which can achieve higher powers than the equivalent sinusoidal waveform for the same force limit.

The simulation permits the nonsinusoidal waveform to have a maximum fundamental a factor of $\frac{4}{\pi}$ above the force limit.
This coincides with the fundamental to peak ratio of a square wave, which is the limiting case both for the bang-singular-bang controller (a piecewise-discontinuous sine-square wave combination that \cite{hendrikx_optimal_2017} finds as the analytically optimal solution), and for the saturated sine controller (a piecewise-continuous sine-square wave combination that \cite{coe_initial_2020} finds as numerically optimal).
In fact, the optimization need not specify the exact nonlinear waveform, since under the describing function filtering hypothesis, only the fundamental matters for simulation purposes.
For hardware implementation purposes, the nonlinear controller can be obtained for the saturated-sine controller by the process outlined in the study \cite{mccabe_force-limited_2024}. 

The appropriate complex scale factor from the optimal unsaturated controller (damping or reactive respectively) is found as described in \Cref{sec:appendix-qp-solution} and will be real for the damping control case and complex in the reactive control case. 

\subsection{Drag Model
\label{sec:appendix-drag}}
As described in \Cref{sec:drag}, a describing function approximation is used to model the form drag.
The relative-velocity computation in \Cref{eq:drag-pressure} uses the standard finite-depth linear-wave velocity phasor evaluated at body draft, with $\vec{T} = [T_{f,2}, T_s]$ collecting the bottom drafts of the float and spar:
\begin{equation}\label{eq:wave-velocity-phasor}
    \vec{\hat{V}}_{wave}(y) = \frac{H}{2}\,\frac{gk}{\omega}\, \exp\{-k\vec{T}-iky\}.
\end{equation}

Strip theory integrates the spatially-dependent infinitesimal drag force vector $d\vec{F}_d$ over the wetted surface area $A_{wet}$.
For a hollow cylinder of outer radius $\vec{R}$ and inner radius $\vec{R}_{in}$,
\begin{equation}\label{eq:drag-force-integral}
    \vec{\hat{F}}_d = \int_{A_{wet}} d\vec{\hat{F}}_d(y) = -\int_{y=-\vec{R}}^{\vec{R}} \vec{w}(y)\, \vec{\hat{p}}_d(y)\, dy,
\end{equation}
where $\vec{w}(y)$ is the local body width perpendicular to wave propagation:
\begin{equation}\label{eq:drag-width}
    \vec{w}(y) = \begin{cases}
        2\sqrt{\vec{R}^2-y^2}, & \vec{R}_{in}<|y|<\vec{R} \\
        2\sqrt{\vec{R}^2-y^2} - 2\sqrt{\vec{R}_{in}^2-y^2}, & 0<|y|<\vec{R}_{in}.
    \end{cases}
\end{equation}

Plugging the describing function \Cref{eq:drag-pressure} into the drag force integral \eqref{eq:drag-force-integral} reveals the quasi-linearized drag force to be the sum of a damping term in phase with the body motion and an excitation term potentially out of phase with the body motion:
\begin{equation}\label{eq:drag-damping-excitation-2}
    \vec{\hat{F}}_d = -\mathbf{B}_{d} \vec{\hat{\dot{\xi}}}
    + \vec{\gamma}_{d} \hat{\zeta}
\end{equation}
The excitation term arises from the use of the relative velocity rather than direct WEC velocity, which is often overlooked in WEC models and has important implications for the phase of the drag force. 
The damping-excitation grouping of the drag terms, as well as the dependence of the relative velocity on the direction of wave propagation $y$, improves on the approach of \cite{quartier_influence_2021} and is discussed more in \Cref{subsec:appendix-drag-others}.
The coefficients in \Cref{eq:drag-damping-excitation} include real diagonal drag damping matrix $\mathbf{B}_d$ and complex drag excitation vector $\vec{\gamma}_{d}$ defined as follows:
\begin{equation}\label{eq:drag-coeffs}
\begin{aligned}
    \mathbf{B}_{d} &= \frac{8}{3\pi}\frac{1}{2} \rho_w \, \textrm{diag}\left(\vec{C}_d |\vec{\hat{\dot{\xi}}}| \int_{y=-\vec{R}}^{\vec{R}} \vec{w}(y) \vec{\alpha}_v(y) dy \right) \\
    \vec{\gamma}_{d} &= \frac{8}{3\pi} \frac{1}{2} \rho_w \vec{C}_d \, |\vec{\hat{\dot{\xi}}}| \frac{gke^{-k\vec{T}}}{\omega} \int_{y=-\vec{R}}^{\vec{R}} \vec{w}(y) \vec{\alpha}_v(y) e^{-iky} dy 
\end{aligned}
\end{equation}
where we use radii vectors $\vec{R}=[D_f/2,D_d/2]$ and $\vec{R}_{in}=[D_{f,in},0]$ to represent the float and spar.
$\vec{\alpha}_v(y)$ is the amplitude ratio of the relative velocity and WEC velocity:
\begin{equation}\label{eq:drag-alpha-definitions}
    \vec{\alpha}_v(y) = \frac{\vec{|\hat{V}}_{rel}(y)|}{|\vec{\dot{\xi}}|} = \sqrt{1 + \vec{r}^{\, 2} \exp\{-2ky\} - 2 \vec{r} \exp\{-ky\}  \sin\angle \vec{\hat{\dot{\xi}}}}
\end{equation}
and $\vec{r} = |\hat{V}_{\text{wave}}|/ |\vec{\dot{\xi}}|$ is the amplitude ratio of the incident wave velocity and WEC velocity.

\subsubsection{Nondimensional Drag Integrals $I_B$ and $I_G$}
The drag coefficients $\mathbf{B}_{d}$ and $\vec{\gamma}_{d}$ are expressed in \Cref{eq:drag-coeffs} as dimensional integrals over the width of the WEC.
Here we present an alternative form of the coefficients in terms of nondimensionalized integral functions, which we refer to as $I_B(r,\theta,\kappa)$ and $I_G(r,\theta,\kappa)$:
\begin{equation}\label{eq:drag-integrals-IB-IG}
\begin{aligned}
    \mathbf{B}_{d} &= \frac{8}{3\pi} \rho_w R^2 C_d |\hat{\dot{\xi}}|
    \left[ 
        I_B(r,\angle\hat{\dot{\xi}},kR) - \alpha^2 I_B(r,\angle\hat{\dot{\xi}},\alpha k R)
    \right]
    \\
    \vec{\gamma}_{d} &= \frac{8}{3\pi} \rho_w R^2 C_d |\hat{\dot{\xi}}| \frac{g k}{\omega} e^{-k\vec{T}}
    \left[ 
        I_G(r,\angle\hat{\dot{\xi}},kR) - \alpha^2 I_G(r,\angle\hat{\dot{\xi}},\alpha kR)
    \right]
\end{aligned}
\end{equation}
where the nondimensional integral functions $I_B$ (real) and $I_G$ (complex) are defined as:
\begin{equation}\label{eq:definition-IB-IG}
\begin{aligned}
    I_B(r,\theta,\kappa) &= \int_{-1}^1 \sqrt{1-x^2}\sqrt{1+r^2e^{-2\kappa x}-2r\sin\theta e^{-\kappa x}} dx
    \\
    I_G(r,\theta,\kappa) &= \int_{-1}^1 \sqrt{1-x^2}\sqrt{1+r^2e^{-2\kappa x}-2r\sin\theta e^{-\kappa x}} e^{-i\kappa x} dx
\end{aligned}
\end{equation}

The integrals must be evaluated numerically, except for special limit cases for which analytical results are available.
\Cref{fig:drag-integrals} shows the results of numerical integration for $I_B$ and $I_G$ across a range of inputs.
\begin{figure}[h!]
    \centering
    \begin{subfigure}[t]{0.32\textwidth}
        \includegraphics[width=\textwidth]{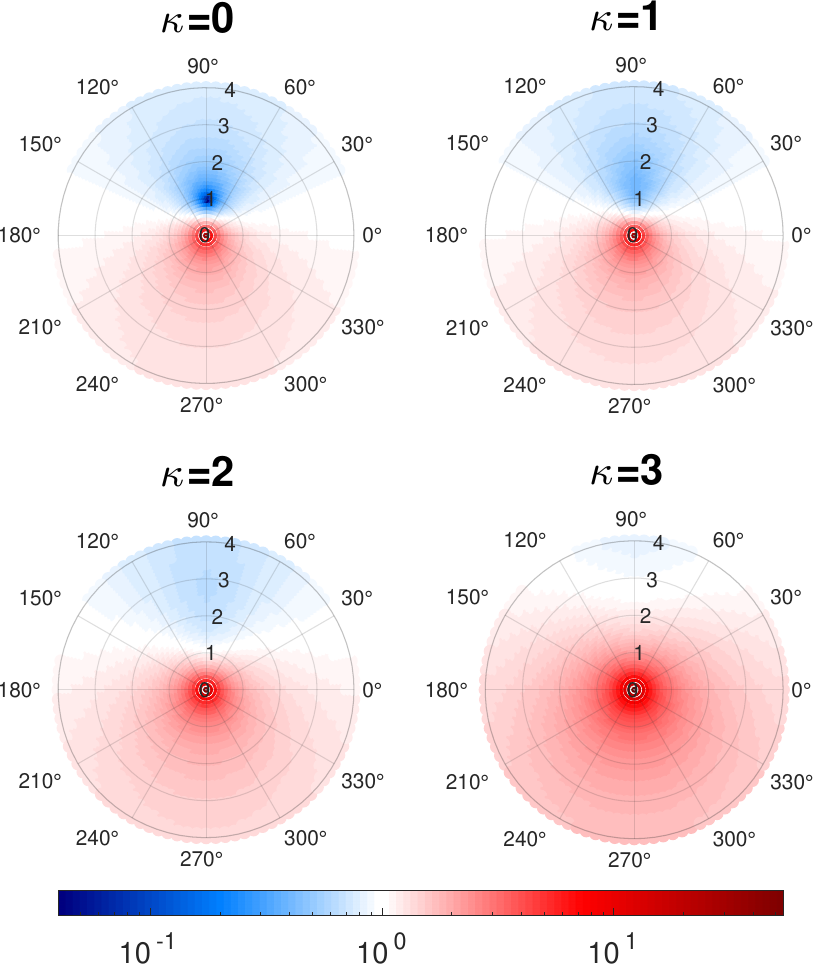}
        \caption{Drag damping integral $I_B$, scaled by $\frac{2}{\pi}$.}
        \label{fig:drag-damping-integral}
    \end{subfigure}
    \hfill
    \begin{subfigure}[t]{0.32\textwidth}
        \includegraphics[width=\textwidth]{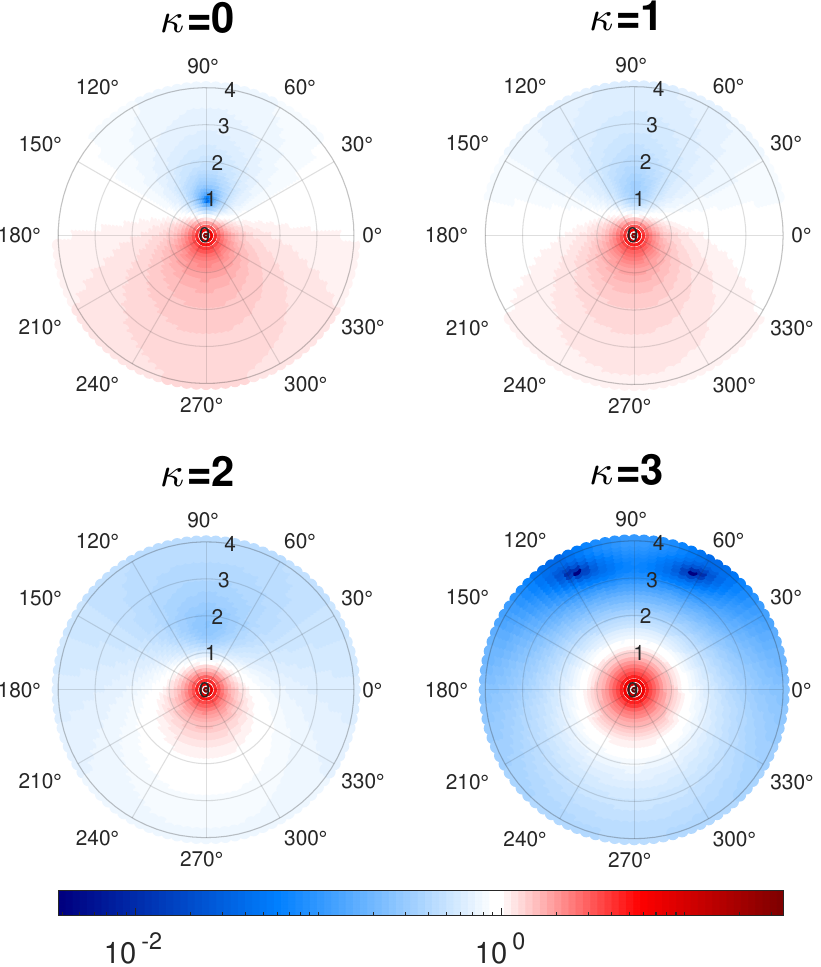}
        \caption{Magnitude of drag excitation integral $|I_G|$, scaled by $\frac{2}{\pi}$.}
        \label{fig:drag-excitation-magnitude-integral}
    \end{subfigure}
    \hfill
    \begin{subfigure}[t]{0.32\textwidth}
        \includegraphics[width=\textwidth]{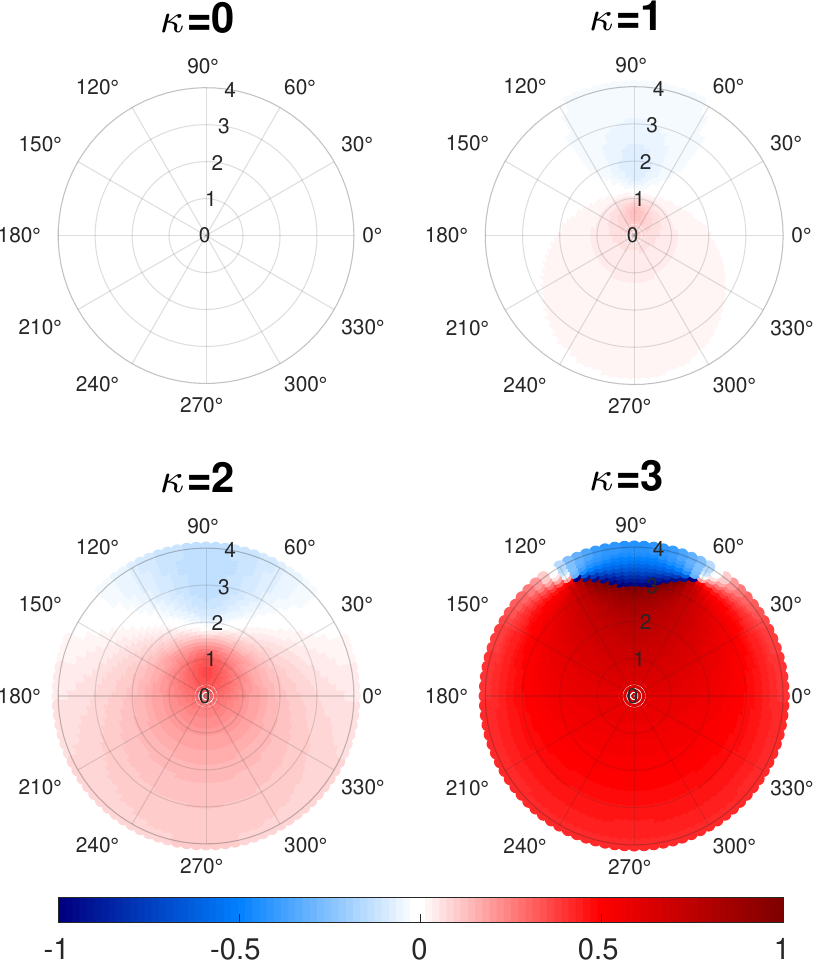}
        \caption{Phase of drag excitation integral $\angle I_G$, scaled by $\frac{1}{\pi}$.}
        \label{fig:drag-excitation-phase-integral}
    \end{subfigure}
    \caption{Nondimensional drag integrals shown in polar coordinates for various values of $\kappa$, where the polar radius represents $1/r$ and the polar angle represents $\theta$.}
    \label{fig:drag-integrals}
\end{figure}

The limit cases in question include $r \rightarrow 0$ (the body velocity dominates the incident wave velocity) and $\kappa \rightarrow 0$ (the body diameter is small with respect to the wavelength), with the following expressions:
\begin{equation}
\begin{aligned}
    \lim_{r\rightarrow 0} I_B(r,\theta,\kappa) = \frac{\pi}{2},
    \qquad
    \lim_{r\rightarrow 0} I_G(r,\theta,\kappa) =
    \begin{cases}
    \frac{\pi J_1(\kappa)}{\kappa} & \textrm{if } \kappa \neq 0 \\
    \frac{\pi}{2} & \textrm{if } \kappa = 0
    \end{cases} \\
    \lim_{\kappa \rightarrow 0} I_B(r,\theta,\kappa) =  \lim_{\kappa \rightarrow 0} I_G(r,\theta,\kappa) = \frac{\pi}{2}\sqrt{1+r^2-2r\sin\theta},
\end{aligned}
\end{equation}
where $J_1$ is the Bessel function of the first kind of order 1.
An important edge case occurs at the origin of the polar plots in \Cref{fig:drag-integrals}, when $r \rightarrow \infty$ (the incident wave velocity dominates the body velocity, so $|\hat{\dot{\xi}}|\rightarrow 0$).
Here, the integrals as defined in \Cref{eq:definition-IB-IG} diverge, and the form of \Cref{eq:drag-integrals-IB-IG} does not apply.
Instead, when $r\rightarrow \infty$, we apply an alternative convention that replaces the WEC speed in \Cref{eq:drag-integrals-IB-IG} with the incident wave speed and uses a primed form of the integrals, $I' = I/r$:
\begin{equation}
\begin{aligned}
    \mathbf{B}_{d} &= \frac{8}{3\pi} \rho_w R^2 C_d |\hat{V}_{wave}|
    \left[ 
        I_B'(r,\angle\hat{\dot{\xi}},kR) - \alpha^2 I_B'(r,\angle\hat{\dot{\xi}},\alpha k R)
    \right]
    \\
    \vec{\gamma}_{d} &= \frac{8}{3\pi} \rho_w R^2 C_d  |\hat{V}_{wave}| \frac{gk}{\omega} e^{-k\vec{T}}
    \left[ 
        I_G'(r,\angle\hat{\dot{\xi}},kR) - \alpha^2 I_G'(r,\angle\hat{\dot{\xi}},\alpha kR)
    \right]
\end{aligned}
\end{equation}
The relevant limits are then:
\begin{equation}
\begin{aligned}
    \lim_{r\rightarrow \infty} I_B'(r,\theta,\kappa) = 0,
    \qquad
    \lim_{r\rightarrow \infty} I_G'(r,\theta,\kappa) = 
    \begin{cases}
    \frac{\pi J_1((1i-1)\kappa)}{(1i-1)\kappa} & \textrm{if } \kappa \neq 0 \\
    \frac{\pi}{2} & \textrm{if } \kappa = 0
    \end{cases}
\end{aligned}
\end{equation}

\subsubsection{Comparison to Other Drag Formulations}
\label{subsec:appendix-drag-others}
The WEC drag model in the study \cite{quartier_influence_2021} also utilizes describing functions with the relative velocity but groups the drag force into damping and stiffness terms rather than damping and excitation terms.
While this representation yields identical results for a floating WEC in the frequency domain, it misleadingly implies that a body fixed at the still water line would experience no drag force from incident waves.
In other words, it fails to account for the edge case of $r \rightarrow \infty$ described above, where the drag force is purely an excitation force and cannot be expressed as a damping or stiffness.
Furthermore, the drag ``damping'' term in the grouping of \cite{quartier_influence_2021} includes not just forces with a dissipative effect on the body but also the portion of nonlinear drag excitation forces that are in phase with the body velocity.
This creates the confusing possibility for negative apparent damping coefficients, leading to false concerns about violating energy conservation if erroneously interpreted as a true damping coefficient.
The damping-excitation grouping chosen in \Cref{sec:drag} avoids this confusion and guarantees positive values of the damping coefficient $\mathbf{B}_d$.

Another improvement over \cite{quartier_influence_2021} is that the present model considers the fact that the incident wave velocity, and thus the relative velocity, differs across the width of the WEC in the direction of wave propagation $y$.
Reference \cite{quartier_influence_2021} more simplistically assumes that the wave velocity at all points equals that at the center of the WEC, which is only accurate when the incident wavelength far exceeds the WEC width.
This means that Ref.~\cite{quartier_influence_2021} captures only the Froude-Krylov-like component of the drag excitation, neglecting the diffraction-like component.
Early MDOcean simulations that used the formulation of \cite{quartier_influence_2021} found this approximation to be unacceptable, producing unstable dynamics (negative drag damping coefficients whose absolute value exceeds the radiation damping) at high frequencies and large wave heights in the operational regime.
The strip-theory approach presented in \Cref{sec:drag} avoids this issue.
Comparisons against WEC-Sim confirm that even without strip theory, assuming that the relative velocity is perfectly in phase with the body velocity (i.e. a drag force of purely damping and in-phase excitation, with no stiffness or out-of-phase excitation) yields more accurate results than the drag formulation used in the study \cite{quartier_influence_2021}.

\subsubsection{Iterative Solution}
With the describing function approximation, the nonlinear time-domain drag equation is now a quasi-linear frequency-domain equation, since the state-dependence of the coefficients $\mathbf{B}_{d}$ and $\vec{\gamma}_{d}$ prevents true linearity.
The solution for states $|\vec{\hat{\dot{\xi}}}|$ and $\angle \vec{\hat{\dot{\xi}}}$ for a given controller can be obtained either through numerical iteration or analytical solution of the nonlinear equation \eqref{eq:eom}.
Iteration is chosen here, the same approach used for a frequency-domain drag simulation of floating offshore wind turbines \cite{hall_open-source_2022}.
MDOcean provides users with two options: to employ a typical nonlinear root-finding algorithm, or to simply use the equation of motion \eqref{eq:eom} to obtain subsequent iterates.
The latter is a form of fixed point iteration where $\vec{\hat{\dot{\xi}}}_k = g(\vec{\hat{\dot{\xi}}}_{k-1})$, where $g(\vec{\hat{\dot{\xi}}})$ is the equation of motion \Cref{eq:eom-freq-domain} and $k$ is the iteration index.
It is therefore guaranteed to converge as long as the dynamics are contracting at the solution, which \cite{chicone_contraction_2006} shows has the following criteria:
\begin{equation}
 \left|
    \frac{\partial}
    {\partial \vec{\hat{\dot{\xi}}}}~\left[
    g\left(
        \vec{\hat{\dot{\xi}}}~
    \right)\right]
\right| < 1
\end{equation}

Five to eight iterations on $|\vec{\hat{\dot{\xi}}}|$ and $\angle \vec{\hat{\dot{\xi}}}$ are typically required to converge all sea states to within 0.01~m and 3 degrees. 
At present (since MDOcean v1.0.0), the solver used for drag iteration can be simultaneously used for the numerical optimal control procedure described in \Cref{sec:optimal-control}, although since v1.2.0 by default the solver is used only for drag convergence and optimal control is done analytically.
Compared to numerical optimal control nested inside the drag solve, combining the solvers in this way has the advantages of code simplicity and reducing the number of dynamics evaluations required for a given simulation, provided that the number of iterations required to converge the coupled drag-control solve is less than the product of the number of iterations required to converge each solve individually.
However, it has the disadvantage that the error signal couples drag nonlinearities with constraint violations, which can potentially lead to slower convergence.

\subsubsection{Optimal Control Condition}
Note that in the quasi-linear formulation, the lack of complete linearity means that the complex-conjugate reactive controller is technically no longer optimal in the unconstrained case.
The optimal load damping $B_l$ is derived by setting
\begin{equation}
\begin{aligned}
0 &= \frac{\partial }{\partial B_l} \left[p_{avg,VI}\right]
= \frac{\partial}{\partial B_l} \left[\frac{1}{2} B_l |\hat{I}|^2\right]
= \frac{\partial }{\partial B_l} \left[\frac{1}{2} B_l \left|\frac{\hat{V}_{s,th}}{Z_{s,th}+B_l+K_l/s}\right|^2 \right]
\end{aligned}
\end{equation}
and the typical simplification $\frac{\partial \hat{V}_{s,th}}{\partial B_l}=\frac{\partial Z_{s,th}}{\partial B_l}=0$ must be replaced with the implicit dependence of the Th\'{e}venin equivalent parameters on $B_l$ because $B_l$ affects $\hat{\dot\xi}$ and therefore $\mathbf{B}_{d}$ and $\vec{\gamma}_{d}$ via \eqref{eq:drag-coeffs}.
In practice, however, the standard complex conjugate controller was observed to produce nearly identical power as the one incorporating the more complicated dependence, so the standard is used for simplicity.

\subsection{Slamming and Submersion Amplitude}
\label{sec:appendix-slam}

In regular waves of height $H$, the conditions for slamming can be derived by comparing the vertical position of the bottom of the WEC, $\xi(t)-\Delta z_{\text{slam}}$, with the wave elevation $\zeta(y,t)$, which depends on the horizontal position coordinate $y$.
To prevent slamming, the criteria
\begin{equation}\label{eq:slamming-time-domain}
\begin{aligned}
    \xi(t) - \Delta z_{slam} &< \zeta(y,t) \\ 
    |\hat{\xi}|\cos(\omega t+\angle \hat{\xi}) - \Delta z_{slam} &< \frac{H}{2}\cos(\omega t - k y)
\end{aligned}
\end{equation}
must be true at all times and over all positions where the body could exit the waves, $-D/2<y<D/2$.
Note that this expression assumes a sinusoidal waveshape but still applies for irregular waves if a wave-by-wave approach \cite{lin_fast_2025} is utilized.
Manipulating \Cref{eq:slamming-time-domain} to ensure the inequality is satisfied for all $t$ yields the criterion:
\begin{equation}
|\hat{\xi}|^2 - H\cos(ky)|\hat{\xi}|\cos\angle\hat{\xi} + H\sin(ky)|\hat{\xi}|\sin\angle\hat{\xi} < \Delta z_{slam}^2-\left(\frac{H}{2}\right)^2
\end{equation}
which is quadratic in $\hat{\xi}$ and therefore already in a form suitable for the constrained optimal control procedure described in \Cref{sec:optimal-control,sec:appendix-qp-solution}, once the appropriate worst-case $y$ is substituted.
The remainder of this section develops an explicit expression for the allowable amplitudes under various scenarios and the corresponding worst-case $y$.
Completing the square results in:
\begin{equation}\label{eq:slamming-squared}
    \left[ |\hat{\xi}|- \frac{H}{2}\cos(\angle\hat{\xi} + ky)\right]^2 < \Delta z_{slam}^2 - (H/2)^2 \sin^2(\angle\hat{\xi} + ky)
\end{equation}
In general, after taking the square root of \Cref{eq:slamming-squared}, both the positive and negative branches must be used.
This yields a maximum and minimum amplitude criteria to prevent slamming, $\xi_{\text{min},\text{slam}}<|\hat{\xi}|<\xi_{\text{max},\text{slam}}$, with the following values:
\begin{equation}\label{eq:slamming-min-max}
\begin{aligned}
    \xi_{max,slam} &= \frac{H}{2} \cos\theta + \sqrt{\Delta z_{slam}^2-\left(\frac{H}{2}\sin\theta\right)^2 } \\
    \xi_{min,slam} &= \frac{H}{2} \cos\theta - \sqrt{\Delta z_{slam}^2-\left(\frac{H}{2}\sin\theta\right)^2 }
\end{aligned}
\end{equation}
where $\theta=\angle\hat{\xi} + ky$ must be evaluated with the value of $y$ that creates the minimum value of $\xi_{\text{max},\text{slam}}$ and the maximum value of $\xi_{\text{min},\text{slam}}$ to ensure lack of slamming across the entire WEC surface.
The relevant value of $\theta$ differs depending on the wave amplitude.

\subsubsection{Case 1: Small Wave Amplitudes}
For the case of $\Delta z_{\text{slam}}>H/2$ (wave amplitudes smaller than the draft), \Cref{eq:slamming-min-max} reduces to only a maximum amplitude criteria $|\hat{\xi}|<\xi_{\text{max},\text{slam}}$ because $\xi_{\text{min},\text{slam}}<0$.
This aligns with the intuition that a body fixed at the still water line would not exit the water when the wave amplitude is less than the draft.
The value of $\theta$ that minimizes $\xi_{\text{max},\text{slam}}$ over the relevant range is 
\begin{equation}\label{eq:slamming-limit-small-waves}
\theta = \pi - \max\left(0,~ \frac{-kD}{2}+|\pi-\angle \xi|\right)
\end{equation}
where the symmetry of \Cref{eq:slamming-min-max} has been used to collapse possible values onto $[0,\pi]$.

This expression reduces to the simple limits $\Delta z_{\text{slam}}-H/2$ and $\Delta z_{\text{slam}}+H/2$ for short wavelengths $(\theta=\pi)$ and long wavelengths $(\theta=0)$ respectively.
These limits are also the minimum and maximum values of $\xi_{\text{max},\text{slam}}$ for any diameter and wave condition where $\Delta z_{\text{slam}}>H/2$.
For intermediate wavelengths, the variable $\theta$ accurately accounts for the effect of body diameter and the phase offset of body motion from the waves.
Intuitively, $\theta$ represents how high up the free surface the bottom of the WEC is allowed to get before the edge exits the water, with $\theta=0$ allowing the body to get all the way to the wave crest (top) and $\theta=\pi$ requiring the float to remain fully below the wave trough (bottom).

\subsubsection{Case 2: Large Wave Amplitudes}
In sufficiently large waves ($H/2>\Delta z_{\text{slam}}$), \Cref{eq:slamming-limit-small-waves} does not apply.
Intuitively, when the wave amplitude exceeds the draft, a minimum amplitude of motion in-phase with the incident wave is required to prevent the wave trough from going below the body surface.
The minimum amplitude arises mathematically when $\xi_{\text{min},\text{slam}}>0$, which occurs exactly for waves meeting the amplitude criterion.
The ``in-phase'' motion requirement is consistent with the observation that in this wave regime, the maximum amplitude in \Cref{eq:slamming-min-max} becomes unsatisfiable ($\xi_{\text{max},\text{slam}}<0$) for WEC phases that are more than 90 degrees out of phase with the incident wave ($\theta \in [\pi/2,3\pi/2]$).

For evaluating \Cref{eq:slamming-min-max}, the value of $\theta$ that produces both minimum $\xi_{\text{max},\text{slam}}$ and maximum $\xi_{\text{min},\text{slam}}$ in this regime is:
\begin{equation}
\theta = \angle\hat{\xi} + \frac{kD}{2}\textrm{sgn}(\pi-\angle\hat{\xi})
\end{equation}
After wrapping the value onto $[0,2\pi]$, it can be mapped to $[0,\pi]$ without changing the value of \Cref{eq:slamming-min-max} via the identity $\cos(\theta)=\cos(\pi-|\pi-\theta|)$.

Notably, in this wave regime it becomes possible for the right hand side of \Cref{eq:slamming-squared} to become negative for certain values of $\angle\hat{\xi}$ and $D$.
This means that those combinations of $\angle\hat{\xi}$ and $D$ are prohibited because slamming would occur regardless of the amplitude $|\hat{\xi}|$.
The following requirement on $D$ ensures that there exist some values of $\angle\hat{\xi}$ without automatic slamming:
\begin{equation}\label{eq:D-slamming}
    kD<2\arcsin\left(\frac{\Delta z_{slam}}{H/2}\right)
\end{equation}
which has upper bound $kD<\pi$.
Once \Cref{eq:D-slamming} is satisfied, the set of acceptable $\angle\hat{\xi}$ that avoid negative values for both $\xi_{\text{max},\text{slam}}$ and the right hand side of \Cref{eq:slamming-squared} are:
\begin{equation}\label{eq:slamming-allowed-angles}
\begin{aligned}
    \angle\hat{\xi} \in &\left[0,~~\arcsin\left(\frac{\Delta z_{slam}}{H/2}\right)-\frac{kD}{2}\right]
    \cup \left[2\pi-\arcsin\left(\frac{\Delta z_{slam}}{H/2}\right)+\frac{kD}{2}, ~~2\pi\right]
\end{aligned}
\end{equation}
At the edge-case $\angle\hat{\xi}$ on the inner boundaries of this interval, only a single amplitude is permitted because the values for $\xi_{\text{max},\text{slam}}$ and $\xi_{\text{min},\text{slam}}$ coincide at $\sqrt{(H/2)^2-\Delta z_{\text{slam}}^2}$.
At any lower amplitude, the WEC would exit the water at the wave trough, and any higher, the WEC would raise out of the water at the wave crest.
The permissible amplitude range can be widened by decreasing $D$ or moving $\angle\hat{\xi}$ towards the closer of $0$ and $2\pi$.
The widest permissible amplitude range occurs when the WEC is perfectly in-phase with the incident wave ($\angle\hat{\xi}=0$) and is small with respect to the wavelength ($kD\approx0$), which yields the limits $H/2-\Delta z_{\text{slam}} < |\hat{\xi}| < H/2+\Delta z_{\text{slam}}$.

\subsubsection{Implementation}
\Cref{fig:slam} visualizes the nondimensional maximum and minimum slamming amplitudes $\xi_{\text{max},\text{slam}}/(H/2)$ and $\xi_{\text{min},\text{slam}}/(H/2)$ as a function of the worst-case phase angle $\theta$ and the draft to wave amplitude ratio $\Delta z_{\text{slam}}/(H/2)$.
This reveals that besides increasing the draft, decreasing $\theta$ can prevent slamming, for example by decreasing the diameter or adjusting $\angle \xi$ via control.
If $\Delta z_{\text{slam}}<H/2|\sin\theta|$, the design is unviable and the slamming limits are undefined, consistent with the negative right hand side of \Cref{eq:slamming-squared} discussed above.
\begin{figure}[htbp]
    \centering
    \includegraphics[width=.8\linewidth]{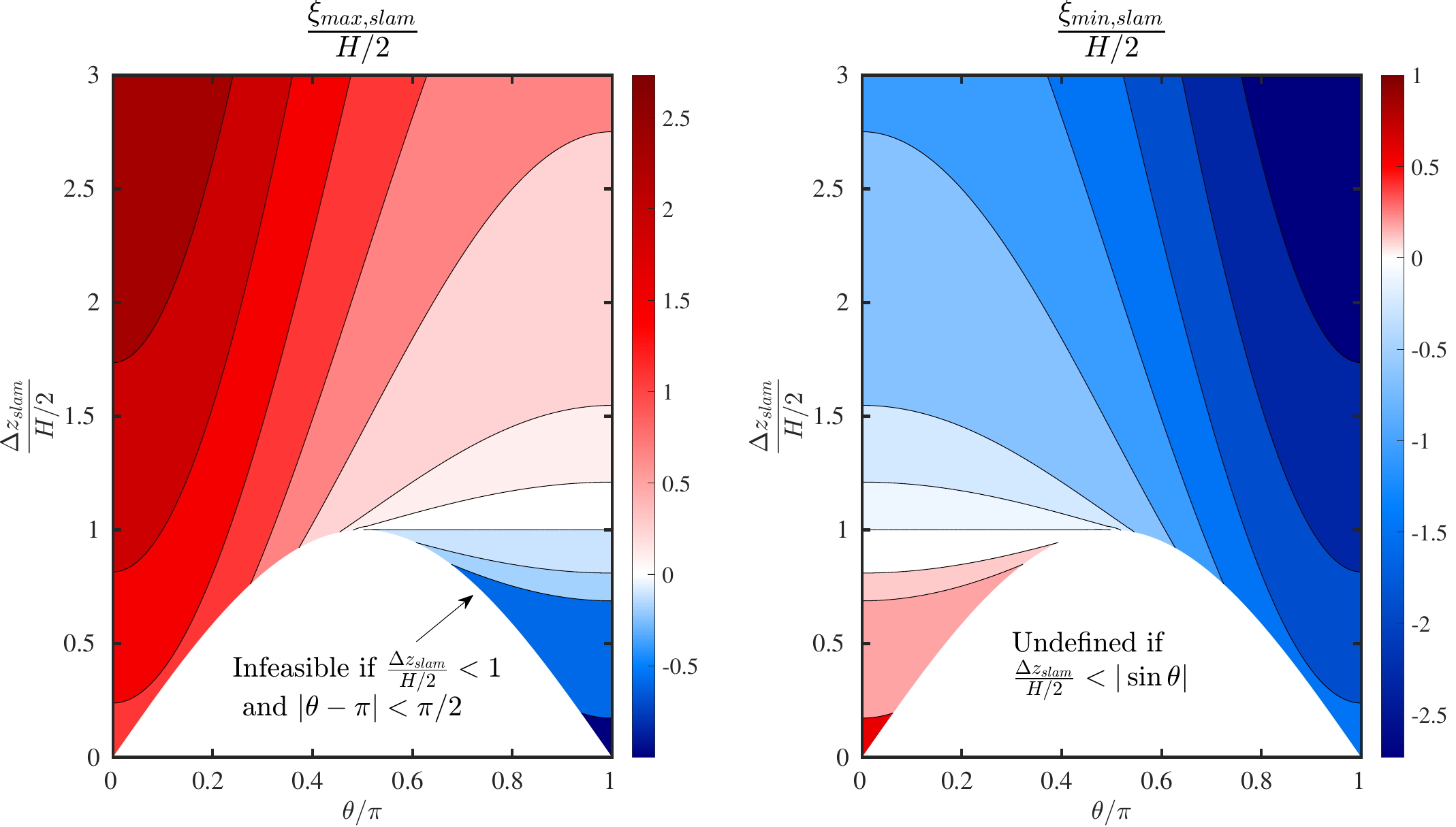}
    \caption{Nondimensional critical slamming amplitude}
    \label{fig:slam}
\end{figure}

Separately, there exists a symmetrical counterpart to the slamming condition that we will call the submersion condition. 
While preventing slamming requires that the bottom of the WEC remains below the free surface, preventing submersion requires that the top of the WEC remains above the free surface. 
Analogous to \eqref{eq:slamming-time-domain}, we obtain:
\begin{equation}\label{eq:submersion}
    \xi(t) + \Delta z_{sub} > \zeta(y,t)
\end{equation}

Conceptually, both the slamming requirement \eqref{eq:slamming-time-domain} and the submersion requirement \eqref{eq:submersion} must be applied to both the float and the spar for a total of 4 free surface constraints for each sea state.
However, the limit on $|\hat{\xi}_f-\hat{\xi}_s|$ described in \Cref{tab:DLCs} already ensures that the top of the spar remains above the top of the float across all dynamic conditions.
Coupled with the fact that the float diameter always exceeds the spar diameter, this means that satisfaction of the float submersion requirement implies automatic satisfaction of the spar submersion requirement.
While this logic works for submersion, it does not apply for slamming because the damping plate diameter may exceed the float diameter.
The remaining three free surface constraints are aggregated into a single constraint for each sea state, using the minimum z-dimension for each body $\Delta z_{\text{surf}}=\min(\Delta z_{\text{sub}}, \Delta z_{\text{slam}})$, where the values for $\Delta z_{\text{sub}}$ and $\Delta z_{\text{slam}}$ were given in \Cref{sec:dynamics}.
This reduces the number of free surface constraints per sea state from four to one.

For operational waves, the vertical dimension $\Delta z$ in \Cref{eq:slam-constraint-quadratic} is set as follows.
For float slamming, $\Delta z = T_{f,1}$ (rather than $T_{f,2}$): this prevents slamming on the slanted underside of the float and additionally maintains constant waterplane area, avoiding unmodeled nonlinearities in the hydrostatic stiffness and Froude–Krylov excitation force.
For spar slamming, $\Delta z = T_s - h_d$ prevents both slamming on the bottom of the spar and surfacing of the damping plate.
For float and spar submersion, $\Delta z$ is set to the above-water heights $h_f - T_{f,2}$ and $h_s - T_s$ respectively.

The storm case slamming constraint is not currently applied.
A future extension should set $\Delta z = T_{f,2}$ in the storm case to prevent slamming on the (now-merged) bottom surface of the float-spar system.

The amplitude limits $h_{fs,\text{clear}}$, $h_{fs,\text{up}}$, $h_{fs,\text{down}}$, $\xi_{f,\text{linear}}$, and $\xi_{s,\text{linear}}$ are sea-state-independent upper limits, while $\xi_{\text{slam}}$ depends on the sea state and can act as both an upper and lower limit on oscillation magnitude.

Relevant model values for the nominal float design under reactive optimal control are plotted for each sea state in \Cref{fig:slamming-validation}.
Comparing the first two subplots for the phase of the float motion $\angle\hat{\xi}_f$ and the slamming angle $\theta$ shows that the slamming angle roughly ``wraps'' the motion angle around a value of $\pi/2$, colored white in the figure.
The third plot shows that the minimum slamming and submersion amplitude is not a concern since it is negative for all sea states and thus will never be an active constraint.
The fourth plot shows the maximum amplitude limit and the fifth depicts the actual amplitude, with amplitudes in violation of the limit indicated with an overlaid black hatching.
A significant fraction of sea states violate the limit and will require adjusted control gains computed via constrained optimal control per \Cref{sec:optimal-control,sec:appendix-qp-solution}.
Significant wave heights in excess of 5.5 meters have negative maximum slamming amplitudes (colored blue), indicating that control adjustments in those waves must not only decrease $|\hat{\xi}_f|$ but also adjust $\angle\hat{\xi}_f$ to reduce $\theta$.
\begin{figure}[htbp]
    \centering
    \includegraphics[width=1.05\linewidth]{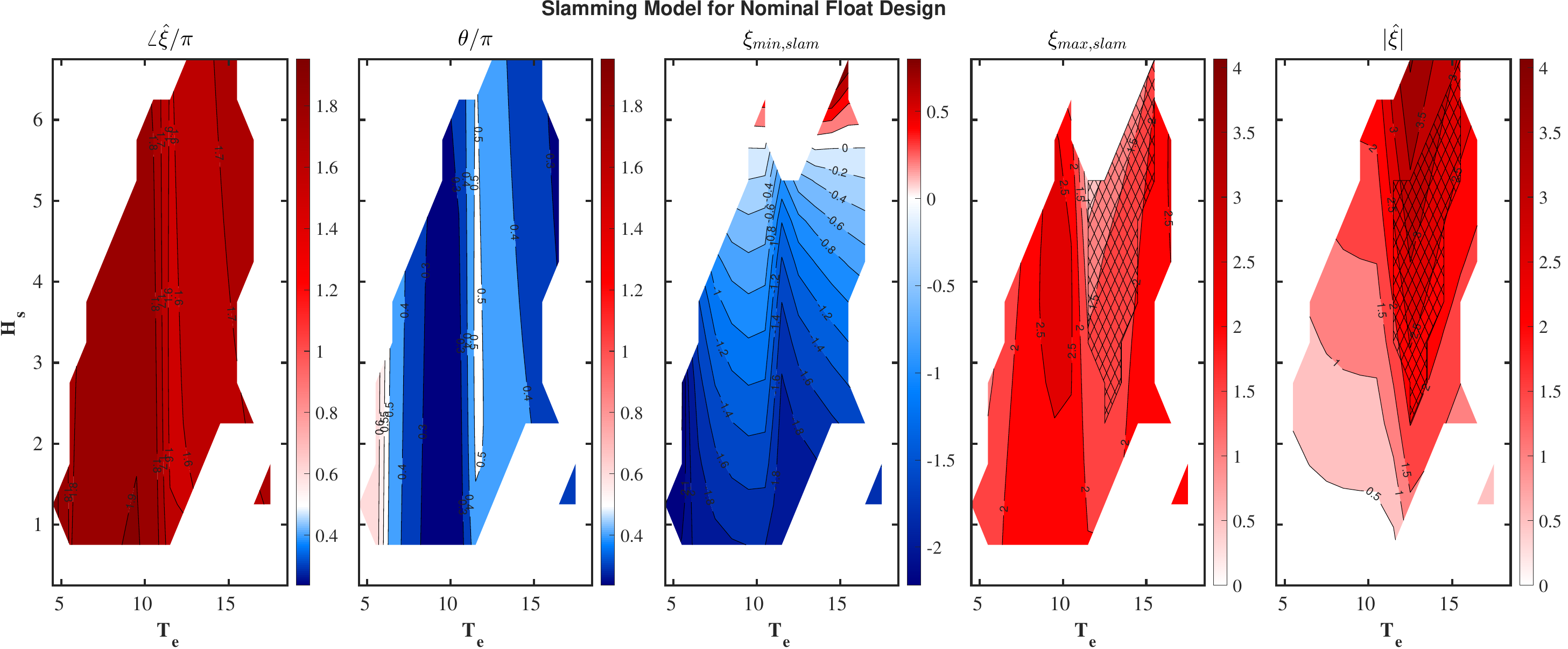}
    \caption{Slamming Model Results for Nominal RM3 Float Design in Operational Sea States}
    \label{fig:slamming-validation}
\end{figure}

\subsubsection{Conservatism of the Fundamental-Amplitude Approximation for Fluid Forces}
\label{sec:appendix-peak-fluid-force}

\Cref{eq:peak-fluid-force} approximates the peak fluid force by the magnitude of its fundamental harmonic phasor.
This approximation can either under- or overestimate the true peak depending on whether the higher-harmonic content (primarily from the drag describing function) interferes constructively or destructively with the fundamental.

The fundamental amplitudes are also used for the peak float and spar displacements $\max_t|\xi_f(t)|$ and $\max_t|\xi_s(t)|$ in both load cases.
The fundamental is more accurate for displacements than for forces because the second-order plant dynamics filter out high-frequency inputs, attenuating drag-induced harmonics in the displacement response.

A future refinement could explicitly reconstruct the time-domain drag force from the describing function and track its peak directly, at the cost of either time-domain integration or additional Fourier-component bookkeeping in the frequency-domain formulation.

\subsection{Dynamic Validation}
\label{sec:appendix-dynamic-validation}

\subsubsection{WEC-Sim}
The drag describing function and MEEM hydrodynamic coefficients have a minor effect assuming a 1-DOF system (9.7\% and 2.7\% error on the average power and maximum amplitude respectively) but a major effect on the 2-DOF system (38.2\% and 28.6\% respectively).
Meanwhile, the error is extremely low in the 2-DOF system enforcing the same hydrodynamic coefficients as WEC-Sim and with zero drag (0.2\% and 1.9\% in power and amplitude respectively).
This indicates that the larger error in the 2-DOF-drag-MEEM case is not an error in the 2-DOF dynamic model itself, but in the way that a 2-DOF model amplifies errors in drag and hydrodynamic coefficients due to the importance of the phase of motion between each DOF.
The remainder of this subsection details the four-quadrant breakdown across drag-on/drag-off and identical/different hydrodynamic coefficient configurations.

A WEC-Sim parallel multiple condition run is performed in accelerator mode for 200s with a 100s ramp time using the ode4 solver and a 10~ms timestep, with the default mass properties and body geometry of the WEC-Sim RM3 tutorial.\footnote{This geometry is not identical to the nominal float design used in the Reference Model Report \cite{RM3} and the rest of this paper, with a shorter spar draft, thicker damping plate, wider float frustum diameter, and assumed deep-water depth regime.
For consistency, MDOcean also uses this alternate geometry for WEC-Sim comparisons only.}
Each of the 210 sea states are simulated individually as the equivalent regular wave.
Body to body interactions (nonzero off-diagonal hydrodynamic coefficients) and state-space computations are turned on.
\Cref{fig:dynamic-validation} compares the results.
The $C_d=0$ simulation with identical hydrodynamic coefficients (top left of \Cref{fig:dynamic-validation}) is conducted to verify correct implementation of the linear dynamics and consistency of sign conventions. 
The validation obtains an error of less than $0.2\%$ for wave periods below 12 seconds.
This aligns with the perfect matching that is expected in the absence of drag non-linearities, and we attribute the tiny error that remains to the discrete-time nature of WEC-Sim's numerical integration and post-processing.
For wave periods above 12 seconds, MDOcean significantly overpredicts the mechanical power.
An additional validation for this case with the spar held fixed (not shown) maintained the excellent $<0.2\%$ accuracy even for $T>12 s$, so we attribute the error to resonant instabilities that occur in the multi-body hydrodynamic interactions in the highly underdamped drag-free dynamics.
Turning on drag ($C_d=1$, top right of \Cref{fig:dynamic-validation}) substantially resolves this inaccuracy, with power errors of $<5\%$ for the most common moderate sea states and error exceeding $40\%$ at the most energetic sea states.
While determining the exact source of the error requires further investigation, it is deemed low-impact enough to justify the model's use in an early-stage design setting.

Moving to the bottom left corner of \Cref{fig:dynamic-validation}, we examine the effect of the error in MDOcean's MEEM hydrodynamic coefficients on its dynamic power production.
The error closely aligns with the error in the identical hydro coefficient case, including the large errors at longer wave periods, confirming that the small error in the MEEM hydro coefficients has minimal effect on the overall dynamic response.
Finally, the bottom right corner of \Cref{fig:dynamic-validation} shows the combined effect of drag and hydro coefficient errors, which is the best representation of MDOcean's total dynamic error in the absence of further model improvements.
Power is underpredicted by up to 40\% at sea states with moderate (10-12s) periods and low wave heights, and overpredicted at both shorter and longer wave periods.
This results in a beneficial positive and negative error cancellation that explains the low error in weighted average power described in \Cref{sec:dynamic-validation}.
Meanwhile, MDOcean underpredicts the peak power by around 25\%.
Error scales more closely with wave period than wave height for periods under 12 seconds, while the opposite is true for periods above 12 seconds.

\begin{landscape}
\begingroup
\begin{figure}[htbp]
\centering
\begin{tabular}{c m{1em} | M{.47\linewidth} | M{.47\linewidth} }
  && \multicolumn{2}{c}{Drag Coefficient}\\
         && $C_d=0$& $C_d=1$\\ \hline
    \multirow{2}{*}[-3em]{\rot{Hydro Coeffs}} &\rot{Identical}& 
    \includegraphics[width=\linewidth]{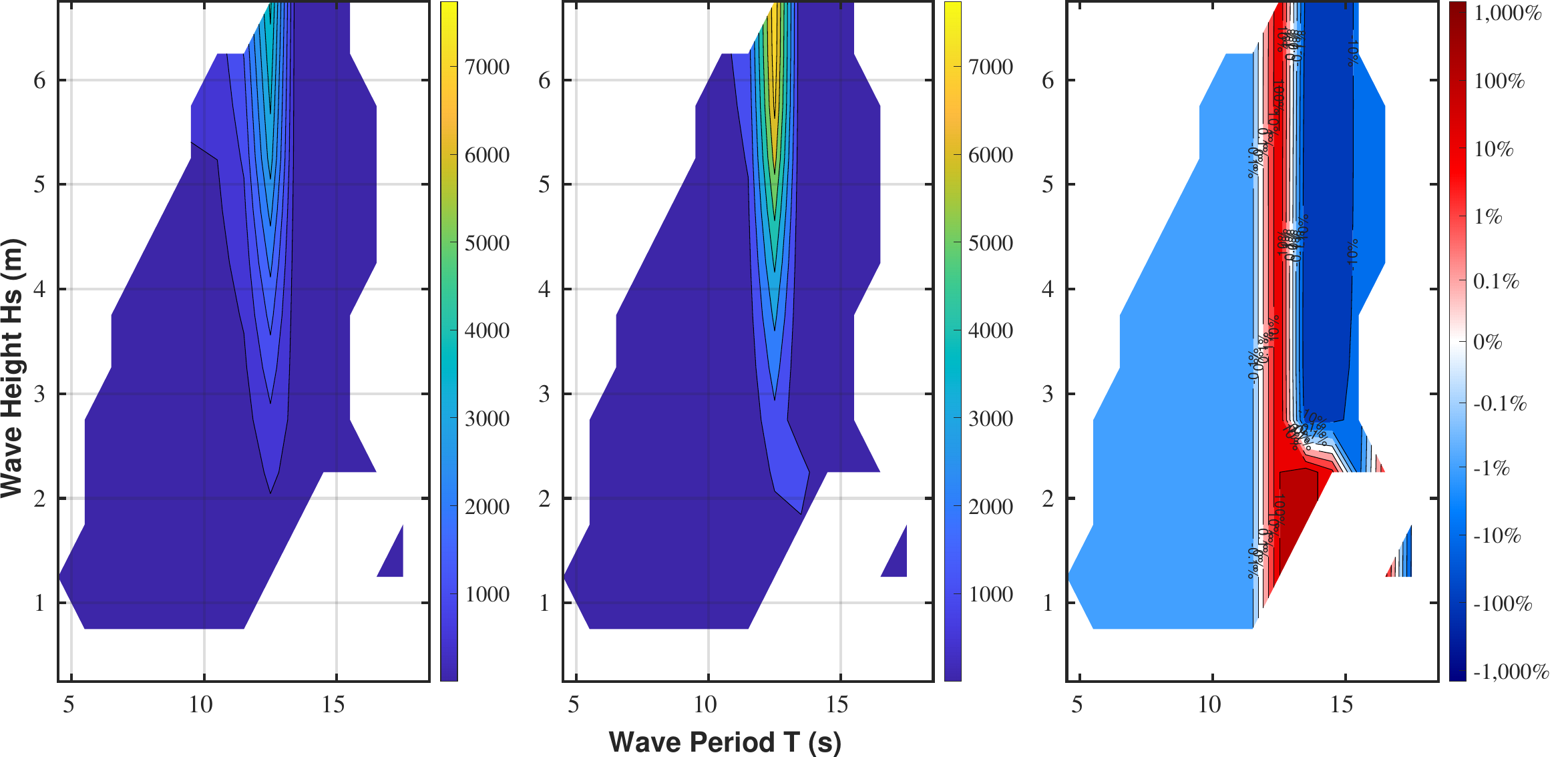} 
  & \includegraphics[width=\linewidth]{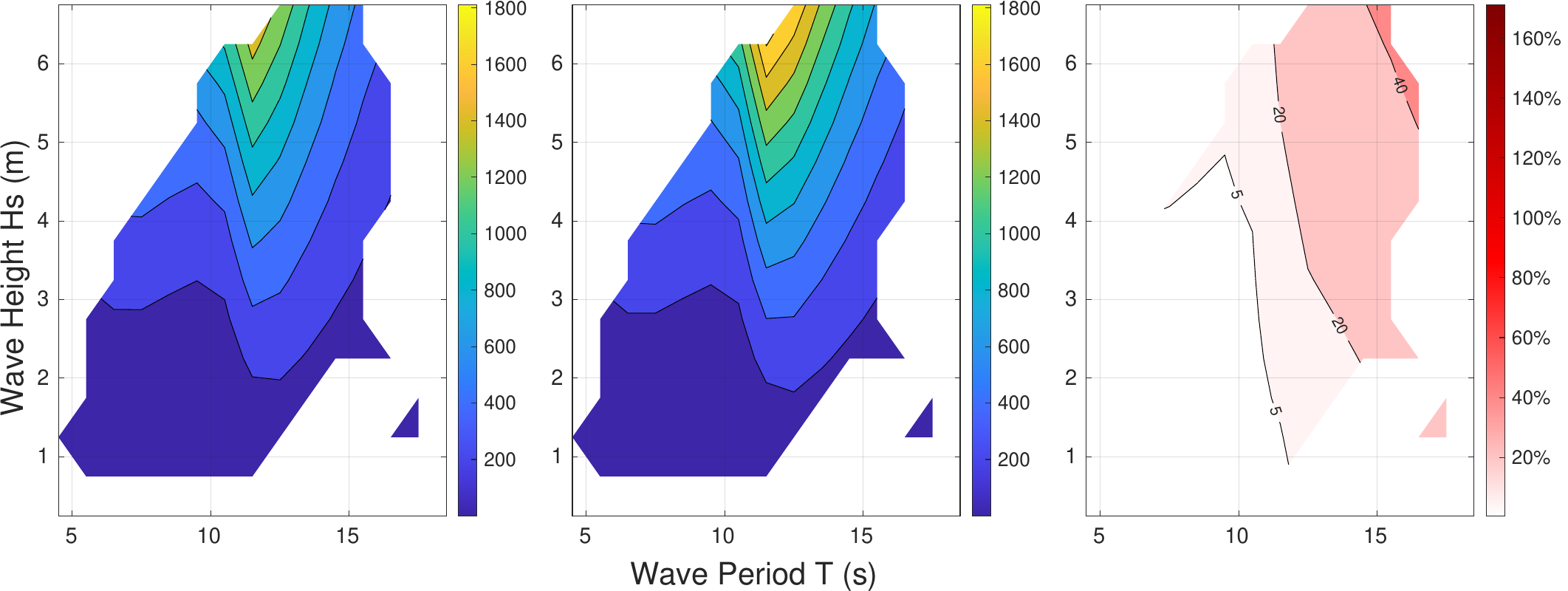}
\\ &\rot{Different} & 
    \includegraphics[width=\linewidth]{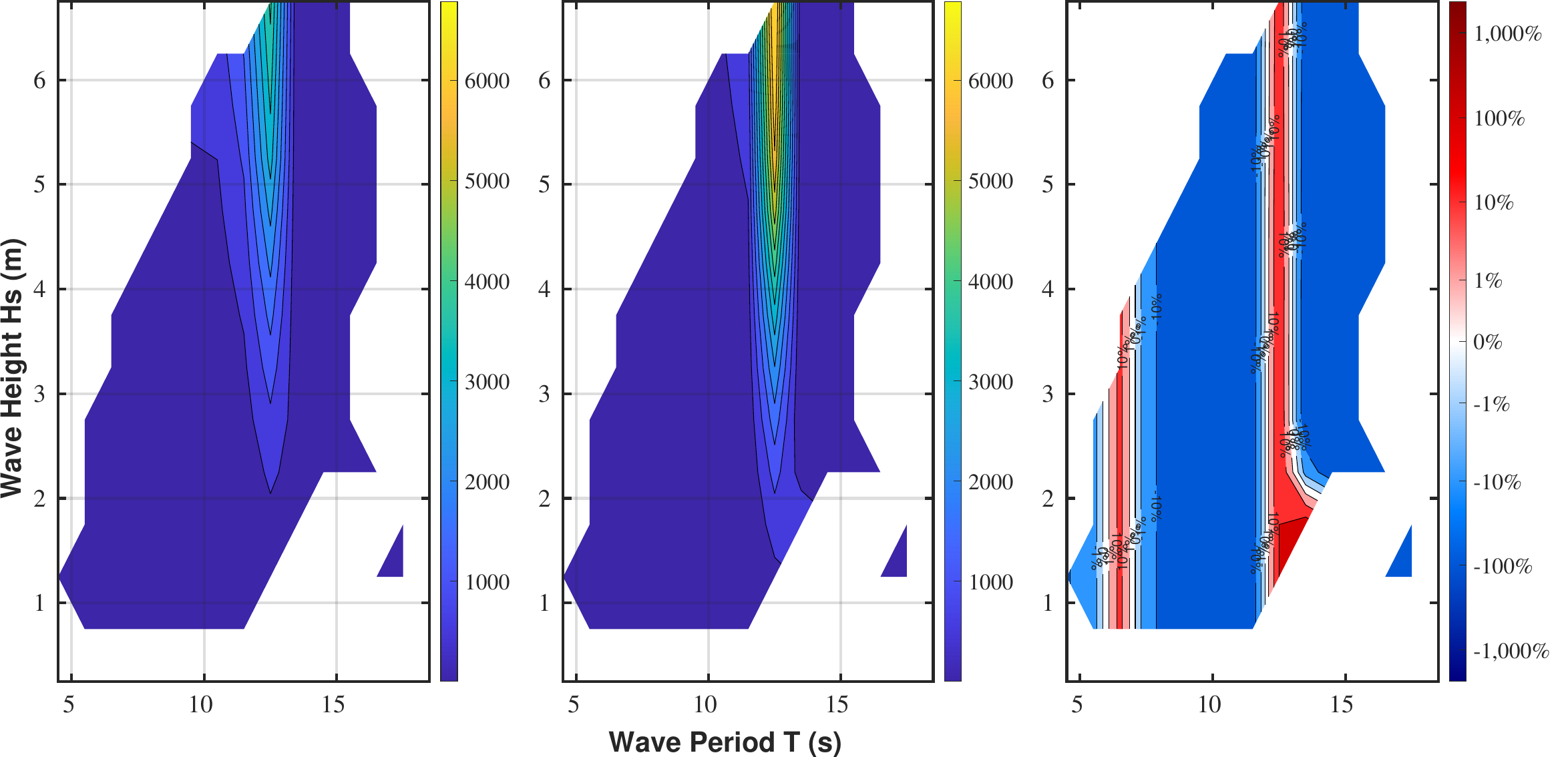} 
  & \includegraphics[width=\linewidth]{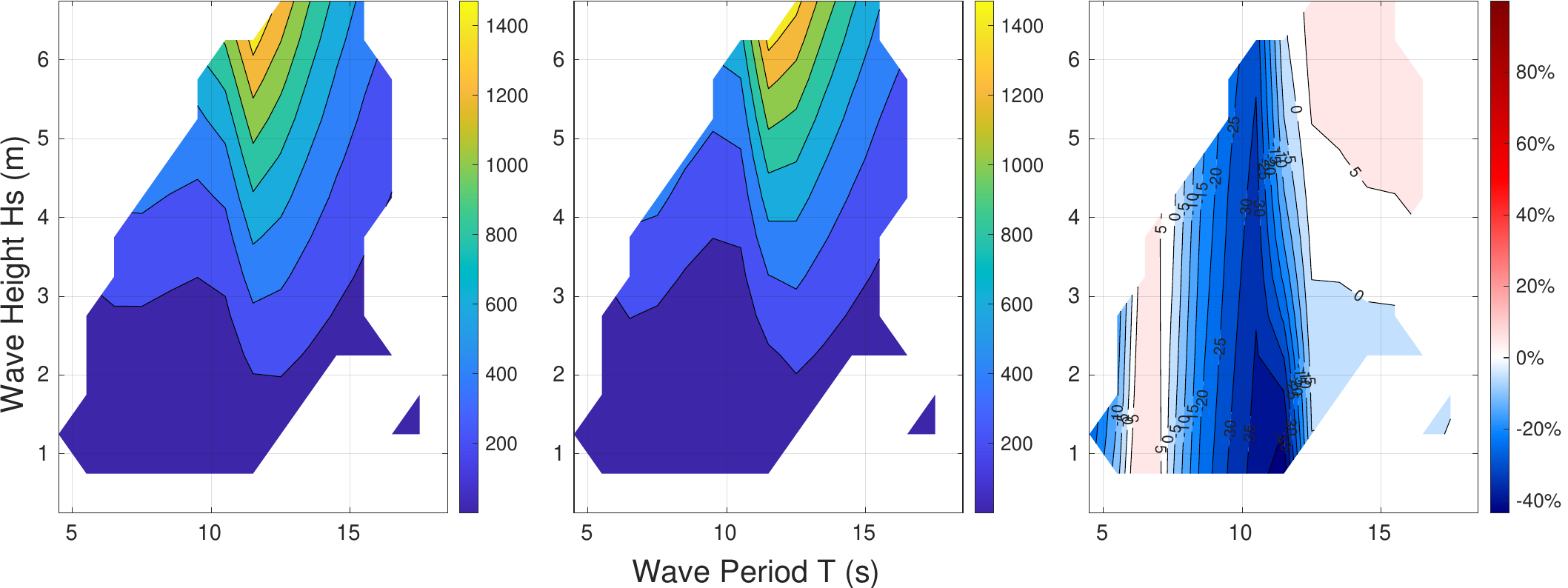} \\
\end{tabular}
\caption{Dynamic validation showing error in unsaturated mechanical power generation for a variety of modeling assumptions}
\label{fig:dynamic-validation}
\fillandplacepagenumber
\end{figure}
\endgroup
\end{landscape}

Besides mechanical power, a variety of other signals including amplitudes, phases, and forces are compared between the models.
While most signals show good agreement between the models, interestingly, large errors are observed in the drag forces, although these errors do not cause significant deviations in the overall system response.
\Cref{fig:drag-validation} compares the drag force on the float and spar in each model using identical hydro coefficients and $C_d=1$.

\begin{landscape}
\begingroup
\begin{figure}[htbp]
\centering
\begin{tabular}{c m{1em} | M{.47\linewidth} | M{.47\linewidth} }
  && \multicolumn{2}{c}{Drag Force}\\
         && Magnitude& Phase\\ \hline
    \multirow{2}{*}[-3em]{\rot{Body}} &\rot{Float}& 
    \includegraphics[width=\linewidth]{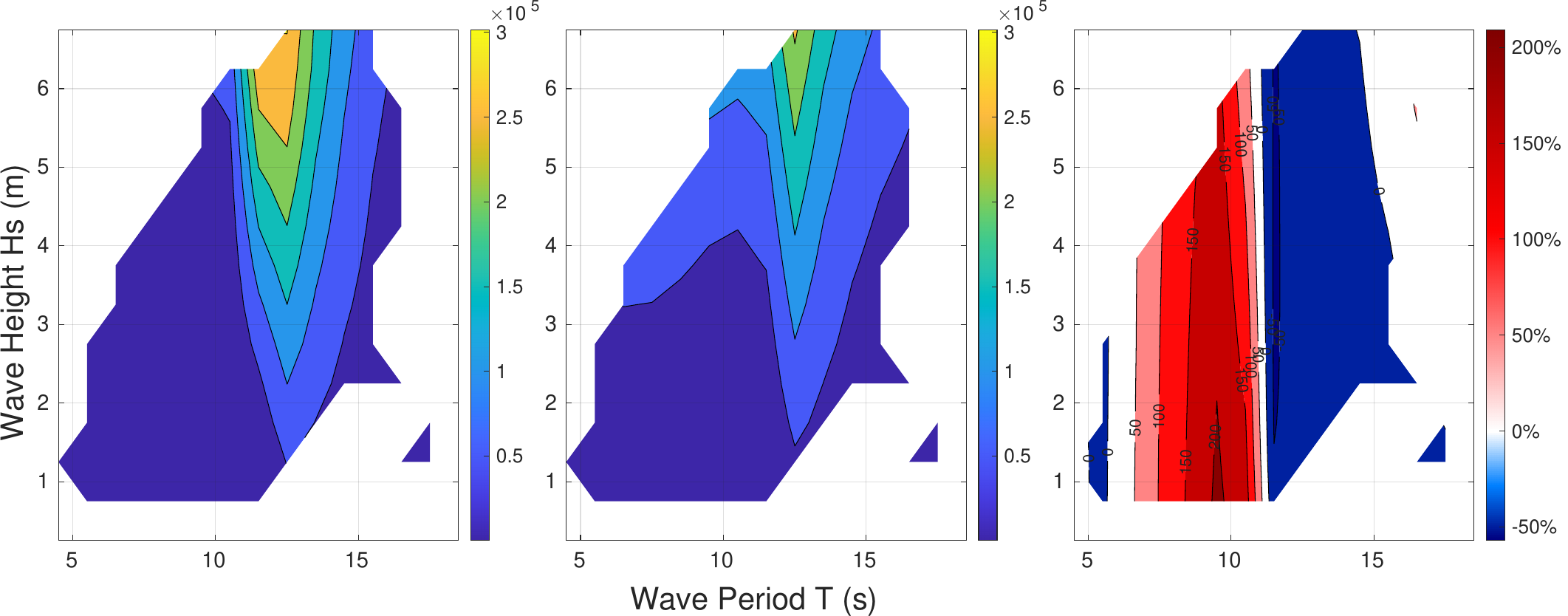} 
  & \includegraphics[width=\linewidth]{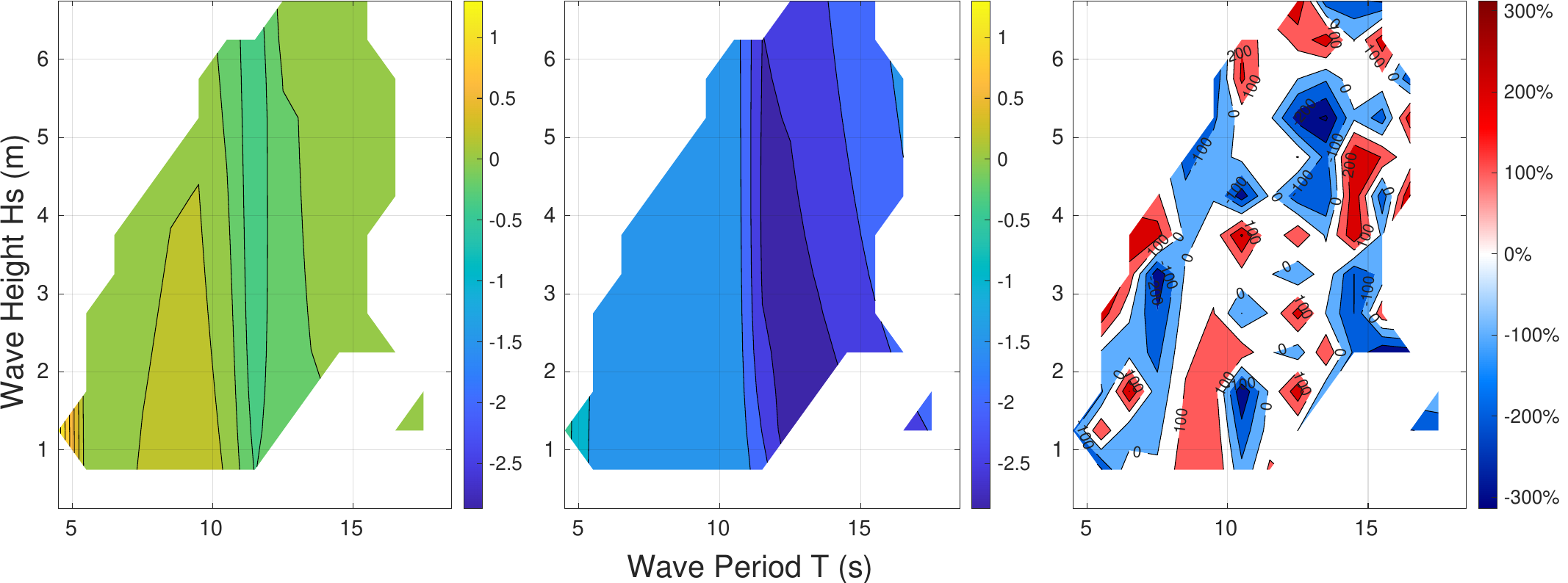}
\\ &\rot{Spar} & 
    \includegraphics[width=\linewidth]{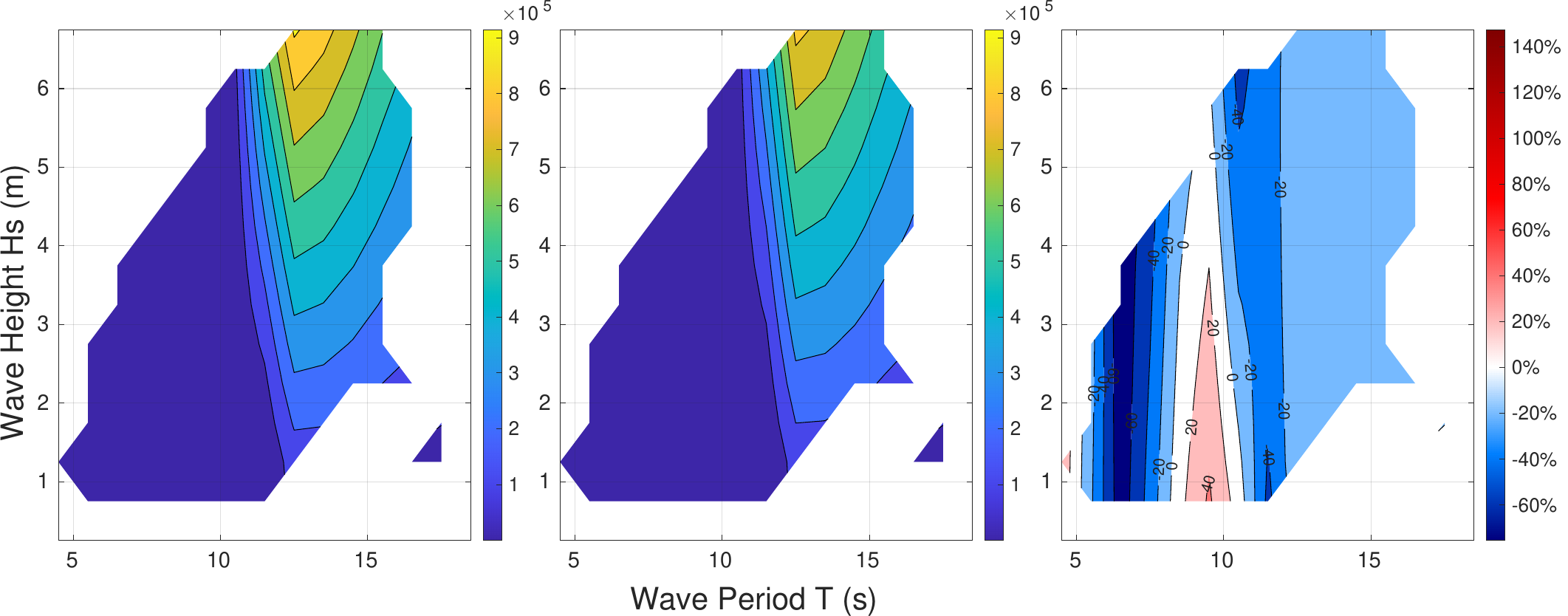} 
  & \includegraphics[width=\linewidth]{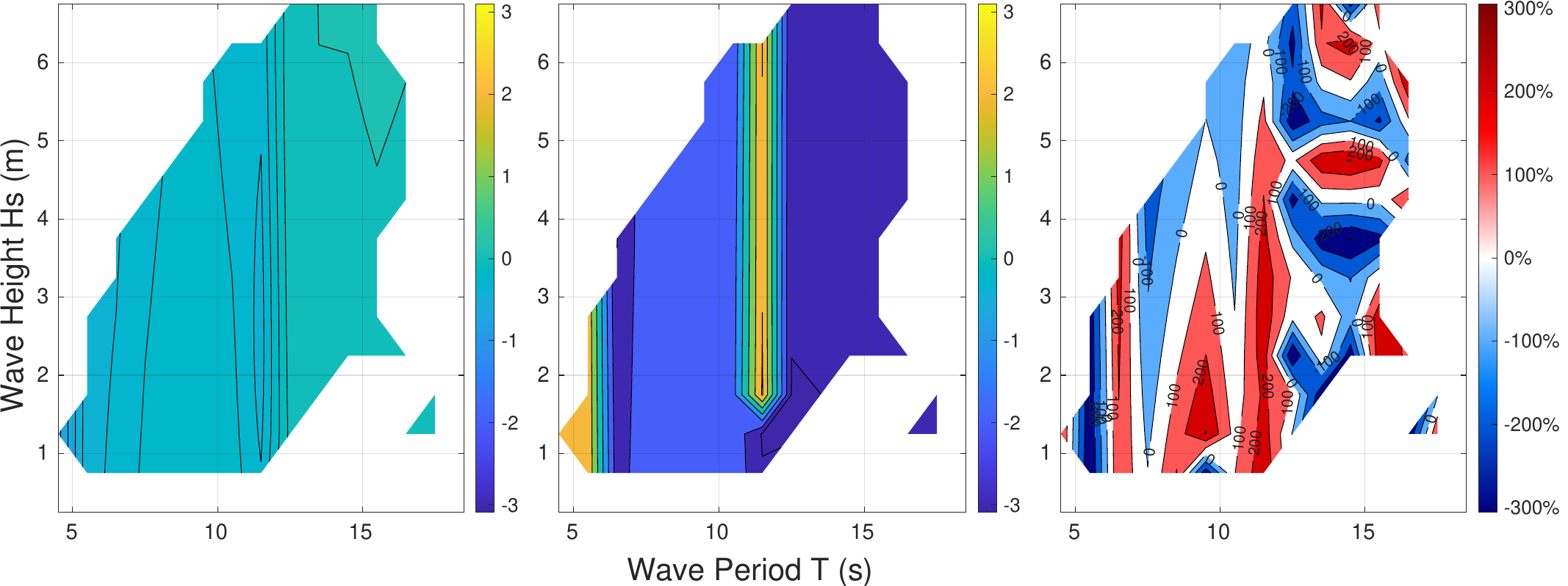} \\
\end{tabular}
\caption{Drag force comparison showing large errors in drag force, despite low errors in amplitude and power}
\label{fig:drag-validation}
\fillandplacepagenumber
\end{figure}
\endgroup
\end{landscape}
The reason for this discrepancy remains unclear, since the expected error of the describing function approximation due to higher harmonics would increase with frequency, whereas the observed error is worse at mid-range frequencies.
Evidently, amplitude and power results are quite sensitive to the presence of drag, but not very sensitive to the exact magnitude and phase of the drag force.
The WEC-Sim simulation uses drag-only Morison elements with a maximum width of one tenth the incident wavelength to ensure accuracy of strip theory, and uses a Fourier transform to extract the fundamental amplitude and phase of the nonlinear drag waveform.
Future work should investigate the source of this error and whether it can be reduced by adjusting the drag model.

We also validate the describing function approximation for the drag force.
The general application of the describing function technique requires that the fundamental frequency $\omega$ contains the vast majority of the energy, resulting in body displacements of the form $\xi(t) \sim \cos(\omega t+\phi)$.
\Cref{fig:wecsim-thd} shows the total harmonic distortion of the float displacement for each sea state in the $C_d=1$ case, which peaks at 1\% in the worst-case sea state and is below 0.5\% for the majority of sea states.
This confirms that higher harmonics have a negligible impact on the system and that the describing function method is a valid approach for modeling drag in this context.
The particular describing function relation used for drag assumes that the drag force has the form $F_d(t) \sim |\cos(\omega t+\phi)|\cos(\omega t+\phi)$.
\Cref{fig:wecsim-drag-waveform} shows the assumed and actual drag force waveform for four representative sea states corresponding to the four corners of the JPD.
The match is excellent in all four cases, confirming that the drag force is indeed well approximated by this form, even in high-amplitude, high-frequency conditions where the nonlinearity is expected to be strongest.

\begin{figure}[htbp]
    \centering
    \includegraphics[width=.6\linewidth]{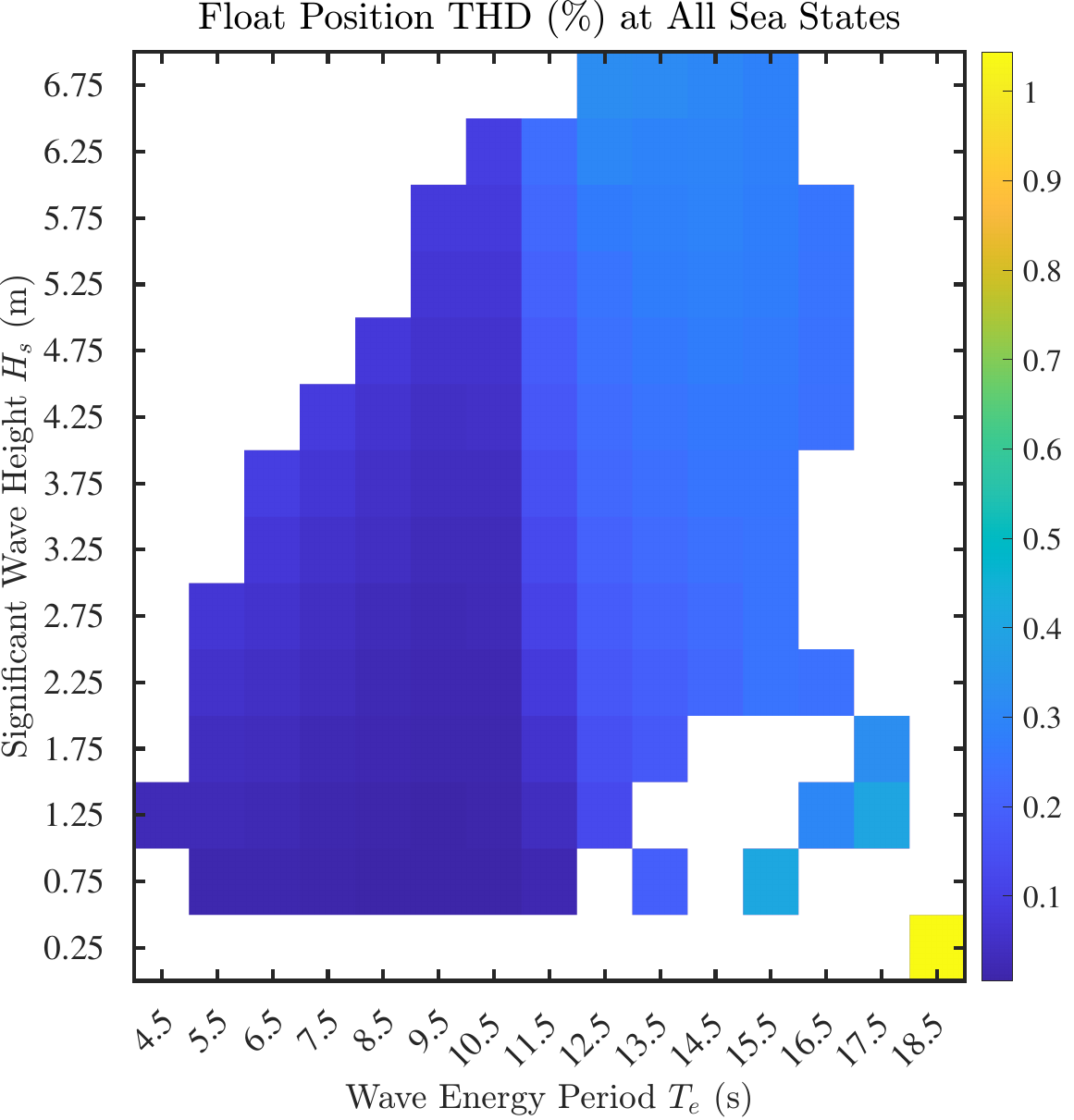}
    \caption{Total Harmonic Distortion of Float Displacement in WEC-Sim Simulations with Drag}
    \label{fig:wecsim-thd}
\end{figure}
\begin{figure}[htbp]
    \centering
    \includegraphics[width=.6\linewidth]{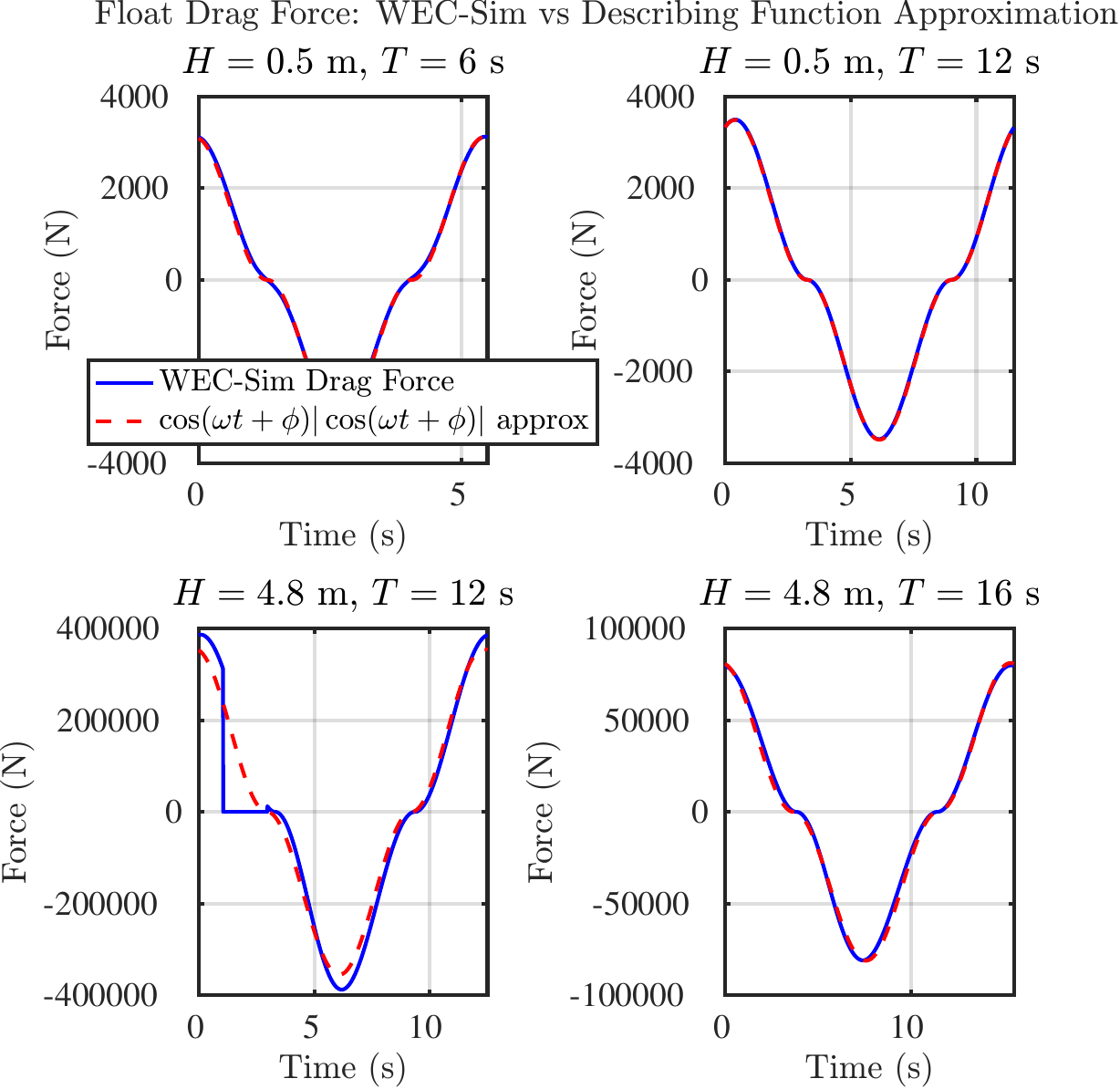}
    \caption{Comparison of Assumed and Actual Drag Force Signal Shape}
    \label{fig:wecsim-drag-waveform}
\end{figure}

\subsubsection{Reference Model Report}
While the original reference model report \cite{RM3} does not document the intermediate values used in dynamics calculations such as response amplitude or hydrodynamic coefficients, 
it does record the mechanical power for each sea state.
\Cref{fig:report-power-validation} compares these values against those predicted by WEC-Sim and MDOcean.
\begin{figure}[htbp]
    \centering
    \includegraphics[width=1.05\linewidth]{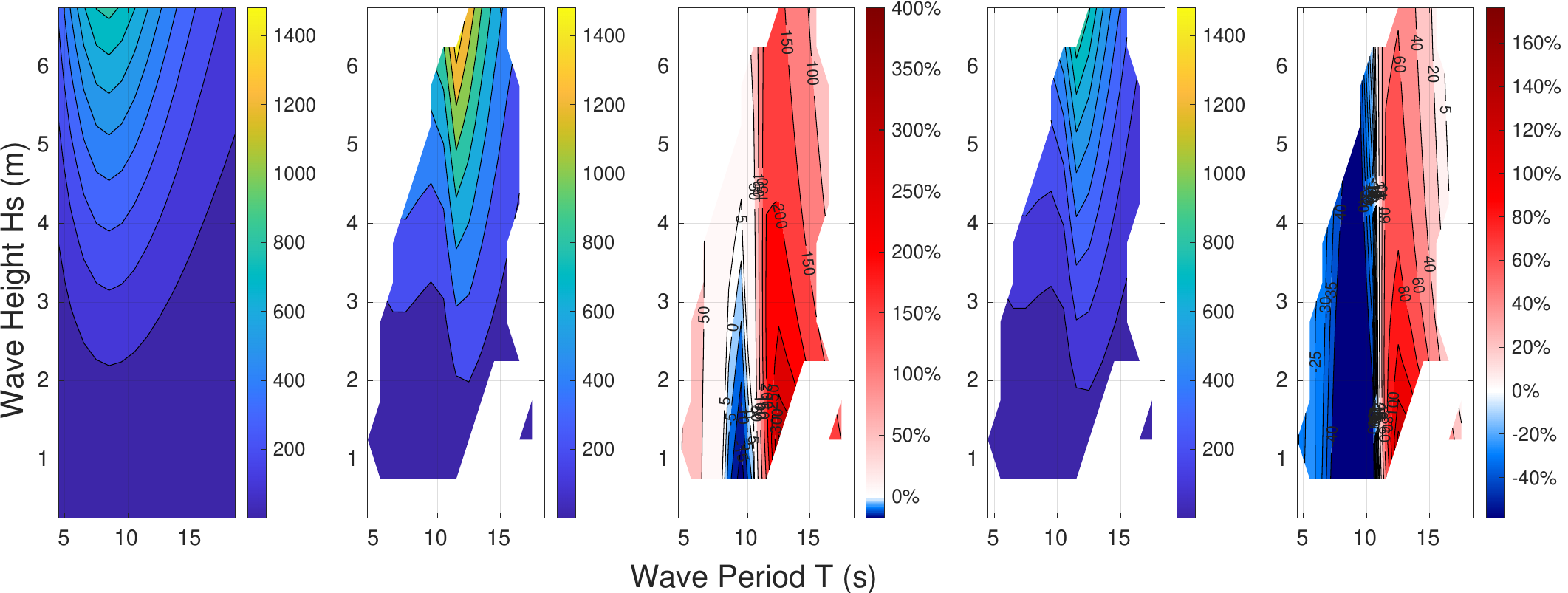}
    \caption{Comparison of mechanical power for RM3 report \cite{RM3} and MDOcean}
    \label{fig:report-power-validation}
\end{figure}
Note that the data is independent of \Cref{fig:dynamic-validation} due to the disparate geometry between the reference model report and provided WEC-Sim hydrodynamic coefficients,
and that the power multiplier mentioned in \Cref{sec:dynamic-validation} has not been applied in the figure.
We observe a rather poor match, with MDOcean underestimating power in the low-period sea states and overestimating in the high-period sea states.
The discrepancy is of minimal concern for the reasons described in \Cref{sec:dynamic-validation}.
Nonetheless, researchers and practitioners should be aware of the inconsistency and exercise caution before using the reference model report's power values for technology benchmarking or model validation.

\subsection{Power Matrix Decomposition}
\label{sec:appendix-power-decomposition}

The efficiency matrices introduced in \Cref{eq:power-matrix-decomposition} are defined as follows.
The wave power density and maximum capture width are independent of design:
\begin{equation}
\mathbf{J}^{H,T}_{wave} = \frac{\rho_w g^2}{64\pi} T_e H_s^2, \qquad
\mathbf{CW}^{H,T}_{max} = \frac{g T_e^2}{4\pi^2}.
\end{equation}
The radiation and drag efficiencies are defined relative to a drag-free simulation $\mathbf{P}_{F\dot{X},\text{no drag}}$ (computed with $C_{d,f}=C_{d,s}=0$, ideal PTO):
\begin{equation}
\mathbf{\eta}^{H,T}_{rad} = \frac{\mathbf{P}_{F\dot{X},\text{no drag}}}{\mathbf{J}^{H,T}_{wave}\,\mathbf{CW}^{H,T}_{max}}, \qquad
\mathbf{\eta}^{H,T}_{drag} = \frac{\mathbf{P}_{F\dot{X},\text{ideal}}}{\mathbf{P}_{F\dot{X},\text{no drag}}}.
\end{equation}
$\mathbf{\eta}^{H,T}_{\text{drag}}$ may exceed 1 at frequencies where the drag-free system is so undamped that the unconstrained controller from \Cref{eq:matched-load,eq:damping-control} would induce instability; in such cases a sub-optimal controller is substituted, which lowers $\mathbf{P}_{F\dot{X},\text{no drag}}$ (the denominator) rather than raising the numerator.
In cases where the unconstrained controller is stable, $\mathbf{\eta}^{H,T}_{\text{drag}} \leq 1$ is guaranteed.

The product of the first four matrices is the theoretical ideal mechanical power:
\begin{equation}
    \mathbf{P}^{H,T}_{F\dot{X},\text{ideal}} = \mathbf{J}^{H,T}_{wave}\,\mathbf{CW}^{H,T}_{max}\,\mathbf{\eta}^{H,T}_{rad}\,\mathbf{\eta}^{H,T}_{drag} = \frac{|\mathbf{\hat{F}}^{H,T}_{th}|^2}{8\Re(\mathbf{Z}^{H,T}_{i,th})}.
\end{equation}
The remaining two efficiency matrices capture the loss to dynamic constraints and to PTO electrical inefficiency:
\begin{equation}
    \mathbf{\eta}^{H,T}_{limits} = \frac{\mathbf{P}^{H,T}_{F\dot{X}}}{\mathbf{P}^{H,T}_{F\dot{X},\text{ideal}}}, \qquad
    \mathbf{\eta}^{H,T}_{elec} = \eta \frac{\mathbf{P}^{H,T}_{IV}}{\mathbf{P}^{H,T}_{F\dot{X}}}.
\end{equation}

Although it would be possible to formulate alternative variants such as $\mathbf{\eta}^{H,T}_{\text{drag},\text{var}} = \mathbf{P}_{IV}/\mathbf{P}_{IV,\text{no drag}}$ that capture the full effect of drag on the saturated system, such matrices would represent independent sensitivities and could not be multiplied together to recover the overall power matrix.
The current decomposition preserves multiplicativity, which is the key advantage of the formulation in \Cref{eq:power-matrix-decomposition}.

\section{Economics Module Details}
\label{sec:appendix-econ}

\subsection{RM3 Cost Breakdown Structure}
\label{sec:appendix-econ-cbs}

The MDOcean economic model is calibrated against the RM3 Cost Breakdown Structure (CBS) of \cite{neary_reference_2014}, reproduced in \Cref{tab:CBS}.
Percentages are shown for both single-device ($N_{WEC}=1$) and array ($N_{WEC}=100$) cases to illustrate the economies-of-scale trend captured by the power-law scaling \Cref{eq:cost-power-law}.

\begin{table}[htbp]
    \centering

    \caption{Cost Breakdown Structure}
    \label{tab:CBS}
\begin{tabular}{>{\raggedright\arraybackslash}p{0.3\linewidth}>{\centering\arraybackslash}p{0.2\linewidth}>{\centering\arraybackslash}p{0.2\linewidth}>{\centering\arraybackslash}p{0.15\linewidth}}
         CBS Category&  Nominal \% for $N_{WEC}=1$& Nominal \% for $N_{WEC}=100$&Scales with design?\\\hline
         1.1 - Development&  26\%&  3\%&No\\
         1.2 - Infrastructure&  6\%&  4\%&No\\
         1.3 - Mooring/Foundation&  3\%&  12\%&No\\
         1.4 - Structural&  17\%&  46\%&$V_{\text{struct}}$\\
         1.5 - Power Take Off&  4\%&  11\%&$F_{\text{max}}, P_{pk,\text{elec}}$\\
         1.6 - Integration \& Profit Margin&  2\%&  6\%&No\\
         1.7 - Installation&  34\%&  10\%&No\\
         1.8 - Decommissioning&  9\%&  9\%&No\\
         1.9 - Contingency&  9\%&  9\%&No\\
 \textbf{Total 1.1-1.9 (CAPEX)}& 93.3\%& 97.6\%&-\\\hline
 2.1 - Insurance& 1.3\%& 0.4\%&No\\
 2.2 - Environmental/Regulatory& 4.1\%& 0.3\%&No\\
 2.3 - Marine Operations& 0.2\%& 0.3\%&No\\
 2.4 - Shoreside Operations& 0.8\%& 0.2\%&No\\
 2.5 - Replacement Parts& 0.3\%& 1.0\%&No\\
 2.6 - Consumables& $<$0.1\%& 0.2\%&No\\
 \textbf{Total 2.1-2.6 (OPEX)}& 6.7\%& 2.4\%&-\\
    \end{tabular}
    \end{table}

\subsection{Power-Law Scaling with Number of Devices}
\label{sec:appendix-econ-power-law}

The per-WEC unit costs and prices in \Cref{eq:unit-cost} ($C_{\text{fixed}}$, $C_{pto,\text{constant}}$, $p_{P}$, $p_{F}$, $p_{s}$) each decrease with the number of devices $N_{WEC}$ according to
\begin{equation}\label{eq:cost-power-law}
    C = C_\infty+( C_1-C_\infty) \cdot (N_{WEC})^{-\beta},
\end{equation}
where $C_\infty$ is the asymptotic large-scale unit cost, $C_1$ is the single-unit cost, and $\beta$ is the rate of decrease.

The values of $C_\infty$, $C_1$, and $\beta$ for each CBS category were obtained by curve-fitting \Cref{eq:cost-power-law} to the CBS estimates for $N_{WEC} \in \{1, 10, 50, 100\}$ \cite{neary_reference_2014}, and are tabulated in \Cref{tab:econ-model-values}.

\begin{table}[h]
    \centering
    \caption{Cost model values for each CBS category}
    \label{tab:econ-model-values}
\begin{tabular}{
        >{\raggedright\arraybackslash}p{0.33\linewidth}
        >{\raggedright\arraybackslash}p{0.1\linewidth}
        >{\raggedright\arraybackslash}p{0.18\linewidth}
        >{\raggedright\arraybackslash}p{0.18\linewidth}}
         CBS Category& Exponent $\beta$&Unit cost at scale $C_{\infty}$&Single-unit cost \newline $C_1$\\\hline
         \textbf{1.1-1.3, 1.6-1.9} - Design-Indep. CAPEX& 0.741& 1.24 \$M&13.92 \$M\\
         \textbf{1.4} - Structural CAPEX & 0.481& 2.387 \$/kg&4.294 \$/kg\\
         \textbf{1.5} - Power Take Off CAPEX& 0.206& Constant: 92.59 \$k Power: 0.4454 \$/W Force: 0.0086 \$/N&Constant: 93.64 \$k Power: 1.355 \$/W Force: 0.0204 \$/N\\
 \textbf{2.1-2.6} - OPEX&0.557& 0&1.193 \$M\\
    \end{tabular}
    \end{table}

\subsection{Linear PTO Cost Scaling}
\label{sec:appendix-econ-pto-scaling}

Linear PTO cost scaling is common for WEC techno-economic analysis.
The RM3 cost numbers \cite{RM3} which MDOcean uses as a baseline are themselves scaled from estimates for a smaller device: the reference model project applies linear scaling with peak power to an unpublished 2011 cost estimate by ReVision Consulting for a 100~kW peak hydraulic PTO (except for the riser cable and control system, whose cost does not scale with power).
The NREL System Advisory Model assumes that hydraulic PTO cost scales slightly less than linearly with power (exponent 0.91), and one study uses regression to obtain separate linear models for the cost of commercial induction and permanent magnet generators as a function of torque capability \cite{nakhai_techno-economic_2022}.

\subsection{Economic Validation}
\label{sec:appendix-econ-validation}
\Cref{fig:econ-nwec-validate} shows the scaling behavior of economic outputs (CAPEX, OPEX, and LCOE) against the number of WECs in the farm, comparing MDOcean results against the \cite{RM3} reference values.
\begin{figure}[htbp]
    \centering
    \includegraphics[width=\linewidth]{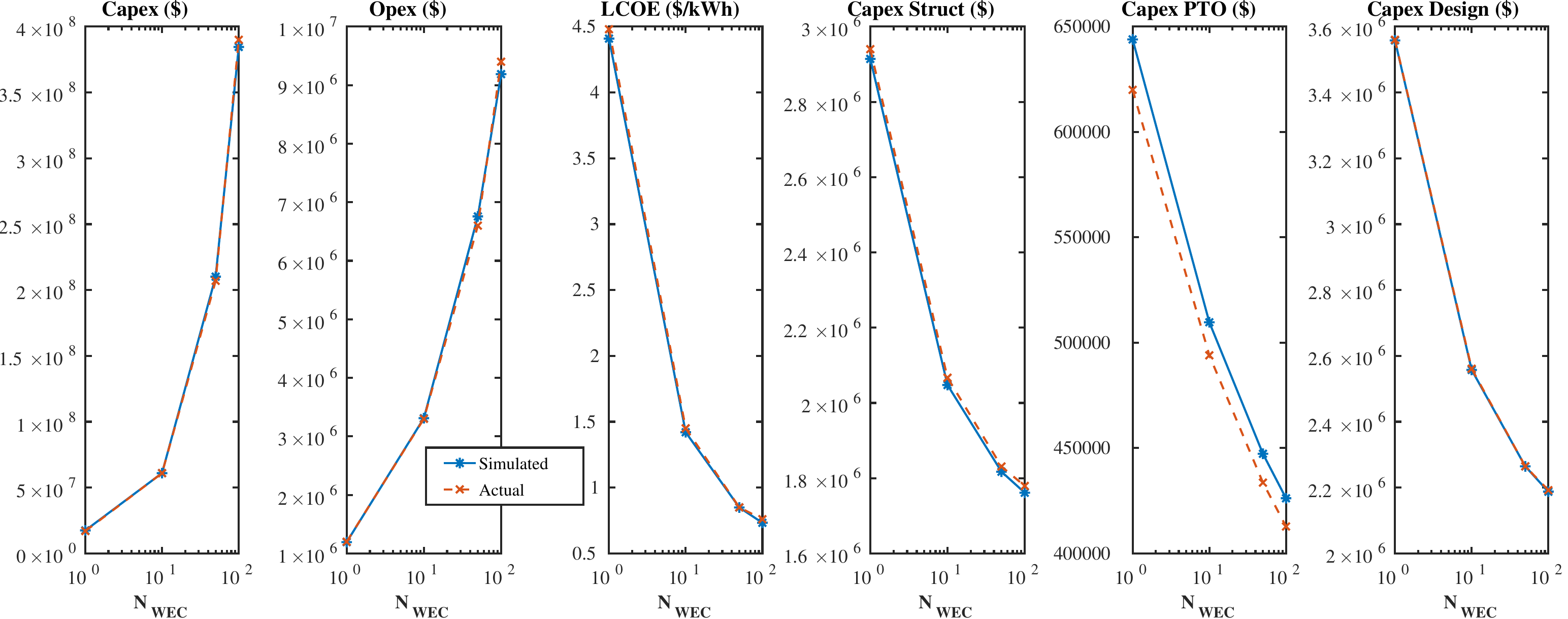}
    \caption{Validation for cost scaling with number of WECs}
    \label{fig:econ-nwec-validate}
\end{figure}

\section{Structures Module Details}
\label{sec:appendix-structures}
Inputs to the structures module include forces, bulk and structural dimensions, and material constants.
It outputs a factor of safety for each limit case.
First, it obtains equivalent section properties based on the plate and stiffener dimensions.
Second, it relates applied loads to stresses and deflection using analytical solutions to structural boundary value problems obtained from the well-known Roark's handbook \cite{young_roarks_2001} and the references therein.
Finally, it utilizes design standards from organizations like Det Norske Veritas, the American Bureau of Shipping, and the American Iron and Steel Institute to develop limit-state expressions for each possible failure mode of each major structural element under each design load case.
For consistency with the reference model report \cite{RM3}, all structural elements are assumed to be welded, although future work should consider the use of pinned joints for certain interfaces to reduce reaction moments and enhance structural efficiency.

\subsection{Equivalent Section Properties for Stiffened Plates}
\label{sec:appendix-equivalent-thickness}
Stiffeners are structural elements, frequently with I or T profiles, that can be welded or riveted to flat plates to provide additional load-carrying capacity and prevent buckling.
Stiffened plates are common in marine and aerospace structural design because they are an efficient way to carry spatially distributed fluid loads.
To analyze these compound sections, equivalent properties can be established that describe the overall bending stiffness of the combined section as if it were a simple uniform section, while still utilizing the maximum distance from the neutral axis of the irregular section to accurately determine stress.
\Cref{fig:stiffened-plate} illustrates the concept of equivalent thickness for a stiffened plate.
\begin{figure}[hbtp]
    \centering
    \includegraphics[width=\linewidth]{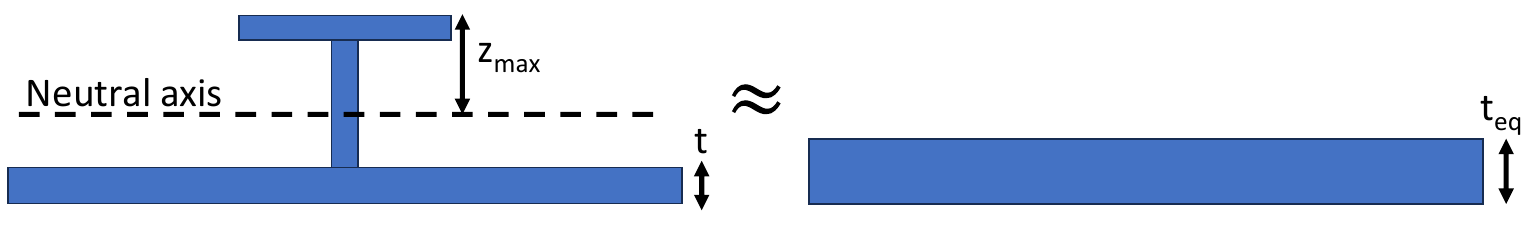}
    \caption{Equivalent thickness for a stiffened plate}
    \label{fig:stiffened-plate}
\end{figure}

Specifically, the equivalent thickness of the plate $t_{eq}$ is derived by equating the stiffened section's second moment of area per unit width with that of a uniform section:
\begin{equation}
    \int z^2dz = \frac{t_{eq}^3}{12}
\end{equation}
Next, deflection of the stiffened plate is equal to the deflection of the equivalent unstiffened plate with flexural rigidity $D_{eq}$:
\begin{equation}
    D_{eq} = \frac{E~  t_{eq}^3}{12(1-\nu^2)}
\end{equation}

The maximum stress $\sigma$ of the stiffened plate is then:
\begin{equation}\label{eq:plate-stress}
    \sigma= 12M z_{\text{max}}/t_{eq}^3
\end{equation}
where $M$ is the maximum moment per unit length and $z_{\text{max}}$ is the maximum distance from the neutral axis of the stiffened section.
This parallels the stress for an unstiffened plate of thickness $t$, which is $\sigma=6M/t^2 = 12 M(t/2)/t^3$.

Note that this equivalent-thickness method of accounting for stiffeners assumes that as a whole, the stiffener-plate system deflects like a single plate element, rather than as a set of multiple plate elements separated by stiffeners.
This is a reasonable assumption when the stiffeners are small and densely spaced with respect to the plate, but breaks down if the stiffeners come to dominate the system.
The so-called effective breadth ratio $\rho$ is used to quantify the validity of this assumption, where the equivalent-thickness method requires $\rho=1$:
\begin{equation}
   \rho = \rho_\lambda \rho_\psi
\quad 
\rho_\lambda = \begin{cases}
        1, & \lambda\leq 0.673 \\
        \frac{1-0.22/\lambda}{\lambda}, & \lambda > 0.673
    \end{cases} \quad
    \rho_\psi = \begin{cases}
        1, & \psi> -0.236 \\
        \frac{1}{2}+\frac{1}{3-\psi}, & \psi \leq -0.236
    \end{cases} 
\end{equation}
The effective breadth ratio depends on the slenderness factor $\lambda$ and load distribution factor $\psi$, found as follows:
\begin{equation}
\begin{aligned}
    \lambda &= \frac{1.052}{\sqrt{k_{\text{buckle}}}} \frac{w}{t} \sqrt{\frac{f_1}{E}} \\
    \psi&=\frac{f_2}{f_1}
    \end{aligned}
\end{equation}
with $t$ referring to the plate unstiffened thickness, $w$ as the distance between stiffeners, $f_1$ as the maximum compression stress along the width, and $f_2$ as the maximum tension stress along the width (or minimum compression stress, if there is no tension).
By convention, compression stress is positive and tension is negative, so $\psi$ ranges from $-\infty$ to $+1$.
The plate buckling coefficient $k_{\text{buckle}}$ is expressed as:
\begin{equation}\label{eq:k-buckle}
  k_{\text{buckle}} = 4+2\left(1-\psi\right)^3+2\left(1-\psi\right)
\end{equation}
This effective breadth procedure follows the American Iron and Steel Institute design manual \cite{american_iron_and_steel_institute_cold-formed_1991}.
Rather than enforcing $\lambda<0.673$, corresponding with $\rho_\lambda=1$, MDOcean more flexibly requires $\lambda<0.809$, corresponding to $\rho_\lambda=0.9$ and thereby capping the error due to insufficient slenderness at around 10\%.
MDOcean also uses $k_{\text{buckle}}=4$, the minimum value that equation \eqref{eq:k-buckle} can take on, conservatively maximizing the slenderness factor $\lambda$.

Broadening the design space to allow for more dominant stiffeners with higher slenderness factors would require modeling the stiffener-plate system not as a plate with the stiffeners absorbed into an equivalent thickness, but as individual stiffeners with the plate absorbed into the effective breadth.
That model has other complexities such as the shear lag phenomenon so is left to future extensions \cite{wierzbicki_lecture_2013,american_iron_and_steel_institute_cold-formed_1991}.


\subsection{Float}
The float is composed of 12 watertight stiffened shells in the shape of trapezoidal prisms.
Each shell consists of a top and a bottom trapezoidal plate, and an inner, an outer, and two side rectangular plates.
All edges are welded.
Rather than model the deflections of each edge and apply compatibility, for simplicity the edges of each plate are conservatively modeled as fixed.
The top and bottom plates are the only ones with external heave loads, arising from the float-spar tubular connection and the hydrodynamic pressure respectively.

The trapezoidal plates are isosceles trapezoids with smaller base $b_1 = \pi D_{f,in}/12$, larger base $b_2=\pi D_f/12$, and perpendicular height $h_0=(D_f-D_{f,in})/2$ (see \Cref{fig:trapezoid}).
The references consulted do not contain structural results for trapezoidal plates, so geometric intuition is used to scale available solutions.
For example, Ref.~\cite{young_roarks_2001} contains models for rectangular plates under perpendicular loading.
They show that the maximum bending stress $\sigma$ scales with the square of the shorter side length, $x_{\text{min}}^2$. 
This makes sense because for fixed-edge plates with distributed loads, it is the square of the minimum distance from an edge to the point of highest deflection, $(x_{\text{min}}/2)^2$, that geometrically sets the maximum curvature, so $\sigma\sim x_{\text{min}}^2$.
For trapezoids, one then expects approximate rectangles ($b_2/b_1\approx1$) to have $x_{\text{min}}=\min(b_1,h_0)$, short trapezoids ($h_0<b_1)$ to have $x_{\text{min}}=h_0$, tall trapezoids ($h_0\gg b_2$) to have $x_{\text{min}}=b_1$, and intermediate trapezoids to have an $x_{\text{min}}$ that depends on the trapezoid slope.
The nominal RM3 design falls into this intermediate case, with $b_1\approx1.7$~m, $b_2\approx5.2$~m, and $h_0\approx6.8$~m.
These dimensions are shown in \Cref{fig:trapezoid}.

\begin{figure}[htbp]
    \centering
    \includegraphics[width=0.5\linewidth]{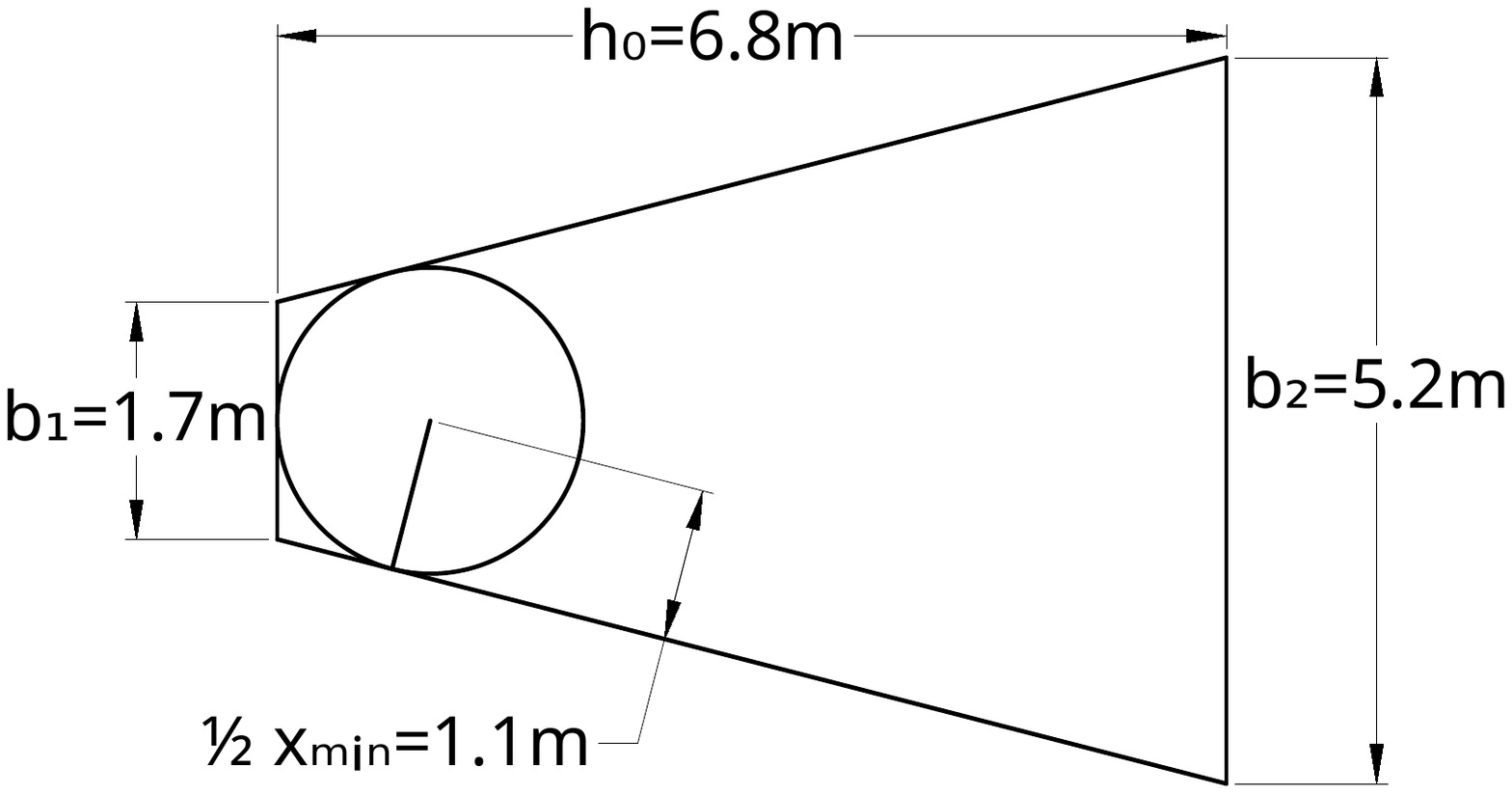}
    \caption{Trapezoidal float plate with inscribed circle used to determine $x_{\text{min}}$}
    \label{fig:trapezoid}
\end{figure}

To approximately capture the dimensions of a trapezoid, which effectively set the deflection curvature and therefore stress, $x_{\text{min}}$ is set to the diameter of the largest circle that can be inscribed in the trapezoid tangent to the shorter base $b_1$.
This inscribed circle scheme satisfies all limit cases mentioned above and can be calculated as
\begin{equation}
    x_{\text{min}} = \begin{cases}b_1 \left( m + \sqrt{1+m^2}\right), & h_0 \geq \sqrt{b_1b_2} \\ h_0, & h_0 < \sqrt{b_1b_2}\end{cases}
\end{equation}
where $m=(b_2-b_1)/(2h_0)$ is the slope of the trapezoid.
The expression is continuous at $h_0=\sqrt{b_1b_2}$, the point where the inscribed circle becomes tangent to base $b_2$ in addition to $b_1$.
Continuity of the model across the design space avoids problems in the gradient-based optimization due to undefined gradients.
The inscribed circle model gives $x_{\text{min}}\approx2.2$~m for the nominal float design.
In addition to the shorter side length $x_{\text{min}}$, stress in a fixed-edge rectangular plate also depends weakly on the longer side length of the plate $x_{\text{max}}$, which for the trapezoid is taken as
\begin{equation}
    x_{\text{max}} = \begin{cases} h_0,& h_0 \geq \sqrt{b_1b_2} \\ \frac{b_1+b_2}{2} + h(1-\sqrt{1+m^2}), & h_0 < \sqrt{b_1 b_2}\end{cases}
\end{equation}
using a similar inscribed circle scheme to maintain continuity.

\subsubsection{Float Bottom Plate}
Under a distributed pressure $q$, the bottom plate maximum moment per unit length is then:
\begin{equation}\label{eq:bottom-plate-moment}
M_{bot}=\beta q x_{\text{min}}^2
\end{equation}
for $\beta$ tabulated as a function of $x_{\text{max}}/x_{\text{min}}$ in reference \cite{timoshenko_theory_1959}, and the bending stress is found by plugging this moment into equation~\eqref{eq:plate-stress}.
While the simulation utilizes the real $\beta$, constant $\beta$ is assumed for the scaling laws of \Cref{eq:struct-scaling} because $\beta$ varies only weakly with $x_{\text{max}}/x_{\text{min}}$.

\subsubsection{Float Top Plate}
The top float plate has equivalent thickness $t_{f,\text{top},eq}$ and the same dimensions and edge fixity as the bottom plate.
Unlike the uniform pressure on the bottom plate, the top plate is subject to the welded tubular float-spar attachment force and moment, modeled as a load distributed over a small circle of radius $r_0$, where $r_0$ is assumed to be at least half the thickness.
Under these conditions, either the stress at the center $\sigma_{\text{cent}}$ or along the longer edge $\sigma_{\text{edge}}$ may dominate:
\begin{equation}\label{eq:top-plate-stress}
    \begin{aligned}
    \sigma_{\text{edge}}&= \frac{3W}{2\pi t_{f,\text{top},eq}^2}\left( (1+\nu)\ln\frac{2x_{\text{min}}}{\pi r_0}+\beta_1\right)\\
        \sigma_{\text{cent}}&=-\beta_2 W/t_{f,\text{top},eq}^2
    \end{aligned}
\end{equation}
using $\beta_1$ and $\beta_2$ tabulated as a function of $x_{\text{max}}/x_{\text{min}}$ in reference \cite{young_roarks_2001}.
The bending moment per unit length is then found as $M_{\text{top}}=\sigma_{\text{edge}}t_{f,\text{top},eq}^2/6$.
However, unlike \Cref{eq:bottom-plate-moment}, it is found that \Cref{eq:top-plate-stress} predicts stresses that do not match well with the finite element results for the nominal RM3 design in references \cite{RM3,previsic_reference_2011}.
It is unknown whether this is an implementation or modeling issue, and remains as future work to address.
In the meantime, the top plate thickness is set by scaling the bottom plate thickness, as described in \Cref{tab:struct-calc-vs-scale}.

\subsubsection{Other Float Plates}
The other float plates besides the top and bottom ones experience only the edge reaction moments that the top and bottom plates apply at their interface, with no additional external loads in heave. 
Assuming that the maximum allowable stress is the same in all float plates, the required side plate thickness can be found by simply scaling the input float bottom plate thickness using the float thickness ratios in the nominal RM3 design.
This prevents the need for a separate structural analysis of the side plates.

\subsection{Damping Plate}
For an annular plate with free outer radius $a$ and fixed inner radius $b$, the nondimensional plate deflection $\overline{\delta}_{\text{plate}}$ and bending moment $\overline{M}_r$ can be calculated as a function of plate aspect ratio $b/a$ for various load cases.
In this case, the plate is loaded by a distributed load (subscript $dis$) equal to $F_{\text{heave}}$, and a concentrated load (subscript $con$) at four equidistant points along the edge, equal to the tubular support reaction force $F_{\text{tube}}$.
Formulas for the concentrated loading are given in reference \cite{boedo_corrected_1998}, and formulas for distributed loading are given as case 2L of table 11.2 in reference \cite{young_roarks_2001}.
The 24 radial plate stiffeners are taken into account with the procedure described in \Cref{sec:appendix-equivalent-thickness}, producing equivalent flexural rigidity $D_{eq}$ and equivalent section modulus $S_{eq}$.
The tube force $F_{\text{tube}}$ is statically indeterminate and is solved with compatibility by equating the plate and tubular support displacements, $\delta_{\text{plate}}=\delta_{\text{tube}}$, where the tube displacement is expressed in terms of its bending stiffness $K_{\text{tube}}$.
This results in the following expression for the radial plate bending stress $\sigma_r$:
\begin{equation}
    \sigma_r =  \frac{F_{\text{heave}}}{S_{eq}} \left( 
    \overline{M}_{r,dis} + \overline{M}_{r,con} \frac{\overline{\delta}_{\text{plate},dis}}{\frac{D_{eq}}{a^2K_{\text{tube}}} - \overline{\delta}_{\text{plate},con}}
    \right)
\end{equation}

The critical buckling stress $\sigma_{\text{buckle}}$ is found according to \cite{american_bureau_of_shipping_requirements_2022}.
The ultimate stress is then $\sqrt{\sigma_Y \sigma_{\text{buckle}}}$  \cite{wierzbicki_lecture_2013-1} and the endurance limit is taken as half of ultimate.
The factor of safety is then the ratio of the maximum (ultimate or endurance, depending on the design load case) stress to the radial plate bending stress.
The procedure is summarized in \Cref{fig:damping-plate-flowchart}.
\begin{figure}[htbp]
    \centering
    \includegraphics[width=1\linewidth]{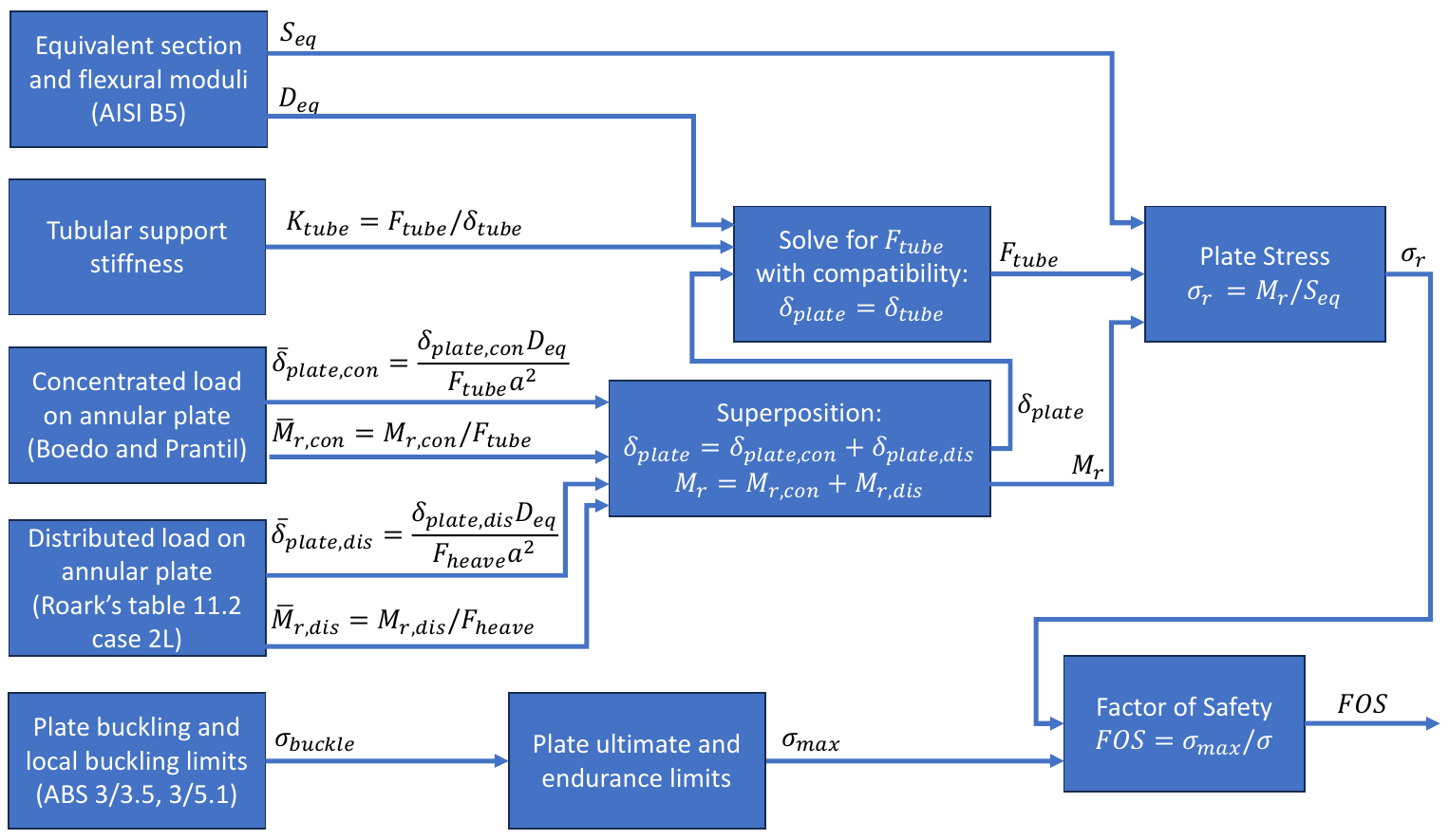}
    \caption{Calculation process flowchart for the damping plate structural assessment}
    \label{fig:damping-plate-flowchart}
\end{figure}

\Cref{fig:damping-plate-struct} shows the spatial distribution of radial bending moment and deflection across the plate radius for the nominal design, computed using the analytical solutions described above.
\begin{figure}[htbp]
    \centering
    \begin{subfigure}[t]{0.48\linewidth}
        \includegraphics[width=\linewidth]{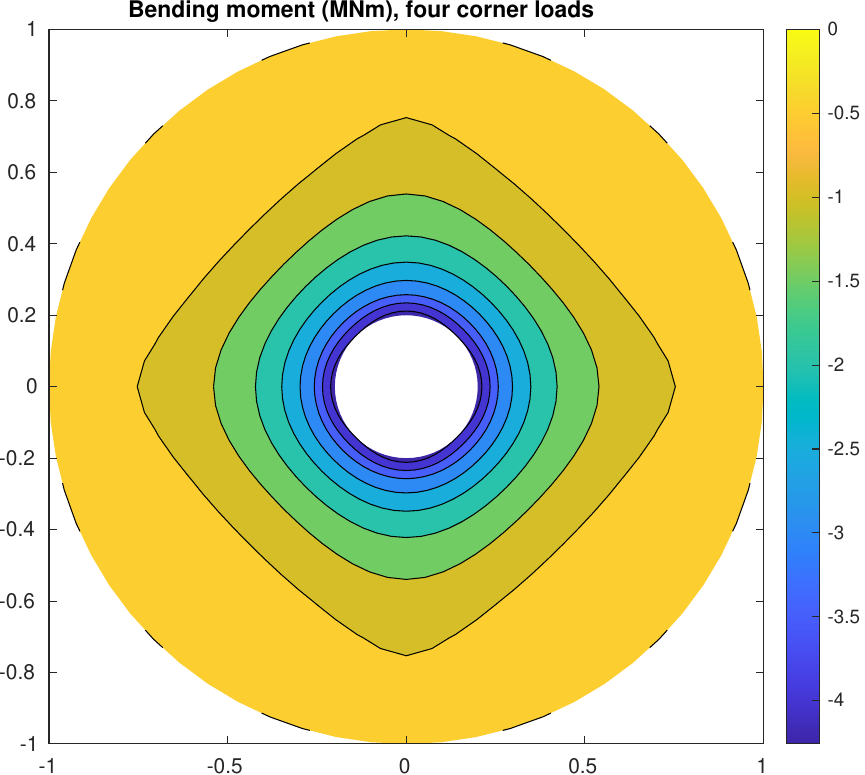}
        \caption{Damping plate moment}
        \label{fig:damping-plate-moment}
    \end{subfigure}
    \hfill
    \begin{subfigure}[t]{0.48\linewidth}
        \includegraphics[width=\linewidth]{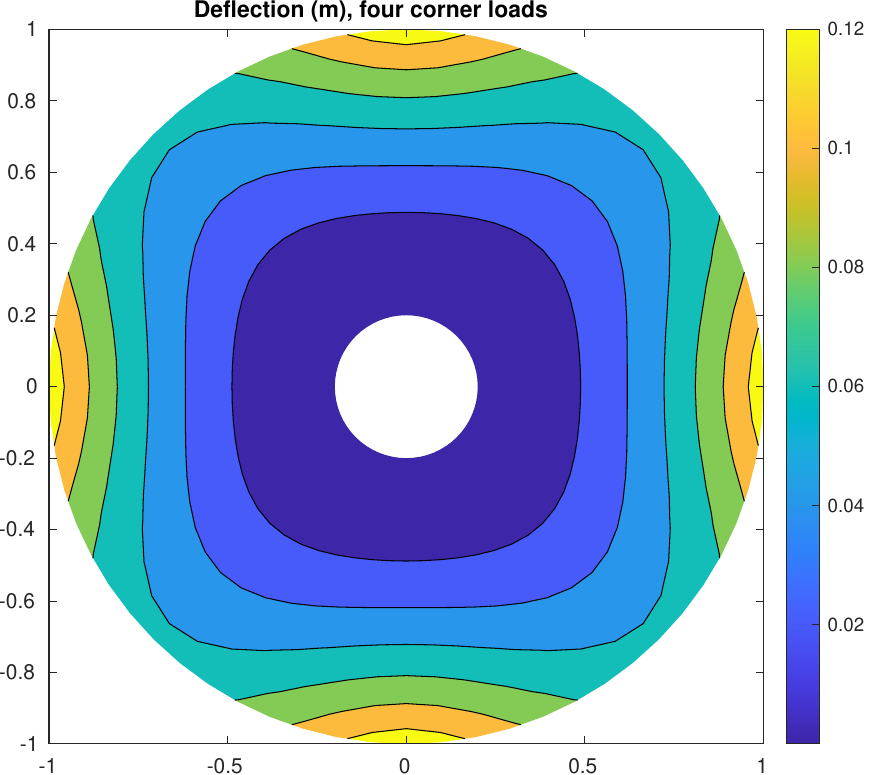}
        \caption{Damping plate deflection}
        \label{fig:damping-plate-deflection}
    \end{subfigure}
    \caption{Damping plate structural analysis: radial bending moment (a) and deflection (b) over the plate radius.}
    \label{fig:damping-plate-struct}
\end{figure}
In \Cref{fig:damping-plate-maxs}, peak stress and deflection are plotted as a function of damping plate aspect ratio, normalized by their values in the nominal design.

\subsection{Column}
As stated in \Cref{sec:structures}, the column's small slenderness ratio means that both axial compression and global Euler buckling must be taken into account.
Meanwhile, the hydrostatic pressure at the bottom creates a substantial compressive hoop stress.

The Euler critical buckling force $F_{\text{crit}}$ is:
\begin{equation}
    F_{\text{crit}} = \frac{\pi^2 E I}{(K_{end} L)^2}
\end{equation}
where $E$ is the Young's modulus, $I$ is the second moment of area, $K_{end}=2$ is the end condition (fixed-free since the hydrostatic rotational stiffness provides a restoring moment at the surface), and $L$ is the beam length, set to $h_s$ the full spar height.
The spar factor of safety $FOS$ to prevent yield and global buckling under the combined loading is \cite{american_bureau_of_shipping_requirements_2022}:
\begin{equation}
    FOS = \frac{\sigma_{s,\text{max}}}{\sigma_s} = \frac{\sigma_Y \frac{\zeta + \sqrt{\zeta^2+4\omega}}{2}}{\frac{F}{A} + q}
\end{equation}
where $\zeta$ and $\omega$ are defined as follows:
\begin{equation}
\begin{aligned}
     \zeta &= 1 - P_r(1 - P_r)~\frac{\sigma_Y}{F_{\text{crit}}/A} - \frac{\sigma_\theta}{\sigma_Y}, \qquad
    \omega = \frac{\sigma_\theta}{2\sigma_Y}  \left(1 - \frac{\sigma_\theta}{2\sigma_Y}\right)
\end{aligned}
\end{equation}
and the hoop stress $\sigma_\theta$ is 
\begin{equation}
     \sigma_\theta = \frac{qD_s}{2~t_{s,r}}
\end{equation}
$P_r$ is a constant and $q$ is the distributed hydrostatic pressure at the bottom of the spar.
These equations assume that the spar is compact, meaning that local buckling of the tube as a plate element is not a concern \cite{american_bureau_of_shipping_requirements_2022}.


\section{Parameters}
\label{sec:appendix-parameters}
\Cref{tab:parameters} lists the constant parameters assumed in this work, including their descriptions, values, and units.

\begin{table}[ht]
\centering
\caption{Selected Parameters}
\label{tab:parameters}
\renewcommand{\arraystretch}{1.4}

\begin{tabular}{c>{\centering\arraybackslash}p{0.35\linewidth}>{\centering\arraybackslash}p{0.3\linewidth}c}
\textbf{Parameter} & \textbf{Description} & \textbf{Value} & \textbf{Units}  \\ \hline
$\rho_w$ & Seawater density& 1000 & kg/$m^3$  \\ 
$g$& Acceleration of gravity& 9.8 & m/$s^2$  \\ 
$h$& Water depth & 100 & m  \\ 
JPD& Wave joint probability distribution & read from file \cite{janzou_sam_2022}  &\%  \\ 
$H_s$ & Wave height \cite{janzou_sam_2022}& 0.25 : 0.5 : 6.75  & m  \\ 
$H_{s,\text{struct}}$ & 100 year wave height \cite{berg_extreme_2011}& [13.4, 18.8, 24.2, 30.1, 24.2, 18.8, 13.4] & m  \\ 
$T$& Wave energy period \cite{janzou_sam_2022}& 4.5 : 1 : 18.5  & s  \\ 
$T_{\text{struct}}$ & 100 year wave peak period \cite{berg_extreme_2011}& [5.57 8.76 12.18 17.26 21.09 24.92 31.70] & s  \\ 
$\sigma_y$ & Material yield strength& 248 & MPa  \\ 
$\rho_m$ & Material density& 7850 & kg/$m^3$  \\ 
$E$& Material Young's modulus& 200 & GPa  \\ 
$\text{cost}_m$ & Material cost& 1.89 & \$/kg \\ 
$FOS_{\text{min}}$ & Minimum factor of safety& 1.5 & -  \\ 
$D_{d_{\text{min}}}$ & Minimum damping plate diameter & 30 & m  \\ 
$FCR$ & Fixed charge rate& 10.8 & \%  \\ 
$N_{WEC}$ & Number of WECs in array& 100 & -  \\ 
$D_{d}/{D_{s}}$ & Normalized damping plate diameter & 5 & -  \\ 
$T_{s}/{D_{s}}$ & Normalized spar draft& 5.83 & -  \\ 
$h_{d}/{D_{s}}$ & Normalized damping plate thickness & 0.004 & -  \\ 
$T_{f}/{h_{f}}$ & Float submergence ratio & 0.5 & -  \\ 
$F_{\text{heave},\text{mult}}$ & Heave force multiplier & 1.65 & - \\ 
\end{tabular}
 \end{table}
\clearpage

}
\printcredits

\bibliographystyle{cas-model2-names}
\bibliography{references,zotero-meem-refs}

@phdthesis{mccabe_dissertation_2026,
	address = {Ithaca, NY},
	type = {{PhD} dissertation},
	title = {Leveraging {Semi}-{Analytical} {Modeling}, {Multidisciplinary} {Design} {Optimization}, and {System} {Value} {Metrics} to {Advance} {Wave} {Energy} {Converter} {Viability}},
	url = {https://calkit.io/symbiotic-engineering/mdocean/publications?path=pubs%2Fdissertation%2FsampleThesis.pdf},
	school = {Cornell University},
	author = {McCabe, Rebecca},
	month = aug,
	year = {2026},
}

@misc{bimali_matrix_2026,
    title = {Matrix structure and convergence behaviour of the matched eigenfunction method for computing heave wave forces on generalized concentric bodies},
    url = {http://arxiv.org/abs/2605.19730},
    doi = {10.48550/arXiv.2605.19730},
    urldate = {2026-06-09},
    publisher = {arXiv},
    author = {Bimali, Yinghui and McCabe, Rebecca and Treacy, Collin and Khanal, Kapil and Lo, En and Haji, Maha},
    month = may,
    year = {2026},
    note = {arXiv:2605.19730 [physics.ao-ph]},
}

@unpublished{mccabe_leveraging_2026,
    title = {Leveraging {Multidisciplinary} {Design} {Optimization} to {Advance} {Wave} {Energy} {Converter} {Viability}},
    author = {McCabe, Rebecca and Dietrich, Madison and Haji, Maha},
    year = {2026},
    note = {Note: In prep.},
}

@book{boyd_convex_2004,
    title = {Convex {Optimization}},
    isbn = {978-0-521-83378-3},
    language = {en},
    publisher = {Cambridge University Press},
    author = {Boyd, Stephen P. and Vandenberghe, Lieven},
    month = mar,
    year = {2004},
    note = {Google-Books-ID: mYm0bLd3fcoC},
}

@article{agrawal_disciplined_2020,
    title = {Disciplined quasiconvex programming},
    volume = {14},
    issn = {1862-4480},
    url = {https://doi.org/10.1007/s11590-020-01561-8},
    doi = {10.1007/s11590-020-01561-8},
    language = {en},
    number = {7},
    urldate = {2026-05-05},
    journal = {Optimization Letters},
    author = {Agrawal, Akshay and Boyd, Stephen},
    month = oct,
    year = {2020},
    pages = {1643--1657},
}

@article{elnagar_pseudospectral_1995,
    title = {The pseudospectral {Legendre} method for discretizing optimal control problems},
    volume = {40},
    issn = {1558-2523},
    url = {https://ieeexplore.ieee.org/document/467672},
    doi = {10.1109/9.467672},
    number = {10},
    urldate = {2026-04-08},
    journal = {IEEE Transactions on Automatic Control},
    author = {Elnagar, G. and Kazemi, M.A. and Razzaghi, M.},
    month = oct,
    year = {1995},
    pages = {1793--1796},
}

@article{hals_comparison_2011,
    title = {A {Comparison} of {Selected} {Strategies} for {Adaptive} {Control} of {Wave} {Energy} {Converters}},
    volume = {133},
    issn = {0892-7219},
    url = {https://doi.org/10.1115/1.4002735},
    doi = {10.1115/1.4002735},
    number = {031101},
    urldate = {2026-04-08},
    journal = {Journal of Offshore Mechanics and Arctic Engineering},
    author = {Hals, Jørgen and Falnes, Johannes and Moan, Torgeir},
    month = mar,
    year = {2011},
}

@article{tan_computationally_2026,
    title = {Computationally efficient spectral-domain wave-to-wire modeling of wave energy converters with geared rotary generators},
    volume = {170},
    issn = {0141-1187},
    url = {https://www.sciencedirect.com/science/article/pii/S0141118726001112},
    doi = {10.1016/j.apor.2026.105028},
    urldate = {2026-04-07},
    journal = {Applied Ocean Research},
    author = {Tan, Jian and Tao, Ji and Tao, Wei and Xi, Chen and Lavidas, George and Shi, Hongda},
    month = may,
    year = {2026},
    pages = {105028},
}

@book{atherton_nonlinear_1982,
    edition = {Student edition},
    title = {Nonlinear {Control} {Engineering}: {Describing} {Function}, {Analysis} and...},
    isbn = {978-0-442-30486-7},
    shorttitle = {Nonlinear {Control} {Engineering}},
    url = {https://www.biblio.com/book/nonlinear-control-engineering-describing-function-analysis/d/1530706447?srsltid=AfmBOopUAO10sh1B9SwmwO78Kn8XmqZsEBUCuyRDsBzWQw9y1AH_QJWz},
    language = {en},
    urldate = {2026-04-07},
    publisher = {Chapman \& Hall},
    author = {Atherton, Derek P},
    year = {1982},
}

@book{ching_quasilinear_2010,
    address = {Cambridge},
    title = {Quasilinear {Control}: {Performance} {Analysis} and {Design} of {Feedback} {Systems} with {Nonlinear} {Sensors} and {Actuators}},
    isbn = {978-1-107-00056-8},
    shorttitle = {Quasilinear {Control}},
    url = {https://www.cambridge.org/core/books/quasilinear-control/7FCDA58513EA21BEEB4C3FA941C818BD},
    doi = {10.1017/CBO9780511976476},
    urldate = {2026-04-07},
    publisher = {Cambridge University Press},
    author = {Ching, ShiNung and Eun, Yongsoon and Gokcek, Cevat and Kabamba, Pierre T. and Meerkov, Semyon M.},
    year = {2010},
}

@book{roberts_random_2003,
    title = {Random {Vibration} and {Statistical} {Linearization}},
    isbn = {978-0-486-43240-3},
    language = {en},
    publisher = {Courier Corporation},
    author = {Roberts, John Brian and Spanos, Pol D.},
    month = dec,
    year = {2003},
    note = {Google-Books-ID: xQ7TtGwEMVoC},
}

@article{vanantwerp_tutorial_2000,
    title = {A tutorial on linear and bilinear matrix inequalities},
    volume = {10},
    issn = {0959-1524},
    url = {https://www.sciencedirect.com/science/article/pii/S0959152499000566},
    doi = {10.1016/S0959-1524(99)00056-6},
    number = {4},
    urldate = {2026-04-07},
    journal = {Journal of Process Control},
    author = {VanAntwerp, Jeremy G. and Braatz, Richard D.},
    month = aug,
    year = {2000},
    pages = {363--385},
}

@article{issoglio_second-order_2026,
    title = {Second-order wave excitation forces in {WEC}-{Sim}/{MOST}: {Implementation}, experimental validation, and code-to-code comparison},
    volume = {353},
    issn = {0029-8018},
    shorttitle = {Second-order wave excitation forces in {WEC}-{Sim}/{MOST}},
    url = {https://www.sciencedirect.com/science/article/pii/S0029801826004397},
    doi = {10.1016/j.oceaneng.2026.124605},
    urldate = {2026-03-20},
    journal = {Ocean Engineering},
    author = {Issoglio, Davide and Shabara, Mohamed A. and Petracca, Ermando and Niosi, Francesco and Keester, Adam and Sirigu, Massimo and Bracco, Giovanni},
    month = apr,
    year = {2026},
    pages = {124605},
}

@inproceedings{mccabe_wec_2025,
    address = {Corvallis, OR},
    title = {{WEC} optimization to maximize grid economic value and avoided emissions},
    url = {https://calkit.io/symbiotic-engineering/mdocean/publications#pubs/UMERC-2025-grid-value/UMERC_2025_grid_value.pdf},
    booktitle = {{UMERC}+{OREC} 2025 {Conference}},
    publisher = {University Marine Energy Research Consortium},
    author = {McCabe, Rebecca},
    month = aug,
    year = {2025},
    pages = {8},
}

@inproceedings{mccabe_open-source_2024,
    address = {Duluth, MN},
    title = {Open-source toolbox for semi-analytical hydrodynamic coefficients  via the matched eigenfunction expansion method},
    url = {https://zenodo.org/records/14504017},
    doi = {10.5281/zenodo.14504017},
    urldate = {2025-01-03},
    booktitle = {{UMERC} ({University} {Marine} {Energy} {Research} {Consortium}) {Conference} 2024},
    publisher = {Zenodo},
    author = {McCabe, Rebecca and Khanal, Kapil and Haji, Maha},
    month = aug,
    year = {2024},
    pages = {6},
}

@techreport{ruehl_next-generation_2023,
    title = {Next-{Generation} {Marine} {Energy} {Software} {Needs} {Assessment}},
    url = {https://www.osti.gov/biblio/2431205},
    doi = {10.2172/2431205},
    language = {English},
    number = {SAND--2023-03906R; NREL/TP-5700-84936},
    urldate = {2026-03-14},
    institution = {Sandia National Laboratories (SNL-NM), Albuquerque, NM (United States); National Renewable Energy Laboratory (NREL), Golden, CO (United States)},
    author = {Ruehl, Kelley Michelle and Leon-Quiroga, Jorge Andres and Michelen Strofer, Carlos Alejandro and Topper, Mathew and Tom, Nathan and Baca, Elena and Ogden, David},
    month = may,
    year = {2023},
}

@unpublished{best_openflash_2026,
    type = {In review},
    title = {{OpenFLASH}: {An} open-source flexible library for analytical and semi-analytical hydrodynamics calculations},
    url = {https://joss.theoj.org/papers/842d9827ef856fd88af7128f707f265b},
    author = {Best, Hope and Khanal, Kapil and McCabe, Rebecca and Jiang, Ruiyang and Treacy, Collin and Haji, Maha},
    year = {2026},
    note = {In review},
}

@incollection{chicone_contraction_2006,
    title = {Contraction},
    isbn = {978-0-387-30769-5},
    url = {https://www.google.com/books/edition/Ordinary_Differential_Equations_with_App/yfY2uGROVrUC},
    language = {en},
    booktitle = {Ordinary {Differential} {Equations} with {Applications}},
    publisher = {Springer Science \& Business Media},
    author = {Chicone, Carmen},
    month = may,
    year = {2006},
    note = {Google-Books-ID: yfY2uGROVrUC},
    pages = {121--144},
}

@inproceedings{philip_damping_2012,
    address = {Rio de Janeiro, Brazil},
    title = {Damping {Characteristics} of {Heave} {Plates} {Attached} to {Spar} {Hull}},
    url = {https://dx.doi.org/10.1115/OMAE2012-83290},
    doi = {10.1115/OMAE2012-83290},
    language = {en},
    urldate = {2025-04-28},
    booktitle = {{ASME} 2012 31st {International} {Conference} on {Ocean}, {Offshore} and {Arctic} {Engineering}},
    publisher = {American Society of Mechanical Engineers Digital Collection},
    author = {Philip, Nimmy Thankom and Nallayarasu, S. and Bhattacharyya, S. K.},
    year = {2012},
    pages = {12},
}

@misc{anderson_re-imagining_2024,
    address = {Duluth, MN},
    type = {Conference presentation},
    title = {Re-{Imagining} the {RM3} through control co-design},
    url = {https://umerc2024.exordo.com/programme/presentation/106},
    urldate = {2024-12-22},
    author = {Anderson, Megan and Gaebele, Daniel and Roach, Aeron and Forbrush, Dominic and Roberts, Jesse and Weber, Jochem},
    month = aug,
    year = {2024},
}

@inproceedings{ogden_hydrochrono_2023,
    title = {{HydroChrono}: {An} {Open}-{Source} {Hydrodynamics} {Package} for {Project} {Chrono}},
    shorttitle = {{HydroChrono}},
    url = {https://research-hub.nrel.gov/en/publications/hydrochrono-an-open-source-hydrodynamics-package-for-project-chro/},
    doi = {10.36688/ewtec-2023-473},
    language = {American English},
    urldate = {2026-02-27},
    booktitle = {15th {European} {Wave} and {Tidal} {Energy} {Conference}},
    author = {Ogden, David and Quinton, Zuriah and Lataillade, Tristan de and Pallud, Maxime},
    year = {2023},
    pages = {10},
}

@inproceedings{mccabe_multidisciplinary_2022,
    title = {Multidisciplinary {Optimization} to {Reduce} {Cost} and {Power} {Variation} of a {Wave} {Energy} {Converter}},
    url = {https://asmedigitalcollection.asme.org/IDETC-CIE/proceedings-abstract/IDETC-CIE2022/86229/1150407},
    doi = {10.1115/DETC2022-90227},
    language = {en},
    urldate = {2023-03-01},
    booktitle = {{ASME} 2022 {International} {Design} {Engineering} {Technical} {Conferences} and {Computers} and {Information} in {Engineering} {Conference}},
    publisher = {American Society of Mechanical Engineers Digital Collection},
    author = {McCabe, Rebecca and Murphy, Olivia and Haji, Maha},
    month = nov,
    year = {2022},
    pages = {10},
}

@inproceedings{chau_inertia_2012,
    title = {Inertia, {Damping}, and {Wave} {Excitation} of {Heaving} {Coaxial} {Cylinders}},
    url = {https://dx.doi.org/10.1115/OMAE2012-83987},
    doi = {10.1115/OMAE2012-83987},
    language = {en},
    urldate = {2023-09-28},
    booktitle = {{ASME} 2012 31st {International} {Conference} on {Ocean}, {Offshore} and {Arctic} {Engineering}},
    publisher = {American Society of Mechanical Engineers Digital Collection},
    author = {Chau, Fun Pang and Yeung, Ronald W.},
    year = {2012},
    pages = {803--813},
}

@mastersthesis{herber_dynamic_2014,
    address = {Urbana, Illinois},
    title = {Dynamic system design optimization of wave energy converters utilizing direct transcription},
    copyright = {Copyright 2014 Daniel Ronald Herber},
    url = {https://hdl.handle.net/2142/49463},
    urldate = {2023-03-03},
    school = {University of Illinois at Urbana-Champaign},
    author = {Herber, Daniel Ronald},
    month = may,
    year = {2014},
}

@techreport{azarm_monotonicity-based_1988,
    address = {College Park, MD},
    type = {Technical report},
    title = {Monotonicity-{Based} {Decomposition} {Methods} for {Design} {Optimization}.},
    url = {http://hdl.handle.net/1903/4743},
    language = {en\_US},
    number = {SRC TR 88-10},
    urldate = {2026-02-27},
    institution = {University of Maryland College Park},
    author = {Azarm, Shapour and Li, Wei-Chu},
    year = {1988},
    pages = {39},
}

@misc{janzou_system_2020,
    address = {Golden, CO},
    title = {System {Advisor} {Model} ({SAM})},
    copyright = {BSD-3-Clause},
    url = {https://github.com/NREL/SAM/},
    urldate = {2022-05-13},
    publisher = {National Renewable Energy Laboratory},
    author = {Janzou, Steve},
    month = nov,
    year = {2020},
    note = {original-date: 2013-01-10T02:52:47Z},
}

@article{wright_coordinate_2015,
    title = {Coordinate descent algorithms},
    volume = {151},
    issn = {1436-4646},
    url = {https://doi.org/10.1007/s10107-015-0892-3},
    doi = {10.1007/s10107-015-0892-3},
    language = {en},
    number = {1},
    urldate = {2026-02-27},
    journal = {Mathematical Programming},
    author = {Wright, Stephen J.},
    month = jun,
    year = {2015},
    pages = {3--34},
}

@article{martins_multidisciplinary_2013,
    title = {Multidisciplinary {Design} {Optimization}: {A} {Survey} of {Architectures}},
    volume = {51},
    issn = {0001-1452},
    shorttitle = {Multidisciplinary {Design} {Optimization}},
    url = {https://arc.aiaa.org/doi/10.2514/1.J051895},
    doi = {10.2514/1.J051895},
    number = {9},
    urldate = {2023-12-04},
    journal = {AIAA Journal},
    publisher = {American Institute of Aeronautics and Astronautics},
    author = {Martins, Joaquim R. R. A. and Lambe, Andrew B.},
    month = sep,
    year = {2013},
    note = {\_eprint: https://doi.org/10.2514/1.J051895},
    pages = {2049--2075},
}

@article{reveyrand_multiport_2018,
    address = {Brive La Gaillarde, France},
    title = {Multiport conversions between {S}, {Z}, {Y}, h, {ABCD}, and {T} parameters},
    copyright = {https://doi.org/10.15223/policy-029},
    url = {https://ieeexplore.ieee.org/document/8430023/},
    doi = {10.1109/INMMIC.2018.8430023},
    urldate = {2026-02-05},
    journal = {2018 International Workshop on Integrated Nonlinear Microwave and Millimetre-wave Circuits (INMMIC)},
    publisher = {IEEE},
    author = {Reveyrand, T.},
    month = jul,
    year = {2018},
    note = {Conference Name: 2018 International Workshop on Integrated Nonlinear Microwave and Millimetre-wave Circuits (INMMIC)
ISBN: 9781538655078},
    pages = {1--3},
}

@book{saadat_power_1999,
    title = {Power system analysis},
    volume = {2},
    url = {https://www.uvic.ca/ecs/ece/assets/docs/current/undergraduate/201901/ece488.pdf},
    urldate = {2026-02-02},
    publisher = {McGraw-hill},
    author = {Saadat, Hadi},
    year = {1999},
}

@misc{weisstein_cassini_2026,
    type = {Text},
    title = {Cassini {Ovals}},
    copyright = {Copyright 1999-2026 Wolfram Research, Inc.  See https://mathworld.wolfram.com/about/terms.html for a full terms of use statement.},
    url = {https://mathworld.wolfram.com/CassiniOvals.html},
    language = {en},
    urldate = {2026-02-02},
    journal = {MathWorld - a Wolfram Resource},
    publisher = {Wolfram Research, Inc.},
    author = {Weisstein, Eric W.},
    month = jan,
    year = {2026},
}

@book{tedrake_underactuated_2024,
    title = {Underactuated {Robotics}: {Algorithms} for {Walking}, {Running}, {Swimming}, {Flying}, and {Manipulation}},
    url = {https://underactuated.csail.mit.edu},
    author = {Tedrake, Russ},
    year = {2024},
    note = {Published: Course Notes for MIT 6.832},
}

@article{coe_co-design_2025,
    title = {Co-design of a wave energy converter through bi-conjugate impedance matching},
    volume = {111},
    issn = {0957-4158},
    url = {https://www.sciencedirect.com/science/article/pii/S0957415825001047},
    doi = {10.1016/j.mechatronics.2025.103395},
    urldate = {2025-12-16},
    journal = {Mechatronics},
    author = {Coe, Ryan G. and Bacelli, Giorgio and Gaebele, Daniel and Keow, Alicia and Forbush, Dominic},
    month = nov,
    year = {2025},
    pages = {103395},
}

@inproceedings{hendrikx_optimal_2017,
    title = {Optimal control of a wave energy converter},
    url = {https://ieeexplore.ieee.org/document/8062556},
    doi = {10.1109/CCTA.2017.8062556},
    urldate = {2025-03-11},
    booktitle = {2017 {IEEE} {Conference} on {Control} {Technology} and {Applications} ({CCTA})},
    author = {Hendrikx, R. W. M. and Leth, J. and Andersen, P. and Heemels, W. P. M. H.},
    month = aug,
    year = {2017},
    pages = {779--786},
}

@article{coe_practical_2021,
    title = {A practical approach to wave energy modeling and control},
    volume = {142},
    issn = {1364-0321},
    url = {https://www.sciencedirect.com/science/article/pii/S1364032121000861},
    doi = {10.1016/j.rser.2021.110791},
    urldate = {2025-02-23},
    journal = {Renewable and Sustainable Energy Reviews},
    author = {Coe, Ryan G. and Bacelli, Giorgio and Forbush, Dominic},
    month = may,
    year = {2021},
    pages = {110791},
}

@incollection{folley_spectral-domain_2016,
    title = {Spectral-{Domain} {Models}},
    isbn = {978-0-12-803210-7},
    url = {https://www.sciencedirect.com/science/article/pii/B9780128032107000049},
    doi = {10.1016/B978-0-12-803210-7.00004-9},
    number = {Chapter 4},
    urldate = {2025-02-16},
    booktitle = {Numerical {Modelling} of {Wave} {Energy} {Converters}},
    publisher = {Academic Press},
    author = {Folley, M.},
    editor = {Folley, Matt},
    month = jan,
    year = {2016},
    pages = {67--80},
}

@article{spanos_efficient_2016,
    title = {Efficient {Dynamic} {Analysis} of a {Nonlinear} {Wave} {Energy} {Harvester} {Model}},
    volume = {138},
    issn = {0892-7219},
    url = {https://doi.org/10.1115/1.4032898},
    doi = {10.1115/1.4032898},
    number = {041901},
    urldate = {2025-02-16},
    journal = {Journal of Offshore Mechanics and Arctic Engineering},
    author = {Spanos, Pol D. and Arena, Felice and Richichi, Alessandro and Malara, Giovanni},
    month = apr,
    year = {2016},
}

@phdthesis{kluger_synergistic_2017,
    type = {Thesis},
    title = {Synergistic design of a combined floating wind turbine - wave energy converter},
    copyright = {MIT theses are protected by copyright. They may be viewed, downloaded, or printed from this source but further reproduction or distribution in any format is prohibited without written permission.},
    url = {https://dspace.mit.edu/handle/1721.1/111692},
    language = {eng},
    urldate = {2025-02-16},
    school = {Massachusetts Institute of Technology},
    author = {Kluger, Jocelyn Maxine},
    year = {2017},
    note = {Accepted: 2017-10-04T14:47:10Z},
}

@article{nie_optimal_2016,
    series = {Selected {Papers} from the {European} {Wave} and {Tidal} {Energy} {Conference} 2015, {Nante}, {France}},
    title = {Optimal causal control of wave energy converters in stochastic waves – {Accommodating} nonlinear dynamic and loss models},
    volume = {15},
    issn = {2214-1669},
    url = {https://www.sciencedirect.com/science/article/pii/S2214166916300169},
    doi = {10.1016/j.ijome.2016.04.004},
    urldate = {2025-02-16},
    journal = {International Journal of Marine Energy},
    author = {Nie, Rudy and Scruggs, Jeff and Chertok, Allan and Clabby, Darragh and Previsic, Mirko and Karthikeyan, Anantha},
    month = sep,
    year = {2016},
    pages = {41--55},
}

@article{neshat_enhancing_2024,
    title = {Enhancing the performance of hybrid wave-wind energy systems through a fast and adaptive chaotic multi-objective swarm optimisation method},
    volume = {362},
    issn = {0306-2619},
    url = {https://www.sciencedirect.com/science/article/pii/S0306261924003386},
    doi = {10.1016/j.apenergy.2024.122955},
    urldate = {2025-02-16},
    journal = {Applied Energy},
    author = {Neshat, Mehdi and Sergiienko, Nataliia Y. and Nezhad, Meysam Majidi and da Silva, Leandro S. P. and Amini, Erfan and Marsooli, Reza and Astiaso Garcia, Davide and Mirjalili, Seyedali},
    month = may,
    year = {2024},
    pages = {122955},
}

@article{da_silva_stochastic_2023,
    title = {Stochastic analysis of the nonlinear dynamics of oscillating water columns: {A} frequency domain approach},
    volume = {139},
    issn = {0141-1187},
    shorttitle = {Stochastic analysis of the nonlinear dynamics of oscillating water columns},
    url = {https://www.sciencedirect.com/science/article/pii/S0141118723002523},
    doi = {10.1016/j.apor.2023.103711},
    urldate = {2025-02-16},
    journal = {Applied Ocean Research},
    author = {da Silva, L. S. P. and Pesce, C. P. and de Oliveira, M. and Sergiienko, N. Y. and Cazzolato, B. and Ding, B.},
    month = oct,
    year = {2023},
    pages = {103711},
}

@article{da_silva_statistical_2020,
    title = {Statistical linearization of the {Morison}’s equation applied to wave energy converters},
    volume = {6},
    copyright = {2020 Springer Nature Switzerland AG},
    issn = {2198-6452},
    url = {https://link.springer.com/article/10.1007/s40722-020-00165-9},
    doi = {10.1007/s40722-020-00165-9},
    language = {en},
    number = {2},
    urldate = {2025-02-16},
    journal = {Journal of Ocean Engineering and Marine Energy},
    publisher = {Springer International Publishing},
    author = {da Silva, Leandro S. P. and Cazzolato, Benjamin S. and Sergiienko, Nataliia Y. and Ding, Boyin and Morishita, Helio M. and Pesce, Celso P.},
    month = may,
    year = {2020},
    note = {Company: Springer
Distributor: Springer
Institution: Springer
Label: Springer
Number: 2},
    pages = {157--169},
}

@article{carmo_slender-body_2025,
    title = {Slender-body approach for computing second-order wave loads in the frequency domain},
    volume = {322},
    issn = {0029-8018},
    url = {https://www.sciencedirect.com/science/article/pii/S0029801825002732},
    doi = {10.1016/j.oceaneng.2025.120558},
    urldate = {2025-02-15},
    journal = {Ocean Engineering},
    author = {Carmo, Lucas and Hall, Matthew},
    month = apr,
    year = {2025},
    pages = {120558},
}

@article{pennock_temporal_2022,
    title = {Temporal complementarity of marine renewables with wind and solar generation: {Implications} for {GB} system benefits},
    volume = {319},
    issn = {0306-2619},
    shorttitle = {Temporal complementarity of marine renewables with wind and solar generation},
    url = {https://www.sciencedirect.com/science/article/pii/S030626192200633X},
    doi = {10.1016/j.apenergy.2022.119276},
    urldate = {2025-01-31},
    journal = {Applied Energy},
    author = {Pennock, Shona and Coles, Daniel and Angeloudis, Athanasios and Bhattacharya, Saptarshi and Jeffrey, Henry},
    month = aug,
    year = {2022},
    pages = {119276},
}

@article{bhattacharya_timing_2021,
    title = {Timing value of marine renewable energy resources for potential grid applications},
    volume = {299},
    issn = {0306-2619},
    url = {https://www.sciencedirect.com/science/article/pii/S030626192100698X},
    doi = {10.1016/j.apenergy.2021.117281},
    urldate = {2025-01-31},
    journal = {Applied Energy},
    author = {Bhattacharya, Saptarshi and Pennock, Shona and Robertson, Bryson and Hanif, Sarmad and Alam, Md Jan E. and Bhatnagar, Dhruv and Preziuso, Danielle and O’Neil, Rebecca},
    month = oct,
    year = {2021},
    pages = {117281},
}

@article{gaebele_tpl_2025,
    title = {From {TPL} assessment to design optimization: {Wave} energy converter control co-design applied to the {RM3}},
    volume = {241},
    issn = {0960-1481},
    shorttitle = {From {TPL} assessment to design optimization},
    url = {https://www.sciencedirect.com/science/article/pii/S0960148124024066},
    doi = {10.1016/j.renene.2024.122338},
    urldate = {2025-01-31},
    journal = {Renewable Energy},
    author = {Gaebele, Daniel T. and Anderson, Megan L. and Roach, Aeron L. and Forbush, Dominic D. and Roberts, Jesse D. and Weber, Jochem},
    month = mar,
    year = {2025},
    pages = {122338},
}

@article{lambe_extensions_2012,
    title = {Extensions to the design structure matrix for the description of multidisciplinary design, analysis, and optimization processes},
    volume = {46},
    issn = {1615-1488},
    url = {https://doi.org/10.1007/s00158-012-0763-y},
    doi = {10.1007/s00158-012-0763-y},
    language = {en},
    number = {2},
    urldate = {2025-01-23},
    journal = {Structural and Multidisciplinary Optimization},
    author = {Lambe, Andrew B. and Martins, Joaquim R. R. A.},
    month = aug,
    year = {2012},
    pages = {273--284},
}

@techreport{previsic_reference_2011,
    type = {Unpublished draft report obtained via personal communication with {Vince} {Neary} of {Sandia} {National} {Lab}},
    title = {Reference {Model} 3 - {Structural} {Design} {Supplement}},
    institution = {ReVision Consulting and US Department of Energy},
    author = {Previsic, Mirko},
    month = sep,
    year = {2011},
    pages = {20},
}

@book{timoshenko_theory_1959,
    series = {Engineering mechanics series},
    title = {Theory of {Plates} and {Shells}},
    isbn = {978-0-07-085820-6},
    url = {https://books.google.com/books?id=rTQFAAAAMAAJ},
    publisher = {McGraw-Hill},
    author = {Timoshenko, S. and Woinowsky-Krieger, S.},
    year = {1959},
    lccn = {58059675},
}

@misc{janzou_sam_2022,
    address = {National Renewable Energy Lab GitHub},
    title = {{SAM} {Wave} {Resource} {Files}},
    url = {https://github.com/NREL/SAM/tree/patch/deploy/wave_resource},
    urldate = {2025-01-12},
    publisher = {SAM (System Advisor Model)},
    author = {Janzou, Steve and Gilman, Paul and Prilliman, Matt},
    month = oct,
    year = {2022},
}

@techreport{berg_extreme_2011,
    title = {Extreme {Ocean} {Wave} {Conditions} for {Northern} {California} {Wave} {Energy} {Conversion} {Device}.},
    url = {https://www.osti.gov/biblio/1113856},
    doi = {10.2172/1113856},
    language = {English},
    number = {SAND2011-9304},
    urldate = {2025-01-13},
    institution = {Sandia National Lab. (SNL-CA), Livermore, CA (United States); Sandia National Lab. (SNL-NM), Albuquerque, NM (United States)},
    author = {Berg, Jonathan Charles},
    month = dec,
    year = {2011},
}

@techreport{livecchi_powering_2019,
    title = {Powering the {Blue} {Economy}: {Exploring} {Opportunities} for {Marine} {Renewable} {Energy} in {Maritime} {Markets}},
    url = {https://www.energy.gov/eere/water/downloads/powering-blue-economy-report},
    institution = {US Department of Energy (DOE)},
    author = {LiVecchi, A and Copping, A and Jenne, D and Gorton, A and Preus, R and Gill, G and Robichaud, R and Green, R and Geerlofs, S and Gore, S and Hume, D and McShane, W and Schmaus, C and Spence, H},
    month = apr,
    year = {2019},
    pages = {207},
}

@misc{neary_reference_2014,
    type = {Spreadsheet},
    title = {Reference {Model} 3 {Cost} {Breakdown} ({RM3}: {Wave} {Point} {Absorber})},
    url = {https://mhkdr.openei.org/submissions/370},
    doi = {10.15473/1819894},
    publisher = {Marine and Hydrokinetic Data Repository},
    author = {Neary, Vincent and Previsic, Mirko and Jenne, Scott and Hallett, Kathleen},
    month = sep,
    year = {2014},
}

@inproceedings{nakhai_techno-economic_2022,
    title = {Techno-{Economic} {Implications} of {Electrical} {Machine} {Scaling} for {Wave} {Energy} {Converters}},
    issn = {0197-7385},
    url = {https://ieeexplore.ieee.org/abstract/document/9977184},
    doi = {10.1109/OCEANS47191.2022.9977184},
    urldate = {2025-01-11},
    booktitle = {{OCEANS} 2022, {Hampton} {Roads}},
    author = {Nakhai, Aryana Y. and McGilton, Ben},
    month = oct,
    year = {2022},
    pages = {1--6},
}

@misc{mcgilton_optimal_2024,
    address = {Rochester, NY},
    type = {{SSRN} {Scholarly} {Paper}},
    title = {On the {Optimal} {Sizing} of {Power} {Take}-{Off} {Systems} for {Wave} {Energy} {Converters}},
    url = {https://papers.ssrn.com/abstract=4886361},
    doi = {10.2139/ssrn.4886361},
    language = {en},
    urldate = {2025-01-11},
    publisher = {Social Science Research Network},
    author = {McGilton, Ben and Nakhai, Aryana and McNally, Jim},
    month = jul,
    year = {2024},
}

@misc{ruehl_wec-simwec-sim_2024,
    title = {{WEC}-{Sim}/{WEC}-{Sim}: v6.1.2},
    shorttitle = {{WEC}-{Sim}/{WEC}-{Sim}},
    url = {https://zenodo.org/records/14549050},
    doi = {10.5281/zenodo.14549050},
    urldate = {2025-01-02},
    publisher = {Zenodo},
    author = {Ruehl, Kelley and Keester, Adam and dforbush2 and Ströfer, Carlos A. Michelén and Topper, Mathew and Lawson, Michael and jtgrasb and Husain, Salman and Leon, Jorge and Ling, Bradley A. and Shabara, Mohamed and Ogden, David and j-vanrij and jhbates and Nguyen, Lily and Jeffalo1 and sedwardsand and ratanakso and emiliofa and crobarcro and agmoore4 and Alves, Erick F. and zmorrell-sand and yuyihsiang and Hall, Matt and gparisella and ashleynchong and SiHeTh and Davies, Ryan and Riley, Pete},
    month = dec,
    year = {2024},
}

@misc{mccabe_mdocean_2024,
    title = {{MDOcean}},
    url = {https://zenodo.org/records/13997244},
    doi = {10.5281/zenodo.13997244},
    urldate = {2025-01-02},
    publisher = {Zenodo},
    author = {McCabe, Rebecca and Dietrich, Madison and Ren, Iris and Murphy, Olivia and Haji, Maha N.},
    month = oct,
    year = {2024},
}

@techreport{gaudin_single_2021,
    type = {Final report},
    title = {From single to multiple wave energy converters: {Cost} reduction through location and configuration optimisation},
    shorttitle = {From single to multiple wave energy converters},
    number = {ARENA 2015 RND086},
    institution = {University of Western Australia},
    author = {Gaudin, C. and David, D. R. and Cai, Y. and Hansen, J. E. and Bransby, M. F. and Rijnsdorp, D. P. and Lowe, R. J. and O’Loughlin, C. D. and Lu, T. and Uzielli, M. and O'Neill, Michael},
    month = oct,
    year = {2021},
}

@techreport{newman_motions_1963,
    address = {Bethesda, MD},
    type = {Research and development report},
    title = {The motions of a spar buoy in regular waves},
    url = {https://apps.dtic.mil/sti/tr/pdf/AD0406333.pdf},
    doi = {10.5962/bhl.title.48348},
    number = {1499},
    institution = {Department of the Navy: David Taylor Model Basin Hydromechanics Laboratory},
    author = {Newman, J. N.},
    month = may,
    year = {1963},
    pages = {38},
}

@book{papalambros_principles_2017,
    edition = {Third},
    title = {Principles of optimal design},
    isbn = {978-1-107-13267-2},
    publisher = {Cambridge University Press, Cambridge},
    author = {Papalambros, Panos Y. and Wilde, Douglass J.},
    year = {2017},
    mrnumber = {3585626},
}

@article{rosati_control_2023,
    title = {Control co-design of power take-off and bypass valve for {OWC}-based wave energy conversion systems},
    volume = {219},
    issn = {0960-1481},
    url = {https://www.sciencedirect.com/science/article/pii/S0960148123014386},
    doi = {10.1016/j.renene.2023.119523},
    urldate = {2024-12-23},
    journal = {Renewable Energy},
    author = {Rosati, Marco and Ringwood, John V.},
    month = dec,
    year = {2023},
    pages = {119523},
}

@article{lin_fast_2025,
    title = {Fast optimal control performance evaluation for wave energy control co-design},
    volume = {239},
    issn = {0960-1481},
    url = {https://www.sciencedirect.com/science/article/pii/S0960148124020421},
    doi = {10.1016/j.renene.2024.121974},
    urldate = {2024-12-22},
    journal = {Renewable Energy},
    author = {Lin, Zechuan and Huang, Xuanrui and Xiao, Xi and Ringwood, John V.},
    month = feb,
    year = {2025},
    pages = {121974},
}

@article{devin_high-dimensional_2024,
    title = {High-dimensional control co-design of a wave energy converter with a novel pitch resonator power takeoff system},
    volume = {312},
    issn = {0029-8018},
    url = {https://www.sciencedirect.com/science/article/pii/S0029801824024624},
    doi = {10.1016/j.oceaneng.2024.119124},
    urldate = {2024-12-22},
    journal = {Ocean Engineering},
    author = {Devin, Michael C. and Gaebele, Daniel T. and Ströfer, Carlos A. Michelén and Grasberger, Jeff T. and Lee, Jantzen and Coe, Ryan G. and Bacelli, Giorgio},
    month = nov,
    year = {2024},
    pages = {119124},
}

@article{cotten_multi-objective_2022,
    title = {Multi-objective optimisation of a sloped-motion, multibody wave energy converter concept},
    volume = {194},
    issn = {0960-1481},
    url = {https://www.sciencedirect.com/science/article/pii/S0960148122006681},
    doi = {10.1016/j.renene.2022.05.030},
    urldate = {2024-12-22},
    journal = {Renewable Energy},
    author = {Cotten, A. and Forehand, D. I. M.},
    month = jul,
    year = {2022},
    pages = {307--320},
}

@article{an_optimal_2024,
    title = {Optimal {Design} of the {Overtopping} {Wave} {Energy} {Converter} {Based} on {Fluid}–{Structure} {Interaction} {Simulation}},
    volume = {116},
    issn = {0749-0208},
    url = {https://doi.org/10.2112/JCR-SI116-117.1},
    doi = {10.2112/JCR-SI116-117.1},
    number = {SI},
    urldate = {2024-12-22},
    journal = {Journal of Coastal Research},
    author = {An, Sung-Hwan and Kim, Geun-Gon and Lee, Jong-Hyun},
    month = jan,
    year = {2024},
    pages = {578--582},
}

@article{garcia-teruel_reliability-based_2021,
    title = {Reliability-based hull geometry optimisation of a point-absorber wave energy converter with power take-off structural reliability objectives},
    volume = {15},
    copyright = {© 2021 The Authors. IET Renewable Power Generation published by John Wiley \& Sons Ltd on behalf of The Institution of Engineering and Technology},
    issn = {1752-1424},
    url = {https://onlinelibrary.wiley.com/doi/abs/10.1049/rpg2.12249},
    doi = {10.1049/rpg2.12249},
    language = {en},
    number = {14},
    urldate = {2024-12-22},
    journal = {IET Renewable Power Generation},
    author = {Garcia-Teruel, Anna and Clark, Caitlyn E.},
    year = {2021},
    note = {\_eprint: https://onlinelibrary.wiley.com/doi/pdf/10.1049/rpg2.12249},
    pages = {3255--3268},
}

@article{ferri_balancing_2014,
    title = {Balancing {Power} {Output} and {Structural} {Fatigue} of {Wave} {Energy} {Converters} by {Means} of {Control} {Strategies}},
    volume = {7},
    copyright = {http://creativecommons.org/licenses/by/3.0/},
    issn = {1996-1073},
    url = {https://www.mdpi.com/1996-1073/7/4/2246},
    doi = {10.3390/en7042246},
    language = {en},
    number = {4},
    urldate = {2024-12-22},
    journal = {Energies},
    publisher = {Multidisciplinary Digital Publishing Institute},
    author = {Ferri, Francesco and Ambühl, Simon and Fischer, Boris and Kofoed, Jens Peter},
    month = apr,
    year = {2014},
    note = {Number: 4},
    pages = {2246--2273},
}

@article{ambuhl_reliability-based_2014,
    title = {Reliability-{Based} {Structural} {Optimization} of {Wave} {Energy} {Converters}},
    volume = {7},
    copyright = {http://creativecommons.org/licenses/by/3.0/},
    issn = {1996-1073},
    url = {https://www.mdpi.com/1996-1073/7/12/8178},
    doi = {10.3390/en7128178},
    language = {en},
    number = {12},
    urldate = {2024-12-22},
    journal = {Energies},
    publisher = {Multidisciplinary Digital Publishing Institute},
    author = {Ambühl, Simon and Kramer, Morten and Sørensen, John Dalsgaard},
    month = dec,
    year = {2014},
    note = {Number: 12},
    pages = {8178--8200},
}

@article{coe_survey_2018,
    title = {A {Survey} of {WEC} {Reliability}, {Survival} and {Design} {Practices}},
    volume = {11},
    copyright = {http://creativecommons.org/licenses/by/3.0/},
    issn = {1996-1073},
    url = {https://www.mdpi.com/1996-1073/11/1/4},
    doi = {10.3390/en11010004},
    language = {en},
    number = {1},
    urldate = {2024-12-22},
    journal = {Energies},
    publisher = {Multidisciplinary Digital Publishing Institute},
    author = {Coe, Ryan G. and Yu, Yi-Hsiang and Van Rij, Jennifer},
    month = jan,
    year = {2018},
    note = {Number: 1},
    pages = {4},
}

@article{zou_practical_2023,
    title = {Practical power absorption assessment limits for generic wave energy converters},
    volume = {277},
    issn = {0029-8018},
    url = {https://www.sciencedirect.com/science/article/pii/S002980182300687X},
    doi = {10.1016/j.oceaneng.2023.114303},
    urldate = {2024-12-22},
    journal = {Ocean Engineering},
    author = {Zou, Shangyan and Robertson, Bryson and Yim, Solomon},
    month = jun,
    year = {2023},
    pages = {114303},
}

@article{garcia-teruel_design_2022,
    title = {Design limits for wave energy converters based on the relationship of power and volume obtained through multi-objective optimisation},
    volume = {200},
    issn = {0960-1481},
    url = {https://www.sciencedirect.com/science/article/pii/S0960148122014033},
    doi = {10.1016/j.renene.2022.09.053},
    urldate = {2024-12-22},
    journal = {Renewable Energy},
    author = {Garcia-Teruel, Anna and Roberts, Owain and Noble, Donald R. and Henderson, Jillian Catherine and Jeffrey, Henry},
    month = nov,
    year = {2022},
    pages = {492--504},
}

@techreport{housner_numerical_2024,
    title = {Numerical {Modeling} and {Optimization} of the {iProTech} {Pitching} {Inertial} {Pump} ({PIP}) {Wave} {Energy} {Converter} ({WEC}) ({Cooperative} {Research} and {Development} {Final} {Report}, {CRADA} {Number}: {CRD}-22-22968)},
    shorttitle = {Numerical {Modeling} and {Optimization} of the {iProTech} {Pitching} {Inertial} {Pump} ({PIP}) {Wave} {Energy} {Converter} ({WEC}) ({Cooperative} {Research} and {Development} {Final} {Report}, {CRADA} {Number}},
    url = {https://www.osti.gov/biblio/2478103},
    doi = {10.2172/2478103},
    language = {English},
    number = {NREL/TP-5000-92188; CRD-22-22968},
    urldate = {2024-12-21},
    institution = {National Renewable Energy Laboratory (NREL), Golden, CO (United States); iProTech, Redwood City, CA (United States)},
    author = {Housner, Stein and Wynn, Nick},
    month = nov,
    year = {2024},
}

@article{abdulkadir_control_2024,
    title = {Control co-design optimization of nonlinear wave energy converters},
    volume = {304},
    issn = {0029-8018},
    url = {https://www.sciencedirect.com/science/article/pii/S002980182401165X},
    doi = {10.1016/j.oceaneng.2024.117827},
    urldate = {2024-12-21},
    journal = {Ocean Engineering},
    author = {Abdulkadir, Habeebullah and Abdelkhalik, Ossama},
    month = jul,
    year = {2024},
    pages = {117827},
}

@article{mi_multi-scale_2025,
    title = {Multi-scale concurrent design of a 100 {kW} wave energy converter},
    volume = {238},
    issn = {0960-1481},
    url = {https://www.sciencedirect.com/science/article/pii/S0960148124019037},
    doi = {10.1016/j.renene.2024.121835},
    urldate = {2024-12-20},
    journal = {Renewable Energy},
    author = {Mi, Jia and Huang, Jianuo and Yang, Lisheng and Ahmed, Alaa and Li, Xiaofan and Wu, Xian and Datla, Raju and Staby, Bill and Hajj, Muhammad and Zuo, Lei},
    month = jan,
    year = {2025},
    pages = {121835},
}

@misc{khanal_multi-objective_2024,
    title = {Multi-{Objective} {Multidisciplinary} {Optimization} of {Wave} {Energy} {Converter} {Array} {Layout} and {Controls}},
    url = {http://arxiv.org/abs/2410.11089},
    doi = {10.48550/arXiv.2410.11089},
    urldate = {2024-12-17},
    publisher = {arXiv},
    author = {Khanal, Kapil and DeGoede, Nate and Vitale, Olivia and Haji, Maha N.},
    month = oct,
    year = {2024},
    note = {arXiv:2410.11089 [eess]},
}

@techreport{wierzbicki_lecture_2013,
    type = {Lecture notes},
    title = {Lecture 11: {Buckling} of {Plates} and {Sections}},
    shorttitle = {2.{080J} {Structural} {Mechanics} {Lecture} 11},
    url = {https://ocw.mit.edu/courses/2-080j-structural-mechanics-fall-2013/resources/mit2_080jf13_lecture11/},
    language = {en},
    number = {11},
    urldate = {2024-12-15},
    institution = {Massachusetts Institute of Technology},
    author = {Wierzbicki, Tomasz},
    year = {2013},
}

@techreport{american_bureau_of_shipping_requirements_2022,
    address = {ABS Plaza 1701 City Plaza Drive Spring, TX 77389 USA},
    type = {Requirement},
    title = {Requirements for {Buckling} and {Ultimate} {Strength} {Assessment} for {Offshore} {Structures}},
    url = {https://ww2.eagle.org/content/dam/eagle/rules-and-guides/current/offshore/126-requirements-for-buckling-and-ultimate-strength-assessment-for-offshore-structures/126-buckling-reqts-july22.pdf},
    language = {en},
    number = {126},
    urldate = {2024-12-15},
    institution = {American Bureau of Shipping},
    author = {{American Bureau of Shipping}},
    month = jul,
    year = {2022},
}

@book{american_iron_and_steel_institute_cold-formed_1991,
    address = {1000 16th Street, NW, Washington, D.C. 20036},
    edition = {1986 Edition with 1989 Addendum},
    title = {Cold-{Formed} {Steel} {Design} {Manual}},
    url = {https://scholarsmine.mst.edu/ccfss-aisi-spec/60/},
    language = {en},
    urldate = {2024-12-15},
    publisher = {W. P. Reyman Associates},
    author = {{American Iron and Steel Institute}},
    year = {1991},
}

@article{paduano_towards_2024,
    title = {Towards standardised design of wave energy converters: {A} high-fidelity modelling approach},
    volume = {224},
    issn = {0960-1481},
    shorttitle = {Towards standardised design of wave energy converters},
    url = {https://www.sciencedirect.com/science/article/pii/S0960148124002064},
    doi = {10.1016/j.renene.2024.120141},
    urldate = {2024-12-15},
    journal = {Renewable Energy},
    author = {Paduano, Bruno and Parrinello, Luca and Niosi, Francesco and Dell’Edera, Oronzo and Sirigu, Sergej Antonello and Faedo, Nicolás and Mattiazzo, Giuliana},
    month = apr,
    year = {2024},
    pages = {120141},
}

@techreport{ove_arup__partners_ltd_structural_2016,
    address = {Scotland},
    title = {Structural {Forces} and {Stresses} for {Wave} {Energy} {Devices}},
    url = {https://tethys-engineering.pnnl.gov/publications/structural-forces-stresses-wave-energy-devices},
    number = {ARP LS2},
    urldate = {2024-12-15},
    institution = {Wave Energy Scotland},
    author = {{Ove Arup \& Partners Ltd} and {Cruz Atcheson Consulting Engineers, Lda.}},
    month = jun,
    year = {2016},
}

@techreport{wierzbicki_lecture_2013-1,
    type = {Lecture notes},
    title = {Lecture 7: {Bending} {Response} of {Plates} and {Optimum} {Design}},
    shorttitle = {2.{080J} {Structural} {Mechanics} {Lecture} 7},
    url = {https://ocw.mit.edu/courses/2-080j-structural-mechanics-fall-2013/resources/mit2_080jf13_lecture7/},
    language = {en},
    number = {7},
    urldate = {2024-12-15},
    institution = {Massachusetts Institute of Technology},
    author = {Wierzbicki, Tomasz},
    year = {2013},
}

@article{boedo_corrected_1998,
    title = {Corrected {Solution} of {Clamped} {Ring} {Plate} with {Edge} {Point} {Load}},
    volume = {124},
    copyright = {Copyright © 1998 American Society of Civil Engineers},
    issn = {0733-9399},
    url = {https://ascelibrary.org/doi/10.1061/%28ASCE%290733-9399%281998%29124%3A6%28696%29},
    doi = {10.1061/(ASCE)0733-9399(1998)124:6(696)},
    language = {EN},
    number = {6},
    urldate = {2024-12-15},
    journal = {Journal of Engineering Mechanics},
    publisher = {American Society of Civil Engineers},
    author = {Boedo, S. and Prantil, V. C.},
    month = jun,
    year = {1998},
    pages = {696--697},
}

@book{young_roarks_2001,
    title = {Roark's {Formulas} for {Stress} and {Strain}},
    isbn = {978-0-07-150181-1},
    language = {en},
    publisher = {McGraw Hill LLC},
    author = {Young, Warren C. and Budynas, Richard G.},
    month = oct,
    year = {2001},
    note = {Google-Books-ID: pummClLoFXEC},
}

@article{mccabe_force-limited_2024,
    series = {15th {IFAC} {Conference} on {Control} {Applications} in {Marine} {Systems}, {Robotics} and {Vehicles} {CAMS} 2024},
    title = {Force-{Limited} {Control} of {Wave} {Energy} {Converters} using a {Describing} {Function} {Linearization}⁎},
    volume = {58},
    issn = {2405-8963},
    url = {https://www.sciencedirect.com/science/article/pii/S2405896324018482},
    doi = {10.1016/j.ifacol.2024.10.093},
    number = {20},
    urldate = {2024-12-09},
    journal = {IFAC-PapersOnLine},
    author = {McCabe, Rebecca and Haji, Maha N.},
    month = jan,
    year = {2024},
    pages = {440--445},
}

@article{hall_open-source_2022,
    title = {An {Open}-{Source} {Frequency}-{Domain} {Model} for {Floating} {Wind} {Turbine} {Design} {Optimization}},
    volume = {2265},
    issn = {1742-6596},
    url = {https://dx.doi.org/10.1088/1742-6596/2265/4/042020},
    doi = {10.1088/1742-6596/2265/4/042020},
    language = {en},
    number = {4},
    urldate = {2024-12-08},
    journal = {Journal of Physics: Conference Series},
    publisher = {IOP Publishing},
    author = {Hall, Matthew and Housner, Stein and Zalkind, Daniel and Bortolotti, Pietro and Ogden, David and Barter, Garrett},
    month = may,
    year = {2022},
    pages = {042020},
}

@article{penalba_review_2016,
    title = {A {Review} of {Wave}-to-{Wire} {Models} for {Wave} {Energy} {Converters}},
    volume = {9},
    copyright = {http://creativecommons.org/licenses/by/3.0/},
    issn = {1996-1073},
    url = {https://www.mdpi.com/1996-1073/9/7/506},
    doi = {10.3390/en9070506},
    language = {en},
    number = {7},
    urldate = {2024-08-23},
    journal = {Energies},
    publisher = {Multidisciplinary Digital Publishing Institute},
    author = {Penalba, Markel and Ringwood, John V.},
    month = jul,
    year = {2016},
    note = {Number: 7},
    pages = {506},
}

@article{quartier_influence_2021,
    title = {Influence of the {Drag} {Force} on the {Average} {Absorbed} {Power} of {Heaving} {Wave} {Energy} {Converters} {Using} {Smoothed} {Particle} {Hydrodynamics}},
    volume = {13},
    copyright = {http://creativecommons.org/licenses/by/3.0/},
    issn = {2073-4441},
    url = {https://www.mdpi.com/2073-4441/13/3/384},
    doi = {10.3390/w13030384},
    language = {en},
    number = {3},
    urldate = {2024-06-25},
    journal = {Water},
    author = {Quartier, Nicolas and Ropero-Giralda, Pablo and M. Domínguez, José and Stratigaki, Vasiliki and Troch, Peter},
    month = jan,
    year = {2021},
    pages = {384},
}

@article{coe_useful_2023,
    title = {Useful {Power} {Maximization} for {Wave} {Energy} {Converters}},
    volume = {16},
    copyright = {http://creativecommons.org/licenses/by/3.0/},
    issn = {1996-1073},
    url = {https://www.mdpi.com/1996-1073/16/1/529},
    doi = {10.3390/en16010529},
    language = {en},
    number = {1},
    urldate = {2024-02-11},
    journal = {Energies},
    author = {Coe, Ryan G. and Bacelli, Giorgio},
    month = jan,
    year = {2023},
    pages = {529},
}

@article{merigaud_geometrical_2023,
    title = {Geometrical {Framework} for {Hydrodynamics} and {Control} of {Wave} {Energy} {Converters}},
    volume = {2},
    url = {https://link.aps.org/doi/10.1103/PRXEnergy.2.023003},
    doi = {10.1103/PRXEnergy.2.023003},
    number = {2},
    urldate = {2024-02-07},
    journal = {PRX Energy},
    author = {Mérigaud, Alexis and Thiria, Benjamin and Godoy-Diana, Ramiro},
    month = may,
    year = {2023},
    pages = {023003},
}

@article{abdulkadir_optimal_2024,
    title = {Optimal {Constrained} {Control} of {Arrays} of {Wave} {Energy} {Converters}},
    volume = {12},
    copyright = {http://creativecommons.org/licenses/by/3.0/},
    issn = {2077-1312},
    url = {https://www.mdpi.com/2077-1312/12/1/104},
    doi = {10.3390/jmse12010104},
    language = {en},
    number = {1},
    urldate = {2024-06-27},
    journal = {Journal of Marine Science and Engineering},
    publisher = {Multidisciplinary Digital Publishing Institute},
    author = {Abdulkadir, Habeebullah and Abdelkhalik, Ossama},
    month = jan,
    year = {2024},
    note = {Number: 1},
    pages = {104},
}

@article{barter_beyond_2023,
    title = {Beyond 15 {MW}: {A} cost of energy perspective on the next generation of drivetrain technologies for offshore wind turbines},
    volume = {344},
    issn = {0306-2619},
    shorttitle = {Beyond 15 {MW}},
    url = {https://www.sciencedirect.com/science/article/pii/S0306261923006360},
    doi = {10.1016/j.apenergy.2023.121272},
    urldate = {2024-06-06},
    journal = {Applied Energy},
    author = {Barter, Garrett E. and Sethuraman, Latha and Bortolotti, Pietro and Keller, Jonathan and Torrey, David A.},
    month = aug,
    year = {2023},
    pages = {121272},
}

@article{faedo_principle_2022,
    title = {On the principle of impedance-matching for underactuated wave energy harvesting systems},
    volume = {118},
    issn = {0141-1187},
    url = {https://www.sciencedirect.com/science/article/pii/S0141118721004223},
    doi = {10.1016/j.apor.2021.102958},
    urldate = {2024-06-02},
    journal = {Applied Ocean Research},
    author = {Faedo, Nicolás and Carapellese, Fabio and Pasta, Edoardo and Mattiazzo, Giuliana},
    month = jan,
    year = {2022},
    pages = {102958},
}

@article{akdemir_opportunities_2023,
    title = {Opportunities for wave energy in bulk power system operations},
    volume = {352},
    issn = {0306-2619},
    url = {https://www.sciencedirect.com/science/article/pii/S0306261923012096},
    doi = {10.1016/j.apenergy.2023.121845},
    urldate = {2024-06-02},
    journal = {Applied Energy},
    author = {Akdemir, Kerem Ziya and Robertson, Bryson and Oikonomou, Konstantinos and Kern, Jordan and Voisin, Nathalie and Hanif, Sarmad and Bhattacharya, Saptarshi},
    month = dec,
    year = {2023},
    pages = {121845},
}

@article{grasberger_control_2024,
    title = {Control co-design and optimization of oscillating-surge wave energy converter},
    volume = {225},
    issn = {0960-1481},
    url = {https://www.sciencedirect.com/science/article/pii/S0960148124002994},
    doi = {10.1016/j.renene.2024.120234},
    urldate = {2024-06-01},
    journal = {Renewable Energy},
    author = {Grasberger, Jeff and Yang, Lisheng and Bacelli, Giorgio and Zuo, Lei},
    month = may,
    year = {2024},
    pages = {120234},
}

@article{son_performance_2016,
    title = {Performance validation and optimization of a dual coaxial-cylinder ocean-wave energy extractor},
    volume = {92},
    issn = {0960-1481},
    url = {https://www.sciencedirect.com/science/article/pii/S0960148116300325},
    doi = {10.1016/j.renene.2016.01.032},
    urldate = {2024-05-24},
    journal = {Renewable Energy},
    author = {Son, Daewoong and Belissen, Valentin and Yeung, Ronald W.},
    month = jul,
    year = {2016},
    pages = {192--201},
}

@book{gelb_multiple-input_1968,
    series = {{McGraw}-{Hill} {Electronic} {Sciences}},
    title = {Multiple-{Input} {Describing} {Functions} and {Nonlinear} {System} {Design}},
    url = {https://www.semanticscholar.org/paper/Multiple-Input-Describing-Functions-and-Nonlinear-Gelb-Velde/8a855f23d04e394328b2978c5843cf9f6d2d8fdc},
    urldate = {2024-04-08},
    publisher = {McGraw-Hill},
    author = {Gelb, Arthur and Vander Velde, Wallace E.},
    year = {1968},
}

@article{bacelli_geometric_2013,
    title = {A geometric tool for the analysis of position and force constraints in wave energy converters},
    volume = {65},
    issn = {0029-8018},
    url = {https://www.sciencedirect.com/science/article/pii/S0029801813001200},
    doi = {10.1016/j.oceaneng.2013.03.011},
    urldate = {2024-01-29},
    journal = {Ocean Engineering},
    author = {Bacelli, Giorgio and Ringwood, John V.},
    month = jun,
    year = {2013},
    pages = {10--18},
}

@article{zou_optimal_2017,
    title = {Optimal control of wave energy converters},
    volume = {103},
    issn = {0960-1481},
    url = {https://www.sciencedirect.com/science/article/pii/S0960148116310059},
    doi = {10.1016/j.renene.2016.11.036},
    urldate = {2024-01-29},
    journal = {Renewable Energy},
    author = {Zou, Shangyan and Abdelkhalik, Ossama and Robinett, Rush and Bacelli, Giorgio and Wilson, David},
    month = apr,
    year = {2017},
    pages = {217--225},
}

@article{chittick_asymmetric_2009,
    title = {An asymmetric suboptimization approach to aerostructural optimization},
    volume = {10},
    issn = {1573-2924},
    url = {https://doi.org/10.1007/s11081-008-9046-2},
    doi = {10.1007/s11081-008-9046-2},
    language = {en},
    number = {1},
    urldate = {2023-12-04},
    journal = {Optimization and Engineering},
    author = {Chittick, Ian R. and Martins, Joaquim R. R. A.},
    month = mar,
    year = {2009},
    pages = {133--152},
}

@inproceedings{sundarrajan_towards_2021,
    title = {Towards a {Fair} {Comparison} between the {Nested} and {Simultaneous} {Control} {Co}-{Design} {Methods} using an {Active} {Suspension} {Case} {Study}},
    issn = {2378-5861},
    url = {https://ieeexplore.ieee.org/abstract/document/9482687},
    doi = {10.23919/ACC50511.2021.9482687},
    urldate = {2023-12-04},
    booktitle = {2021 {American} {Control} {Conference} ({ACC})},
    author = {Sundarrajan, Athul K. and Herber, Daniel R.},
    month = may,
    year = {2021},
    pages = {358--365},
}

@article{bacelli_numerical_2015,
    title = {Numerical {Optimal} {Control} of {Wave} {Energy} {Converters}},
    volume = {6},
    issn = {1949-3037},
    url = {https://ieeexplore.ieee.org/abstract/document/6987295},
    doi = {10.1109/TSTE.2014.2371536},
    number = {2},
    urldate = {2023-11-27},
    journal = {IEEE Transactions on Sustainable Energy},
    author = {Bacelli, Giorgio and Ringwood, John V.},
    month = apr,
    year = {2015},
    note = {Conference Name: IEEE Transactions on Sustainable Energy},
    pages = {294--302},
}

@article{coe_maybe_2021,
    title = {Maybe less is more: {Considering} capacity factor, saturation, variability, and filtering effects of wave energy devices},
    volume = {291},
    issn = {0306-2619},
    shorttitle = {Maybe less is more},
    url = {https://www.sciencedirect.com/science/article/pii/S0306261921002701},
    doi = {10.1016/j.apenergy.2021.116763},
    urldate = {2023-11-27},
    journal = {Applied Energy},
    author = {Coe, Ryan G. and Ahn, Seongho and Neary, Vincent S. and Kobos, Peter H. and Bacelli, Giorgio},
    month = jun,
    year = {2021},
    pages = {116763},
}

@article{herber_nested_2018,
    title = {Nested and {Simultaneous} {Solution} {Strategies} for {General} {Combined} {Plant} and {Control} {Design} {Problems}},
    volume = {141},
    issn = {1050-0472},
    url = {https://doi.org/10.1115/1.4040705},
    doi = {10.1115/1.4040705},
    number = {011402},
    urldate = {2023-10-30},
    journal = {Journal of Mechanical Design},
    author = {Herber, Daniel R. and Allison, James T.},
    month = oct,
    year = {2018},
}

@article{roberts_bringing_2021,
    title = {Bringing {Structure} to the {Wave} {Energy} {Innovation} {Process} with the {Development} of a {Techno}-{Economic} {Tool}},
    volume = {14},
    copyright = {http://creativecommons.org/licenses/by/3.0/},
    issn = {1996-1073},
    url = {https://www.mdpi.com/1996-1073/14/24/8201},
    doi = {10.3390/en14248201},
    language = {en},
    number = {24},
    urldate = {2023-03-07},
    journal = {Energies},
    publisher = {Multidisciplinary Digital Publishing Institute},
    author = {Roberts, Owain and Henderson, Jillian Catherine and Garcia-Teruel, Anna and Noble, Donald R. and Tunga, Inès and Hodges, Jonathan and Jeffrey, Henry and Hurst, Tim},
    month = jan,
    year = {2021},
    note = {Number: 24},
    pages = {8201},
}

@inproceedings{caio_tackling_2019,
    title = {Tackling the {Wave} {Energy} {Paradox} - {Stepping} {Towards} {Commercial} {Deployment}},
    url = {https://onepetro.org/ISOPEIOPEC/proceedings-abstract/ISOPE19/All-ISOPE19/21575},
    language = {en},
    urldate = {2023-03-03},
    publisher = {OnePetro},
    author = {Caio, Andrea and Davey, Thomas and McNatt, Cameron},
    month = jun,
    year = {2019},
}

@article{mccabe_constrained_2013,
    title = {Constrained optimization of the shape of a wave energy collector by genetic algorithm},
    volume = {51},
    issn = {0960-1481},
    url = {https://www.sciencedirect.com/science/article/pii/S0960148112006258},
    doi = {10.1016/j.renene.2012.09.054},
    language = {en},
    urldate = {2023-03-01},
    journal = {Renewable Energy},
    author = {McCabe, A. P.},
    month = mar,
    year = {2013},
    pages = {274--284},
}

@article{giannini_wave_2022,
    title = {Wave energy converters design combining hydrodynamic performance and structural assessment},
    volume = {249},
    issn = {0360-5442},
    url = {https://www.sciencedirect.com/science/article/pii/S0360544222005448},
    doi = {10.1016/j.energy.2022.123641},
    language = {en},
    urldate = {2023-03-01},
    journal = {Energy},
    author = {Giannini, Gianmaria and Rosa-Santos, Paulo and Ramos, Victor and Taveira-Pinto, Francisco},
    month = jun,
    year = {2022},
    pages = {123641},
}

@article{trueworthy_wave_2020,
    title = {The {Wave} {Energy} {Converter} {Design} {Process}: {Methods} {Applied} in {Industry} and {Shortcomings} of {Current} {Practices}},
    volume = {8},
    copyright = {http://creativecommons.org/licenses/by/3.0/},
    issn = {2077-1312},
    shorttitle = {The {Wave} {Energy} {Converter} {Design} {Process}},
    url = {https://www.mdpi.com/2077-1312/8/11/932},
    doi = {10.3390/jmse8110932},
    language = {en},
    number = {11},
    urldate = {2023-03-01},
    journal = {Journal of Marine Science and Engineering},
    publisher = {Multidisciplinary Digital Publishing Institute},
    author = {Trueworthy, Ali and DuPont, Bryony},
    month = nov,
    year = {2020},
    note = {Number: 11},
    pages = {932},
}

@article{al_shami_parameter_2019,
    title = {A parameter study and optimization of two body wave energy converters},
    volume = {131},
    issn = {0960-1481},
    url = {https://www.sciencedirect.com/science/article/pii/S0960148118307833},
    doi = {10.1016/j.renene.2018.06.117},
    language = {en},
    urldate = {2022-02-14},
    journal = {Renewable Energy},
    author = {Al Shami, Elie and Wang, Xu and Zhang, Ran and Zuo, Lei},
    month = feb,
    year = {2019},
    pages = {1--13},
}

@techreport{RM3,
    address = {Albuquerque, New Mexico},
    title = {Methodology for {Design} and {Economic} {Analysis} of {Marine} {Energy} {Conversion} ({MEC}) {Technologies}},
    shorttitle = {{RM3}},
    url = {https://energy.sandia.gov/wp-content/gallery/uploads/SAND2014-9040-RMP-REPORT.pdf},
    language = {en},
    number = {SAND2014-9040},
    institution = {Sandia National Laboratories},
    author = {Neary, Vincent S and Previsic, Mirko and Jepsen, Richard A and Lawson, Michael J and Yu, Yi-Hsiang and Copping, Andrea E and Fontaine, Arnold A and Hallett, Kathleen C and Murray, Dianne K},
    month = mar,
    year = {2014},
    pages = {262},
}

@book{newman,
    title = {Marine {Hydrodynamics}},
    shorttitle = {newman},
    url = {https://direct.mit.edu/books/book/2693/Marine-Hydrodynamics},
    language = {en},
    urldate = {2022-01-26},
    author = {Newman, J. N.},
    month = aug,
    year = {1977},
}

@article{coe_initial_2020,
    title = {Initial conceptual demonstration of control co-design for {WEC} optimization},
    volume = {6},
    issn = {2198-6452},
    url = {https://doi.org/10.1007/s40722-020-00181-9},
    doi = {10.1007/s40722-020-00181-9},
    language = {en},
    number = {4},
    urldate = {2021-10-20},
    journal = {Journal of Ocean Engineering and Marine Energy},
    author = {Coe, Ryan G. and Bacelli, Giorgio and Olson, Sterling and Neary, Vincent S. and Topper, Mathew B. R.},
    month = nov,
    year = {2020},
    pages = {441--449},
}

@article{garcia-sanz_control_2019,
    title = {Control {Co}-{Design}: {An} engineering game changer},
    volume = {1},
    issn = {2578-0727},
    shorttitle = {Control {Co}-{Design}},
    url = {https://onlinelibrary.wiley.com/doi/abs/10.1002/adc2.18},
    doi = {10.1002/adc2.18},
    language = {en},
    number = {1},
    urldate = {2021-10-20},
    journal = {Advanced Control for Applications},
    author = {Garcia-Sanz, Mario},
    year = {2019},
    pages = {e18},
}

@article{nguyen_theoretical_2024,
    title = {Theoretical modeling of a bottom-raised oscillating surge wave energy converter structural loadings and power performances},
    volume = {149},
    issn = {0141-1187},
    url = {https://www.sciencedirect.com/science/article/pii/S0141118724001536},
    doi = {10.1016/j.apor.2024.104031},
    urldate = {2026-05-19},
    journal = {Applied Ocean Research},
    author = {Nguyen, Nhu and Davis, Jacob and Tom, Nathan and Thiagarajan, Krish},
    month = aug,
    year = {2024},
    pages = {104031},
}

@inproceedings{chau_inertia_2010,
    address = {Harbin, China},
    title = {Inertia and {Damping} of {Heaving} {Compound} {Cylinders}},
    url = {https://www.academia.edu/73219479/Inertia_and_Damping_of_Heaving_Compound_Cylinders_Fun},
    language = {en},
    urldate = {2023-09-28},
    booktitle = {25th {International} {Workshop} on {Water} {Waves} and {Floating} {Bodies}},
    author = {Chau, Fun Pang and Yeung, Ronald W.},
    month = jan,
    year = {2010},
    pages = {4},
}

@article{khanal_fully_2025,
    title = {Fully differentiable boundary element solver for hydrodynamic sensitivity analysis of wave-structure interactions},
    volume = {163},
    issn = {0141-1187},
    url = {https://www.sciencedirect.com/science/article/pii/S0141118725002937},
    doi = {10.1016/j.apor.2025.104707},
    urldate = {2026-03-20},
    journal = {Applied Ocean Research},
    author = {Khanal, Kapil and Ströfer, Carlos A. Michelén and Ancellin, Matthieu and Haji, Maha N.},
    month = oct,
    year = {2025},
    pages = {104707},
}

@article{edwards_optimisation_2022,
    title = {Optimisation of the geometry of axisymmetric point-absorber wave energy converters},
    volume = {933},
    issn = {0022-1120, 1469-7645},
    url = {https://www.cambridge.org/core/journals/journal-of-fluid-mechanics/article/optimisation-of-the-geometry-of-axisymmetric-pointabsorber-wave-energy-converters/EE6FAE12B0F2F9C4DC607C301406411F},
    doi = {10.1017/jfm.2021.993},
    language = {en},
    urldate = {2023-10-07},
    journal = {Journal of Fluid Mechanics},
    publisher = {Cambridge University Press},
    author = {Edwards, Emma C. and Yue, Dick K.-P.},
    month = feb,
    year = {2022},
    pages = {A1},
}

@book{chatjigeorgiou_analytical_2018,
    address = {Cambridge},
    title = {Analytical {Methods} in {Marine} {Hydrodynamics}},
    isbn = {978-1-107-17969-1},
    url = {https://www.cambridge.org/core/books/analytical-methods-in-marine-hydrodynamics/FA575866CF4838EE370460746C304B55},
    doi = {10.1017/9781316838983},
    urldate = {2023-09-28},
    publisher = {Cambridge University Press},
    author = {Chatjigeorgiou, Ioannis K.},
    year = {2018},
}

@article{mavrakos_second-order_2009,
    title = {Second-order hydrodynamic effects on an arrangement of two concentric truncated vertical cylinders},
    volume = {22},
    issn = {0951-8339},
    url = {https://www.sciencedirect.com/science/article/pii/S0951833908000580},
    doi = {10.1016/j.marstruc.2008.12.003},
    number = {3},
    urldate = {2025-02-03},
    journal = {Marine Structures},
    author = {Mavrakos, Spyros A. and Chatjigeorgiou, Ioannis K.},
    month = jul,
    year = {2009},
    pages = {545--575},
}

@article{kokkinowrachos_behaviour_1986,
    title = {Behaviour of vertical bodies of revolution in waves},
    volume = {13},
    issn = {0029-8018},
    url = {https://www.sciencedirect.com/science/article/pii/0029801886900375},
    doi = {10.1016/0029-8018(86)90037-5},
    number = {6},
    urldate = {2023-12-12},
    journal = {Ocean Engineering},
    author = {Kokkinowrachos, Konstantin and Mavrakos, Spyridon and Asorakos, Sampson},
    month = jan,
    year = {1986},
    pages = {505--538},
}

@article{yeung_added_1981,
    title = {Added mass and damping of a vertical cylinder in finite-depth waters},
    volume = {3},
    issn = {0141-1187},
    url = {https://www.sciencedirect.com/science/article/pii/0141118781901012},
    doi = {10.1016/0141-1187(81)90101-2},
    number = {3},
    urldate = {2023-09-28},
    journal = {Applied Ocean Research},
    author = {Yeung, Ronald W.},
    month = jul,
    year = {1981},
    pages = {119--133},
}

@article{cong_novel_2020,
    title = {A novel solution to the second-order wave radiation force on an oscillating truncated cylinder based on the application of control surfaces},
    volume = {204},
    issn = {0029-8018},
    url = {https://www.sciencedirect.com/science/article/pii/S0029801820303243},
    doi = {10.1016/j.oceaneng.2020.107278},
    urldate = {2025-02-03},
    journal = {Ocean Engineering},
    author = {Cong, Peiwen and Teng, Bin and Chen, Lifen and Gou, Ying},
    month = may,
    year = {2020},
    pages = {107278},
}

@article{mciver_added_1991,
    title = {The added mass of bodies heaving at low frequency in water of finite depth},
    volume = {13},
    issn = {0141-1187},
    url = {https://www.sciencedirect.com/science/article/pii/S0141118705800367},
    doi = {10.1016/S0141-1187(05)80036-7},
    number = {1},
    urldate = {2025-04-27},
    journal = {Applied Ocean Research},
    author = {McIver, P. and Linton, C. M.},
    month = feb,
    year = {1991},
    pages = {12--17},
}

@article{olaya_hydrodynamic_2015,
    title = {Hydrodynamic {Coefficient} {Computation} for a {Partially} {Submerged} {Wave} {Energy} {Converter}},
    volume = {40},
    issn = {1558-1691},
    url = {https://ieeexplore.ieee.org/document/6895317},
    doi = {10.1109/JOE.2014.2344951},
    number = {3},
    urldate = {2025-01-10},
    journal = {IEEE Journal of Oceanic Engineering},
    author = {Olaya, Sébastien and Bourgeot, Jean-Matthieu and Benbouzid, Mohamed El Hachemi},
    month = jul,
    year = {2015},
    note = {Conference Name: IEEE Journal of Oceanic Engineering},
    pages = {522--535},
}

@article{mavrakos_hydrodynamic_2004,
    title = {Hydrodynamic coefficients in heave of two concentric surface-piercing truncated circular cylinders},
    volume = {26},
    issn = {0141-1187},
    url = {https://www.sciencedirect.com/science/article/pii/S0141118705000076},
    doi = {10.1016/j.apor.2005.03.002},
    number = {3},
    urldate = {2023-12-13},
    journal = {Applied Ocean Research},
    author = {Mavrakos, Spyros A.},
    month = may,
    year = {2004},
    pages = {84--97},
}

\end{document}